\documentclass[10pt,a4paper,oneside]{report}
\usepackage{mathrsfs}
\usepackage[T1]{fontenc}
\usepackage{charter}

\usepackage[utf8]{inputenc}   

\usepackage[english]{babel} 

\newcommand{\acknowledgments}{@undefined} 
\addto\captionsenglish{\renewcommand{\acknowledgments}{Acknowledgments}}

\addto\captionsportuguese{\renewcommand{\acknowledgments}{Agradecimentos}}
\addto\captionsportuguese{\renewcommand{\nomname}{Lista de S\'{i}mbolos}} 

\newcommand{\coverThesis}{@undefined} 
\newcommand{\coverCoSupervisor}{@undefined} 
\newcommand{\coverClassification}{@undefined} 
\newcommand{\coverFunding}{@undefined} 

\newcommand{\coverExaminationCommittee}{@undefined} 
\newcommand{\coverSupervisor}{@undefined} 
\addto\captionsenglish{\renewcommand{\coverThesis}{\FontMb Thesis approved in public session to obtain the PhD Degree in}}
\addto\captionsenglish{}
\addto\captionsenglish{\renewcommand{\coverExaminationCommittee}{Jury}}
\addto\captionsenglish{}
\addto\captionsenglish{\renewcommand{\coverSupervisor}{\FontMb Supervisor}}
\addto\captionsenglish{\renewcommand{\coverCoSupervisor}{\FontMb Co-Supervisor}}
\addto\captionsenglish{\renewcommand{\coverClassification}{\FontMb Jury final classification: Pass with Distinction}}
\addto\captionsenglish{}
\addto\captionsenglish{\renewcommand{\coverFunding}{\FontMb Funding Institution}}

\addto\captionsportuguese{\renewcommand{\coverThesis}{Disserta\c{c}\~{a}o para obten\c{c}\~{a}o do Grau de Mestre em}}
\addto\captionsportuguese{}
\addto\captionsportuguese{\renewcommand{\coverExaminationCommittee}{J\'{u}ri}}
\addto\captionsportuguese{}
\addto\captionsportuguese{\renewcommand{\coverSupervisor}{Orientador}}
\addto\captionsportuguese{}

\def\FontLb{
  \usefont{T1}{phv}{b}{n}\fontsize{16pt}{16pt}\selectfont}

\def\FontMb{
  \usefont{T1}{phv}{b}{n}\fontsize{14pt}{14pt}\selectfont}
\def\FontSn{
  \usefont{T1}{phv}{m}{n}\fontsize{12pt}{12pt}\selectfont}

\usepackage{geometry}	
\usepackage{setspace}

\usepackage{enumerate}
\usepackage{stackengine}
\usepackage[pdftex]{graphicx}

\usepackage{amsmath}  
\usepackage{amsthm}   
\usepackage{amsfonts} %
\usepackage{amssymb}
\usepackage{afterpage}
\usepackage{subfigmat}

\usepackage{slashed}

\usepackage{dcolumn}
\newcolumntype{d}{D{.}{.}{-1}} 
\newcolumntype{e}{D{E}{E}{-1}} 

\usepackage{nomencl}
\makenomenclature
\RequirePackage{ifthen} 
\ifthenelse{\equal{\languagename}{english}}%
    { 
    \renewcommand{\nomgroup}[1]{%
      \ifthenelse{\equal{#1}{A}}{%
        \item[\textbf{Abbreviations}]}{%
        \ifthenelse{\equal{#1}{G}}{%
          \item[\textbf{Greek symbols}]}{%
          \ifthenelse{\equal{#1}{S}}{%
            \item[\textbf{Subscripts}]}{%
            \ifthenelse{\equal{#1}{T}}{%
              \item[\textbf{Superscripts}]}{}}}}}%
    }{
    \renewcommand{\nomgroup}[1]{%
      \ifthenelse{\equal{#1}{R}}{%
        \item[\textbf{Simbolos romanos}]}{%
        \ifthenelse{\equal{#1}{G}}{%
          \item[\textbf{Simbolos gregos}]}{%
          \ifthenelse{\equal{#1}{S}}{%
            \item[\textbf{Subscritos}]}{%
            \ifthenelse{\equal{#1}{T}}{%
              \item[\textbf{Sobrescritos}]}{}}}}}%
    }%

\usepackage{rotating}

\usepackage{hyperref}
\hypersetup{colorlinks,allcolors=black}
\usepackage[figure,table]{hypcap}

\usepackage{cite}

\usepackage{multirow}
\usepackage{tikz}

\usepackage[compat=1.1.0]{tikz-feynman}
\tikzfeynmanset{warn luatex=false}
\usepackage{cancel}

\usepackage{stackengine}
\usepackage{booktabs}

\usepackage{pdfpages}

\newcommand{\degree}{\ensuremath{^\circ\,}} 

\DeclareMathOperator{\diag}{diag}
\DeclareMathOperator{\im}{Im}
\DeclareMathOperator{\re}{Re}

\newcommand{\cV}{\mathcal{V}}
\newcommand{\cM}{\mathcal{M}}
\newcommand{\cD}{\mathcal{D}}
\newcommand{\bzero}{\textbf{0}}
\newcommand{\id}{\mathds{1}}
\newcommand{\N}{\mathbb{N}_0}

\providecommand{\cM}{\mathcal{M}}
\providecommand{\cV}{\mathcal{V}}
\providecommand{\om}{\overline{m\hspace{-.2ex}}\hspace{.2ex}}
\providecommand{\oM}{\overline{\!M}\mkern-2mu}
\providecommand{\oYd}{\overline{Y}\!_d}
\providecommand{\oYu}{\overline{Y}\!_u}

\providecommand{\oK}{\overline{\!K}\,}

\providecommand{\mtrx}[1]{\begin{pmatrix} #1 \end{pmatrix}}

\def\bm#1{\mathchoice                             
  {\mbox{\boldmath$\displaystyle#1$}}%
  {\mbox{\boldmath$#1$}}%
  {\mbox{\boldmath$\scriptstyle#1$}}%
  {\mbox{\boldmath$\scriptscriptstyle#1$}}}

\usepackage{tikz}
\usepackage{blindtext}

\usepackage{easybmat}
\usepackage{multirow,bigdelim}

\usepackage{appendix}
\usepackage{verbatim}
\usepackage{makecell}

\usepackage{float}
\usepackage{tikz}
\usetikzlibrary{fit,matrix}

\usepackage{stackengine}

\newcommand\brabar{\scalebox{.3}{(}\raisebox{-2pt}{\bf --}\scalebox{.3}{)}}

\usepackage{listings}
\usepackage{mathtools}
\usepackage{dsfont}
\usepackage{arydshln}
\usepackage{mathrsfs}

\usepackage{units}
\usepackage{amsmath}
\usepackage{amssymb}
\usepackage{amsfonts}
\usepackage{xcolor}
\usepackage[abbreviations,symbols,nogroupskip,nonumberlist,automake]{glossaries-extra}

\usepackage{enumitem}

\newcommand\myunder[2]{\mathrlap{\smash{\underbrace{\phantom{%
    \begin{matrix} #2 \end{matrix}}}_{\mbox{$#1$}}}}#2}
\newcommand\myrightbrace[2]{%
\left.\vphantom{\begin{matrix} #1 \end{matrix}}\right\}#2}

\makeglossaries

\glsxtrnewsymbol[description={number of isosinglet $q$-type vector-like quarks ($q=u,d$)}]{nq}{\ensuremath{n_q}}
\glsxtrnewsymbol[description={quark mixing matrix, of dimension $(3+n_u)\times (3+n_d)$}]{V}{\ensuremath{V}}
\glsxtrnewsymbol[description={$3\times 3$ CKM matrix (upper-left block of $V$, not necessarily unitary)}]{VCKM}{\ensuremath{V_\textrm{\normalfont CKM}}}
\glsxtrnewsymbol[description={exact matrix parameterizing deviations from unitarity}]{X}{\ensuremath{X}}
\glsxtrnewsymbol[description={approximate matrix parameterizing deviations from unitarity}]{Theta}{\ensuremath{\Theta}}
\glsxtrnewsymbol[description={matrices controlling FCNC ($q=u,d$)}]{Fq}{\ensuremath{F^q}}
\glsxtrnewsymbol[description={$q$-type quark mass matrix ($q=u,d$)}]{Mq}{\ensuremath{\mathcal{M}_q}}
\glsxtrnewsymbol[description={first 3 rows of $\mathcal{V}^q_\chi$ ($q=u,d$; $\chi=L,R$)}]{A}{\ensuremath{A^q_\chi}}
\glsxtrnewsymbol[description={last $n_q$ rows of $\mathcal{V}^q_\chi$ ($q=u,d$; $\chi=L,R$)}]{B}{\ensuremath{B^q_\chi}}
\glsxtrnewsymbol[description={$(3+n_q)$-dimensional unitary diagonalization matrix ($q=u,d$; $\chi=L,R$)}]{mV}{\ensuremath{\mathcal{V}^q_\chi}}

\begin{document}

\pagestyle{plain}

\pagenumbering{roman}

\thispagestyle {empty}


    \begin{tikzpicture}[remember picture,overlay]
    \node[anchor=north west,yshift=-1.5pt,xshift=1pt]%
        at (current page.north west)
        {\includegraphics[scale=0.25]{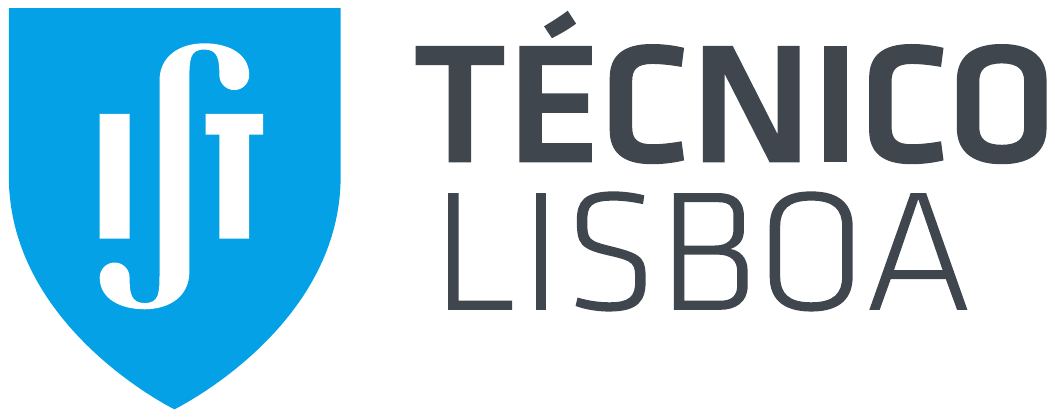}};
    \end{tikzpicture}

\begin{center}
\vspace{1.0cm}
{\FontLb UNIVERSIDADE DE LISBOA} \\
{\FontLb INSTITUTO SUPERIOR TÉCNICO} \\
\vspace{2.0cm}
{\FontLb  Non-Unitary $3 \times 3$ Mixing in Majorana Neutrinos and Vector-like Quark Models }\\ 
\vspace{2.5cm}
{\FontMb Pedro Manuel Ferreira Pereira} \\ 
\vspace{2.0cm}

\vspace{1.0cm} 
\vspace{0.3cm}
{\FontSn %
\begin{tabular}{c}
\coverSupervisor: Doctor \normalfont Maria Margarida Nesbitt Rebelo da Silva \\ 
\coverCoSupervisor: Doctor \normalfont Gustavo da Fonseca Castelo Branco \\        
\end{tabular} } \\

\vspace{2cm}

{\FontSn \coverThesis} \\
\vspace{0.3cm}
{\LARGE Physics} \\ 
\vspace{1.0cm}
{\FontSn \coverClassification} \\
\vspace{1.0cm}
\vspace{0.3cm}
\vspace{1.0cm}
\vspace{1.0cm}
{\FontMb 2024} \\ 

\end{center}

\cleardoublepage

\thispagestyle {empty}


    \begin{tikzpicture}[remember picture,overlay]
    \node[anchor=north west,yshift=-1.5pt,xshift=1pt]%
        at (current page.north west)
        {\includegraphics[scale=0.25]{Figures/IST_A_CMYK_POS.pdf}};
    \end{tikzpicture}

\begin{center}
\vspace{1.0cm}
{\FontLb UNIVERSIDADE DE LISBOA} \\
{\FontLb INSTITUTO SUPERIOR TÉCNICO \\
\vspace{1.0cm}
{\FontLb  Non-Unitary $3 \times 3$ Mixing in Majorana Neutrinos and Vector-like Quark Models \\ 
\vspace{0.5cm}
{\FontMb Pedro Manuel Ferreira Pereira} \\ 
\vspace{0.5cm}
{\FontSn %
\normalsize
\begin{tabular}{c}
\coverSupervisor: Doctor \normalfont Maria Margarida Nesbitt Rebelo da Silva \\ 
\coverCoSupervisor: Doctor \normalfont Gustavo da Fonseca Castelo Branco \\        
\end{tabular} } \\

\vspace{0.5cm}

{\FontSn \coverThesis} \\
\vspace{0.3cm}
{\LARGE Physics} \\ 
\vspace{0.8cm}
{\FontSn \coverClassification} \\

\vspace{0.5cm} }
{\FontMb \coverExaminationCommittee} \\
\vspace{0.3cm} }
{
\normalsize 
\begin{tabular}{@{}p{\textwidth}@{}}
\textbf{Chairperson:} \textbf{Doctor} Mário João Martins Pimenta, Instituto Superior Técnico, Universidade de Lisboa \\ 
\textbf{Member of the Committee:} \\
\vspace{-0.7cm}
\begin{list}{}{\setlength{\leftmargin}{1cm}\setlength{\itemindent}{0cm}\setlength{\itemsep}{-0.1cm}\setlength{\parsep}{-0.0cm}}
    \item[]\textbf{Doctor} Francisco Jose Botella Olcina, Institut de Física Corpuscular, Universitat de València,
    \item[] Espanha
    \item[]\textbf{Doctor} Mário João Martins Pimenta, Instituto Superior Técnico, Universidade de Lisboa
    \item[]\textbf{Doctor} António Joaquim Onofre Abreu Ribeiro Gonçalves, Escola de Ciências, Universidade do
    \item[] Minho
    \item[]\textbf{Doctor} Maria Margarida Nesbitt Rebelo da Silva, Instituto Superior Técnico, Universidade de
    \item[] Lisboa
    \item[]\textbf{Doctor} Ricardo Jorge Gonzalez Felipe, Instituto Superior de Engenharia de Lisboa - Instituto
    \item[] Politécnico de Lisboa
    \item[]\textbf{Doctor} Ivo de Medeiros Varzielas, Instituto Superior Técnico, Universidade de Lisboa
\end{list}
\end{tabular}
}
\vspace{0.5cm}
{\FontSn \coverFunding} \\
\vspace{0.1cm}
{\large FCT: Fundação para a Ciência e a Tecnologia, Fellowship SFRH/BD/145399/2019} \\ 
\vspace{0.5cm}
{\FontMb 2024} \\ 
\end{center}

\cleardoublepage


\null\vskip5cm%
\begin{flushleft}

"\textit{Às vezes, quando ergo a cabeça estonteada dos livros em que escrevo as contas alheias e a ausência de vida própria, sinto uma náusea física, que pode ser de me curvar, mas que transcende os números e a desilusão. A vida desgosta-me como um remédio inútil. E é então que eu sinto com visões claras como seria fácil o afastamento deste tédio se eu tivesse a simples força de o querer deveras afastar.\\
    ~~    \\
Vivemos pela acção, isto é, pela vontade. Aos que não sabemos querer — sejamos génios ou mendigos — irmana-nos a impotência. De que me serve citar-me génio se resulto ajudante de guarda-livros? Quando Cesário Verde fez dizer ao médico que era, não o Sr. Verde empregado no comércio, mas o poeta Cesário Verde, usou de um daqueles verbalismos do orgulho inútil que suam o cheiro da vaidade. O que ele foi sempre, coitado, foi o Sr. Verde empregado no comércio. O poeta nasceu depois de ele morrer, porque foi depois de ele morrer que nasceu a apreciação do poeta.\\
    ~~    \\
Agir, eis a inteligência verdadeira. Serei o que quiser. Mas tenho que querer o que for. O êxito está em ter êxito, e não em ter condições de êxito. Condições de palácio tem qualquer terra larga, mas onde estará o palácio se o não fizerem ali?}"\footnote{Livro do Desassossego por Bernardo Soares.Vol.I. Fernando Pessoa. (Recolha e transcrição dos textos de Maria Aliete Galhoz e Teresa Sobral Cunha. Prefácio e Organização de Jacinto do Prado Coelho.) Lisboa: Ática, 1982.  - 85.
}\\
    ~~    \\
A minha mãe,\\ 
Que sempre me fez acreditar que seria o que quisesse.\\
Que nunca leu Piaget,\\
Nem nenhum desses supostos entendidos em psicologia do desenvolvimento,\\
E que sempre soube mais que eles todos juntos. \\

\end{flushleft}

\vfill\newpage

\cleardoublepage


\section*{\acknowledgments}

\addcontentsline{toc}{section}{\acknowledgments}

Gostaria de agradecer a todos os meus amigos e familiares que, de alguma forma, contribuíram para o sucesso deste doutoramento e para a conclusão desta tese. Sem o seu apoio emocional nada teria sido possível.\par
Deixo também um agradecimento muito especial aos meus orientadores por todos estes anos de orientação e aconselhamento. Estendo esse agradecimento a todas as pessoas que tive o prazer de conhecer graças à Física e a todos com que tive o prazer de colaborar. Obrigado.\par
Este trabalho teve o apoio financeiro da Fundação para a Ciência e Tecnologia (FCT) através da bolsa de Doutoramento SFRH/BD/145399/2019. 
\cleardoublepage


\section*{Resumo}

\addcontentsline{toc}{section}{Resumo}
Matrizes de mistura $3 \times 3$ não unitárias estão presentes em muitas extensões do Modelo Padrão. Uma das extensões mais simples é a adição de $n_R$ neutrinos de direita sem impor a Conservação do Número Leptónico ($n_R\nu$SM). Na região do espaço de parâmetros onde a conhecida aproximação de seesaw é válida, os desvios de unitariedade da matriz de mistura leptónica  $3 \times 3$ são desprezáveis, existindo três neutrinos leves e dois ($n_R=2$) ou três ($n_R=3$) neutrinos pesados com massas próximas da Escala GUT. 
Apesar desta diferença de escala de massa, os acoplamentos de Yukawa são de ordem 1, fornecendo uma explicação natural para a pequenez das massas dos neutrinos leves - daí o nome seesaw.\par
Outras regiões do espaço de parâmetros, com neutrinos pesados mais leves e maiores desvios da unitariedade da matriz de mistura leptónica $3 \times 3$, são ainda permitidas pela experiência. 
De facto, sempre que as massas dos neutrinos pesados são muitas ordens de magnitude abaixo da escala GUT, há implicações fenomenológicas muito interessantes a baixas energias.
Para analisar tais regiões do espaço de parâmetros com precisão, é preciso uma parametrização exata, pois as aproximações válidas no cenário seesaw comum falham.
Esta tese consistirá num estudo e classificação de regiões do espaço de parâmetros do 3$\nu$SM à luz de uma parametrização exata recém-desenvolvida. \par
Além disso, também incluirá um estudo de modelos com vector-like quarks, outra extensão simples do SM, que contém uma matriz de mistura de quarks $3 \times 3$ não-unitária. O estudo das regiões mais interessantes do espaço de parâmetros dos modelos com vector-like quarks também beneficia do uso da parametrização exata mencionada anteriormente. Essas regiões são ricas em fenomenologia e podem explicar certas questões em aberto como o problema da unitariedade CKM/anomalia do ângulo de Cabibbo.\par
Esta tese também inclui um capítulo onde se discute as implicações experimentais destes modelos.\par

\vfill

\textbf{\Large Palavras-chave: Modelo Padrão, Neutrinos, Vector-like Quarks, Matriz de Mistura, Desvios de Unitariedade.}   
\cleardoublepage


\section*{Abstract}

\addcontentsline{toc}{section}{Abstract}
Non-unitary $3 \times 3$ mixing matrices are present in many extensions of the Standard Model. One of the most simple extensions is the addition of $n_R$ right-handed neutrinos without imposing Lepton Number Conservation ($n_R\nu$SM). In the region of the parameter space where the usual seesaw approximation is valid, deviations from $3 \times 3$ unitarity of the leptonic mixing matrix are negligible, while one has three light neutrinos and two ($n_R=2$) or three ($n_R=3$) heavy neutrinos with masses close to the GUT scale. This gap between the mass scales occurs while having order 1 Yukawa couplings, providing a natural explanation for the smallness of light neutrino masses - hence the name seesaw.\par
Other regions of the parameter space, with lighter heavy neutrinos - with an eV or keV mass, for instance - and bigger deviations from unitarity of the leptonic mixing matrix, are still allowed by experiment. In fact, whenever heavy neutrino masses are many orders of magnitude below the GUT scale there are interesting phenomenological implications at low energies. To analyse such regions with precision, one needs an exact parameterisation, since approximations that were valid in the usual seesaw case now fail. This thesis will consist of a study and classification of regions of the 3$\nu$SM parameter space in the light of a newly-developed exact parameterisation. \par
Furthermore, it will also include a study of models with vector-like quarks, another simple extension of the SM, that contains a $3 \times 3$ non-unitary quark mixing matrix. Studying the most interesting regions of the parameter space of vector-like quarks models also benefits from the use of the aforementioned exact parameterisation. These regions are phenomenologically rich and may explain certain open questions like the CKM unitarity problem/Cabibbo angle anomaly. \par
A chapter discussing the experimental implications of these models is also included. \par

\vfill

\textbf{\Large Keywords: Standard Model, Neutrinos, Vector-like Quarks, Mixing Matrix, Deviations from Unitarity.}
\cleardoublepage

\section*{List of Publications}

\addcontentsline{toc}{section}{List of Publications}

\vfil
This thesis is based on the following publications, listed in chronological order:
\begin{itemize}
\item N. R. Agostinho, G. C. Branco, P. M. F. Pereira, M. N. Rebelo and J. I. Silva-Marcos, \textit{Can one have significant deviations from leptonic $3\times 3$ unitarity in the framework of type I seesaw mechanism?}, Eur. Phys. J. \textbf{C78} (2018) no.11, 895 doi:10.1140/epjc/s10052-018-6347-2 [arXiv:1711.06229 [hep-ph]] \footnote{Strictly speaking, this paper constitutes the research conducted for my Master's degree.}

\item G. C. Branco, J. T. Penedo, P. M. F. Pereira, M. N. Rebelo and J. I. Silva-Marcos, \textit{Type-I Seesaw with eV-Scale Neutrinos}, JHEP \textbf{07} (2020), 164 doi:10.1007/JHEP07(2020)164 [arXiv:1912.05875 [hep-ph]].

\item  G. C. Branco, J. T. Penedo, P. M. F. Pereira, M. N. Rebelo and J. I. Silva-Marcos, \textit{Addressing the CKM unitarity problem with a vector-like up quark}, JHEP \textbf{07} (2021), 099
doi:10.1007/JHEP07(2021)099 [arXiv:2103.13409 [hep-ph]].
\end{itemize}

\hspace{-0.5cm}The research performed by the author during his PhD studies has also led to the publication of the Review
\begin{itemize}
    \item J. M. Alves, G. C. Branco, A. L. Cherchiglia, J. T. Penedo, P. M. F. Pereira, C. C. Nishi, M. N. Rebelo and J. I. Silva-Marcos, \textit{Vector-like Singlet Quarks: a Roadmap}, Phys. Rept. \textbf{1057} (2024) 1-69 doi:10.1016/j.physrep.2023.12.004 [arXiv: 2304.10561 [hep-ph]]
\end{itemize}

\hspace{-0.5cm}as well as to the following contribution to conference proceedings:
\begin{itemize}
    \item P. M. F. Pereira, \textit{Non-Unitary Mixing Matrices in Neutrino and Vector-like Quark Models}, in Proceedings of the Corfu Summer Institute “School and Workshops on Elementary Particle Physics and Gravity” (CORFU2021) Corfu, Greece, 29 August-8 September 2021, [arXiv: 2205.05101].
\end{itemize}
\vfil 
\cleardoublepage


%
\tableofcontents
\cleardoublepage 

%
\phantomsection
\addcontentsline{toc}{section}{\listtablename}
\listoftables
\cleardoublepage 

%
\phantomsection
\addcontentsline{toc}{section}{\listfigurename}
\listoffigures
\cleardoublepage 

%
%
%
%
%
%
%
\nomenclature[a]{SM}{Standard Model}
\nomenclature[a]{WB}{Weak Basis}
\nomenclature[a]{WBT}{Weak Basis Transformation}
\nomenclature[a]{LHC}{Large Hadron Collider}
\nomenclature[a]{QFT}{Quantum Field Theory}
\nomenclature[a]{SSB}{Spontaneous Symmetry Breaking}
\nomenclature[a]{CKM}{Cabibbo-Kobayashi-Maskawa}
\nomenclature[a]{PMNS}{Pontecorvo-Maki-Nakagawa-Sakata}
\nomenclature[a]{VLQ}{Vector-like Quark}
\nomenclature[a]{FCNC}{Flavour-Changing Neutral Coupling}
\nomenclature[a]{VEV}{Vacuum Expectation Value}
\nomenclature[a]{BR}{Branching Ratio} 
\nomenclature[a]{CPV}{CP Violation}
\nomenclature[a]{LFV}{Lepton Flavour Violation}
\nomenclature[a]{LFNU}{Lepton Flavour Non-Universality}
\nomenclature[a]{LBL}{Long Baseline}
\nomenclature[a]{SBL}{Short Baseline}
\nomenclature[a]{DU}{Deviations from Unitarity}
\nomenclature[a]{BAU}{Baryon Asymmetry of the Universe}
\nomenclature[a]{NP}{New Physics}
\nomenclature[a]{$n_R$$\nu$SM}{SM with the addition of $n_R$ right-handed neutrinos}
\nomenclature[a]{$n_u$SM$n_d$}{SM with the addition of $n_u$ up-type VLQs and $n_d$ down-type VLQs}

\nomenclature[s]{$\mu, \nu, \sigma, \rho$}{Lorentz indices.}
\nomenclature[s]{$i,j,k, l$}{Mass state indices.}
\nomenclature[s]{$\alpha, \beta, \gamma, \delta$}{Flavour indices.}
\nomenclature[s]{$u, v, w $}{Color indices.}
\nomenclature[s]{$a, b, c $}{Gauge group Generators indices.}

\nomenclature[t]{T}{Transpose.}
\nomenclature[t]{*}{Conjugate.}
\nomenclature[t]{$\dagger$}{Conjugate Transpose.}

%
\phantomsection
\addcontentsline{toc}{section}{\nomname}
\printnomenclature
\cleardoublepage



%
\setcounter{page}{1}
\pagenumbering{arabic}


\chapter{Introduction and Motivation}
\label{chapter:introduction}

The very precise measurements of the quark mixing matrix - the CKM matrix \cite{PhysRevLett.10.531,10.1143/PTP.49.652} and its (apparent) consistency with unitarity \cite{pdg} was one of the main motivations for the prediction of the top quark, which was discovered in 1995 \cite{PhysRevLett.74.2626,PhysRevLett.74.2632}. Since then, the unitarity of the mixing matrices of the Standard Model (SM) is almost taken for granted.

In the leptonic sector, with massless \textbf{neutrinos} and only the three known charged leptons - $e$, $\mu$ and $\tau$, the leptonic mixing matrix was forced to be the identity matrix, as it could be rotated away via weak-basis transformations (WBT).

With the discovery of \textbf{neutrino oscillations} \cite{PhysRevLett.81.1158,PhysRevLett.89.011301}, which implies that at least two have non-zero mass \cite{b.pontecorvo1947,b.pontecorvo1958,b.pontecorvo1968,s.m.bilenkyb.pontecorvo1976}, there was finally room for a \textbf{non-trivial leptonic mixing matrix} - the PMNS matrix \cite{ziromakimasaminakagawashoichisakata1962}. The study of this matrix was supported by the work of B. Pontecorvo and his peers \cite{b.pontecorvo1947,b.pontecorvo1958,b.pontecorvo1968,s.m.bilenkyb.pontecorvo1976,b.pontecorvo21958}, which during decades questioned the hypothesis of all neutrinos being massless. These neutrinos are Dirac particles \cite{palashpal,Branco:1988ex} if right-handed neutrinos are added to the SM and lepton number Conservation is imposed, yielding a PMNS that is a unitary $3 \times 3$ matrix.

However, if one does not impose lepton number conservation - the approach with the least number of assumptions - the \textbf{Majorana mass term} is allowed and neutrinos are Majorana particles. In this case, the $3 \times 3$ PMNS is non-unitary in general, as there will be \textbf{extra neutrino states}.

Furthermore, current oscillation experiments, both short and long baseline, measure the entries of the PMNS matrix with very low precision, while many results one finds in the literature are obtained assuming the unitarity of PMNS. Hence, there is no experimental indication of a unitary PMNS and a non-unitary PMNS is still allowed \cite{Parke:2015goa,Miranda:2019ynh}.

Thus, considering the Majorana neutrinos hypothesis - which given the number of papers in the literature that use it,  should be what the HEP community believes is more likely to be true - a \textbf{leptonic $3 \times 3$ non-unitary mixing matrix} has to be incorporated in the SM, which spoils the presumed feature that all the mixing matrices in the SM are unitary matrices.

This would be enough to motivate this thesis but recently a new anomaly was discovered in the quark sector - known as the \textbf{CKM unitarity problem} or the \textbf{Cabibbo angle anomaly} \cite{Czarnecki:2019mwq,Seng:2020wjq,Hayen:2020cxh,Shiells:2020fqp,Belfatto:2019swo,Belfatto:2021jhf}.

This anomaly stems from new measurements of $|V_{us}|$ and $|V_{ud}|$ that indicate that \textbf{the unitarity of the first row of the CKM matrix may be violated} -
$|V_{ud}|^2+|V_{us}|^2+|V_{ub}|^2 < 1$ - at two or three standard deviations. 
This deficit results from new theory calculations of the SM radiative corrections to
$\beta$-decay processes~\cite{Seng:2018yzq,Seng:2018qru}.
The fact that a unitary CKM matrix has been so successful until this anomaly made an appearance shows that deviations from $3 \times 3$ unitarity, if present, should be small. One must keep in mind that, until now, a large number of experimental data both on quark mixing and CP violation is consistent with a unitary CKM matrix.

It is then curious that an extension of the SM almost as simple, and in many ways similar, to adding right-handed neutrinos can provide an answer to this anomaly. The addition of particles known as \textbf{vector-like quarks} has the notable feature of leading to a \textbf{naturally suppressed violation of $3 \times 3$ unitarity} as well as to naturally suppressed flavour changing neutral couplings (FCNC) at tree level. 

Since this thesis deals with unitarity, it is relevant to \textbf{stress the differences} between \textbf{Unitarity in the QFT sense} and what is meant by \textbf{unitarity of the mixing matrices of the SM}.
Unitarity in the QFT sense usually refers to quantities being bounded from above - see ref. \cite{jorgec.romão2016} where the absence of the Higgs results in some cross sections growing with energy ad infinitum - due to the unitarity of the $S$ matrix, a matrix connecting sets of asymptotically free particle states (the in-states and the out-states) in the Hilbert space of physical states.
Unitarity is essentially the condition that the time evolution of a quantum state according to a given evolution equation is mathematically represented by a unitary operator (i.e. there is conservation of probability) - in QFT, this operator is the $S$ matrix. Unitarity is usually taken as an axiom or basic postulate of quantum mechanics and there is no experimental evidence of its violation \cite{jorgec.romão2016,jorgec.romão2018,peskinschroeder1995,Schwartz:2014sze}. The thing to keep in mind is that, \textbf{in this thesis, what is being discussed is the violation of the unitarity of what the HEP community, at the time of writing of this thesis, defines as the mixing matrices of the SM in the leptonic and quark sectors}, which are  $3 \times 3$ matrices. This does not violate the unitarity of the $S$ matrix of the SM + massive neutrinos or SM + vector-like quarks. The complete mixing matrices \textbf{must} be unitary and probability is conserved when one takes into account the new extra states and the new decays the standard particles may have. It just might be a possibility that what until now was considered to be the full mixing matrix is actually a $3 \times 3$ non-unitary part of a bigger matrix that is unitary.

\section{Thesis Outline}

Although more significant in the quark sector \cite{Miranda:2019ynh,Belfatto:2019swo}, the experimental hints of deviations from unitarity of the mixing matrices on both fermionic sectors of the SM show that studying models with $3 \times 3$ non-unitary mixing matrices is a relevant topic.\par
In the published work that this thesis is based on, an exact parameterisation of these non-unitary mixing matrices was developed, such that one could study more exotic regions of the parameter space of these models without recurring to approximations, which usually fail out of these standard areas.\par

This work is organized as follows. In the rest of this section a brief
review of the SM will be made, in order to fix notation and introduce concepts. In
chapter 2, the Standard Model + $n_R$ right-handed neutrinos ($n_R\nu$SM) will be covered, using the parameterisation first presented in ref. \cite{Agostinho:2017wfs}. Results from refs. \cite{Agostinho:2017wfs,Branco:2019avf}
are discussed, including some unpublished work regarding CP Violating phases and the classification of the whole parameter space of $n_R\nu$SM.

In chapter 3, the Standard Model + $n_u$ up-type vector-like quarks + $n_d$ down-type vector-like quarks ($n_u$SM$n_d$) will be covered, using notations and definitions from refs. \cite{Pereira:2022pqu,Alves:2023ufm}. Some emphasis will be placed on models with only up vector-like quarks $n_d=0$, ($n_u$SM$0$), with a discussion of results from ref. \cite{Branco:2021vhs}. 
Models with only down vector-like quarks $n_u=0$, ($0$SM$n_d$) will also be examined, using results from ref. \cite{Alves:2023ufm}.

Chapter 4 will include the most important results of these models and the experimental signals they can leave on current and future experiments, as discussed in refs \cite{Agostinho:2017wfs,Branco:2019avf,Alves:2023ufm,Branco:2021vhs}. To conclude, a final chapter with the main
conclusions of the published works and some future prospects is included.   

\section{Brief Summary of the Standard Model of Particle Physics}
\label{sec:summary_SM}

This section consists of a brief review of the SM and its main features and is directed to someone who is not familiar with the topic. For a more complete treatment, with emphasis on the technical subjects, refer to refs. \cite{jorgec.romão2018,peskinschroeder1995,Schwartz:2014sze,Branco:1999fs,paullangacker2009,tapeichenglingfongli1984,s.weinberg1995}. \par
The SM is a Quantum Field Theory (QFT) that introduces fundamental interactions (strong, weak and electromagnetic) such that the Lagrangian is locally invariant under Gauge Symmetries - where every field is in a representation of the Gauge Symmetry Group. Thus, the SM is a Gauge Theory. In QFT, particles are seen as excitations (quanta) of a field that permeates the universe. Going from a Classical Field Theory to a QFT does not involve a modification of the Lagrangian or of the field equations, but rather a reinterpretation of the field variables.\par

The SM particles currently known can be divided into Bosons and Fermions. In the Bosons category there is a scalar (spin $0$) boson - $H$ - the Higgs particle, responsible for the mass generation mechanism, and 4 kinds of gauge bosons with spin $1$: $W^\pm$, $Z$ and $\gamma$ (the photon) - responsible for electro-weak interactions - and $G$ (the gluons), responsible for strong interactions. Fermions can be divided into Quarks and Leptons and there are three families of each. Each family of Quarks has an up (electric charge $+2/3$) and down (electric charge $-1/3$) type quark with 3 colours each, while for Leptons each family has a charged lepton (electric charge $-1$) and the correspondent neutrino (electric charge $0$). Although a fermion is any particle with a half-integer spin - $1/2,3/2,5/2,...$ - fundamental fermions with spins different from $1/2$ were never discovered in nature. Hence, all SM fermions have spin $1/2$. \par

\begin{figure}[h!]
  \centering
  \includegraphics[width=0.5\textwidth]{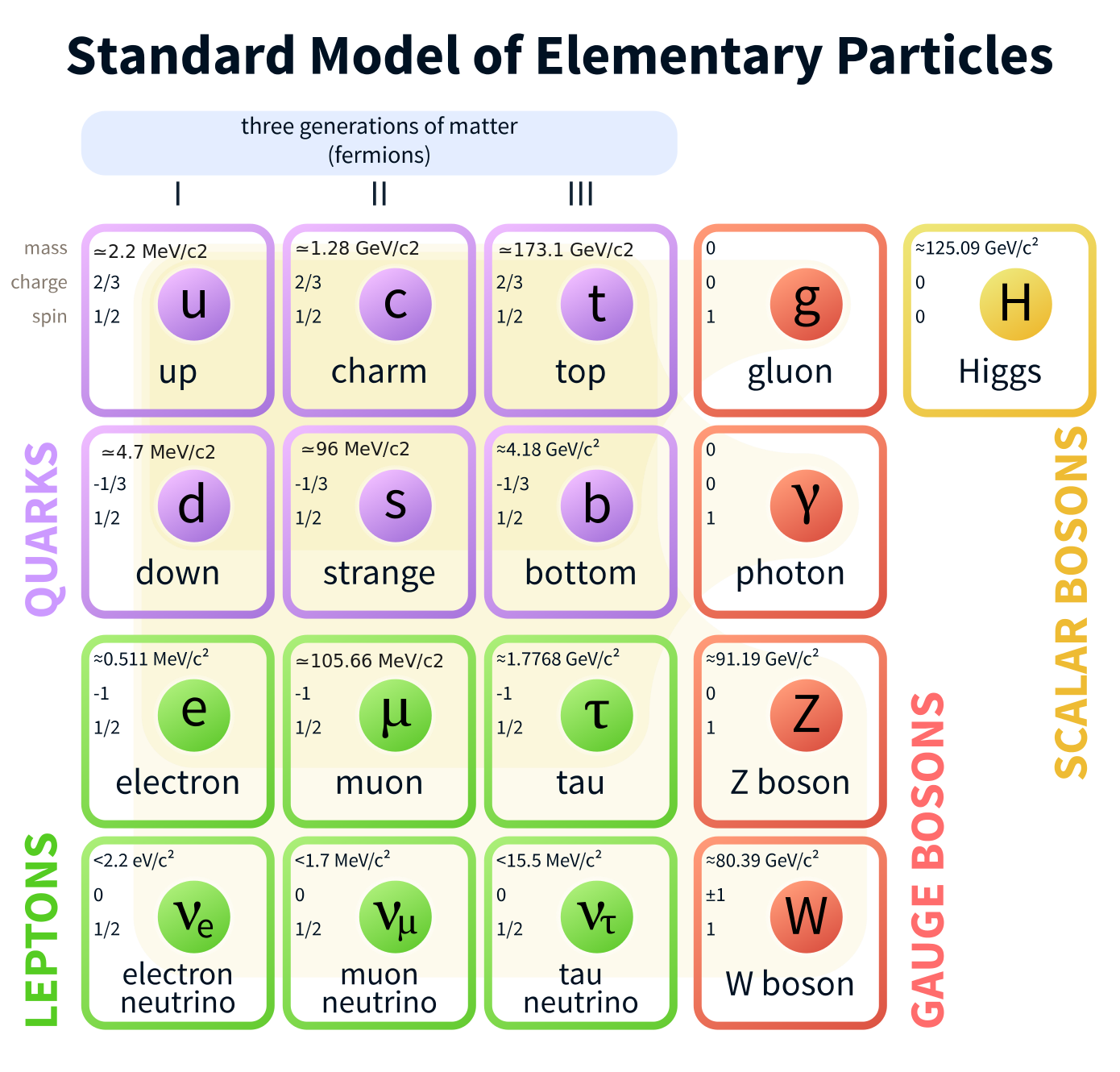} 
  \caption{List of SM particles. Until 2012, all of them but the Higgs ($H$) were experimentally discovered. Image is not property of this thesis' author. Licensed under Creative Commons Attribution 3.0 Unported license, one is free to to copy, distribute, transmit and adapt the work. Numerical values taken from Particle Data Group Booklet 2016.}
  \label{fig:SMparticles}
\end{figure}
Fig. \ref{fig:SMparticles}, illustrates all the SM particles. It is clear that the only characteristic that distinguishes particles with the same quantum numbers and from different families is their mass. Each of these fundamental fermions can be thought as a unique particle species and is usually referred to as a flavour. Hence, flavour physics is the study of how these different particle species mix with each other and flavour changing interactions are interactions where one species turns into another - one down quark transforms into a charm quark, for instance. \par
In gauge theories, masses cannot be introduced in the free theory because they would break the required local gauge invariance. Hence, masses are introduced via Spontaneous Symmetry Breaking (SSB) of the SM Group
\begin{equation}
    G_{SM}=SU(3)_c \times SU(2)_L \times U(1)_Y ~.
    \label{eq:GSM}
\end{equation}
 The massless Lagrangian for this theory is locally invariant under transformations of this group.

One can define the construction of the SM into two parts: Before SSB, where every field is massless and every weak interaction is diagonal - this is named the interaction or flavour basis -, and after SSB, where some fields become massive and mixing occurs - named mass or physical basis.

The quantum numbers of the particle spectrum of the SM in the interaction basis \footnote{Note that in the interaction basis $G^+$ and $G^0$ compose the Higgs doublet and $B$, $W^1$, $W^2$ and $W^3$ are linear combinations of the physical bosons $W^\pm$, $Z$ and $\gamma$.} are given in [Table \ref{tab:smqnumbers}].
\begin{table}[h!]
\centering
\caption{Quantum Numbers and Representations of every SM Particle \label{tab:smqnumbers}}
\resizebox{\textwidth}{!} {
\begin{tabular}{|c|c|c|c|c|c|c|c|c|c|c|c|c|c|}
\hline
\textbf{Field}                          & \textbf{$l^{\alpha}_L$} & \textbf{$l^{\alpha}_R$} & \textbf{$\nu^{\alpha}_L$}  & \textbf{$u^{\alpha}_L$} & \textbf{$u^{\beta}_R$} & \textbf{$d^{\beta}_L$} & \textbf{$d^{\beta}_R$} & \textbf{$G^{+}$} & \textbf{$ G^0$} & \textbf{$G$} & \textbf{$B$} & \textbf{$W^{1,2}$} & \textbf{$W^3$} \\ \hline
$\bm{T_3}$                  & -1/2          & 0             & 1/2             & 1/2           & 0             & -1/2          & 0             & 1/2                            & -1/2                           & 0          & 0              & $\pm 1$          & 0          \\ \hline
\textbf{Y}                     & -1/2          & -1            & -1/2         & 1/6           & 2/3           & 1/6           & -1/3          & 1/2                            & 1/2                            & 0          & 0              & 0          & 0          \\ \hline
\textbf{Q}                     & -1            & -1            & 0                & 2/3           & 2/3           & -1/3          & -1/3          & 1                              & 0                              & 0          & 0              & $\pm 1$          & 0          \\ \hline
$\bm{SU(3)_c}$ \textbf{Rep.}                     &           \textbf{1}  & \textbf{1}         & \textbf{1}                          & \textbf{3}           & \textbf{3}           & \textbf{3}          & \textbf{3}          & \textbf{1}                             & \textbf{1}                              & \textbf{8}          & \textbf{1}             & \textbf{1}          & \textbf{1}         \\ \hline

\textbf{Lorentz-Poincare Rep.} & \textbf{(1/2,0) }      & \textbf{(0,1/2) }        & \textbf{(1/2,0) }        & \textbf{(1/2,0) }         & \textbf{(0,1/2) }         & \textbf{(1/2,0) }         & \textbf{(0,1/2) }         & \textbf{(0,0) }                          & \textbf{(0,0) }                            & \textbf{(1/2,1/2) }    & \textbf{(1/2,1/2) }     & \textbf{(1/2,1/2) }    & \textbf{(1/2,1/2) }    \\ \hline
\end{tabular}
}
\end{table}

Where $\alpha=1,2,3$ is the leptonic generation index  and $\beta=1,2,3$ is the quark generation index- one can define that to the top quark ($t_{R,L}$) corresponds $u^3_{R,L}$ and to the muon ($\mu_{R,L}$) corresponds $l^2_{R,L}$, and so on. $Y$ is defined by the relation $Q=Y+T_3$, which will be explained later. \par

The quantum numbers of the particle spectrum of the SM are chosen such that they are in the correct representation of the gauge group and that the conserved quantum number (after SSB) - electric charge $Q$ - has the correct value for each one.
$Y$ is named Weak Hypercharge and is the Quantum Number corresponding to $U(1)_Y$, $T_3$ is named Weak Isospin and is the quantum number corresponding to $SU(2)_L$.
The Quantum Number corresponding to $SU(3)_c$ is colour. Quarks can have three colours (red, green and blue) and anti-quarks can have three anti-colours (anti-red, anti-green and anti-blue). They are in the triplet representation of $SU(3)_c$. Gluons can have the 8 independent combinations of these 6 (3+3) colours and anti-colours. However, every object observed in nature at low energy is colourless (a singlet of $SU(3)_c$,   like an electron or a proton), which means quarks and gluons aren't asymptotic states and don't have a spectral representation, only hadrons (combinations of these) do. This happens due to a special property of QCD: Asymptotic freedom \cite{h.davidpolitzer1973,davidj.grossfrankwilczek1973}.\par

The chirality of a particle also plays a major role, albeit being a particularly abstract concept. The chirality of a particle is positive (negative) if the particle transforms in a right-handed (left-handed) representation of the Poincaré group.
A related and easier to understand concept is helicity. Mathematically, helicity is the sign of the projection of the spin vector onto the momentum vector: “left” is negative, “right” is positive.
Although similar, Chirality and Helicity only match if a particle is massless \cite{peskinschroeder1995,Schwartz:2014sze}. 
Understandably, helicity is an easier concept to understand in classical terms - and actually is a constant of motion, since it commutes with the Hamiltonian, and thus, in the absence of external forces, is time-invariant. Nonetheless, one should use chirality when working in particle physics, since it is Lorentz invariant. Often times in the literature (and in this thesis) one refers to left-handed or right-handed particles, it should be understood as negative or positive chirality particles. 
Thus, if one considers the chirality state of a particle, positive and negative chirality states could be considered different particles, since they have different interactions (the SM is a chiral QFT, more on that later). In a massless SM without Yukawa couplings, this would be the case, as one wouldn't have other choice but to identify particles based on their interactions. One wouldn't have 3 families of Quarks and Leptons, but just one, with the double amount of fields - 1 Dirac field can be decomposed into two Weyl fields (chiral massless fields) \cite{palashpal}:\par
\begin{table}[h!]
\centering
\resizebox{0.2\textwidth}{!} {
\begin{tabular}{|l|l|l|}
\hline
 & \textbf{$P_L$} & \textbf{$P_R$} \\ \hline
\textbf{$l$} & $l_L $     & $l_R$
                                        \\ \hline
\textbf{$\overline{l}$} & $\overline{l}_R$       &   $\overline{l}_L $                                              \\ \hline
\end{tabular}
}
\caption{The two particle/anti-particle states of a massless SM without Yukawa couplings. $P_L$ and $P_R$ are chirality projection operators - $P_L l_L = l_L$, $P_L l_R = 0$, and so on. \label{tab:chiralparticles} }
\end{table}\par
This happens because the SM is a chiral theory - it treats differently particles with different chirality. For instance, considering again the massless SM without Yukawa couplings, this time with the addition of positive chirality neutrinos, $\nu_R$. In this model, the $\nu_R$ would be a sterile particle, as it doesn't interact with anything, while the $\nu_L$ would interact via weak interaction. In the case of the SM, which is a theory with massive particles with only Dirac fermions, a particle is considered a state with definite mass which is a superposition of the negative and positive chirality states - a Dirac mass term in the SM Lagrangian can be interpreted as a interaction term between the positive chirality state and the negative chirality state.
Thus, SSB bounds the two chirality states into what is defined as a particle state. \par
To obtain a Lagrangian that is locally invariant under transformations of the group defined in eq. \eqref{eq:GSM}, one must introduce gauge fields and correspond to each one of them a generator of the group. The SM group has $12$ generators - $8$ $SU(3)$ bosons ($G^a$), $3$ $SU(2)$ bosons ($W^b$) and $1$ $U(1)$ boson ($B$).\par
Gauge bosons are in the adjoint representation of their corresponding gauge group (octet for $SU(3)_c$ and triplet for $SU(2)_L$) and in the vector representation of the Lorentz-Poincare group, quarks and leptons are in the fundamental or anti-fundamental representation of the Lorentz-Poincare Group, while the negative chirality ones are in in the fundamental representation of $SU(2)_L$  (doublet):
\begin{equation}
    \Psi^{\alpha}_L= \begin{pmatrix}\nu^{\alpha}_L \\ l^{\alpha}_L \end{pmatrix} ~,~ \Upsilon^{\beta}_L= \begin{pmatrix} u^{\beta}_L \\ d^{\beta}_L  \end{pmatrix} ~.
\end{equation}
Note that $\alpha$ and $\beta$ are flavour indices.
The positive chirality ones are in the singlet representation of $SU(2)_L$:
\begin{equation}
    l^{\alpha}_R ~,~ u^{\beta}_R ~,~ d^{\beta}_R ~.
\end{equation}
 Regarding $SU(3)_c$, Leptons are in the singlet representation - don't take part in coloured interactions - while quarks are in the triplet representation:
 \begin{equation}
     Q^{\gamma}= \begin{pmatrix} q^{\gamma}_{r} \\ q^{\gamma}_{g} \\ q^{\gamma}_{b} \end{pmatrix} ~,
     \label{tripletsu3}
 \end{equation}
 where $\gamma=1,2,..,6$ is the quark flavour index - implicitly defining the up quark ($u$) as the triplet $Q^1$ with entries $q^1_i$, the down quark ($d$) as the triplet $Q^2$ with entries $q^2_i$, the charm quark ($c$) as the triplet $Q^3$ with entries $q^3_i$, and so on. Note that $i=r,g,b$ is a colour index and that the order is irrelevant.\par

The gauge fields are necessarily bosons because to get gauge invariance it is necessary to add fields that transform like the derivative - creating what is known as the Covariant Derivative

 \begin{equation}
    D_{\mu}^q=\partial_{\mu} + ig_s \sum_{a=1}^{8} G_{\mu}^{a}\frac{\lambda^a}{2} ~,~
      D_{\mu}^L=\partial_{\mu}+ig \sum_{b=1}^{3} W_{\mu}^{b}\frac{\sigma^b}{2} +ig' Y B_{\mu} ~,~
         D_{\mu}^R=\partial_{\mu}+ig' Y B_{\mu} ~.
         \label{eq:covderivint}
\end{equation}

 where the $g_s$ are the couplings of each interaction,$\frac{\lambda^a}{2}$ are the 8 generators of SU(3) (Gell-Mann matrices) and $T^b= \frac{\sigma^b}{2}$ are the 3 generators of SU(2) (Pauli matrices). The used sign notation is consistent with \cite{jorgec.romaojoaop.silva2012},  taking all $\eta_i=1$. Since the derivative transforms like a vector under the Lorentz-Poincare group, gauge fields must have the same behaviour, and thus, be vector fields, with integer non-zero spin (vector bosons).
This covariant derivative generates the interactions between the gauge bosons and the fermionic fields:

\begin{equation}
\begin{split}
    \mathcal{L}_{\textrm{Fermion}}= &\sum_{\gamma} \overline{Q^{\gamma}} i \gamma^{\mu} D_{\mu}^q Q^{\gamma} + \sum_{\alpha} \left( \overline{\Psi^{\alpha}_L} i \gamma^{\mu} D_{\mu}^L \Psi^{\alpha}_L + \overline{l^{\alpha}_R} i \gamma^{\mu} D_{\mu}^R l^{\alpha}_R  \right) \\
    & +\sum_{\beta} \left( \overline{\Upsilon^{\beta}_L} i \gamma^{\mu} D_{\mu}^L \Upsilon^{\beta}_L  +  \overline{u^{\beta}_R} i \gamma^{\mu} D_{\mu}^R u^{\beta}_R + \overline{d^{\beta}_R} i \gamma^{\mu} D_{\mu}^R d^{\beta}_R \right) ~,
    \label{eq:lfermion}
    \end{split}
\end{equation}
where the indices $\alpha$, $\beta$ and $\gamma$ have the same meaning as in Table \ref{tab:smqnumbers} and in eq. \eqref{tripletsu3}.

The gauge boson self-interactions come from their kinetic terms, which are of the form:
\begin{equation}
    \mathcal{L}_{\textrm{kin}}=-\frac{1}{4}G^{a \mu \nu} G_{a\mu \nu} -\frac{1}{4}W^{b \mu \nu} W_{b\mu \nu}  -\frac{1}{4}B^{ \mu \nu} B_{\mu \nu} ~,
    \label{eq:lkin}
\end{equation}
where
\begin{equation}
G^a_{\mu \nu}=\partial_{\mu} G^a_{\nu}-\partial_{\nu} G^a_{\mu}- g_s f^{acd}  G_{c \mu}  G_{d \nu} ~,
    W^b_{\mu \nu}=\partial_{\mu} W^b_{\nu}-\partial_{\nu} W^b_{\mu}- g f^{bcd}  W_{c \mu}  W_{d \nu} ~, B_{\mu \nu}=\partial_{\mu} B_{\nu}-\partial_{\nu} B_{\mu} 
    \label{eq:lkin2}
\end{equation}

and $a=1,...,8$ is summed implicitly and runs over the number of gauge bosons of $SU(3)$ and $b=1,2,3$ is summed implicitly and runs over the number of gauge bosons of $SU(2)$. $f^{abc}$ are the structure constants for the gauge group - for $SU(2$) $f^{abc}=\epsilon^{abc}$. For abelian gauge groups - like $U(1)$ - these are zero. 
These terms generate self-interactions (in the gluon case) but also interactions between different gauge bosons.
After electroweak symmetry breaking the physical states $Z$, $\gamma$, $W^+$, $W^-$ are revealed to be linear combinations of the gauge fields $B$, $W^1$, $W^2$, $W^3$. It is easy to see from eqs. \eqref{eq:lkin} \eqref{eq:lkin2} that this generates triple and quartic interactions between $Z$, $\gamma$, $W^+$ and $W^-$. \par
When electroweak symmetry breaking happens the mass of fundamental particles is generated, and here the Higgs is the protagonist. The existence of the Higgs is one of the simplest ways of generating mass for fundamental particles without explicitly breaking the gauge symmetry while keeping the unitarity of the SM \cite{johnm.cornwalldavidn.levingeorgetiktopoulos1973,jorgec.romão2016}. 
This 'spontaneous' symmetry breaking means that the vacuum of the theory at a certain point in time - spontaneously - stops having the same symmetry as the Lagrangian.
The Higgs mechanism spontaneously breaks $G_{SM}$ into $SU(3)_C \times U(1)_Q$, in order for this to happen, a scalar doublet of $SU(2)_L$ - $\phi$ - is added to the theory:
\begin{equation}
\phi= \begin{pmatrix}G^+ \\ G^0 \end{pmatrix}  ~,~  V(\phi) = -\mu^2(\phi^\dagger \phi)+\lambda(\phi^\dagger \phi)^2 ~,~   \mathcal{L}_{\textrm{Higgs}} = (D_{\mu}^L \phi)^{\dagger}\cdot (D^{\mu L} \phi) - V(\phi) ~,
\end{equation}
where $V(\phi)$ is the most general renormalizable potential that can be added to the Lagrangian. Considering $<\phi>$ the value of $\phi$ for which $V(\phi)$ is minimal, if $\mu^2>0$ and $\lambda>0$, the potential has an absolute minimum for $<\phi> \neq \begin{pmatrix} 0 \\ 0 \end{pmatrix}$. From the minimization equation one gets $<\phi^\dagger \phi>=\frac{\mu^2}{2\lambda}$. All other SM fields with spin different from zero are compelled to have a zero vacuum expectation value, by Lorentz invariance.  This vev can be parameterised in the following way:
\begin{equation}
<\phi>= \frac{1}{\sqrt{2}}\begin{pmatrix}0 \\v \end{pmatrix}  ,
\end{equation}  which gives
\begin{equation}
    <\phi^\dagger \phi>=\frac{\mu^2}{2\lambda} = \frac{v^2}{2} \implies v^2= \frac{\mu^2}{\lambda} ~.
\end{equation}

To get a proper spectrum it's useful to write the $\phi$ field as a perturbation around its vev, taking into account all its degrees of freedom:

\begin{equation}
\phi= <\phi> + \begin{pmatrix}G^+ \\ \frac{H+iG^0}{\sqrt{2}} \end{pmatrix} ~.
\label{eq:higgsgeneral}
\end{equation}
One can parameterise three degrees of freedom in the form of a global $SU(2)_L$ transformation:
\begin{equation}
\phi=  e^{i\frac{\sigma^b}{2} \omega_b}  \begin{pmatrix}0 \\ \frac{v+H}{\sqrt{2}} \end{pmatrix} ~,
\end{equation}
 and then use the freedom to apply a global $SU(2)_L$ transformation to absorb them. This is known as going to the unitary gauge:
 
 \begin{equation}
\phi \xrightarrow{} e^{-i\frac{\sigma^b}{2} \omega_b} \phi = \begin{pmatrix}0 \\ \frac{v+H}{\sqrt{2}} \end{pmatrix} = <\phi> + \begin{pmatrix}0 \\ \frac{H}{\sqrt{2}} \end{pmatrix} ~.
\label{eq:higgsunit}
\end{equation}

In this gauge, the $H$ field parameterises the deviations from the value of $\phi$ that minimizes the potential. It will correspond to the Higgs field.
Note that the vacuum $<\phi>$ is the kind of vacuum we need for SSB because it's not invariant under $SU(2)_L$ transformations anymore:

\begin{equation}
    e^{i\frac{\sigma^b}{2} \theta_b} \cdot <\phi> \neq <\phi> \implies \frac{\sigma^b}{2} \cdot <\phi> \neq \begin{pmatrix}0 \\0 \end{pmatrix} ~.
\end{equation} 

However, it's invariant under the combination that represents electric charge, given by $Q=T_3 + Y $ \cite{nishijima}, the Gell-Mann-Nishijima relation, where $Y$ is a diagonal matrix in flavour space:

\begin{equation}
Q <\phi > =  \left(\frac{\sigma^3}{2} +Y \cdot \id_{2\times2}\right) \cdot <\phi> =  \begin{pmatrix}  1/2+Y & 0 \\ 0 & -1/2 + Y \end{pmatrix} \cdot \frac{1}{\sqrt{2}}\begin{pmatrix}0 \\v \end{pmatrix} = \begin{pmatrix}0 \\0 \end{pmatrix} = \bf{0} 
\end{equation} 

which is zero if $Y \phi=\frac{1}{2} \phi$. This proves that in the SM with only one Higgs doublet it is impossible to break electric charge conservation.

Thus, this vacuum  spontaneously breaks part of the electroweak gauge symmetry, after which only one neutral Higgs scalar, $H$, remains in the physical particle spectrum. However note that the number of scalars in the theory is not constrained and there can be more than one Higgs-like particle if one introduces more than one Higgs-like doublet. This itself is another topic of theoretical and experimental research.
The Covariant derivative acting on $\phi$ (kinetic term on $\mathcal{L}_{Higgs}$) generates electroweak mixing among the bosons. After a basis rotation, the gauge bosons mass matrices - a collection of all terms that are order two in the gauge boson fields - become diagonal, revealing the masses of the gauge bosons. This identifies the physical states and gives a relation for their couplings  
\begin{equation}
    Z_{\mu}=\cos{\theta_\textrm{W}}W^3_{\mu} - \sin{\theta_\textrm{W}}B_{\mu}~,~     A_{\mu}=\cos{\theta_\textrm{W}} B_{\mu} + \sin{\theta_\textrm{W}}W^3_{\mu} ~,~  W^{\pm}_{\mu}=\frac{1}{\sqrt{2}}(W^1_{\mu}  \mp i W^2_{\mu}) ~,
\end{equation}
\begin{equation}
    sin^2(\theta_\textrm{W})=\frac{g'^2}{g^2+g'^2} ~,~~m_W=m_Z \cos{\theta_\textrm{W}} ~,~ e=g \sin{\theta_\textrm{W}} = g' \cos{\theta_\textrm{W}} ~,
\end{equation}
\begin{equation}
    m_W^2=\frac{1}{4}v^2g^2 ~,~     m_Z^2=\frac{1}{4}v^2(g^2+g'^2) ~,~ m_H^2=2 \mu^2 ~.
    \label{eq:electroweakmwmz}
\end{equation}

Note that $A_\mu$ is the traditional name for the $\gamma$ (the photon) field. The initial vacuum was identically zero and was left invariant under the 4 generators of $SU(2)_L \times U(1)_Y$. Now, it is left invariant under one combination of two of these generators. From the Goldstone Theorem one should have $3\times (4-1)$ Nambu-Goldstone bosons (massless scalar bosons), but there are $3$ massive gauge bosons instead. In a pictoric language it is said the gauge bosons "eat" the Nambu-Goldstone bosons. If the broken symmetry was global (and not gauge/local) one would have 3 massless scalars \cite{goldstonewein,goldstone} - $G^\pm$ (2, real and imaginary part) and $G^0$ (1).\par

After SSB one can go the physical basis and write the covariant derivative in terms of the physical bosons
\begin{equation}
    D_{\mu}=\partial_{\mu}+i \sum_{a=1}^{8} G_{\mu}^{a}\frac{\lambda^a}{2} + i \frac{g}{\sqrt{2}} (W_{\mu}^+ T_+ + W_{\mu}^- T_-) + i \frac{g}{\cos{\theta_\textrm{W}}} (T_3-Q \sin^2{\theta_\textrm{W}}) Z_{\mu} +      ieQ A_{\mu} ~,
\end{equation}
where $T_\pm = T_1 \pm i T_2 = \frac{\sigma_1}{2} \pm i \frac{\sigma_2}{2} = \sigma_\pm$.

It's useful to rewrite eq. \eqref{eq:covderivint}\footnote{$D^q_{\mu}$ stays the same after electroweak symmetry breaking. $SU(3)_c$ is not broken.}:
\begin{equation}
    D_{\mu}^L=  \partial_{\mu} + i \frac{g}{\sqrt{2}} (W_{\mu}^+ T_+ + W_{\mu}^- T_-) + i \frac{g}{\cos{\theta_\textrm{W}}} (T_3-Q \sin^2{\theta_\textrm{W}}) Z_{\mu} +      ieQ A_{\mu}  ~,
\end{equation}
 \begin{equation}
D_{\mu}^R=  \partial_{\mu}+i (Q-T_3) \frac{g}{\cos{\theta_\textrm{W}}} (\sin{\theta_\textrm{W}}\cos{\theta_\textrm{W}} A_{\mu} - \sin^2{\theta_\textrm{W}} Z_{\mu}) ~.
\end{equation}

To generate Dirac fermion masses one needs to create an  $SU(2)_L \times U(1)_Y$ invariant term, using $\phi$ and the fermion fields, that after SSB generates their mass terms. Using Table \ref{tab:smqnumbers} it is easy to see that a term like

\begin{equation}
  Y^l_{\delta \sigma} \begin{pmatrix} \overline{\nu^{\delta}_L} &  \overline{l^{\delta}_L} \end{pmatrix}   \phi l^{\sigma}_R = Y^l_{\delta \sigma}   \begin{pmatrix} \overline{\nu^{\delta}_L} &  \overline{l^{\delta}_L} \end{pmatrix}  \begin{pmatrix} G^+ \\ v + H +iG^0 \end{pmatrix} l^{\sigma}_R \xrightarrow[\text{~}]{\text{SSB}}  Y^l_{\delta \sigma} \frac{v}{\sqrt{2}} \overline{l^{\delta}_L} l^{\sigma}_R ~,
  \label{eq:ssbmassmatrix}
\end{equation}
is invariant because it has $Y=1/2+1/2-1=0$, hence being a $SU(2)_L \times U(1)_Y$ singlet.
After SSB this generates a mass term proportional to $v$. The proportionality constant is $Y^l_{\delta \sigma}=\frac{ \sqrt{2}m^l_{\delta \sigma}}{v}$, where $m^l_{\delta \sigma}$ is the $(\delta,\sigma)$ entry of the Dirac mass matrix for the charged leptons. $Y^l_{\delta \sigma}$ are known  as the Yukawa couplings, and they parameterise the couplings between the Higgs field and fermions before SSB. Their origin and possible connection to other couplings is an object of study \cite{Saldana-Salazar:2016hxb,Kubo:1995gb}. This method also works for other fermions. However, for up quarks (the ones with $T_3=\frac{1}{2}$) one needs to define the adjoint doublet $\Tilde{\phi}= \begin{pmatrix} v+ H -i G^{0*} \\ -G^- \end{pmatrix}$ which has $Y=-\frac{1}{2}$. After SSB, one has $<\Tilde{\phi}>= \frac{1}{\sqrt{2}}\begin{pmatrix}v \\0 \end{pmatrix}  $.

Thus, after SSB one gets a Dirac mass matrix for up quarks, down quarks and for charged leptons:
\begin{equation}
    \mathcal{L}_{\textrm{Yukawa}}= - \overline{\Psi^{\delta}_L} Y^l_{\delta \sigma} \phi l^{\sigma}_R -  \overline{\Upsilon^{\delta}_L} Y^d_{\delta \sigma} \phi  d^{\sigma}_R -  \overline{\Upsilon^{\delta}_L} Y^u_{\delta \sigma} \Tilde{\phi} u^{\sigma}_R + h.c. ~.
    \label{eq:lyukawa}
\end{equation}
\par
Going from the interaction basis to the mass basis requires the diagonalisation of these mass matrices, with the new rotated states becoming the physical states. This generates mixing - mass states being linear combinations of interaction states - which makes the interaction Lagrangian in eq. \eqref{eq:lfermion} no longer diagonal in the fields. \par
In the SM, this only happens in the quark sector, as neutrinos are massless and it is possible to remove the mixing in the leptonic sector.

Let us rewrite eq. \eqref{eq:lyukawa} after SSB, including only the mass terms
\begin{equation}
    \mathcal{L}_{\textrm{FMass}} = - \overline{l_L^0} M^l l_R^0 -  \overline{d_L^0} M^d  d_R^0 -  \overline{u_L^0} M^u  u_R^0 + h.c. ~,
     \label{eq:massmatrixintro}
\end{equation}
where $M = \frac{v}{\sqrt{2}} Y$.
Thus, in the SM, a general Dirac mass matrix is a $3 \times 3$ matrix. Since in the SM there are no $\nu_R$, there are only mass matrices for up and down quarks as well as for charged leptons. After one diagonalises these mass matrices one finds new states - mass states that correspond to the physical particles.\par
The diagonalisation of a mass matrix cannot be performed with just one diagonalising matrix. One needs a unitary bi-diagonalisation (or as mathematicians call it - Singular Value Decomposition), because one needs to diagonalise a matrix while rotating two different fields - one with positive chirality and another with negative chirality - cf. eq. \eqref{eq:massmatrixintro}. Thus, the matrices acting on them are, in general, different.  Then, one has:
\begin{equation}
      {U^{l}_L}^{\dagger} M^l U^{l}_R = d_l ~,~ {U^{u}_L}^{\dagger} M^u U^{u}_R = d_u ~,~ {U^{d}_L}^{\dagger} M^d U^{d}_R = d_d ~,
   \label{eq:bidiagonalisation}
\end{equation}
where $d_u$, $d_d$ and $d_l$ are  diagonal $3 \times 3$ matrices with positive real entries, which contain the masses of the quarks and charged leptons, respectively, in the diagonal.
From eq. \eqref{eq:massmatrixintro} one notes that to go to this mass basis, the fields must transform as
\begin{equation}
   l^{\delta 0}_{R,L}=({U^{l}_{R,L}})^{\delta k} l^{k}_{R,L} ~,~   u^{\delta 0}_{R,L}=({U^{u}_{R,L}})^{\delta k} u^{k}_{R,L} ~,~   d^{\delta 0}_{R,L}=({U^{d}_{R,L}})^{\delta k} d^{k}_{R,L} ~,
\label{eq:weakbasis}
\end{equation}
where the left hand side corresponds to interaction states ($\delta$=$e$, $\mu$, $\tau$, for the charged leptons, for instance) and the right hand side corresponds to mass states ($k=1, 2, 3$).\par

It is now possible to write the Lagrangian in eq. \eqref{eq:lfermion} in the physical basis, highlighting the terms that contain interactions with the physical electroweak gauge bosons. For charged interactions with leptons,

\begin{equation}
    \mathcal{L}^l_{W}=- \frac{g}{\sqrt{2}} \left[W^+_{\mu} \overline{\nu^j_L}  \delta_{ij}\gamma^{\mu} l^i_L + W^-_{\mu} \overline{l^i_L}  \delta_{ij} \gamma^{\mu} \nu^j_L     \right] ~,
    \label{eq:wcurrentssml}
\end{equation}
where the matrix $U^l_L$ was rotated away by an unphysical redefinition of the $\nu_L$ fields. And for leptonic interactions with the photon and $Z$ boson,
\begin{equation}
   \mathcal{L}^l_{A,Z}= -\frac{g}{2 \cos{\theta_\textrm{W}}} \left[   Z_{\mu} ( \overline{\nu^i_L}
 \delta_{ij} \gamma^{\mu} \nu^j_L - \overline{l^i_L} \delta_{ij} \gamma^{\mu} l^j_L)  \right] -\left[( \frac{g \sin^2{\theta_\textrm{W}} } {\cos{\theta_\textrm{W}}} Z_{\mu} - e A_{\mu}) (\overline{l^i_L} \delta_{ij} \gamma^{\mu} l^j_L +\overline{l^i_R} \delta_{ij} \gamma^{\mu} l^j_R  ) \right] ~,
 \label{eq:neutralcurrentssml}
\end{equation}
where $e$ is the modulus of the electron charge. For charged interactions of quarks, one has
\begin{equation}
    \mathcal{L}^q_{W}=- \frac{g}{\sqrt{2}} \left[W^+_{\mu} \overline{u^j_L}  V_{\text{CKM}} \gamma^{\mu} d^i_L + W^-_{\mu} \overline{d^i_L}  V_{\text{CKM}}^\dag \gamma^{\mu} u^j_L     \right] ~,
    \label{eq:wcurrentssmq}
\end{equation}
where the matrix $V_{\text{CKM}}=U^{u \dagger}_L U^d_L$ is the unitary $3 \times 3$ quark mixing matrix. The existence of this matrix implies that the otherwise diagonal charged interactions become non-diagonal, giving origin to what is known as mixing. In the quark sector, the mixing matrix is known as $V_{\textrm{CKM}}$. 
As for the neutral interactions among quarks and the $Z$ and $\gamma$
\begin{equation}
\begin{split}
   \mathcal{L}^q_{A,Z}= &-\frac{g}{2 \cos{\theta_\textrm{W}}} \left[   Z_{\mu} ( \overline{u^i_L}
 \delta_{ij} \gamma^{\mu} u^j_L - \overline{d^i_L} \delta_{ij} \gamma^{\mu} d^j_L)  \right] \\
 &-\left[( -\frac{g \sin^2{\theta_\textrm{W}} } {\cos{\theta_\textrm{W}}} Z_{\mu} + e A_{\mu}) \left(\frac{2}{3}\right)(\overline{u^i_L} \delta_{ij} \gamma^{\mu} u^j_L +\overline{u^i_R} \delta_{ij} \gamma^{\mu} u^j_R) \right] \\
  &-\left[( \frac{g \sin^2{\theta_\textrm{W}} } {\cos{\theta_\textrm{W}}} Z_{\mu} - e A_{\mu}) \left(\frac{1}{3}\right)(\overline{d^i_L} \delta_{ij} \gamma^{\mu} d^j_L +\overline{d^i_R} \delta_{ij} \gamma^{\mu} d^j_R) \right] ~,
 \label{eq:neutralcurrentssmq}
\end{split}
\end{equation}
which shows that neutral currents remain diagonal, which means that there are no flavour Changing Neutral Currents (FCNC) at tree level in the SM - in any sector. Beyond tree level they are highly suppressed - cf. $K^0_L \rightarrow \mu^{+} \mu^{-}$ - this suppression stems from what is known as the GIM mechanism \cite{PhysRevD.2.1285}.\par
The quark mixing matrix $V_{\textrm{CKM}} \equiv V$ is complex but some of its phases have no physical meaning, due to the fact that one has the freedom to rephase mass eigenstates, $u'_i=e^{i \phi_i} u_i$ and $d'_j=e^{i \phi_j} d_j$ , transforming the entries of the mixing matrix:
\begin{equation}
    V'_{ij}=e^{i(\phi_j-\phi_i)} V_{ij} ~.
    \label{eq:reph}
\end{equation}
Thus, it's useful to look for rephasing invariants such as the the moduli of each entry, $|V_{ij}|$, or the quartets:
\begin{equation}
    Q_{ijkl}=V_{ij}V_{kl}V^{*}_{il}V^{*}_{kj} ~,
\end{equation}
because it can be proved that invariants of higher order can always be written as combinations of quartets and the moduli.\par
The fact that V is complex, in general, means that CP Violation can exist. Performing a CP transformation is performing a Charge transformation - transforming a particle in its anti-particle - followed by a Parity transformation - flipping the sign of the spatial coordinate, which means changing the chirality of a field, since axial vectors get an extra sign under parity transformations. One can check section 13.2 of \cite{Branco:1999fs} to see how SM fields transform under CP transformations. Performing a CP transformation to the SM Lagrangian after SSB, one obtains the condition for CP invariance of the SM:
\begin{equation}
    V^{*}_{ij}=e^{i(\xi_W + \xi_j-\xi_i)} V_{ij} ~,
    \label{eq:cpinv}
\end{equation}
where $\xi_\alpha$ are spurious CP phases that arise from the transformation. One can make eq. \eqref{eq:cpinv} always true for a single entry of $V$, however, if one considers all of the entries, one is forced to conclude that all quartets must be real and, thus, all other rephasing invariants. Hence, CP violation stems from the non-removable phases of $V_{\textrm{CKM}}$.
Thus, it is of uttermost importance to determine how many physical CP violating phases might exist in $V$. Due to the rephasing invariance, for $N_f$ generations ($V$ is an $N_f  \times N_f$ unitary matrix parametrized by $N_f^2$ parameters) one can remove $2N_f-1$ phases, making the total number of parameters $(N_f-1)^2$.  $N_f(N_f-1)/2$ of these parameters will be angles, while $(N_f-1)(N_f-2)/2$ will be phases. If one takes $N_f=3$, one obtains that there is only one phase. This is a CP violating phase, Kobayashi and  Maskawa \cite{makotokobayashitoshihidemaskawa1973} arrived to the above conclusion, proving that only for $N_f \geq 3$ one has CP Violating phases in the quark sector. For $N_f=3$ the imaginary part of all quartets are equal, up to a sign 
\begin{equation}
    J=|Im[Q]| ~.
    \label{eq:jarlskog}
\end{equation}
From the unitarity constraints on the entries of V one can define what is known as unitarity triangles - cf. section 13.5 and 13.6 of \cite{Branco:1999fs}. From these one gets a remarkable geometrical interpretation to $J$ - it is twice the area of any of the six possible unitarity triangles.\par
For an extensive treatment of CP Violation refer to ref. \cite{Branco:1999fs}. The standard parametrization \cite{pdg} of the \textrm{CKM} mixing matrix is 
\begin{eqnarray}
V=\left(
\begin{array}{ccc}
c_{12} c_{13} & s_{12} c_{13} & s_{13} e^{-i \delta}  \\
-s_{12} c_{23} - c_{12} s_{23} s_{13}   e^{i \delta}
& \quad c_{12} c_{23}  - s_{12} s_{23}  s_{13} e^{i \delta} \quad 
& s_{23} c_{13}  \\
s_{12} s_{23} - c_{12} c_{23} s_{13} e^{i \delta}
& -c_{12} s_{23} - s_{12} c_{23} s_{13} e^{i \delta}
& c_{23} c_{13} 
\end{array}\right) ~,
\label{eq:upmns}
\end{eqnarray}
where $c_{ij} = \cos \theta_{ij}$, $s_{ij} = \sin \theta_{ij}$, and $\delta$ is a Dirac-type CP violating phase. In this parameterisation,
\begin{equation}
    J = c_{12}c_{23}c^2_{13}s_{12}s_{23}s_{13}\sin{\delta} ~.
\end{equation}
\par
However, note that $\delta$ is not a rephasing invariant and is only meaningful under this parametrization. It is possible to parametrize the mixing matrix using a non optimal number of parameters with every single one of them being a rephasing invariant \cite{pdg}.
The current best-fit values for the quark mixing matrix are:\par
\begin{equation}
  V_{\textrm{CKM}} = \left( 
  \begin{array}{ccc}
0.97373 \pm 0.00031 & 0.2243 \pm 0.0008 & 0.00382 \pm 0.00020\\
0.221 \pm 0.004 & 0.975 \pm 0.006 & 0.0408 \pm 0.0014\\
0.0086 \pm 0.0002 & 0.0415 \pm 0.0009& 1.014 \pm 0.029
\end{array}\right) ~.
\end{equation} \par
The unitarity of $V_{\textrm{CKM}}$ is heavily constrained \cite{pdg}, contrarily to the leptonic sector where sizable deviations from unitarity of the mixing matrix are not ruled out. \par
This difference stems from the fact that there are many hadron decay processes which enable the direct measurement of individual $V_{\textrm{CKM}}$ entries. However, in the leptonic sector this is not possible. There is not enough precision to detect neutrino mass states (their mass scale is too small) in leptonic weak decays, so in each process one can only know the produced interaction state, which is a linear combination of the mass states, with certainty. Currently, one of the most reliable ways to get information regarding the mixing matrix are oscillation experiments and, even in those, one only has access to the first row and the last column of the $3 \times 3$ mixing matrix. Furthermore, what is measured in those cases are combinations of the entries of the mixing matrix, and not individual entries like in the quark case \cite{Parke:2015goa}.\par
In chapter \ref{chapter:framework_neutrinos}, in the discussion of the $n_R\nu$SM - the SM with right-handed neutrinos-, the emergence of a non-trivial mixing matrix in the leptonic sector will be studied and the steps are very similar to what has been done here for the quark sector. The emergence of a $(3+n_u) \times (3+n_d)$ unitary mixing matrix in the $n_u$SM$n_d$ - the SM with the addition of $n_u$ up vector-like quarks and $n_d$ down vector-like quarks - will also be discussed in chapter \ref{chapter:framework_quarks}.
There are only two pieces left to have the SM Lagrangian completely defined:
\begin{equation}
    \mathcal{L}_{\textrm{SM}}=    \mathcal{L}_{\textrm{Fermion}}+     \mathcal{L}_{\textrm{Kin}} +     \mathcal{L}_{\textrm{Higgs}} +     \mathcal{L}_{\textrm{Yukawa}} +     \mathcal{L}_{\textrm{GF}} +     \mathcal{L}_\textrm{{Ghost}} ~.
    \label{eq:LSM}
\end{equation}
The result in eq. \eqref{eq:higgsunit} is gauge dependent. A gauge independent formulation of the SM should use eq. \eqref{eq:higgsgeneral}. However, with this definition, $\mathcal{L}_{Higgs}$ will generate mixed quadratic terms in fields, with the three Goldstone bosons $G^0$ and $G^\pm$, that complicate the definition of the gauge boson propagators.
Using the gauge independent eq. \eqref{eq:higgsgeneral} in $\mathcal{L}_{Fermion}$ also introduces interactions between fermions and the unphysical Goldstone bosons. These should be taken into account when performing calculations in a general gauge. To  cancel the mixed quadratic terms that arise from $\mathcal{L}_{Higgs}$, one should include a new term in the SM Lagrangian:
\begin{equation}
        \mathcal{L}_{\textrm{GF}}=-\frac{1}{2 \xi_G} F_G^a F_{G a} -\frac{1}{2 \xi_A} F_A^2 -\frac{1}{2 \xi_Z} F_Z^2 -\frac{1}{\xi_W} F_- F_+ ,
\end{equation}
where $F_G^a=\partial^{\mu} G_{\mu}^a$, $F_A=\partial^{\mu} A_{\mu}$ , $F_Z=\partial^{\mu} Z_{\mu} +\xi_Z m_Z G^0$, $F_{\pm}=\partial^{\mu} W_{\mu}^{\pm} \pm i \xi_W m_W G^\pm$. $\mathcal{L}_{GF}$ are, actually, the gauge breaking terms in $\mathcal{L}_{SM}$.

The last piece is the Ghost Lagrangian. Faddeev-Popov Ghosts are unphysical particles that violate the Spin-Statistics Theorem. In theories like the SM they are bosonic (spin 0) with anti-comutation relations (fermionic). Every gauge boson correspond to a non-Abelian Gauge Group will have a Ghost - 8 for $SU(3)$ , 4 for $SU(2) \times U(1)$. These ghost fields are necessary to achieve a linear gauge fixing condition like in $\mathcal{L}_{GF}$, generating gauge field propagators with transverse and longitudinal component, thus, invertible. If the SM gauge group was abelian there would be no need for Faddeev-Popov Ghosts \cite{jorgec.romão2018}.
\begin{equation}
        \mathcal{L}_{\textrm{Ghost}}= \sum_{i=1}^4 \left [  \overline{c}_+ \frac{\partial (\delta F_+)}{\partial \alpha_i} +  \overline{c}_- \frac{\partial (\delta F_-)}{\partial \alpha_i}  +  \overline{c}_Z \frac{\partial (\delta F_Z)}{\partial \alpha_i}  + \overline{c}_A \frac{\partial (\delta F_A)}{\partial \alpha_i}  \right] c_i+ \sum_{a,b=1}^8  \overline{\omega}^a \frac{\partial (\delta F^a_G)}{\partial \beta_b}  \omega^b ,
\end{equation}
where $\alpha=1,2,3,4$ and $\beta=1,...,8$ are parameters of the correspondent gauge transformations.
For a more technical treatment of Faddeev-Popov Ghosts refer to Appendix A of \cite{jorgec.romaojoaop.silva2012}, sections 16.2 and 21.1 of \cite{peskinschroeder1995} and chapter 6 of \cite{jorgec.romão2018}. Since ghosts don't couple to matter fields, their contribution to one-loop corrections of physical processes involving fermions in external lines is zero, hence explaining ghosts further goes beyond the scope of this work. \par
Adding the final two terms in eq. \eqref{eq:LSM}, one can check that the Unitary gauge corresponds to $\xi_{Z,W} \xrightarrow{} \infty$ and $ \xi_{A,G} \xrightarrow{} 1$ and that the Goldstone bosons correspondent to massive gauge bosons - $G^0$ and $G^\pm$ -  acquire gauge dependent masses. The unitary gauge - used in eq. \eqref{eq:higgsunit} - only contains physical particles because having infinite masses means a decoupling from the theory. The formulation of the SM in a general gauge will be needed in later parts of this work.\par
To conclude, one final note regarding the SM. The SM  has 19 free parameters: 9 fermion masses, 3 quark mixing angles, 1 CP violating  phase in the quark sector, 1 Higgs mass, 1 Higgs vaccum expectation value ($v$), 1 QCD Vacuum angle ($\Theta_{\text{QCD}}$), 1 weak mixing angle\footnote{Instead of considering $\theta_\textrm{W}$ one may consider the $U(1)_Y$ gauge coupling ($g^\prime$) or the fine-structure constant ($\alpha$), as they are all proportional to each other.} ($\theta_\textrm{W}$), 1 $SU(2)_L$ gauge coupling ($g$) and 1 $SU(3)_c$ gauge coupling ($g$). These are the parameters that need to be fitted to experimental data. \par
Hence, the SM does not predict fermion masses nor gives an explanation to the number of generations of these. The only known thing regarding the generation number is that the number of generations of leptons and quarks must be equal in order to cancel gauge anomalies that appear at one-loop corrections - cf. Chapter 19 section 19.4 and chapter 20 section "Anomaly Cancellation" of \cite{peskinschroeder1995}.\par
Furthermore, the SM does not include gravity nor particles that can be dark matter candidates. Also, it does not explain dark energy. \par
In what regards CP Violation, it was not seen in strong sector (QCD) while there is nothing that inhibits it. In addition, the CP Violation detected in the electroweak sector is not large enough to explain the matter-antimatter asymmetry observed in the universe.\par
There are also some theoretical-inspired shortcomings that haunt the SM. One is the hierarchy problem, related to the fine-tuning that needs to happen in higher order calculations to achieve a Higgs mass near the electroweak scale ($v$) and the fact that the gauge couplings do not unify at high-energy, unlike what happens in some GUT models. \par
All these drawbacks lead to the belief that the SM is not the final theory of everything (TOE) but just a low-energy effective theory of it. The SM can be summarized in its Feynman Rules, which can be found in ref. \cite{jorgec.romaojoaop.silva2012}. In this work, these were used setting all $\eta_i=1$.

\cleardoublepage

\chapter{The Standard Model + $n_R$ right-handed neutrinos ($n_R\nu$SM)}
\label{chapter:framework_neutrinos}
This chapter includes the study of a model where $n_R$ neutrinos are added to the SM without imposing Lepton Number Conservation, allowing for Majorana mass terms. This extension of the SM will be treated exactly using the parameterisation first devised in \cite{Agostinho:2017wfs}, in the more succinct form of \cite{Pereira:2022pqu}.

In a WB where the charged leptons are diagonal and their masses are in the diagonal of the matrix $d_l$, the neutrino Dirac mass matrix is labeled $m$, and is a $3 \times n_R$ matrix and the symmetric Majorana mass matrix, $M$, is an $n_R \times n_R$ matrix, the leptonic mass Lagrangian is given by
\begin{equation}
\begin{aligned}
\mathcal{L}_{m} \,&=\,
-\left[\frac{1}{2}n_{L}^T C^{*} \cM^* n_{L}
+\overline{l_{L}} d_{l}l_{R}\right]+\text{h.c.}\,, 
\label{eq:Lmnu}
\end{aligned}
\end{equation}
where $n_L^{T} = (\nu_L^{T}\,\,\, \overline{\nu_R}\, C^T)$,
$n_L = (\nu_L\,, \,\,C\, \overline{\nu_R}^T)$ is a column vector and $C$ is the charge conjugation matrix. The $(3+n_R) \times (3+n_R)$ neutrino mass matrix is 
\begin{align}
\renewcommand{\arraystretch}{1}
\setlength{\extrarowheight}{6pt}
\mathcal{M} \,=\,
\left(\begin{array}{c;{2pt/2pt}c}
\,\textbf{0}\,\, & \,m\,
\\[1.5mm] \hdashline[2pt/2pt]
{\,\,\,m^T\,\,\,}\! &
{\,\,\,M\,\,\,}\!\\[1mm]
\end{array}\right) \,.
\label{eq:massmatrixnu}
\end{align}
And its diagonalisation equation, using the Takagi decomposition, a special case of singular value decomposition for symmetric matrices, is
\begin{equation}
  \cV^T \cM^* \cV = \cD ~,
  \label{eq:diagneutrino}
\end{equation}
where $\cD$ is a diagonal matrix and has the masses of the 3 light neutrinos and the $n_R$ heavy neutrinos in the diagonal and $\cV$ is a unitary  $(3+n_R) \times (3+n_R)$ matrix that can be written as
 \begin{equation}
\cV \,=\,
\setlength{\extrarowheight}{1.2pt}
    \left(\begin{array}{c}
     { }\\[-4mm]
      \qquad A \qquad 
      \\[2mm] \hdashline[2pt/2pt]
       { }\\[-4mm]
      \qquad B\qquad 
      \\[2mm]
    \end{array}\right) 
\,,
\label{eq:vneutrinos}
\end{equation}
which combined with eq. \eqref{eq:massmatrixnu} enables the rewriting of the diagonalisation equations in eq. \eqref{eq:diagneutrino} as
 \begin{equation}
 \begin{split}
&\bzero = A \,\cD\, A^T\,, \quad\\
&m = A \,\cD\, B^T\,, \quad \\
  &M   = B \,\cD\, B^T\,.
 \end{split}
 \label{eq:mneutrinos}
\end{equation}
These are exact equations for the mass matrices that will be very useful when $A$ and $B$ are properly defined.
The leptonic charged current Lagrangian in the physical basis takes the form of
 \begin{equation}
   \mathcal{L}_W^\nu=-\frac{g}{\sqrt{2}}
\overline{l}_L \,
V\,    \gamma^\mu
\begin{pmatrix}
n_L \\[2mm] N_L 
\end{pmatrix} 
    W_\mu
    \,+\,\text{h.c.} ~,
    \label{eq:lW}
 \end{equation}
 where $n$ ($N$) are light (heavy) neutrino mass states and
\begin{equation}
 V=A  ~,
 \label{eq:mixingmatrix_neutrino}
\end{equation}
plays the role of the physical mixing matrix. Note that $V$ is a $3 \times (3+n_R)$ matrix.
The neutral interaction - involving the $Z$ and the Higgs boson - Lagrangians can be written as
\begin{equation}
\begin{split}
\label{eq:lZ}
    \mathcal{L}_Z^\nu&=- \frac{g}{2\cos{\theta_W}}  Z_\mu [ \left( 
\begin{array}{cc}
\overline{n_L} &  \overline{N_L}
\end{array}%
\right) ~F~  \gamma^\mu\left( 
\begin{array}{cc}
n_L  \\ 
N_L%
\end{array}%
\right)] +\,\text{h.c.} ~,\\
\end{split}
\end{equation}
and
\begin{equation}
 \mathcal{L}_H^\nu \,=\,
 -\frac{h}{v} \bigg[\begin{pmatrix} \overline{n}_L & \overline{N}_L \end{pmatrix}\,
    F~\cD\, \begin{pmatrix} n_L^c \\[2mm] N_L^c \end{pmatrix} 
 \, \bigg] 
 \,+\,\text{h.c.}
    \,,
\end{equation}
where
   \begin{equation}
        F=A^\dag A ~.
        \label{eq:Fdef}
   \end{equation}
 Note that this matrix $F$ is only the identity matrix if $A$ is unitary on the left, which will be proved to be only possible when the heavy neutrinos are completely decoupled from the light ones, being completely sterile. Actually, $A$ is only unitary on the right. Hence, $F$ controls the Flavour Changing Neutral Couplings (FCNCs) of this model. Given the unitarity constraints, it is possible to write the matrix $\cV$ defined in eq. \eqref{eq:massmatrixnu} as

   \begin{align}
   \label{eq:calVfull}
\renewcommand{\arraystretch}{1.2}
\mathcal{V}\,=\,
\renewcommand{\arraystretch}{1}
\setlength{\extrarowheight}{6pt}
    \left(\begin{array}{c;{2pt/2pt}c}
      \,\,\,K\,\,\,\,\, & \,\,\,K\,X^\dag\,
      \\[1.5mm] \hdashline[2pt/2pt]
     {- \oK \,{X}\,\, }\! &
{\quad \,\,\oK\quad\,\,}\!\\[1mm]
    \end{array}\right) ~,
\\[-4mm] \nonumber
\end{align}
for a non-singular $3 \times 3$  general complex matrix $K$, a non-singular $n_R \times n_R$  general complex matrix ~$\oK$ and a $n_R \times 3$ matrix, $X$, that will be defined next. This parameterisation is completely general and exact, under the aforementioned conditions, and it was first introduced in ref. \cite{Agostinho:2017wfs}. From eq. \eqref{eq:vneutrinos} one can identify
\begin{equation}
    A = ( K~~KX^\dag ) ~,~ B = ( -\oK X~~\oK ) ~,
    \label{eq:ABdef}
\end{equation}
where the unitarity of $\mathcal{V}$ yields
\begin{align}
\mathcal{V} \mathcal{V}^\dag= \begin{pmatrix}
 \,A\, \\[2mm] 
B
\end{pmatrix} \begin{pmatrix}
\,{A}^\dagger & {B}^\dagger\,
\end{pmatrix} 
\,&=\,
\begin{pmatrix}
\,A\,{A}^\dagger & A\,{B}^\dagger\,\\[2mm] 
\,B\,{A}^\dagger & B\,{B}^\dagger\,
\end{pmatrix} 
\,=\,
\begin{pmatrix}
\,\id_{3}\, & \,0\, \\[2mm] 
\,0\, & \,\id_{n_R}\,
\end{pmatrix}\,, 
 \label{eq:unitAB02n}
\\[2mm]
\mathcal{V}^\dag \mathcal{V}= \begin{pmatrix}
\,{A}^\dagger & {B}^\dagger\,
\end{pmatrix} \begin{pmatrix}
\,A\, \\[2mm] 
B
\end{pmatrix}
\,&=\,
\,
{A}^\dagger\,A+ 
{B}^\dagger\,B
\,
\,=\,
\,
\id_{3+n_R}
\,.
 \label{eq:unitAB2n}
\end{align}

Note that by eq. \eqref{eq:unitAB2n}, $A$ is only unitary on the left if $\oK$ and $X$ are exactly zero. This corresponds to the case of having heavy fully sterile neutrinos, decoupled from the rest of the theory. Using eq. \eqref{eq:ABdef} and the unitarity relations in eqs. \eqref{eq:unitAB02n} and \eqref{eq:unitAB2n}, eq. \eqref{eq:mneutrinos} becomes

 \begin{equation}
 \begin{aligned}
& \bzero = d + X^\dag D X^* ~,\\
&m \,=\, K\, X^\dagger D \left( \oK ^{-1}\right)^* ~,\\
  &M \,=\, \oK \, (D+X\,d\,X^T)\,\oK ^T\,.
 \end{aligned}
 \label{eq:mneutrino}
\end{equation}
The last two equations are exact equations for the $3 \times n_R$ Dirac mass matrix, $m$, and the $n_R \times n_R$ Majorana mass matrix, $M$.
The equation involving the null matrix has the solution
\begin{align}
X\,=\pm\,i\,\sqrt{D^{-1}}\,O_{c}\,\sqrt{d}\,,
\label{eq:Xdef}
\end{align}
where $O_c$ is an orthogonal complex matrix and $d$ ($D$) is a diagonal matrix with the masses of the light (heavy) neutrinos in the diagonal. This unique solution defines the matrix $X$. From the unitarity relations of $\mathcal{V}$ one can also obtain the following definitions
 \begin{equation}
 \begin{split}
&K=U_{K}(\id_3+X^\dag X)^{-1/2} ~,\\
&\oK=U_{\oK}(\id_{n_R}+X X^\dag)^{-1/2}~,\\
&K_{PMNS}= K~,
 \end{split}
 \label{eq:kdef}
\end{equation}
where $U_K$ and $U_{\oK}$ are unitary matrices, since the unitarity relations only define $K$
 and $\oK$ up to a unitary matrix on the left. $K$ will play the role of the PMNS mixing matrix and is only unitary when $X \xrightarrow[]{} \bzero$.\\
 
Furthermore, using this parameterisation is also possible to rewrite the matrix that controls FCNCs, defined in eq. \eqref{eq:Fdef}, as
 \begin{equation}
F= \begin{pmatrix}
(\id_3+X^\dag X)^{-1} &(\id_3+X^\dag X)^{-1}X^\dag\\ 
X (\id_3+X^\dag X)^{-1} &X (\id_3+X^\dag X)^{-1} X^\dag\\ 
\end{pmatrix} ~.
\label{eq:Fdef2}
\end{equation}
The complete form of the  $(3+n_R) \times (3+n_R)$ unitary matrix $\cV$ is written as
 \begin{equation}
  \mathcal{V} =       \begin{pmatrix}
U_{K}(\id_3+X^\dag X)^{-1/2} &U_{K}(\id_3+X^\dag X)^{-1/2}X^\dag\\ 
- U_{\oK}(\id_{n_R}+X X^\dag)^{-1/2}X &U_{\oK}(\id_{n_R}+X X^\dag)^{-1/2}\\
\end{pmatrix} ~.
\label{eq:calV}
\end{equation}  
With all the relevant matrices defined in a general and exact way, it is now possible to proceed to the study of the model in interesting regions of the parameter space.
Note that parameterisations with a similar structure, in the leptonic sector, existed in the literature prior to this work \cite{Korner:1992zk,Grimus:2000vj}, but are either approximations or a special case of this one.\par
In the rest of this chapter and remaining of this thesis, in the context of neutrinos and whenever there is no risk of confusion, $M$ should refer to the scale of the masses of the heavy neutrinos, while $m$ refers to the scale of the masses of the light neutrinos. Where there is risk of confusion, it will be specifically stated if one is referring to the mass matrix or to the mass of a specific species of neutrino.


 \section{The Quasi-Decoupled Parameter Space: $M \sim$ GeV/TeV} \label{sec:quasi}
 In ref. \cite{Agostinho:2017wfs} one tried to argue that a region of the parameter space where all the heavy neutrinos have a GeV-scale mass and the deviations from unitarity (DU) of the $3 \times 3$ leptonic mixing matrix match the most stringent experimental bounds at the time \cite{Fernandez-Martinez:2016lgt}, was allowed and still provided a reasonable explanation to the smallness of light neutrinos masses, given that the Yukawa couplings were $\mathcal{O}(1)$. \par

In the literature \cite{pdg} the PMNS matrix is usually parameterised as 
\begin{eqnarray}
U_{\textrm{PMNS}}=\left(
\begin{array}{ccc}
c_{12} c_{13} & s_{12} c_{13} & s_{13} e^{-i \delta}  \\
-s_{12} c_{23} - c_{12} s_{23} s_{13}   e^{i \delta}
& \quad c_{12} c_{23}  - s_{12} s_{23}  s_{13} e^{i \delta} \quad 
& s_{23} c_{13}  \\
s_{12} s_{23} - c_{12} c_{23} s_{13} e^{i \delta}
& -c_{12} s_{23} - s_{12} c_{23} s_{13} e^{i \delta}
& c_{23} c_{13} 
\end{array}\right) \cdot  P ~,
\label{eq:Vstd}
\end{eqnarray}
with $P$ given by
\begin{equation}
P=\mathrm{diag} \ (1,e^{i\alpha_{21}}, e^{i\alpha_{31}}) ~,
\end{equation}
where $c_{ij} = \cos \theta_{ij}$, $s_{ij} = \sin \theta_{ij}$ 
and $\delta$ is a Dirac-type 
CP violating phase, while $\alpha_{21}$, $\alpha_{31}$ denote Majorana 
phases. The PMNS matrix has this form in the framework of a low-energy effective theory, where the deviations from unitarity of the $3 \times 3$ mixing matrix are negligible and the heavy neutrinos are so heavy that it is possible to integrate them out. In this effective theory, only mass terms involving $\nu_L$ exist, preventing the rephasing of the neutrino field, leading to the presence of two additional Majorana phases.  \par

\subsection{Status of Neutrino Physics and DU for $M>M_W$}
When it is necessary to go beyond this level of approximation, the consensus in the literature \cite{Gluza:2002vs,Antusch:2006vwa,Fernandez-Martinez:2007iaa,Antusch:2014woa,Fernandez-Martinez:2016lgt,Blennow:2016jkn} dictates that one should write the $3 \times 3$ leptonic mixing matrix as the product of a
hermitian matrix by a unitary matrix:
\begin{equation}
K = (\id_3 - \eta ) U_K ~,
\label{eq:khu}
\end{equation}
where $U_K$ is a unitary matrix and $\eta $ is a hermitian matrix with small entries, that are proportional to the deviations from unitarity. 
Note that under the previous definition of the $3 \times 3$ leptonic mixing matrix in eq. \eqref{eq:kdef}, the matrices $U_K$ and $\eta$ can be written as
\begin{equation}
    \eta  = \id_3 - U_{K}(\id_3+X^\dag X)^{-1/2} U_{K}^\dag \approx \frac{1}{2} U_K \left(X^\dag X\right) U_K^\dag ~.
\end{equation}

In ref. \cite{Agostinho:2017wfs}, the matrix $U_K$ is then fixed making use of the, at the time of publication, most updated best-fit values obtained from 
a global analysis based on the assumption of unitarity, present in Table \ref{tab:bestfit1}. These were the values used in the published results.

\begin{table}[h]

\centering
\caption[Neutrino oscillation parameter summary from 2018. 
For $\Delta m^2_{31}$, 
$\sin^2 \protect\theta_{23}$ , $\sin^2 \protect\theta_{13}$, and $\protect
\delta$ the upper (lower) row corresponds to normal (inverted) neutrino mass
hierarchy.]{Neutrino oscillation parameter summary from \cite{deSalas:2017kay}. 
For $\Delta m^2_{31}$, 
$\sin^2 \protect\theta_{23}$ , $\sin^2 \protect\theta_{13}$, and $\protect
\delta$ the upper (lower) row corresponds to normal (inverted) neutrino mass
hierarchy.}
\label{reps}
\begin{tabular}{|c|c|c|}

\hline
Parameter & Best fit & $1 \sigma $ range \\ \hline
$\Delta m^2_{21}$ $[10^{-5} eV^2 ] $ & 7.56 & 7.37 -- 7.75\\ 
\hline
$|\Delta m^2_{31}|$ $[10^{-3} eV^2 ]$ (NO) & 2.55 & 2.41 -- 2.59 \\ 
\hline
$|\Delta m^2_{31}|$ $[10^{-3} eV^2 ]$ (IO)  & 2.49 & 2.45 -- 2.53 \\ 
\hline
$\sin^2 \theta_{12}$ & 0.321 & 0.305 -- 0.339 \\ 
\hline
$\sin^2 \theta_{23}$ (NO) & 0.430  & 0.412 -- 0.450 \\ 
\hline
$\sin^2 \theta_{23}$ (IO) & 0.596 & 0.576 -- 0.614 \\ 
\hline
$\sin^2 \theta_{13}$ (NO)& 0.02155 & 0.02080 --0.02245 \\ 
\hline
$\sin^2 \theta_{13}$ (IO) & 0.02140 & 0.02055 -- 0.02222 \\ 
\hline
$\delta$ (NO) & 1.40 $\pi $ & 1.20 --1.71 $\pi$ \\ 
\hline
$\delta$ (IO) & 1.44 $\pi$ & 1.70 --1.21 $\pi$ \\ \hline
\end{tabular}
\label{tab:bestfit1}
\end{table}
For comparison, at the time of writing of this thesis, the most updated best-fit values obtained from 
a global analysis based on the assumption of unitarity are present in Table \ref{tab:bestfit2}. These values were not used in any of the published results.

\begin{table}[h]
\centering
\caption[Neutrino oscillation parameter summary from 2020.
For $\Delta m^2_{31}$, 
$\sin^2 \protect\theta_{23}$ , $\sin^2 \protect\theta_{13}$, and $\protect
\delta$ the upper (lower) row corresponds to normal (inverted) neutrino mass
hierarchy.]{Neutrino oscillation parameter summary from \cite{deSalas:2020pgw}. 
For $\Delta m^2_{31}$, 
$\sin^2 \protect\theta_{23}$ , $\sin^2 \protect\theta_{13}$, and $\protect
\delta$ the upper (lower) row corresponds to normal (inverted) neutrino mass
hierarchy.}
\label{reps}
\begin{tabular}{|c|c|c|}

\hline
Parameter & Best fit & $1 \sigma $ range \\ \hline
$\Delta m^2_{21}$ $[10^{-5} eV^2 ] $ & 7.50 & 7.30 -- 7.72\\ 
\hline
$|\Delta m^2_{31}|$ $[10^{-3} eV^2 ]$ (NO) & 2.55 & 2.52 -- 2.57 \\ 
\hline
$|\Delta m^2_{31}|$ $[10^{-3} eV^2 ]$ (IO)  & 2.45 & 2.42 -- 2.47 \\ 
\hline
$\sin^2 \theta_{12}$ & 0.318 & 0.302 -- 0.340 \\ 
\hline
$\sin^2 \theta_{23}$ (NO) & 0.574  & 0.560 -- 0.588 \\ 
\hline
$\sin^2 \theta_{23}$ (IO) & 0.578 & 0.568 -- 0.561 \\ 
\hline
$\sin^2 \theta_{13}$ (NO)& 0.02200 & 0.02138 --0.02269 \\ 
\hline
$\sin^2 \theta_{13}$ (IO) & 0.02225 & 0.02155 -- 0.02289 \\ 
\hline
$\delta$ (NO) & 1.08 $\pi $ & 0.96 --1.21 $\pi$ \\ 
\hline
$\delta$ (IO) & 1.58 $\pi$ & 1.42 --1.73 $\pi$ \\ \hline
\end{tabular}
\label{tab:bestfit2}
\end{table}

According to \cite{Fernandez-Martinez:2016lgt}, it is advantageous, from a phenomenological perspective, to represent $K$ with a unitary matrix on the right, as it is not possible to determine which physical light neutrino is produced in experiments. Therefore, one must sum over all neutrino indices. As a result, most observables are dependent on $K K^\dagger$, which depends on the following combination
\begin{equation}
\left( K K^\dagger \right)_{\alpha \beta} = \delta_{\alpha \beta} - 
2 \eta_{\alpha \beta} + \mathcal{O}(\eta^2_{\alpha \beta}) ~.
\label{eq:etadef}
\end{equation}
A fit of 28 observables, including the W boson mass and the effective mixing weak angle $\theta_W$, was used to derive global constraints on the matrix $\eta$ in ref.~\cite{Fernandez-Martinez:2016lgt}. The fit considered several ratios of $Z$ fermionic decays, the invisible width of the $Z$, ratios of weak decays that constrain EW universality, weak decays that constrain CKM unitarity, and some radiative lepton flavour violating (LFV) processes.
 \begin{eqnarray}
\left| 2 \eta_{\alpha \beta} \right| \leq \left(
\begin{array}{ccc}
2.5 \times 10^{-3} & 2.4 \times 10^{-5}  & 2.7 \times 10^{-3}    \\
2.4 \times 10^{-5} & 4.0 \times 10^{-4}  & 1.2 \times 10^{-3}   \\
2.7 \times 10^{-3}  & 1.2 \times 10^{-3}  & 5.6 \times 10^{-3}  
\end{array} \right) ~.
\label{eq:hen}
\end{eqnarray}
It is crucial to emphasize that this result only holds when all heavy neutrinos possess masses above the electroweak (EW) scale \cite{Blennow:2016jkn}. In other words, if $M$ represents the mass of the heavy neutrinos, it is required that all of them satisfy $M > M_W$, where $M_W$ denotes the mass of the $W$ boson. This is the case in ref. \cite{Agostinho:2017wfs}.
\subsection{Benchmarks and Perturbativity Remarks}
Before examining the benchmarks that demonstrate the feasibility of having GeV heavy neutrinos with substantial deviations from the unitarity of the $3 \times 3$ leptonic mixing matrix while still having $\mathcal{O}(1)$ Yukawa couplings, it is crucial to understand the theoretical basis behind this possibility. By combining equations \eqref{eq:mneutrino}, \eqref{eq:Xdef}, and \eqref{eq:kdef}, one can reformulate the $3 \times 3$ Dirac mass matrix for neutrinos as

\begin{equation}
    \begin{aligned}
  m \,=\      V\,\sqrt{\left(\id+X^\dagger X\right)^{-1}}\,X^\dagger D\, \sqrt{\id+X^* X^T} U_{\oK}^T\,, \\
    \end{aligned}
    \label{eq:mdirac}
\end{equation}
which, using eq. \eqref{eq:Xdef} becomes
\begin{equation}
\begin{aligned}
m \,=\, - i \,
U_K\,\sqrt{\left(\id_3+X^\dagger X\right)^{-1}}
\, \sqrt{d} \, O_{c}^\dagger \sqrt{D}\,
\sqrt{\id_{n_R}+X^* X^T}~U_{\oK}^T\,.
\end{aligned}
\label{eq:mdiracWB}
\end{equation}

\begin{figure}[h!]

\includegraphics[width=1.\linewidth]{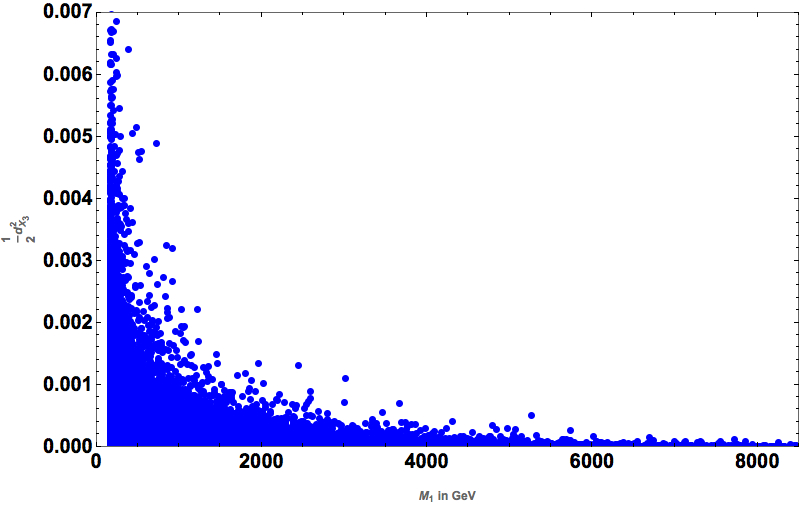}
  
\caption{Maximum deviations from unitarity as a function of $M_1$, generated under 
the condition that Tr$( mm^\dagger)\leq m_t^2$ and $|\eta_{12}|\leq 10^{-4}$.  }
\label{fig:sub1}
 \end{figure}

Taking $U_{\oK}=\id_{n_R}$, since this matrix is unphysical at low energy, the equation for $m$ in first order approximation in $X$ (or $Oc$) takes the form
\begin{equation}
    \begin{aligned}
m \,\approx\, - i \,
U_K\,\, \sqrt{d} \, O_{c}^\dagger \sqrt{D}.
\end{aligned}
\end{equation}
It is evident that for $\mathcal{O}(1)$ Yukawa couplings and fixed light neutrino masses, there is a seesaw relationship between the magnitude of deviations from unitarity, the singular values of $O_c$, and the scale of heavy neutrino masses in $D$. This relationship becomes clearer when one defines the singular value decomposition of $X$
\begin{equation}
    d_X = W^\dag X U ~,
\end{equation}
and calculates the quantity
\begin{equation}
Tr\left[ mm^{\dagger}\right] =
Tr\left[ \left( \sqrt{\left( \id_3,+d_{X}^{2}\right) }\
\right) ^{-1}d_{X}\ W^{\dagger }\ D\ W^{\ast }\ \left( \id_3%
\,+d_{X}^{2}\right) \ W^{T}\ D\ W\ d_{X}\ \left( \sqrt{\left( {1\>\!\!\!
\mathrm{I}}\,+d_{X}^{2}\right) }\ \right) ^{-1}\right] ~,
\label{mmt}
\end{equation}
which to a very good approximation yields
\begin{equation}
Tr\left[ mm^{\dagger}\right] =
Tr\left[ d_{X}\ W^{\dagger }D^{2}W\ d_{X}\right] ~.
\end{equation}
This can be rewritten as
\begin{equation}
\begin{array}{l}
Tr\left[ mm^{\dagger}\right] =
d_{X_{1}}^{2}\left( M_{1}^{2}\left\vert W_{11}\right\vert
^{2}+M_{2}^{2}\left\vert W_{21}\right\vert ^{2}+M_{3}^{2}\left\vert
W_{31}\right\vert ^{2}\right) + \\ 
d_{X_{2}}^{2}\left( M_{1}^{2}\left\vert W_{12}\right\vert
^{2}+M_{2}^{2}\left\vert W_{22}\right\vert ^{2}+M_{3}^{2}\left\vert
W_{32}\right\vert ^{2}\right) + \\ 
d_{X_{3}}^{2}\left( M_{1}^{2}\left\vert W_{13}\right\vert
^{2}+M_{2}^{2}\left\vert W_{23}\right\vert ^{2}+M_{3}^{2}\left\vert
W_{33}\right\vert ^{2}\right) ~.
\end{array}
\label{mmt1}
\end{equation}
It can be demonstrated that, through the use of the properties of orthogonal complex matrices, only the largest of the $d_{X_i}$, represented by $d_{X_3}$, can have a substantial value (for example, $d_{X_3} \approx 10^{-2}$) while the other two are negligible. As a result, in good approximation,
\begin{equation}
Tr\left[ mm^{\dagger}\right] =
d_{X_{3}}^{2}\left( M_{1}^{2}\left\vert W_{13}\right\vert
^{2}+M_{2}^{2}\left\vert W_{23}\right\vert ^{2}+M_{3}^{2}\left\vert
W_{33}\right\vert ^{2}\right) ~, \label{mmt2}
\end{equation}
and using the unitarity of $W$
\begin{equation}
Tr\left[ mm^{\dagger}\right] =
d_{X_{3}}^{2}M_{1}^{2}\left( 1+\left( \frac{M_{2}^{2}}{M_{1}^{2}}-1\right)
\left\vert W_{23}\right\vert ^{2}+\left( \frac{M_{3}^{2}}{M_{1}^{2}}
-1\right) \left\vert W_{33}\right\vert ^{2}\right) ~,
\label{mmt3}
\end{equation}
which, with the choice $M_{3}\geq M_{2}\geq M_{1}$, leads to
\begin{equation}
d_{X_{3}}^{2}M_{1}^{2}\leq Tr\left[ mm^{\dagger }\right] =\sum_{i,j} \left\vert
m_{ij}\right\vert ^{2}  \sim \left(\frac{v}{\sqrt{2}}\right)^2 \approx m_t^2 ~, \label{dim}
\end{equation}
for $\mathcal{O}(1)$ Yukawa couplings.
This seesaw between the deviations from unitarity and the scale of the heavy neutrino mass is easy to see if one plots $\frac{1}{2}d_{X_3}^2$ against $M_1$, as in fig. \ref{fig:sub1}.

It should now be clear that by decreasing the mass scale of the heavy neutrinos while keeping the Yukawa couplings $\mathcal{O}(1)$, the deviations from $3 \times 3$ unitarity can be increased.\par

The benchmarks for NO (IO) of the light neutrino masses, where the 3 heavy neutrinos have masses in the range of $500-1500$~GeV and the $\eta$ matrix is consistent with the most stringent upper bounds from \cite{Fernandez-Martinez:2016lgt} as given in equation \eqref{eq:hen} for $\textrm{Tr}(mm^\dag) \sim (0.03 - 0.5)~m_t^2$, can be found in Table \ref{tab:bench1} (\ref{tab:bench2}).

\begin{table}[h!]
\caption{Examples of deviations from unitarity, with light neutrino masses of $m_1=0.005050$ eV, $m_2=0.01005$ eV, and $m_3=0.05075$ eV for the case of NO, are expressed by the Hermitian matrix $\eta$. Three different hierarchies for the heavy Majorana neutrino masses $M_i$ are considered. For each hierarchy, two examples are given where the choice of the matrix $X$ is varied. The third column shows the Dirac-type neutrino mass matrix expressed in GeV units, the fifth column displays the Majorana-type neutrino mass matrix expressed in units of the top quark mass, and the fourth column displays the trace of $mm^\dagger$ in units of the squared top quark mass.}
\begin{center}
\label{tab:bench1}
\setlength{\tabcolsep}{0.5pc} 
\resizebox{\textwidth}{!}{\begin{tabular}{|c|c|c|c|c|}
\hline
\textbf{Heavy Neutrino Masses }& \textbf{$|\eta|$} &\textbf{$|m| (GeV)$ }& \textbf{$\textrm{Tr}(mm^{\dagger}) (m_t^2)$}& \textbf{$|M| (m_t)$ } \\
\hline
$\begin{pmatrix}
	M_1= 3 \ m_t\\
	M_2= (3 + 1 \times 10^{-7}) \ m_t\\\
	M_3=9\ m_t\
\end{pmatrix}$ & 
$\begin{pmatrix}
1.12\times 10^{-3} & 8.92\times 10^{-6} & 1.36\times 10^{-3} \\
. & 7.11\times 10^{-8} & 1.08\times 10^{-5} \\
 . & . & 1.65\times 10^{-3}
\end{pmatrix}$

 & 
$\begin{pmatrix}
24.6 & 6.88\times 10^{-11} & 2.71\times 10^{-5} \\
 0.196 & 3.43 \times 10^{-10} & 1.69\times 10^{-4} \\
 29.9 & 3.03\times 10^{-10} & 1.49 \times 10^{-4} 
\end{pmatrix}$ \
& 0.0497
&

$\begin{pmatrix}
9.99\times 10^{-8} & 2.99 & 7.68\times 10^{-6} \\
 2.99 & 1.99 \times 10^{-11} & 7.66\times 10^{-6} \\
 7.68 \times 10^{-6}  & 7.66\times 10^{-6} & 8.99
\end{pmatrix}$ \
\\
\hline

$\begin{pmatrix}
	M_1= 3 \ m_t\\
	M_2= (3 + 1\times 10^{-7})\ m_t\\\
	M_3=9\ m_t\
\end{pmatrix}$ & 
$\begin{pmatrix}
5.43\times 10^{-7} & 9.57\times 10^{-6} & 3.18\times 10^{-5} \\
. & 1.69\times 10^{-4} & 5.61\times 10^{-4} \\
 . & . & 1.86\times 10^{-3}
\end{pmatrix}$

 &

$\begin{pmatrix}
0.542 & 1.21\times 10^{-10} & 4.92\times 10^{-5} \\
 9.57 & 3.69 \times 10^{-10} & 1.85\times 10^{-4} \\
 31.78 & 3.41\times 10^{-10} & 1.32 \times 10^{-4} 
\end{pmatrix}$ \
&  0.0366
&

$\begin{pmatrix}
1.00\times 10^{-7} & 2.99 & 8.72\times 10^{-6} \\
 2.99 & 2.40 \times 10^{-11} & 8.70\times 10^{-6} \\
 8.72 \times 10^{-6}  & 8.70\times 10^{-6} & 8.99
\end{pmatrix}$ \
\\
\hline

$\begin{pmatrix}
	M_1= 3 \ m_t\\
	M_2= (3 + 1\times 10^{-7})\ m_t\\\
	M_3= 100\ m_t\
\end{pmatrix}$ &

$\begin{pmatrix}
1.12\times 10^{-3} & 8.92\times 10^{-6} & 1.36\times 10^{-3} \\
. & 7.11\times 10^{-8} & 1.08\times 10^{-5} \\
 . & . & 1.65\times 10^{-3}
\end{pmatrix}$
 &

$\begin{pmatrix}
24.6 & 6.88\times 10^{-11} & 5.40\times 10^{-5} \\
 0.196 & 3.43 \times 10^{-10} & 5.61\times 10^{-4} \\
 29.9 & 3.03\times 10^{-10} & 6.11 \times 10^{-4} 
\end{pmatrix}$ \
&  0.0497
&

$\begin{pmatrix}
1.00\times 10^{-7} & 2.99 & 3.73\times 10^{-5} \\
 2.99 & 1.99 \times 10^{-11} & 3.72\times 10^{-5} \\
 3.73 \times 10^{-5}  & 3.72\times 10^{-5} & 99.9
\end{pmatrix}$ \
\\
\hline

$\begin{pmatrix}
	M_1= 3 \ m_t\\
	M_2= (3+1 \times 10^{-7})\ m_t\\\
	M_3=100\ m_t\
\end{pmatrix}$ &

$\begin{pmatrix}
5.43\times 10^{-7} & 9.57\times 10^{-6} & 3.18\times 10^{-5} \\
. & 1.69\times 10^{-4} & 5.61\times 10^{-4} \\
 . & . & 1.86\times 10^{-3}
\end{pmatrix}$
 &

$\begin{pmatrix}
0.542 & 1.21\times 10^{-10} & 1.64\times 10^{-4} \\
 9.57 & 3.69 \times 10^{-10} & 6.02\times 10^{-4} \\
 31.8 & 3.41\times 10^{-10} & 5.72 \times 10^{-4} 
\end{pmatrix}$ \
& 0.0366
&

$\begin{pmatrix}

1.00\times 10^{-7} & 2.99 & 4.23\times 10^{-5} \\
 2.99 & 2.40 \times 10^{-11} & 4.22\times 10^{-5} \\
 4.23 \times 10^{-5}  & 4.22\times 10^{-5} & 99.9
\end{pmatrix}$ \
\\
\hline

$\begin{pmatrix}
	M_1= 3 \ m_t\\
	M_2= 9\ m_t\\\
	M_3=(9+1\times 10^{-7})\ m_t\
\end{pmatrix}$ &

$\begin{pmatrix}
1.12\times 10^{-3} & 8.92\times 10^{-6} & 1.36\times 10^{-3} \\
. & 7.11\times 10^{-8} & 1.08\times 10^{-5} \\
 . & . & 1.65\times 10^{-3}
\end{pmatrix}$
 &

$\begin{pmatrix}
73.8 & 6.89\times 10^{-11} & 2.73\times 10^{-5} \\
 0.589 & 3.43 \times 10^{-10} & 9.70\times 10^{-5} \\
 89.6 & 3.03\times 10^{-10} & 1.33\times 10^{-4} 
\end{pmatrix}$ \
& 0.447
&

$\begin{pmatrix}
1.00\times 10^{-7} & 8.97 & 2.01\times 10^{-5} \\
 8.97 & 1.85\times 10^{-11} & 8.85\times 10^{-6} \\
 2.01 \times 10^{-5}  & 8.85\times 10^{-6} & 2.99
\end{pmatrix}$ \
\\
\hline

$\begin{pmatrix}
	M_1= 3 \ m_t\\
	M_2= 9\ m_t\\\
	M_3=(9+ 1\times 10^{-7})\ m_t\
\end{pmatrix}$ &

$\begin{pmatrix}
5.23\times 10^{-7} & 9.22\times 10^{-6} & 3.06\times 10^{-5} \\
. & 1.62\times 10^{-4} & 5.41\times 10^{-4} \\
 . & . & 1.80\times 10^{-3}
\end{pmatrix}$
 &

$\begin{pmatrix}
1.60 & 1.23\times 10^{-10} & 2.85\times 10^{-5} \\
 28.2 & 3.75 \times 10^{-10} & 1.02\times 10^{-4} \\
 93.6 & 3.47\times 10^{-10} & 1.29 \times 10^{-4} 
\end{pmatrix}$ \
& 0.317
&

$\begin{pmatrix}
1.00\times 10^{-7} & 8.98 & 2.40\times 10^{-5} \\
 8.98 & 2.49 \times 10^{-11} & 1.04\times 10^{-5} \\
 2.40 \times 10^{-5}  & 1.04\times 10^{-5} & 2.99
\end{pmatrix}$ \
\\
\hline

\end{tabular}}
\end{center}
\end{table}

\begin{table}[h!]
\caption{Examples of deviations from unitarity for the case of IO are provided for light neutrino masses of $m_1=0.05064 ~ eV$\ , \ $m_2=0.05138 ~ eV$\ , \ $m_3=0.00864 ~ eV$. These deviations are quantified by the Hermitian matrix $\eta$. Three different hierarchies are considered for the heavy Majorana neutrino masses $M_i$, and for each hierarchy, two examples are given by varying the choice of the matrix $X$. The third column shows the Dirac-type neutrino mass matrix in GeV units, and the fifth column displays the Majorana-type neutrino mass matrix in units of the top quark mass. The fourth column presents the trace of the product $mm^\dagger$ in units of the squared top quark mass.}
\begin{center}
\label{tab:bench2}
\setlength{\tabcolsep}{0.5pc} 
\resizebox{\textwidth}{!}{\begin{tabular}{|c|c|c|c|c|}
\hline
\textbf{Heavy Neutrino Masses }& \textbf{$|\eta|$} &\textbf{$|m| (GeV)$ }& \textbf{$\textrm{Tr}(mm^{\dagger}) (m_t^2)$}& \textbf{$|M| (m_t)$ } \\
\hline
$\begin{pmatrix}
	M_1= 3 \ m_t\\
	M_2= (3 + 1 \times 10^{-7}) \ m_t\\\
	M_3=9\ m_t\
\end{pmatrix}$ &

$\begin{pmatrix}
1.18\times 10^{-3} & 1.00\times 10^{-5} & 1.20\times 10^{-3} \\
. & 8.50\times 10^{-8} & 1.02\times 10^{-5} \\
 . & . & 1.22\times 10^{-3}
\end{pmatrix}$
&

$\begin{pmatrix}
25.2 & 5.41\times 10^{-10} & 1.67\times 10^{-4} \\
 0.215 & 2.51 \times 10^{-10} & 1.59\times 10^{-4} \\
 25.8 & 3.22\times 10^{-10} & 1.64 \times 10^{-4} 
\end{pmatrix}$ \
& 0.0431
&

$\begin{pmatrix}
1.00\times 10^{-7} & 2.99 & 1.24\times 10^{-7} \\
 2.99 & 5.01 \times 10^{-11} & 1.23\times 10^{-7} \\
 1.24 \times 10^{-7}  & 1.23\times 10^{-7} & 8.99
\end{pmatrix}$ \
\\
\hline

$\begin{pmatrix}
	M_1= 3 \ m_t\\
	M_2= (3 + 1\times 10^{-7})\ m_t\\\
	M_3=9\ m_t\
\end{pmatrix}$ &

$\begin{pmatrix}
1.33\times 10^{-7} & 5.36\times 10^{-6} & 1.09\times 10^{-5} \\
. & 2.16\times 10^{-4} & 4.38\times 10^{-4} \\
 . & . & 8.85\times 10^{-4}
\end{pmatrix}$

 &

$\begin{pmatrix}
0.269 & 9.83\times 10^{-10} & 1.63\times 10^{-4} \\
 10.8 & 3.19 \times 10^{-10} & 1.17\times 10^{-4} \\
 21.91 & 2.47\times 10^{-10} & 9.96 \times 10^{-6} 
\end{pmatrix}$ \
&  0.0198
&

$\begin{pmatrix}
1.00\times 10^{-7} & 2.99 & 7.49\times 10^{-6} \\
 2.99 & 4.04 \times 10^{-11} & 7.48\times 10^{-6} \\
 7.49 \times 10^{-6}  & 7.48\times 10^{-6} & 8.99
\end{pmatrix}$ \
\\
\hline

$\begin{pmatrix}
	M_1= 3 \ m_t\\
	M_2= (3 + 1\times 10^{-7})\ m_t\\\
	M_3= 100\ m_t\
\end{pmatrix}$ &

$\begin{pmatrix}
1.16\times 10^{-3} & 9.90\times 10^{-6} & 1.19\times 10^{-3} \\
. & 8.43\times 10^{-8} & 1.01\times 10^{-5} \\
 . & . & 1.21\times 10^{-3}
\end{pmatrix}$
 &

$\begin{pmatrix}
25.1 & 5.43\times 10^{-10} & 5.56\times 10^{-4} \\
 0.214 & 2.52 \times 10^{-10} & 5.31\times 10^{-4} \\
 25.6 & 3.24\times 10^{-10} & 5.47 \times 10^{-4} 
\end{pmatrix}$ \
&  0.0428
&

$\begin{pmatrix}
1.00\times 10^{-7} & 2.99 & 3.20\times 10^{-7} \\
 2.99 & 5.05 \times 10^{-11} & 3.19\times 10^{-7} \\
 3.20 \times 10^{-7}  & 3.19\times 10^{-7} & 100
\end{pmatrix}$ \
\\
\hline

$\begin{pmatrix}
	M_1= 3 \ m_t\\
	M_2= (3+1 \times 10^{-7})\ m_t\\\
	M_3=100\ m_t\
\end{pmatrix}$ &

$\begin{pmatrix}
1.33\times 10^{-7} & 5.36\times 10^{-6} & 1.09\times 10^{-5} \\
. & 2.16\times 10^{-4} & 4.38\times 10^{-4} \\
 . & . & 8.85\times 10^{-4}
\end{pmatrix}$
 &

$\begin{pmatrix}
0.269 & 9.83\times 10^{-10} & 5.44\times 10^{-4} \\
 10.8 & 3.19 \times 10^{-10} & 4.31\times 10^{-4} \\
 21.9 & 2.47\times 10^{-10} & 1.13 \times 10^{-4} 
\end{pmatrix}$ \
& 0.0198
&

$\begin{pmatrix}
1.00\times 10^{-7} & 3.00 & 3.63\times 10^{-5} \\
 3.00 & 4.04 \times 10^{-11} & 3.63\times 10^{-5} \\
 3.63 \times 10^{-5}  & 3.63\times 10^{-5} & 99.9
\end{pmatrix}$ \
\\
\hline

$\begin{pmatrix}
	M_1= 3 \ m_t\\
	M_2= 9\ m_t\\\
	M_3=(9+1\times 10^{-7})\ m_t\
\end{pmatrix}$ &

$\begin{pmatrix}
1.11\times 10^{-3} & 9.46\times 10^{-6} & 1.14\times 10^{-3} \\
. & 8.05\times 10^{-8} & 9.65\times 10^{-6} \\
 . & . & 1.16\times 10^{-3}
\end{pmatrix}$
 &

$\begin{pmatrix}
73.7 & 5.56\times 10^{-10} & 9.63\times 10^{-5} \\
 0.626 & 2.58 \times 10^{-10} & 9.20\times 10^{-5} \\
 75.2 & 3.31\times 10^{-10} & 9.48\times 10^{-5} 
\end{pmatrix}$ \
& 0.367
&

$\begin{pmatrix}
1.00\times 10^{-7} & 8.98 & 1.68\times 10^{-7} \\
 8.98 & 5.29\times 10^{-11} & 6.42\times 10^{-8} \\
 1.68 \times 10^{-7}  & 6.42\times 10^{-8} & 2.99
\end{pmatrix}$ \
\\
\hline

$\begin{pmatrix}
	M_1= 3 \ m_t\\
	M_2= 9\ m_t\\\
	M_3=(9+ 1\times 10^{-7})\ m_t\
\end{pmatrix}$ &

$\begin{pmatrix}
1.32\times 10^{-7} & 5.34\times 10^{-6} & 1.08\times 10^{-5} \\
. & 2.15\times 10^{-4} & 4.35\times 10^{-4} \\
 . & . & 8.80\times 10^{-4}
\end{pmatrix}$
 &

$\begin{pmatrix}
0.804 & 9.85\times 10^{-10} & 9.54\times 10^{-5} \\
 32.4 & 3.20 \times 10^{-10} & 1.27\times 10^{-4} \\
 65.5 & 2.48\times 10^{-10} & 1.25 \times 10^{-4} 
\end{pmatrix}$ \
& 0.177
&

$\begin{pmatrix}
1.00\times 10^{-7} & 8.99 & 2.43\times 10^{-5} \\
 8.99 & 4.06 \times 10^{-11} & 2.08\times 10^{-5} \\
 2.43 \times 10^{-5}  & 2.08\times 10^{-5} & 2.99
\end{pmatrix}$ \
\\
\hline

\end{tabular}}
\end{center}
\end{table}

Note that at least two of the heavy neutrinos must be quasi-degenerate. This is due to the light neutrino mass one-loop corrections. In order for these not to be too large, a lepton-number-like $U(1)$ symmetry must be present and softly broken. This will be discussed in more detail in the next section.\par
\clearpage
\pagebreak
 \section{The Non-Decoupled Parameter Space: $M \sim$ eV/keV}\label{sec:non-dec}
 In this section, a study of a region of the parameter space where some heavy neutrinos have eV and keV mass is done, based in ref. \cite{Branco:2019avf}. This region is interesting because of the anomalies found in Short-baseline experiments, that may be hinting at the existence of an eV almost sterile neutrino \cite{LSND:1996ubh,LSND:2001aii,Armbruster:2002mp,MiniBooNEDM:2018cxm,Dentler:2019dhz,deGouvea:2019qre,PalomaresRuiz:2005vf,Alekseev:2018efk,Ko:2016owz,Abdurashitov:2005tb,Laveder:2007zz,Giunti:2006bj,Giunti:2019aiy,Diaz:2019fwt}. One exciting feature of this region of the parameter space is also that the deviations from $3 \times 3$ unitarity can be much larger than the region studied in the previous section, avoiding the bounds from ref. \cite{Fernandez-Martinez:2016lgt}, as discussed in \cite{Blennow:2016jkn}, which enables the detection of its effects in current and future neutrino oscillation experiments.\par
In the presence of deviations from unitarity, the probabilities of neutrino oscillations are altered, as documented in refs. \cite{Antusch:2006vwa,Blennow:2016jkn}. If $n$ heavy neutrinos can be produced in the beam used in oscillation experiments, the calculation of oscillation probabilities involves a $3 \times (3 + n)$ matrix, $\Theta$, which is a submatrix of $\mathcal{V}$ defined in equation \eqref{eq:calV},
\begin{equation}
\Theta\,=\,
\left( \begin{array}{cc}
K & (KX^\dag)_{3\times n}
\end{array} \right)\,.
\label{eq:wdef}
\end{equation}
$\Theta$ is the matrix defined by the first $3+n$ columns of $A$, defined in eq. \eqref{eq:ABdef}.
The probability of transition between flavour (anti-)neutrinos
\stackon[-.7pt]{$\nu$}{\brabar}$_\alpha$ and 
\stackon[-.7pt]{$\nu$}{\brabar}$_\beta$,
or of survival for a given flavour ($\alpha = \beta$),
with $\alpha, \beta = e, \mu, \tau$, can be shown to take the form
\begin{equation}
\begin{aligned}
P_{\stackon[-.7pt]{$\nu$}{\brabar}_\alpha \rightarrow \stackon[-.7pt]{$\nu$}{\brabar}_\beta}(L,E)
\,=\,\frac{1}{(\Theta\Theta^\dagger)_{\alpha\alpha}(\Theta\Theta^\dagger)_{\beta\beta}}
\Bigg[ 
\left|(\Theta\Theta^\dagger)_{\alpha\beta}\right|^2
&-   4 \sum_{i>j}^{3+n}\,\re
\left(\Theta_{\alpha i}^*\,\Theta_{\beta i}\,\Theta_{\alpha j}\,\Theta_{\beta j}^*\right)
\sin^2 \Delta_{ij} \\
&\pm 2 \sum_{i>j}^{3+n}\,\im
\left(\Theta_{\alpha i}^*\,\Theta_{\beta i}\,\Theta_{\alpha j}\,\Theta_{\beta j}^*\right)
\sin 2 \Delta_{ij}
\Bigg] \,,
\end{aligned}
\label{eq:probability}
\end{equation}%
where the plus or minus sign in the second line refers to neutrinos or anti-neutrinos, respectively.
Here, $L$ denotes the source-detector distance, $E$ is the (anti-)neutrino energy,
and one has defined
$\begin{aligned}
\Delta_{ij} \,\equiv \, \frac{\Delta m^2_{ik}\, L}{4E}
\,\simeq\, 1.27\,\frac{\Delta m_{ij}^{2}[\text{eV}^{2}]\,L[\text{km}] }{ E[\text{GeV}]}\,,
\end{aligned}$
with mass-squared differences $\Delta m_{ij}^2 \equiv m_i^2 - m_j^2$, as usual.

The equation \eqref{eq:probability} is a formula for the probability of neutrino oscillation, which takes into account deviations from a unitary mixing matrix. When $n=3$, the equation reduces to the usual formula for unitary mixing matrices. The normalization term in the equation will be close to unity in all cases, and the term proportional to $|(\Theta\Theta^\dagger)_{\alpha\beta}|^2$ is referred to as the "zero-distance" term, which is a result of deviations from $3\times3$ unitarity. This term will also be negligible. The equation can be simplified for specific cases, such as short-baseline (SBL) and long-baseline (LBL) experimental setups. However, matter effects must be considered in LBL experiments.

The SBL anomalies is the name given to the consistent measurements, in the last few decades, of anomalous excesses/deficits of $\nu_e$/$\overline{\nu}_e$ at $L/E \sim 1~\textrm{m/MeV}$ from $\nu_\mu /$ $\overline{\nu}_\mu$  and $\nu_e /$ $\overline{\nu}_e$  sources.
It first started with the LSND experiment \cite{PhysRevD.64.112007,PhysRevLett.75.2650}, which observed an excess of $\overline{\nu}_e$ at the detector, from a $\overline{\nu}_\mu$ source. Given the $L/E$ settings of the detector, this result pointed towards a large $\Delta m^2$, if interpreted as a two neutrino oscillation $\overline{\nu}_\mu \xrightarrow{} \overline{\nu}_e$.  Some years after the LSND results came the Reactor anomaly \cite{Mueller:2011nm,Huber:2011wv} and the Radioactive Source anomalies  \cite{Dentler:2018sju}. These experiments operated in nuclear reactors, NEOS and DANSS, or used radioactive $\nu_e$ sources, GALLEX and SAGE, and measured a deficit of $\nu_e$ or $\overline{\nu}_e$ in the detector.
Recently, the MiniBOONE experiment, which has about the same $L/E$ as LSND, but has different energy, beam and detector systematics and different event signatures and backgrounds, obtained results consistent with the ones from LSND \cite{MiniBooNE:2018esg,MiniBooNE:2020pnu}.
Combining all other experimental constraints together with the results from these experiments seems to indicate a preference for $(3+1)$ scenario, where one has 3 light neutrinos and 1 $eV$ mass neutrino \cite{Giunti:2019aiy}. Although this is the case, it is important to note that a closer examination on the data set reveals tension between the appearence and disappearence datasets and the neutrino and anti-neutrino datasets. \cite{Acero:2022wqg}. \par

\subsection{Deviations from Unitarity and Loop Corrections}
\label{sec:devunit_non}
These anomalies may be explained if the spectra of heavy neutrinos includes at least one eV mass neutrino and the deviations from $3 \times 3$ unitarity (DU) are sizeable. These DU need to be larger than in the previous studied region of the parameter space. Thus, they must violate the experimental bounds from ref. \cite{Fernandez-Martinez:2016lgt}, cf. eq. \eqref{eq:hen}. This can be done without much effort because the bounds from ref. \cite{Fernandez-Martinez:2016lgt} are only valid when all heavy neutrinos have a mass larger than the W boson mass, $M_W$. Hence, constraints on the entries of $\eta$ depend on the mass scale of the new neutrinos.\par
Therefore, to limit the parameter space of spectra with light heavy neutrinos, different sets of experimental constraints must be considered based on the spectrum. For light heavy neutrinos with eV-scale masses, the strongest constraints come from oscillation experiments~\cite{Blennow:2016jkn}. At the time of writing of ref. \cite{Branco:2019avf}, the most stringent bounds came from BUGEY-3~\cite{Declais:1994su}, MINOS~\cite{MINOS:2016viw}, NOMAD~\cite{NOMAD:2001xxt,NOMAD:2003mqg} and Super-Kamiokande~\cite{Super-Kamiokande:2014ndf}. In those cases, observables constrain directly the entries of $K X^\dag$, and not just the product $KX^\dag X X^{\dagger}$. To take constraints from these experiments, the relevant exclusion curves in the $\sin^{2}2\vartheta_{\alpha \beta}$ -- $\Delta m^{2}$ planes  are considered and translated into constraints on the elements of the mixing matrix block $K X^\dag$. 
For keV or GeV-TeV light heavy neutrinos, $\beta$-decay experiments (see e.g.~\cite{Drewes:2016upu} and references within) and from LHC searches for heavy Majorana neutrinos~\cite{Antusch:2015mia,
Deppisch:2015qwa, Das:2015toa, Das:2017nvm, Das:2017zjc,CMS2018iaf, CMS2018jxx} should be taken into account. Another crucial experimental input, also taken into account
in this analysis, is the limit on the $\mu \rightarrow e\gamma $ branching ratio
obtained by the MEG Collaboration, 
$BR(\mu \rightarrow e\gamma) < 4.2\times 10^{-13}$ ($90\%$~CL)~\cite{MEG:2016leq}, as it is one of the strictest bounds on lepton flavour violation. For the MeV-GeV intermediate mass range, constraints can be found in~\cite{Chrzaszcz:2019inj}.\par
For more details regarding new physics, effects on physical observables, the best experimental constraints and the most stringent bounds on a given spectra of heavy almost-sterile neutrinos see section \ref{sec:res_neutrinos}.\par
Deviations from $3 \times 3$ unitarity can cause problems when calculations are done at one loop. At one loop, the $\mathbf{0}_{3 \times 3}$ matrix in the upper-left corner of $\mathcal{M}$, defined in eq. \eqref{eq:massmatrixnu}, is no longer a null matrix and is corrected. These corrections can be very significant and increase the light neutrino masses at tree level by orders of magnitude. The one-loop corrected matrix is given by

\begin{equation}
\delta M_{L}\,=\,\delta M_{L}^{Z}+\delta M_{L}^{H}~, 
\end{equation}
where $\delta M_{L}^{Z}$ and $\delta M_{L}^{H}$ represent contributions
depending on the $Z$ and Higgs boson masses, $m_Z$ and $m_H$, respectively. 
Explicitly, one has (see also Appendix A of Ref.~\cite{AristizabalSierra:2011mn}):
\begin{equation}
\begin{aligned}
\delta M_{L}^{Z} &\,=\,
\frac{3}{32\pi ^{2}\, v^2 }\,
\left( \begin{array}{cc} K & K X^\dag \end{array} \right)\,
\frac{\mathcal{D}^{3}}{\mathcal{D}^{2}/{m_{Z}^{2}} - \id}\,
\log \left( \frac{\mathcal{D}^{2}}{m_{Z}^{2}} \right)\,
\left( \begin{array}{c} K^T \\ (K X^\dag)^T \end{array} \right)\,, \\[2mm]
\delta M_{L}^{H} &\,=\,
\frac{1}{32\pi ^{2}\, v^2 }\,
\left( \begin{array}{cc} K & K X^\dag \end{array} \right)\,
\frac{\mathcal{D}^{3}}{\mathcal{D}^{2}/{m_{H}^{2}} - \id}\,
\log \left( \frac{\mathcal{D}^{2}}{m_{H}^{2}} \right)\,
\left( \begin{array}{c} K^T \\ (K X^\dag)^T \end{array} \right)\,, 
\label{dM1}
\end{aligned}
\end{equation}
in a generic weak basis, with $v \simeq 174$ GeV being the Higgs VEV
and with $\mathcal{D}$, $K$ and $K X^\dag$ given in Eqs.~\eqref{eq:diagneutrino} and \eqref{eq:calVfull}.
This result can be cast in a simple form:
\begin{align}
\delta M_{L}\,=\,
K\,f(d)\,K^T + K X^\dag\,f(D)\,(K X^\dag)^T\,, 
\end{align}
where naturally $f$ is applied element-wise to diagonal matrices,
with
\begin{align}
f(m) \,\equiv \, \frac{m^3}{(4\pi\,v)^2} \left(\frac{3\log(m/m_Z)}{m^2/m_Z^2 -1} + \frac{\log(m/m_H)}{m^2/m_H^2 -1}\right)\,.
\end{align}

The following section describes the steps taken to obtain numerical results that are accurate to one-loop corrections. The tree-level form of $\mathcal{M}$ in Eq.\eqref{eq:massmatrixnu} can either be derived from a specific model or constructed using the matrices $d$, $D$, $V$, and $O_c$, as shown in Table \ref{tab:parameters}. These matrices, along with $m_l$, encompass the entire physical parameter space.
\begin{table}[h!]
\centering
\renewcommand{\arraystretch}{1.2}
\begin{tabular}{lcccccc}
\toprule 
 & $\,\,\,\, m_l \,\,\,\,$ & $\,\,\,\, d \,\,\,\,$ & $\,\,\,\, D \,\,\,\,$ & $\,\,\,\, O_c \,\,\,\,$ & $\,\,\,\, V \,\,\,\,$ & Total \\
\midrule
Moduli \quad& 3 & 3 & 3 & 3 & 3 & 15 = 9 + 6 \\[1mm] 
Phases \quad& $-3$ & 0 & 0 & 3 & 6 & 6 \\ 
\bottomrule
\end{tabular}
\caption{
The counting of physical parameters in a type-I seesaw scenario with three sterile neutrinos involves 15 variables. These include 9 lepton masses (3 for charged leptons and 6 for neutrinos) and 6 mixing angles. There are also 6 physical phases, taking into account that rephasing the charged leptons can eliminate 3 phases from the matrix $V$. It's worth mentioning that in the basis under consideration, $m_l$ is real and diagonal.}
\label{tab:parameters}
\end{table}

It is then straightforward to obtain $K$ and $X$.
One can then compute $\delta M_L$ and proceed to diagonalise
\begin{align}
\mathcal{M}^{\text{1-loop}} \simeq \left( 
\begin{array}{cc}
\delta M_L & m \\ 
m^{T} & M%
\end{array}\right)\,,
\label{eq:oneloop}
\end{align}%
obtaining the loop-corrected $K$, $X$, $d$ and $D$. It is reasonable to neglect one-loop corrections to $m$ and $M$~\cite{Grimus:2002nk,Grimus:2018rte}.

The mass loop-corrections can be controlled by incorporating a lepton-number-like symmetry and softly breaking it. Light sterile neutrinos can emerge naturally in a seesaw framework with an approximately conserved lepton number~\cite{Shaposhnikov:2006nn, Kersten:2007vk, Ibarra:2010xw}. An exact $U(1)_L$ symmetry imposes specific textures on the mass matrices $m$ and $M$, which can be slightly altered when the symmetry is only approximate, allowing for non-zero Majorana neutrino masses and non-trivial mixing.

\subsection{Benchmarks}
\label{sec:benc_non-dec}

The goal of this subsection is to examine the parameter space where at least one mostly-sterile neutrino is light, with a mass of $\mathcal{O}$(eV), to establish a connection to the Short-Baseline (SBL) anomalies. In order to achieve Yukawa couplings that are of order one, one must have substantial deviations from $3 \times 3$ unitarity, as explained in the end of section \ref{sec:quasi}. It is interesting to note that these large deviations from unitarity are necessary to connect to the SBL anomalies. All of this is only possible in the presence of a softly broken lepton-number-like $U(1)$ symmetry. The choice of lepton charges should then be such that, in the limit of exact conservation:

\begin{enumerate}
\item  $M$ has zero determinant.
\item Not all entries of $m$ are small. One needs a couple of $\mathcal{O}(1)$ Yukawa couplings.
\end{enumerate}
The available choices for $U(1)_L$ charges are restricted by these conditions. The study in the manner of reference \cite{Branco:1988ex} is carried out in a specific `symmetry' weak basis where the lepton charge vectors $\lambda_\nu$ and $\lambda_L$ are assigned to the three right-handed neutrino singlets and the three lepton doublets respectively.
As $\oK$ is not included in the Lagrangian and thus not a physical matrix, in many cases it is convenient to set $U_{\oK}$ equal to the identity matrix. Nevertheless, when textures are considered, it is typical to have $U_{\oK} \neq \id$ as specified in equation (\ref{eq:kdef}).
In the exact conservation limit, there are only four non-trivial choices of $U(1)_L$ charges for permutations that result in an $M$ matrix with a zero determinant: 
$\lambda_\nu=(1,1,0)$, $\lambda_\nu=(1,-1,-1)$,
$\lambda_\nu=(1,1,1)$ and $\lambda_\nu=(0,0,1)$. 
Out of these four options, the choice of $\lambda_\nu=(1,1,1)$ is not feasible as it leads to an $M$ matrix equal to zero, and $\lambda_\nu=(0,0,1)$ is disregarded since the requirement for controlled loop corrections within the framework effectively reduces it to the case with $\lambda_\nu=(1,-1,-1)$.
The remaining two options $\lambda_\nu=(1,1,0)$ and $\lambda_\nu=(1,-1,-1)$ are used in what follows.
The choice of $\lambda_L$ is determined based on the conditions that the seesaw mechanism is effective for all light neutrinos and that all left-handed neutrinos are able to couple with the right-handed ones, as stated in~\cite{Branco:1988ex}.
The numerical benchmarks in the study described in section \ref{sec:quasi} and based on ref. \cite{Agostinho:2017wfs} were obtained with the $U(1)_L$ charge assignment $\lambda_\nu = (1,-1,0)$ and $\lambda_L=(1,1,1)$. However, this leads to $\det ,M \neq 0$ in the symmetric limit.
Given these considerations, the following study is divided into three possibilities: \textbf{Ia} , \textbf{Ib}  and \textbf{II}. \par
For each, an exploration of the parameter space of qualitatively similar seesaw structures will be done, providing an illustrative and representative numerical benchmark in the end. By approaching it in this manner, the possibility of a significant active-sterile mixing only within a finely tuned region of parameter space is ruled out.
As anticipated in the beginning of this section, section \ref{sec:non-dec},  approximate forms
of the transition probabilities of muon to electron (anti-)neutrinos,
$P_{\stackon[-.7pt]{$\nu$}{\brabar}_\mu \rightarrow \stackon[-.7pt]{$\nu$}{\brabar}_e}$,
obtained from Eq.~\eqref{eq:probability} while having in mind SBL and LBL setups,
for each of the three scenarios will be provided.
Considering the most recent global fits available at the time~\cite{DeSalas:2018rby, Esteban:2018azc}, which strongly disfavor a light neutrino mass spectrum with inverted ordering (IO) compared to one with normal ordering (NO) beyond a $3\sigma$ significance level, the numerical examples are confined to the case of normal ordering (NO).

Before proceeding, it is important to note that the three scenarios of interest bear some resemblance to the commonly considered $3+1+1$ (case \textbf{Ia} ), $3+2$ (case \textbf{Ib} ), and $3+1$ (case \textbf{II}) schemes, as discussed in ref.~\cite{Giunti:2019aiy}. While the connection to the latter schemes is not exact, especially in the case of scenario \textbf{Ib}  which does not have a typical $3+2$ spectrum, it could still be beneficial to consider quantities defined within those schemes in the analysis, as outlined in~\cite{Gariazzo:2015rra}.
\begin{align}
\sin^2 2 \vartheta^{(k)}_{\mu e}
\,\equiv\, 4\, \big|\Theta_{\mu k}\big|^2 \,\big|\Theta_{e k}\big|^2 
\,,
\label{eq:sinmue}
\end{align}
with $k=4$ in the$3+1$ case, while $k=4,5$ for the other two cases.
According to the global fit to SBL data of Ref.~\cite{Gariazzo:2015rra},
explaining the observed anomalies requires
$\Delta m^2_{41} \in [0.87,\,2.04]$ eV$^2$ and
$\sin^2 2 \vartheta^{(4)}_{\mu e} \in [6.5\times 10^{-4},\, 2.6 \times 10^{-3}]$ ($99.7\%$ CL)
in the$3+1$ scheme. This result may also be of relevance in the$3+1+1$ scheme.
These intervals are taken as guidelines in the numerical explorations, since they encompassed the consensus of global-fits at the time of writing of ref. \cite{Branco:2019avf}. Thus, the sterile neutrino parameter space is constrained through the conservative bounds
$\sum_i |R_{\alpha i}|^2 < 0.1$ ($\alpha = e,\mu,\tau$),
and via the constraints of \cite{NOMAD:2001xxt, NOMAD:2003mqg,Drewes:2016upu}
on mixing matrix elements corresponding to
large mass-squared differences $\Delta m^2 \sim 10\text{ eV}^2 - 1\text{ keV}^2$,
as anticipated in subsection~\ref{sec:devunit_non}.

\subsubsection{Case I: \texorpdfstring{$\lambda_\nu = (1,1,0)$}{lambda nu = (1,1,0)}}
\label{sec:caseI}
For this case, the only sensible choice for the doublet charges is
$\lambda_L = (0,0,0)$.
The mass matrices in the symmetric limit read:
\begin{align}
m\,=\,\left( 
\begin{array}{ccc}
0 & 0 & a \\ 
0 & 0 & b \\ 
0 & 0 & c%
\end{array}%
\right) \,,\quad
M\,=\,\left( 
\begin{array}{ccc}
0 & 0 & 0 \\ 
0 & 0 & 0 \\ 
0 & 0 & M_3%
\end{array}%
\right) \,.
\label{eq:MI}
\end{align}
Breaking the symmetry generates the light neutrino masses, two (mainly) sterile states with masses $M_1$ and $M_2$ that can be significantly smaller than $M_3$, and a heavy sterile with a mass close to $M_3$.

As expected, some Yukawa couplings remain at $\mathcal{O}(1)$, which can be explained by the dependence of the Dirac mass matrix $m$ on the sterile masses contained in $D$, as seen in Eq.~\eqref{eq:mdirac}.

This scenario is further divided into two subcases: a scenario with a hierarchy of $M_2 \gg M_1$ ({\bf case \textbf{Ia}}), which may occur in a stepwise symmetry breaking scenario, or a scenario with a single new light-sterile scale, with $M_1 \sim M_2$ ({\bf case \textbf{Ib}}).\par

\clearpage

{\bfseries\LARGE Case Ia: \texorpdfstring{$M_1\ll M_2\ll M_3$}{M1 << M2 << M3}\par}

\vskip 8mm
\begin{table}[h!]
\renewcommand{\thetable}{\arabic{table}a}
\centering
\renewcommand{\arraystretch}{1.2}
\begin{tabular}{lr}
\toprule
 & {\bf Case Ia} numerical benchmark \\
\midrule
\addlinespace
$m$ (GeV) &
$\begin{bmatrix*}[r]
 ( 2.11 -5.58\, i)\times 10^{-11} & ( 1.29 +1.65\, i)\times 10^{-9} &  11.2 -10.9\, i \\ 
 ( 0.85 +2.22\, i)\times 10^{-10} & (-5.29 +3.99\, i)\times 10^{-9} &  10.4 + 0.4\, i\\ 
 (-0.26 +1.98\, i)\times 10^{-10} & (-4.51 -1.05\, i)\times 10^{-9} & -10.5 -34.6\, i
\end{bmatrix*}$ \\
\addlinespace
$M$ (GeV)  & 
$\begin{bmatrix*}[r]
 8.93\times 10^{-10} & 4.45\times 10^{-11} & 1.28\times 10^{-13} \\
 4.45\times 10^{-11} & 1.00\times 10^{-6} & 6.22\times 10^{-11} \\
 1.28\times 10^{-13} & 6.22\times 10^{-11} & 5.00\times 10^{14} 
\end{bmatrix*}$
\\
\addlinespace
\midrule
\addlinespace
$K$  &
$\begin{bmatrix*}[r]
-0.797 +0.071 \, i &  0.578 +0.006 \, i & -0.115+0.096 \, i \\
 0.293 -0.086 \, i &  0.575 +0.027 \, i &  0.719+0.010 \, i \\
-0.516 -0.004 \, i & -0.570 +0.020 \, i &  0.606
\end{bmatrix*}$
\\
\addlinespace
$KX^\dag$  &
$\begin{bmatrix*}[r]
 0.024 -0.057 \, i & ( 1.29 +1.65 \, i)\times 10^{-3} & (-2.24+2.18 \, i) \times 10^{-14} \\
 0.093 +0.223 \, i & (-5.29 +3.99 \, i)\times 10^{-3} & (-2.08+0.08 \, i) \times 10^{-14} \\
-0.026 +0.199 \, i & (-4.51 -1.05 \, i)\times 10^{-3} & (-2.10+6.92 \, i) \times 10^{-14}
\end{bmatrix*}$
\\
\addlinespace
$X$  &
$\begin{bmatrix*}[r]
-0.003-0.015 \, i & 0.102 +0.023 \, i & 0.050 -0.317 \, i \\
(-5.12+1.72 \, i)\times 10^{-4}  & ( 0.46 -4.33 \, i)\times 10^{-3}  & (-7.30-2.18 \, i)\times 10^{-3} \\
( 0.23+5.33 \, i)\times 10^{-14} & (-3.44 +2.75 \, i)\times 10^{-14} & ( 0.36-4.41 \, i)\times 10^{-14}
\end{bmatrix*}$
\\
\addlinespace
$O_c$ (tree level) \!\!\!\!\!\! \!\!\!\!\!\!\!\!\!\!\!\!& 
$\begin{bmatrix*}[r]
-0.53 +0.12 \, i &  0.22 -1.12 \, i & -1.41 -0.22 \, i \\
 0.22 +0.56 \, i & -1.50 -0.13 \, i & -0.30 +1.03 \, i \\
 1.00 -0.06 \, i &  0.23 +0.25 \, i & -0.14 -0.01 \, i
\end{bmatrix*}$
\\
\addlinespace
\midrule
\addlinespace
Masses &
$\begin{matrix*}[l]
m_1 \simeq 1.06\times 10^{-3}\text{ eV}\,,\quad & m_2 \simeq 8.48\times 10^{-3}\text{ eV} \,,\quad & m_3 \simeq 5.02\times 10^{-2}\text{ eV} \,,\,\\ 
M_1 \simeq 1.00\text{ eV}\,,\,               & M_2 \simeq 1.00\text{ keV} \,,\,              & M_3 \simeq 5.00\times 10^{14}\text{ GeV}
\end{matrix*}$ \\
\addlinespace
\midrule
\addlinespace
$3\nu$ $\Delta m^2$ &
$\Delta m^2_\odot = \Delta m^2_{21} \simeq 7.08 \times 10^{-5}\text{ eV}^2\,,\,
\quad\,\,\,\,
\Delta m^2_\text{atm} = \Delta m^2_{31} \simeq 2.52 \times 10^{-3}\text{ eV}^2$
\\
\addlinespace
$3\nu$ mixing angles \!\!\!\!\!\!\!\!\!\!\!\!& 
$\sin^2 \theta_{12} \simeq 0.344\,,\,\quad\,\,\,\,
\sin^2 \theta_{23} \simeq 0.585\,,\,\quad\,\,\,\,
\sin^2 \theta_{13} \simeq 0.0236$
\\
\addlinespace
$3\nu$ CPV phases \!\!\!\!\!\! & 
$\delta \simeq 1.21 \pi\,,\,\quad\,\,\,\,
\alpha_{21} \simeq 0.06 \pi\,,\,\quad\,\,\,\,
\alpha_{31} \simeq 0.06 \pi$
\\
\addlinespace
\midrule
\addlinespace
$\sin^2 2 \vartheta^{(i)}_{\mu e}$ & 
$\sin^2 2 \vartheta_{\mu e}^{(4)} \simeq 8.8\times 10^{-4}\,,\,\quad\,\,\,\,
\sin^2 2 \vartheta_{\mu e}^{(5)} \simeq 7.7\times 10^{-10}$
\\
\bottomrule
\end{tabular}
\caption[Numerical benchmark for case \textbf{Ia}. The ordering of light neutrinos is NO.
From the input matrices $m$ and $M$, and taking into account one-loop corrections, the other quantities here listed follow.
It should be noted that $O_c$ of Eq.~\eqref{eq:Xdef} is only defined at tree level.
Values for the mixing angles and CPV phases of the $3\nu$-framework in the standard parametrisation are extracted by identifying the unitary matrix $V$ with a unitary $3\times 3$ PMNS mixing matrix.]{Numerical benchmark for case \textbf{Ia}. The ordering of light neutrinos is NO.
From the input matrices $m$ and $M$, and taking into account one-loop corrections, the other quantities here listed follow.
It should be noted that $O_c$ of Eq.~\eqref{eq:Xdef} is only defined at tree level.
Values for the mixing angles and CPV phases of the $3\nu$-framework in the standard parametrisation~\cite{pdg} are extracted by identifying the unitary matrix $V$ with a unitary $3\times 3$ PMNS mixing matrix.}
\label{tab:Ia}
\end{table}

The numerical results for scenario \textbf{Ia} are presented in Table~\ref{tab:Ia}, including the one-loop correction from Eq.\eqref{eq:oneloop}. This scenario features three mostly-active light neutrinos, as well as three mostly-sterile neutrinos with masses of approximately 1 eV for $M_1$, 1 keV for $M_2$, and a few orders of magnitude below the grand unification scale for $M_3$, with a value of approximately $10^{14}$ GeV. The keV-scale neutrino may serve as a potential dark matter candidate, as suggested in Refs.\cite{Drewes:2016upu, Boyarsky:2018tvu}.


For case \textbf{Ia}, there are two kinematically accessible almost-sterile heavy neutrinos, meaning $n=2$ in Eq.\eqref{eq:probability}. In the context of a LBL experiment like DUNE~\cite{DUNE:2016evb}), the expression of Eq.~\eqref{eq:probability} applied to the transition probability of muon to electron (anti-)neutrinos
can, in this case, be approximated by:
\begin{equation}
\begin{aligned}
P^\text{LBL}_{\stackon[-.7pt]{$\nu$}{\brabar}_\mu \rightarrow \stackon[-.7pt]{$\nu$}{\brabar}_e}
\,\simeq\,
&\frac{1}{(\Theta\Theta^\dagger)_{\mu\mu}(\Theta\Theta^\dagger)_{ee}}
\Bigg[ 
\left|(\Theta\Theta^\dagger)_{\mu e}\right|^2
\\ &
-   4 \sum_{i>j}^3\,\re
\left(\Theta_{\mu i}^*\,\Theta_{e i}\,\Theta_{\mu j}\,\Theta_{e j}^*\right)
\sin^2 \Delta_{ij} 
\pm 2 \sum_{i>j}^{3}\,\im
\left(\Theta_{\mu i}^*\,\Theta_{e i}\,\Theta_{\mu j}\,\Theta_{e j}^*\right)
\sin 2 \Delta_{ij}
\\ &
-   4 \,\cdot\,\frac{1}{2}\,\re
\left(\Theta_{\mu 4}^*\,\Theta_{e 4}\,\sum_{j=1}^3\,\Theta_{\mu j}\,\Theta_{e j}^*\right)
-   4 \,\cdot\,\frac{1}{2}\,\re
\left(\Theta_{\mu 5}^*\,\Theta_{e 5}\,\sum_{j=1}^4\,\Theta_{\mu j}\,\Theta_{e j}^*\right)
\Bigg] \,,
\label{eq:LBL1Ia}
\end{aligned}
\end{equation}
where terms depending on $\Delta_{4j},\,\Delta_{5j} \gg 1$ 
have been replaced by their averaged versions
($\sin^2 \Delta_{ij} \to 1/2$, $\sin 2\Delta_{ij} \to 0$).
%

%

The loss of unitarity and a zero-distance effect are indicated by the normalization and the first term in this equation, while the effects of the two lightest mostly-sterile states on oscillations are explicitly represented by the last two terms.

If one is in a situation similar to that of the numerical benchmark of  Table~\ref{tab:Ia}, for which $|(\Theta\Theta^\dagger)_{\mu\mu}(\Theta\Theta^\dagger)_{ee} - 1|$
and $|(\Theta\Theta^\dagger)_{\mu e}|^2$ are negligible, this expression can be further approximated by:
\begin{align}
P^\text{LBL}_{\stackon[-.7pt]{$\nu$}{\brabar}_\mu \rightarrow \stackon[-.7pt]{$\nu$}{\brabar}_e}
\,\simeq\,
P^{\text{LBL, }3\nu}_{\stackon[-.7pt]{$\nu$}{\brabar}_\mu \rightarrow \stackon[-.7pt]{$\nu$}{\brabar}_e}
+ \frac{1}{2}\sin^2 2 \vartheta^{(4)}_{\mu e}
 \,,
\label{eq:LBL2Ia}
\end{align}
where it was defined a $3\nu$-framework transition probability which, however, incorporates the effects of deviations of $K$ from unitarity,
\begin{align}
P^{\text{LBL, }3\nu}_{\stackon[-.7pt]{$\nu$}{\brabar}_\mu \rightarrow \stackon[-.7pt]{$\nu$}{\brabar}_e}
\,\equiv\,
-   4 \sum_{i>j}^3\,\re
\left(\Theta_{\mu i}^*\,\Theta_{e i}\,\Theta_{\mu j}\,\Theta_{e j}^*\right)
\sin^2 \Delta_{ij} 
\pm 2 \sum_{i>j}^{3}\,\im
\left(\Theta_{\mu i}^*\,\Theta_{e i}\,\Theta_{\mu j}\,\Theta_{e j}^*\right)
\sin 2 \Delta_{ij}
\,,
\label{eq:3nu}
\end{align}
and have used the definition of Eq.~\eqref{eq:sinmue},
the unitarity of the full $6\times 6$ mixing matrix, and the fact that
$|\Theta_{\alpha 4}|^2 (= |(KX^\dag)_{\alpha 1}|^2) \gg |\Theta_{\alpha 5}|^2 (= |(KX^\dag)_{\alpha 2}|^2) \gg |(KX^\dag)_{\alpha 3}|^2$.

%
In an SBL experiment (e.g.~MicroBooNE~\cite{MicroBooNE:2015bmn}),
the relevant form of Eq.~\eqref{eq:probability} for
$\stackon[-.7pt]{$\nu$}{\brabar}_\mu \rightarrow \stackon[-.7pt]{$\nu$}{\brabar}_e$
transitions is:
\begin{equation}
\begin{aligned}
P^\text{SBL}_{\stackon[-.7pt]{$\nu$}{\brabar}_\mu \rightarrow \stackon[-.7pt]{$\nu$}{\brabar}_e}
\,\simeq\,
&\frac{1}{(\Theta\Theta^\dagger)_{\mu\mu}(\Theta\Theta^\dagger)_{ee}}
\Bigg[ 
\left|(\Theta\Theta^\dagger)_{\mu e}\right|^2
-   4 \,\cdot\,\frac{1}{2}\,\re
\left(\Theta_{\mu 5}^*\,\Theta_{e 5}\,\sum_{j=1}^4\,\Theta_{\mu j}\,\Theta_{e j}^*\right)
\\ &
-   4 \,\re
\left(\Theta_{\mu 4}^*\,\Theta_{e 4}\,\sum_{j=1}^3\,\Theta_{\mu j}\,\Theta_{e j}^*\right)
\sin^2 \Delta_{41} 
\pm 2 \,\im
\left(\Theta_{\mu 4}^*\,\Theta_{e 4}\,\sum_{j=1}^{3}\,\Theta_{\mu j}\,\Theta_{e j}^*\right)
\sin 2 \Delta_{41}
\Bigg] \,,
\label{eq:SBL1Ia}
\end{aligned}
\end{equation}
with $\Delta_{41}\simeq \Delta_{42}\simeq \Delta_{43}$, and
where terms depending on $\Delta_{5j} \gg 1$ 
have been replaced by their averaged versions
($\sin^2 \Delta_{5j} \to 1/2$, $\sin 2\Delta_{5j} \to 0$).
In this context, one is sensitive to oscillations due to the scale of the mass-squared differences $\Delta m^2_{4j}$ with $j=1,2,3$, whereas the oscillations pertaining to smaller mass-squared differences have not yet had a chance to develop.

%
Finally, if one is in a condition similar to that of
the numerical benchmark, this expression can be simply approximated by:
\begin{align}
P^\text{SBL}_{\stackon[-.7pt]{$\nu$}{\brabar}_\mu \rightarrow \stackon[-.7pt]{$\nu$}{\brabar}_e}
\,\simeq\, \sin^2 2 \vartheta^{(4)}_{\mu e}\,\sin^2 \Delta_{41} \,,
\label{eq:SBL2Ia}
\end{align}
where once again one has taken into account
the unitarity of the full mixing matrix and the fact that
$|(KX^\dag)_{\alpha 1}|^2 \gg  |(KX^\dag)_{\alpha 2}|^2 \gg |(KX^\dag)_{\alpha 3}|^2$.

\begin{figure}[h!]
\centering
\includegraphics[width=1.0\linewidth]{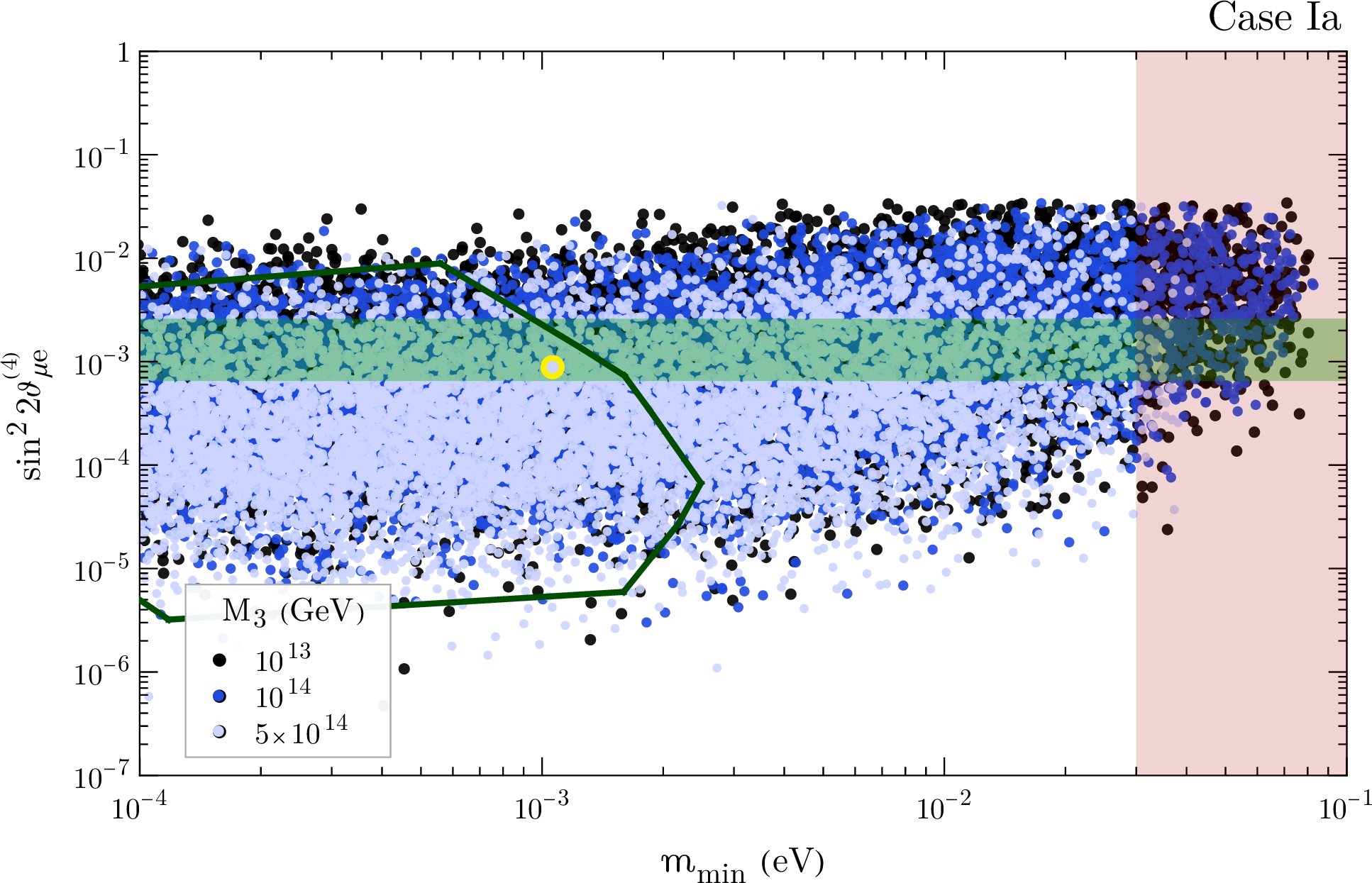}
\caption[Active-sterile mixing measure $\sin^2 2\vartheta_{\mu e}^{(4)}$ 
{\it versus} the lightest neutrino mass $m_\text{min}$ from a scan of the case-\textbf{Ia} parameter space,
with NO ($m_\text{min} = m_1$). The heavy spectrum at tree level has
$M_1 = 1$ eV and $M_2 = 1$ keV, while three values of
the heaviest mass are considered, $M_3 = 10^{13}\, (10^{14})\, \{5\times 10^{14}\}$ GeV,
corresponding to the black (dark blue) \{light blue\} points in the scatter plot. The horizontal green band shows the $99.7\%$ CL interval, while the vertical red exclusion band is obtained by combining
the most stringent bound on the sum of light neutrino masses from cosmology, $\sum_i m_i < 0.12$ eV ($95\%$ CL), with the $3\sigma$ ranges of mass-squared differences. The dark green contour delimits the region inside which loop-stable points have been found (see text), while the benchmark of Table~\ref{tab:Ia} is marked in yellow.]{Active-sterile mixing measure $\sin^2 2\vartheta_{\mu e}^{(4)}$ 
{\it versus} the lightest neutrino mass $m_\text{min}$ from a scan of the case-\textbf{Ia} parameter space,
with NO ($m_\text{min} = m_1$). The heavy spectrum at tree level has
$M_1 = 1$ eV and $M_2 = 1$ keV, while three values of
the heaviest mass are considered, $M_3 = 10^{13}\, (10^{14})\, [5\times 10^{14}]$ GeV,
corresponding to the black (dark blue) [light blue] points in the scatter plot.
The horizontal green band shows the $99.7\%$ CL interval of Ref.~\cite{Gariazzo:2015rra},
while the vertical red exclusion band is obtained by combining
the most stringent bound on the sum of light neutrino masses from cosmology,
$\sum_i m_i < 0.12$ eV ($95\%$ CL)~\cite{Vagnozzi:2017ovm,Aghanim:2018eyx},
with the $3\sigma$ ranges of mass-squared differences.
The dark green contour delimits the region inside which loop-stable points have been found (see text),
while the benchmark of Table~\ref{tab:Ia} is marked in yellow.}
\label{fig:Ia}
\end{figure}

To further study the parameter space of case \textbf{Ia}, numerical seesaw structures were created by setting the values of the unitary part $V$ of the mixing matrix $K$, the mostly-active and mostly-sterile masses in $d$ and $D$, and by scanning the complex orthogonal matrix $O_c$ which is represented as a combination of three complex rotations and a sign that corresponds to its determinant. The goal is to find seesaw structures that are similar to the benchmark, for which was set (at tree level) $M_1 = 1$ eV and $M_2 = 1$ keV, while considering three different values for the heaviest neutrino mass, $M_3 \in {10^{13}, 10^{14}, 5\times 10^{14}}$ GeV. The minimum lightest neutrino mass, $m_\text{min}$, is examined in the range of $[10^{-4}, 0.1]$ eV, while the remaining elements of $d$ are set by determining the solar and atmospheric mass differences.
The $3\nu$ mixing angles and the Dirac CPV phase in $V$, as well as the three-neutrino mass-squared differences, are selected to be the central values obtained from the global fit in Ref.~\cite{Esteban:2018azc}.
It is important to emphasize that, as with the numerical benchmark in Table~\ref{tab:Ia}, the $3\nu$ mixing angles and CPV phases for three neutrinos obtained while identifying $V$ with a unitary $3\times 3$ mixing matrix may differ slightly from the mixing angles and CPV phases derived from the full $6\times 6$ mixing matrix $\mathcal{V}$, due to deviations from unitarity.
In Figure~\ref{fig:Ia}, values of $\sin^2 2\vartheta_{\mu e}^{(4)}$ from Eq.~\eqref{eq:sinmue} are displayed against the lightest neutrino mass for the numerical examples of case \textbf{Ia}. Only the points for which $\textrm{Tr}\left[m,m^\dagger\right] \in [0.01,1],v^2$ are displayed.%
\footnote{By selecting suitable values for $M_3$ and $O_c$, it is possible to avoid extremely small Yukawa couplings.}

The horizontal green band in the figure marks the range of $\sin^2 2\vartheta_{\mu e}^{(4)}$ that is preferred by the global fit from Ref.\cite{Gariazzo:2015rra}, which was cited earlier in the section. The dark green contour outlines the area where relatively loop-stable points can be found. These are points that, even after the implementation of the one-loop correction from Eq.\eqref{eq:oneloop}, still have three-neutrino mass-squared differences and mixing angles (extracted from $V$) within the $3\sigma$ ranges of the fit~\cite{Esteban:2018azc}. The areas outside of the dark green lines are not necessarily ruled out, as it is possible to imagine a scenario where the incorrect tree-level values lead to acceptable values after corrections.
As illustrated in the figure, increasing the magnitude of $M_3$ results in a decrease in the magnitude of light neutrino masses, thus preventing excessively large values of $m_\text{min}$.
The approximations used in deriving the oscillation formulae
of Eqs.~\eqref{eq:LBL2Ia} and~\eqref{eq:SBL2Ia} hold for all the plotted points.

Some quantities of potential phenomenological relevance, unrelated to 
neutrino oscillations, include 
the effective electron neutrino mass in $\beta$-decay, $m_\beta$,
the absolute value of the effective neutrino Majorana mass controlling the rate
of neutrinoless double beta ($(\beta\beta)_{0\nu}$-)decay, $|m_{\beta\beta}|$,
and the $\mu \to e \gamma$ branching ratio, $BR(\mu \rightarrow e\gamma)$.
For all stable numerical examples pertaining to case \textbf{Ia} that stable under loop corrections, the branching ratio of $\mu$ to electron gamma decay is so small as to be unobservable ($BR(\mu \rightarrow e\gamma) \ll 10^{-30}$). The effective electron neutrino mass in beta decay ($m_\beta$) and the magnitude of the effective Majorana mass that governs the rate of neutrinoless double beta decay ($|m_{\beta\beta}|$) are still too low to be measured by current or near-future experiments, with bounds of $m_\beta < 9.4$ meV and $|m_{\beta\beta}| < 6.7$ meV. The calculation of $|m_{\beta\beta}|$ has taken into account the effects of the eV- and keV-scale
neutrinos.

In scenarios where active-sterile mixing is relatively high, such as this one, experiments like KATRIN in the future may be able to detect the presence of sterile neutrinos with masses around $\mathcal{O}$(eV)~\cite{Riis:2010zm}. The sensitivity of these experiments is determined by $|(KX^\dag){e1}|^2 = |\mathcal{V}{e4}|^2$, which for the loop-stable numerical examples of this case, is found to be limited to $|(KX^\dag)_{e4}|^2 \lesssim 0.02$.
Sterile neutrinos with masses of $\mathcal{O}$(keV) may be detected through kink-like signatures in next-generation beta decay experiments, even if the mixing between active and almost-sterile neutrinos is small ($|(KX^\dag){e2}|^2= |\mathcal{V}{e5}|^2 \sim 10^{-6}$)~\cite{Mertens:2014nha}.
\clearpage
{\bfseries\LARGE Case Ib: \texorpdfstring{$M_1\sim M_2\ll M_3$}{M1 \textasciitilde{} M2 << M3}\par}
\vskip 8mm
\begin{table}[h!]
\addtocounter{table}{-1}
\renewcommand{\thetable}{\arabic{table}b}
\centering
\renewcommand{\arraystretch}{1.2}
\begin{tabular}{lr}
\toprule
 & {\bf Case Ib} numerical benchmark \\
\midrule
\addlinespace
$m$ (GeV) &
$\begin{bmatrix*}[r]
 ( 0.46 - 2.57\, i) \times 10^{-10} & ( 2.37 + 0.54\, i) \times 10^{-10} & 11.24 - 2.72\, i \\
 (-5.50 - 1.04\, i) \times 10^{-10} & ( 0.68 - 6.20\, i) \times 10^{-10} &  8.90 -27.50\, i \\
 (-3.69 + 1.78\, i) \times 10^{-10} & (-1.60 - 4.45\, i) \times 10^{-10} & -1.85 + 0.43\, i\end{bmatrix*}$ \\
\addlinespace
$M$ (GeV)  & 
$\begin{bmatrix*}[r]
 2.88\times 10^{-9}  & 8.24\times 10^{-11} & 1.41\times 10^{-11}\\
 8.24\times 10^{-11} & 2.87\times 10^{-9}  & 1.42\times 10^{-11}\\
 1.41\times 10^{-11} & 1.42\times 10^{-11} & 1.00\times 10^{14}
\end{bmatrix*}$
\\
\addlinespace
\midrule
\addlinespace
$K$  &
$\begin{bmatrix*}[r]
-0.799 +0.137 \, i &  0.558 +0.001 \, i &  0.116 -0.071 \, i \\
 0.272 -0.172 \, i &  0.582 -0.036 \, i & -0.695 +0.014 \, i \\
-0.480 +0.099 \, i & -0.560 +0.141 \, i & -0.620 -0.019 \, i
\end{bmatrix*}$
\\
\addlinespace
$KX^\dag$  &
$\begin{bmatrix*}[r]
0.039 +0.077\, i &  0.067 -0.040\, i & (-1.12 +0.27\, i)\times 10^{-13} \\
0.156 -0.105\, i & -0.097 -0.170\, i & (-0.89 +2.75\, i)\times 10^{-13} \\
0.061 -0.140\, i & -0.115 -0.071\, i & ( 1.85 -0.43\, i)\times 10^{-14}
\end{bmatrix*}$
\\
\addlinespace
$X$  &
$\begin{bmatrix*}[r]
-0.003 +0.009\, i & 0.073 -0.064 \, i & -0.168 -0.196 \, i \\
-0.009 -0.005\, i & 0.049 +0.078 \, i &  0.170 -0.185 \, i \\
(1.40 -5.37 \, i)\times 10^{-14} & (-1.47 -1.74 \, i)\times 10^{-13} & (0.37 +2.24 \, i)\times 10^{-13}
\end{bmatrix*}$
\\
\addlinespace
$O_c$ (tree level) \!\!\!\!\!\! \!\!\!\!\!\!\!\!\!\!\!\!& 
$\begin{bmatrix*}[r]
-1.06 -0.51 \, i & -1.10 -1.29 \, i & -1.51 +1.30 \, i \\
 0.75 -1.08 \, i &  1.40 -0.83 \, i & -1.46 -1.35 \, i \\
 0.91 +0.31 \, i & -0.60 +0.44 \, i &  0.32 -0.05 \, i
\end{bmatrix*}$
\\
\addlinespace
\midrule
\addlinespace
Masses &
$\begin{matrix*}[l]
m_1 \simeq 0.24\times 10^{-3}\text{ eV}\,,\quad & m_2 \simeq 8.76\times 10^{-3}\text{ eV} \,,\quad & m_3 \simeq 5.00\times 10^{-2}\text{ eV} \,,\,\\ 
M_1 \simeq 3.00\text{ eV}\,,\,               & M_2 \simeq 3.16\text{ eV} \,,\,              & M_3 \simeq 1.00\times 10^{14}\text{ GeV}
\end{matrix*}$ \\
\addlinespace
\midrule
\addlinespace
$3\nu$ $\Delta m^2$ &
$\Delta m^2_\odot = \Delta m^2_{21} \simeq 7.66 \times 10^{-5}\text{ eV}^2\,,\,
\quad\,\,\,\,
\Delta m^2_\text{atm} = \Delta m^2_{31} \simeq 2.50 \times 10^{-3}\text{ eV}^2$
\\
\addlinespace
$3\nu$ mixing angles \!\!\!\!\!\!\!\!\!\!\!\!& 
$\sin^2 \theta_{12} \simeq 0.327\,,\,\quad\,\,\,\,
 \sin^2 \theta_{23} \simeq 0.562\,,\,\quad\,\,\,\,
 \sin^2 \theta_{13} \simeq 0.0232$
\\
\addlinespace
$3\nu$ CPV phases \!\!\!\!\!\! & 
$\delta \simeq 1.26 \pi\,,\,\quad\,\,\,\,
\alpha_{21} \simeq 0.11 \pi\,,\,\quad\,\,\,\,
\alpha_{31} \simeq 0.22 \pi$
\\
\addlinespace
\midrule
\addlinespace
$\sin^2 2 \vartheta^{(i)}_{\mu e}$ & 
$\sin^2 2 \vartheta_{\mu e}^{(4)} \simeq 1.1\times 10^{-3}\,,\,\quad\,\,\,\,
 \sin^2 2 \vartheta_{\mu e}^{(5)} \simeq 9.2\times 10^{-4}$
\\
\bottomrule
\end{tabular}
\caption{The same as Table~\ref{tab:Ia} for case \textbf{Ib}.}
\label{tab:Ib}
\end{table}
The numerical data for the benchmark corresponding to this case is given in Table~\ref{tab:Ib}.
In addition to the three mainly active neutrinos with low levels of activity, the spectrum consists of three mainly sterile neutrinos, whose masses are roughly equal to $M_1 \sim M_2 \sim 3$ eV, with a difference of $M_2^2 - M_1^2$ being approximately 1 eV$^2$. The mass of the third neutrino, $M_3$, is around $10^{14}$ GeV.


In the spectrum of scenario \textbf{Ib}, $n=2$ in equation~\eqref{eq:probability}. When applied to the transition probability of muon to electron (anti-)neutrinos in a Long-Baseline (LBL) context, the expression in equation~\eqref{eq:probability} can be approximated using the same formula~\eqref{eq:LBL1Ia} as in scenario \textbf{Ia}.
%
%
Once again,
the last two terms in that equation explicitly show the effects
of the two lightest mostly-sterile states in oscillations.
In a scenario similar to the benchmark outlined in Table~\ref{tab:Ib}, where $|(\Theta\Theta^\dagger){\mu\mu}(\Theta\Theta^\dagger){ee} - 1|$ and $|(\Theta\Theta^\dagger)_{\mu e}|^2$ are considered to be negligible, the expression can be simplified further to:
\begin{align}
P^\text{LBL}_{\stackon[-.7pt]{$\nu$}{\brabar}_\mu \rightarrow \stackon[-.7pt]{$\nu$}{\brabar}_e}
\,\simeq\,
P^{\text{LBL, }3\nu}_{\stackon[-.7pt]{$\nu$}{\brabar}_\mu \rightarrow \stackon[-.7pt]{$\nu$}{\brabar}_e}
\,+\, \frac{1}{2}
\Big[
 \sin^2 2 \vartheta^{(4)}_{\mu e}
+\sin^2 2 \vartheta^{(5)}_{\mu e}
+4\, \re
\left(\Theta_{\mu 4}^*\,\Theta_{e 4}\,\Theta_{\mu 5}\,\Theta_{e 5}^*\right)
\Big] \,,
\label{eq:LBL2Ib}
\end{align}
where the unitarity of the full $6\times 6$ mixing matrix was used,
and the fact that $|(KX^\dag)_{\alpha 1}|^2  \sim |(KX^\dag)_{\alpha 2}|^2 \gg |(KX^\dag)_{\alpha 3}|^2$.
This means that $|(KX^\dag){\alpha 2}|^2$ (and thus $\sin^2 2\vartheta^{(5)}{\mu e}$) cannot be ignored in comparison to $|(KX^\dag){\alpha 1}|^2$ (and $\sin^2 2\vartheta^{(4)}{\mu e}$), as was done in the previous scenario.

%
In a SBL context,
the relevant form of Eq.~\eqref{eq:probability} for
$\stackon[-.7pt]{$\nu$}{\brabar}_\mu \rightarrow \stackon[-.7pt]{$\nu$}{\brabar}_e$
transitions in case \textbf{Ib} is:
\begin{equation}
\begin{aligned}
P^\text{SBL}_{\stackon[-.7pt]{$\nu$}{\brabar}_\mu \rightarrow \stackon[-.7pt]{$\nu$}{\brabar}_e}
\,\simeq\,
&\frac{1}{(\Theta\Theta^\dagger)_{\mu\mu}(\Theta\Theta^\dagger)_{ee}}
\Bigg[ 
\left|(\Theta\Theta^\dagger)_{\mu e}\right|^2
\\ &
-   4 \,\cdot\, \frac{1}{2} \,\re
\left(\Theta_{\mu 4}^*\,\Theta_{e 4}\,\sum_{j=1}^3\,\Theta_{\mu j}\,\Theta_{e j}^*\right)
-   4 \,\cdot\, \frac{1}{2} \,\re
\left(\Theta_{\mu 5}^*\,\Theta_{e 5}\,\sum_{j=1}^3\,\Theta_{\mu j}\,\Theta_{e j}^*\right)
\\ &
-   4 \,\re
\left(\Theta_{\mu 5}^*\,\Theta_{e 5}\,\Theta_{\mu 4}\,\Theta_{e 4}^*\right)
\sin^2 \Delta_{54} 
\pm 2 \,\im
\left(\Theta_{\mu 5}^*\,\Theta_{e 5}\,\Theta_{\mu 4}\,\Theta_{e 4}^*\right)
\sin 2 \Delta_{54}
\Bigg] \,,
\label{eq:SBL1Ib}
\end{aligned}
\end{equation}
where terms depending on the large $\Delta_{4j}$ and $\Delta_{5j}$ ($j=1,2,3$)
have been replaced by their averaged versions.

It can be seen that this scenario does not conform to the typical$3+2$ scenario, outlined in reference~\cite{Gariazzo:2015rra}, as the mass-squared differences $\Delta m^2_{4j}$ and $\Delta m^2_{5j}$ for $j=1,2,3$ are around 10 eV$^2$. As a result, oscillations caused by the mass-squared difference of $\Delta m^2_{54} \sim 1$ eV$^2$ can be detected, while oscillations arising from larger differences are averaged out and those driven by smaller mass-squared differences are underdeveloped.
%
%
If one is in a condition similar to that of
the numerical benchmark, this expression can be approximated by:
\begin{equation}
\begin{aligned}
P^\text{SBL}_{\stackon[-.7pt]{$\nu$}{\brabar}_\mu \rightarrow \stackon[-.7pt]{$\nu$}{\brabar}_e}
\,\simeq\, 
 \frac{1}{2} \Big( \sin^2 2 \vartheta^{(4)}_{\mu e} +\sin^2 2 \vartheta^{(5)}_{\mu e} \Big) 
&+   4 \,\re\left(\Theta_{\mu 4}^*\,\Theta_{e 4}\,\Theta_{\mu 5}\,\Theta_{e 5}^*\right)\cos^2 \Delta_{54}\\
&\mp 2 \,\im\left(\Theta_{\mu 4}^*\,\Theta_{e 4}\,\Theta_{\mu 5}\,\Theta_{e 5}^*\right)\sin 2 \Delta_{54}
\,,
\label{eq:SBL2Ib}
\end{aligned}
\end{equation}
where once again it was taken into account the unitarity of the full mixing matrix and the fact that
$|(KX^\dag)_{\alpha 1}|^2 \sim  |(KX^\dag)_{\alpha 2}|^2 \gg |(KX^\dag)_{\alpha 3}|^2$.
It is important to note that in this scenario, unlike the typical$3+2$ case, oscillations are dependent on the square of the cosine of the corresponding $\Delta_{ij}$.

\begin{figure}[t]
\centering
\includegraphics[width=1.0\linewidth]{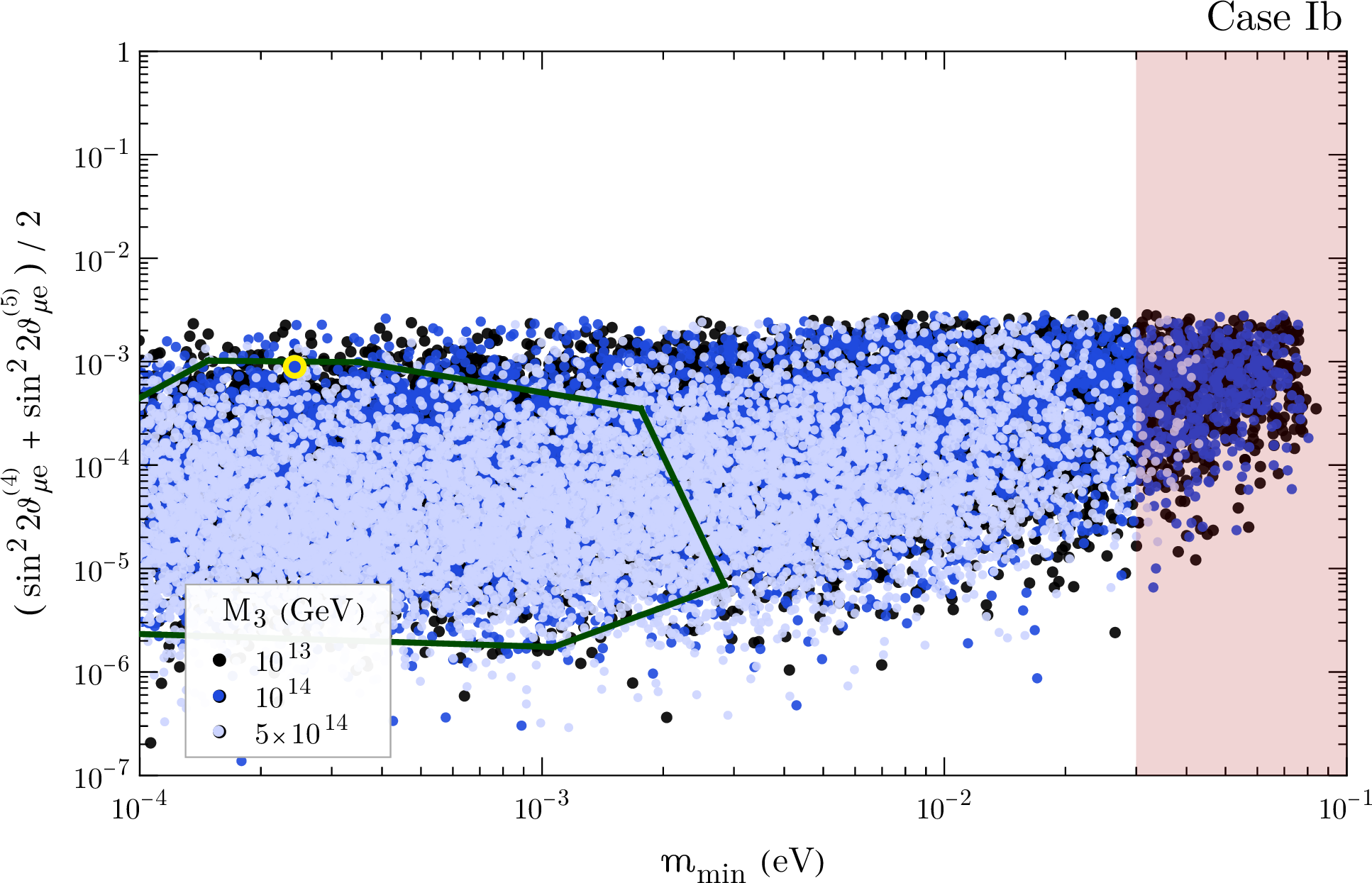}
\caption{The average of $\sin^2 2\vartheta_{\mu e}^{(4)}$ and $\sin^2 2\vartheta_{\mu e}^{(5)}$ 
{\it versus} the lightest neutrino mass $m_\text{min}$, from a scan of the case-\textbf{Ib} parameter space,
with NO ($m_\text{min} = m_1$). The heavy spectrum at tree level has
$M_1 = 3.00$ eV and $M_2 = 3.16$ eV, while three values of
the heaviest mass are considered, $M_3 = 10^{13}\, (10^{14})\, [5\times 10^{14}]$ GeV,
corresponding to the black (dark blue) [light blue] points in the scatter plot.
The vertical red band corresponds to the cosmological constraint, as in Figure~\ref{fig:Ia}.
The dark green contour delimits the region inside which loop-stable points have been found,
while the benchmark of Table~\ref{tab:Ib} is marked in yellow.}
\label{fig:Ib}
\end{figure}
To examine the parameter space of case \textbf{Ib} more thoroughly, numerical seesaw structures similar to the benchmark have been generated using the method outlined in the discussion of case \textbf{Ia}. The lightest two neutrino masses were specified at tree level as $M_1 = 3.00$ eV and $M_2 = 3.16$ eV, and three different values were considered for the heaviest neutrino mass: $M_3 \in {10^{13},~10^{14},~5\times 10^{14}}$ GeV.
In Figure~\ref{fig:Ib}, the average of $\sin^2 2\vartheta_{\mu e}^{(4)}$ and $\sin^2 2\vartheta_{\mu e}^{(5)}$ is plotted against the lightest neutrino mass for the numerical examples found in case \textbf{Ib}.
The latter quantity represents the expected order of magnitude of signals in SBL and LBL experiments. Only the points where $\textrm{Tr}\left[m,m^\dagger\right]$ falls within the range $[0.01,1],v^2$ are included.
As in the previous case, the dark green contour outlines the area where relatively stable points can be found with loop corrections. An increase in the value of $M_3$ will result in a decrease of the light neutrino masses.
The approximations used in deriving the oscillation formulae of Eqs.~\eqref{eq:LBL2Ib} and~\eqref{eq:SBL2Ib}
are valid for all the plotted points.

For all numerical examples pertaining to case \textbf{Ib} which are stable under loop corrections,
$BR(\mu \rightarrow e\gamma) \ll 10^{-30}$ is unobservably small,
while one finds $m_\beta < 9.3$ meV and $|m_{\beta\beta}| < 4.6$ meV,
still out of reach of present and near-future experiments.
In the computation of $|m_{\beta\beta}|$, the effects of both eV-scale
neutrinos are important and have been taken into account.
One additionally finds the bounds $|(KX^\dag)_{e4}|^2,\,|(KX^\dag)_{e5}|^2 \lesssim 0.01$
for the loop-stable numerical examples of this case.

\subsubsection{Case II: \texorpdfstring{$\lambda_\nu = (1,-1,-1)$}{lambda nu = (1,-1,-1)}, \texorpdfstring{$M_1\ll M_2\sim M_3$}{M1 << M2 \textasciitilde{} M3}}
\label{sec:caseII}
For this case, one is instead led to
$\lambda_L = (1,1,1)$. In the exact conservation limit,
the mass matrices are given by: 
\begin{align}
m\,=\,\left( 
\begin{array}{ccc}
a & 0 & 0 \\ 
b & 0 & 0 \\ 
c & 0 & 0%
\end{array}%
\right)\,,\quad
M\,=\,\left( 
\begin{array}{ccc}
0 & A & B \\ 
A & 0 & 0 \\ 
B & 0 & 0%
\end{array}%
\right)\,.
\label{eq:MII}
\end{align}
In this limit, there are two degenerate neutrinos with mass $\sqrt{|A|^2+|B|^2}$ and opposite CP parities, forming a single heavy Dirac particle.

Breaking the symmetry will generate light neutrino masses and another massive sterile state with a mass that can be much smaller than $|A|$ and $|B|$. Additionally, it will lift the mass degeneracy for the Dirac neutrino, resulting in a pseudo-Dirac neutrino pair~\cite{Wolfenstein:1981kw,Petcov:1982ya}.

As noted in~\cite{Ibarra:2011xn}, a strong mass degeneracy implies a symmetry in the $KX^\dag$ block of the mixing matrix, specifically $(KX^\dag){\alpha 2}\simeq \pm i (KX^\dag){\alpha 3}$ ($\alpha = e,\mu,\tau$).
Such a relation has a crucial impact in reducing the effect of large masses $M_2$ and $M_3$ on the one-loop correction $\delta M_L$ as seen in Eq.~\eqref{eq:oneloop}. This relation indicates a proximity to the limit of lepton number conservation, even if $(KX^\dag)_{\alpha 2}$ and $(KX^\dag)_{\alpha 3}$ are not extremely suppressed, which allows for relatively large Yukawa couplings, in spite of the fact that $M_2 \simeq M_3$ are not as large as the $M_3$ in case I. This is evident from the form of the Dirac mass matrix $m$, defined in eq.~\eqref{eq:mdirac}. The mass of the pseudo-Dirac pair can be in the TeV range \cite{Ibarra:2010xw,Ibarra:2011xn,Dinh:2012bp,Cely:2012bz,Penedo:2017knr}, as the lightest neutrino masses are safeguarded by approximate lepton number conservation. The same symmetry and related effects are also observed in the examples discussed in section \ref{sec:quasi}.\\

\begin{table}[h!]
\addtocounter{table}{-1}
\renewcommand{\thetable}{\arabic{table}c}
\centering
\renewcommand{\arraystretch}{1.2}
\begin{tabular}{lr}
\toprule
 & {\bf Case II} numerical benchmark \\
\midrule
\addlinespace
$m$ (GeV) &
$\begin{bmatrix*}[r]
-4.15 + 0.47 \,i & ( 4.51 - 1.49 \,i)\times 10^{-9} & (-1.59 + 0.13 \,i)\times 10^{-9} \\
 3.98 + 6.17 \,i & (-5.04 - 4.64 \,i)\times 10^{-9} & ( 1.52 + 2.31 \,i)\times 10^{-9} \\
 1.53 + 6.58 \,i & (-1.90 - 2.68 \,i)\times 10^{-9} & ( 0.59 + 2.59 \,i)\times 10^{-9}
 \end{bmatrix*}$ \\
\addlinespace
$M$ (GeV)  & 
$\begin{bmatrix*}[r]
2.18\times 10^{-6} &  1390               & 2.96               \\
1390               & -2.19\times 10^{-6} & 5.52\times 10^{-7} \\
2.96               &  5.52\times 10^{-7} & 3.33\times 10^{-9}
\end{bmatrix*}$
\\
\addlinespace
\midrule
\addlinespace
$K$  &
$\begin{bmatrix*}[r]
 0.825 +0.061 \,i &  0.536 +0.027 \,i & -0.092 +0.108 \,i \\
-0.302 +0.113 \,i &  0.581 -0.017 \,i &  0.728 -0.052 \,i \\
 0.455 +0.054 \,i & -0.599 +0.075 \,i &  0.651 +0.002 \,i
 \end{bmatrix*}$
\\
\addlinespace
$KX^\dag$  &
$\begin{bmatrix*}[r]
 0.063 -0.056 \,i & ( 2.11 -0.24 \,i)\times 10^{-3} & (-0.24 -2.11 \,i)\times 10^{-3} \\
-0.066 -0.147 \,i & (-2.03 -3.13 \,i)\times 10^{-3} & (-3.13 +2.03 \,i)\times 10^{-3} \\
-0.021 -0.036 \,i & (-0.79 -3.35 \,i)\times 10^{-3} & (-3.35 +0.79 \,i)\times 10^{-3} 
\end{bmatrix*}$
\\
\addlinespace
$X$  &
$\begin{bmatrix*}[r]
 0.042 +0.014 \,i & 0.007 +0.099 \,i & -0.069 +0.140 \,i \\
( 1.48 +0.64 \,i)\times 10^{-3} & ( 2.30 +0.66 \,i)\times 10^{-4} & (-2.11 +4.89 \,i)\times 10^{-3} \\
(-0.64 +1.48 \,i)\times 10^{-3} & (-0.66 +2.30 \,i)\times 10^{-4} & (-4.89 -2.11 \,i)\times 10^{-3}
\end{bmatrix*}$
\\
\addlinespace
$O_c$ (tree level) \!\!\!\!\!\! \!\!\!\!\!\!\!\!\!\!\!\!& 
$\begin{bmatrix*}[r]
-0.21 +0.62 \,i & -1.02 +0.06 \,i & 0.62 + 0.31 \,i \\
(-1.11 +2.54 \,i)\times 10^4 & (-1.05 +2.42 \,i)\times 10^3 & ( 2.55 +1.12 \,i)\times 10^4 \\
(-2.54 -1.11 \,i)\times 10^4 & (-2.42 -1.05 \,i)\times 10^3 & (-1.12 +2.55 \,i)\times 10^4 
\end{bmatrix*}$
\\
\addlinespace
\midrule
\addlinespace
Masses &
$\begin{matrix*}[l]
m_1 \simeq 4.65\times 10^{-3}\text{ eV}\,,\quad & m_2 \simeq 9.47\times 10^{-3}\text{ eV} \,,\quad & m_3 \simeq 5.01\times 10^{-2}\text{ eV} \,,\,\\ 
M_1 \simeq 1.00\text{ eV}\,,\,               & M_2 \simeq 1390\text{ GeV} \,,\,              & M_3 \simeq 1390\text{ GeV}
\end{matrix*}$ \\
\addlinespace
\midrule
\addlinespace
$3\nu$ $\Delta m^2$ &
$\Delta m^2_\odot = \Delta m^2_{21} \simeq 6.80 \times 10^{-5}\text{ eV}^2\,,\,
\quad\,\,\,\,
\Delta m^2_\text{atm} = \Delta m^2_{31} \simeq 2.48 \times 10^{-3}\text{ eV}^2$
\\
\addlinespace
$3\nu$ mixing angles \!\!\!\!\!\!\!\!\!\!\!\!& 
$\sin^2 \theta_{12} \simeq 0.298\,,\,\quad\,\,\,\,
 \sin^2 \theta_{23} \simeq 0.563\,,\,\quad\,\,\,\,
 \sin^2 \theta_{13} \simeq 0.0212$
\\
\addlinespace
$3\nu$ CPV phases \!\!\!\!\!\! & 
$\delta \simeq 1.32 \pi\,,\,\quad\,\,\,\,
\alpha_{21} \simeq 1.99 \pi\,,\,\quad\,\,\,\,
\alpha_{31} \simeq 0.02 \pi$
\\
\addlinespace
\midrule
\addlinespace
$\sin^2 2 \vartheta^{(i)}_{\mu e}$ & 
$\sin^2 2 \vartheta_{\mu e}^{(4)} \simeq 7.4\times 10^{-4}$
\\
\bottomrule
\end{tabular}
\caption{The same as Table~\ref{tab:Ia} for case \textbf{II}.
For this benchmark, $M_3 - M_2 \simeq 7.6$ eV $\ll M_{2,3}$.}
\label{tab:II}
\end{table}
The numerical information for the benchmark scenario of case \textbf{II} is provided in Table~\ref{tab:II}.
In this case, apart from the three light mostly-active neutrinos, there is also a mostly-sterile neutrino with a mass around $1$ eV and a pair of quasi-degenerate neutrinos with masses close to $1$ TeV.
The light neutrino mass spectrum in this scenario is similar to that of the conventional $\nu$MSM~\cite{Asaka:2005an,Asaka:2005pn}. However, in the $\nu$MSM, the lightest neutrino has a mass of about 1 keV, making it a warm dark matter candidate, while the two heavier neutrinos with masses around 1 GeV generate the baryon asymmetry in the universe. In contrast, in the scenario under consideration, the lightest neutrino has a mass around 1 eV and the two heavier neutrinos have masses around 1 TeV.

From Table~\ref{tab:II} one sees that the symmetry in the last two columns of $KX^\dag$ is tied to an analogous symmetry in the last two rows of $X$ and of $O_c$.
The latter can be understood from Eqs.~\eqref{eq:calV} and \eqref{eq:Xdef}.

The symmetry observed in the last two columns of $KX^\dag$ is also reflected in the last two rows of $X$ and $O_c$, as seen in Table~\ref{tab:II}. This is due to the relationship between $X$, $O_c$, and $KX^\dag$ as described in Eqs.~\eqref{eq:calV} and \eqref{eq:Xdef}.


For the case \textbf{II} spectrum, one should consider $n=1$ in the expression in Eq.\eqref{eq:probability}. In a long-baseline experiment, the transition probability of muon to electron (anti-)neutrinos can be approximately expressed as described in Eq.\eqref{eq:probability}
\begin{equation}
\begin{aligned}
P^\text{LBL}_{\stackon[-.7pt]{$\nu$}{\brabar}_\mu \rightarrow \stackon[-.7pt]{$\nu$}{\brabar}_e}
\,\simeq\,
&\frac{1}{(\Theta\Theta^\dagger)_{\mu\mu}(\Theta\Theta^\dagger)_{ee}}
\Bigg[ 
\left|(\Theta\Theta^\dagger)_{\mu e}\right|^2
-   4 \,\cdot\,\frac{1}{2}\,\re
\left(\Theta_{\mu 4}^*\,\Theta_{e 4}\,\sum_{j=1}^3\,\Theta_{\mu j}\,\Theta_{e j}^*\right)
\\ &
-   4 \sum_{i>j}^3\,\re
\left(\Theta_{\mu i}^*\,\Theta_{e i}\,\Theta_{\mu j}\,\Theta_{e j}^*\right)
\sin^2 \Delta_{ij} 
\pm 2 \sum_{i>j}^{3}\,\im
\left(\Theta_{\mu i}^*\,\Theta_{e i}\,\Theta_{\mu j}\,\Theta_{e j}^*\right)
\sin 2 \Delta_{ij}
\Bigg] \,,
\label{eq:LBL1II}
\end{aligned}
\end{equation}
where terms depending on $\Delta_{4j} \gg 1$ 
have been replaced by their averaged versions.
%
%
If the conditions are similar to those of the benchmark in Table~\ref{tab:II} where $|(\Theta\Theta^\dagger){\mu\mu}(\Theta\Theta^\dagger){ee} - 1|$ and $|(\Theta\Theta^\dagger)_{\mu e}|^2$ are negligible, the expression for the muon to electron (anti-)neutrino transition probability in a LBL context can be simplified to:
\begin{align}
P^\text{LBL}_{\stackon[-.7pt]{$\nu$}{\brabar}_\mu \rightarrow \stackon[-.7pt]{$\nu$}{\brabar}_e}
\,\simeq\,
P^{\text{LBL, }3\nu}_{\stackon[-.7pt]{$\nu$}{\brabar}_\mu \rightarrow \stackon[-.7pt]{$\nu$}{\brabar}_e}
+ \frac{1}{2}\sin^2 2 \vartheta^{(4)}_{\mu e}
+ 4\,\re\left(\Theta_{\mu 4}^*\,\Theta_{e 4}\,(KX^\dag)_{\mu 2}\,(KX^\dag)_{e 2}^*\right)
 \,.
\label{eq:LBL2II}
\end{align}
In this case, the unitarity of the full $6\times 6$ mixing matrix has been utilized, as well as the approximate symmetry $(KX^\dag){\alpha 2}\simeq i, (KX^\dag){\alpha 3}$. Additionally, if $|\Theta_{\alpha 4}|^2 = |(KX^\dag){\alpha 1}|^2 \gg |(KX^\dag){\alpha 2}|^2 \simeq |(KX^\dag)_{\alpha 3}|^2$, the last term in the expression can be ignored, resulting in Eq.~\eqref{eq:LBL2Ia} from case \textbf{Ia}.

%
In a SBL context,
the relevant form of Eq.~\eqref{eq:probability} for
$\stackon[-.7pt]{$\nu$}{\brabar}_\mu \rightarrow \stackon[-.7pt]{$\nu$}{\brabar}_e$
transitions in case \textbf{II} is:
\begin{equation}
\begin{aligned}
P^\text{SBL}_{\stackon[-.7pt]{$\nu$}{\brabar}_\mu \rightarrow \stackon[-.7pt]{$\nu$}{\brabar}_e}
\,\simeq\,
\frac{1}{(\Theta\Theta^\dagger)_{\mu\mu}(\Theta\Theta^\dagger)_{ee}}
\Bigg[ 
\left|(\Theta\Theta^\dagger)_{\mu e}\right|^2
&-   4 \,\re
\left(\Theta_{\mu 4}^*\,\Theta_{e 4}\,\sum_{j=1}^3\,\Theta_{\mu j}\,\Theta_{e j}^*\right)
\sin^2 \Delta_{41} 
\\ &
\pm 2 \,\im
\left(\Theta_{\mu 4}^*\,\Theta_{e 4}\,\sum_{j=1}^{3}\,\Theta_{\mu j}\,\Theta_{e j}^*\right)
\sin 2 \Delta_{41}
\Bigg] \,,
\label{eq:SBL1II}
\end{aligned}
\end{equation}
with $\Delta_{41}\simeq \Delta_{42}\simeq \Delta_{43}$.

One is sensitive to oscillations due to the mass-squared differences $\Delta m^2_{4j}$ with $j=1,2,3$, while the oscillations related to smaller mass-squared differences have not yet emerged.
%
%
In a situation resembling the numerical benchmark, this expression can be estimated as:
\begin{equation}
\begin{aligned}
P^\text{SBL}_{\stackon[-.7pt]{$\nu$}{\brabar}_\mu \rightarrow \stackon[-.7pt]{$\nu$}{\brabar}_e}
\,&\simeq\, 
\left[
\sin^2 2 \vartheta^{(4)}_{\mu e} + 8\,\re\left(\Theta_{\mu 4}^*\,\Theta_{e 4}\,(KX^\dag)_{\mu 2}\,(KX^\dag)_{e 2}^*\right)
\right]\sin^2 \Delta_{41} \\
&\mp 4\,\im\left(\Theta_{\mu 4}^*\,\Theta_{e 4}\,(KX^\dag)_{\mu 2}\,(KX^\dag)_{e 2}^*\right)\sin 2 \Delta_{41}
\,,
\label{eq:SBL2II}
\end{aligned}
\end{equation}
where once again the unitarity of the full mixing matrix has been taken into account,
as well as the relation $(KX^\dag)_{\alpha 2}\simeq i\, (KX^\dag)_{\alpha 3}$. 
If also $|(KX^\dag)_{\alpha 1}|^2\gg |(KX^\dag)_{\alpha 2}|^2 \simeq |(KX^\dag)_{\alpha 3}|^2$,
then the two terms containing $(KX^\dag)_{\alpha 2}$ in this equation
can be neglected and one recovers Eq.~\eqref{eq:SBL2Ia} of case \textbf{Ia}.

\begin{figure}[t]
\centering
\includegraphics[width=1.0\linewidth]{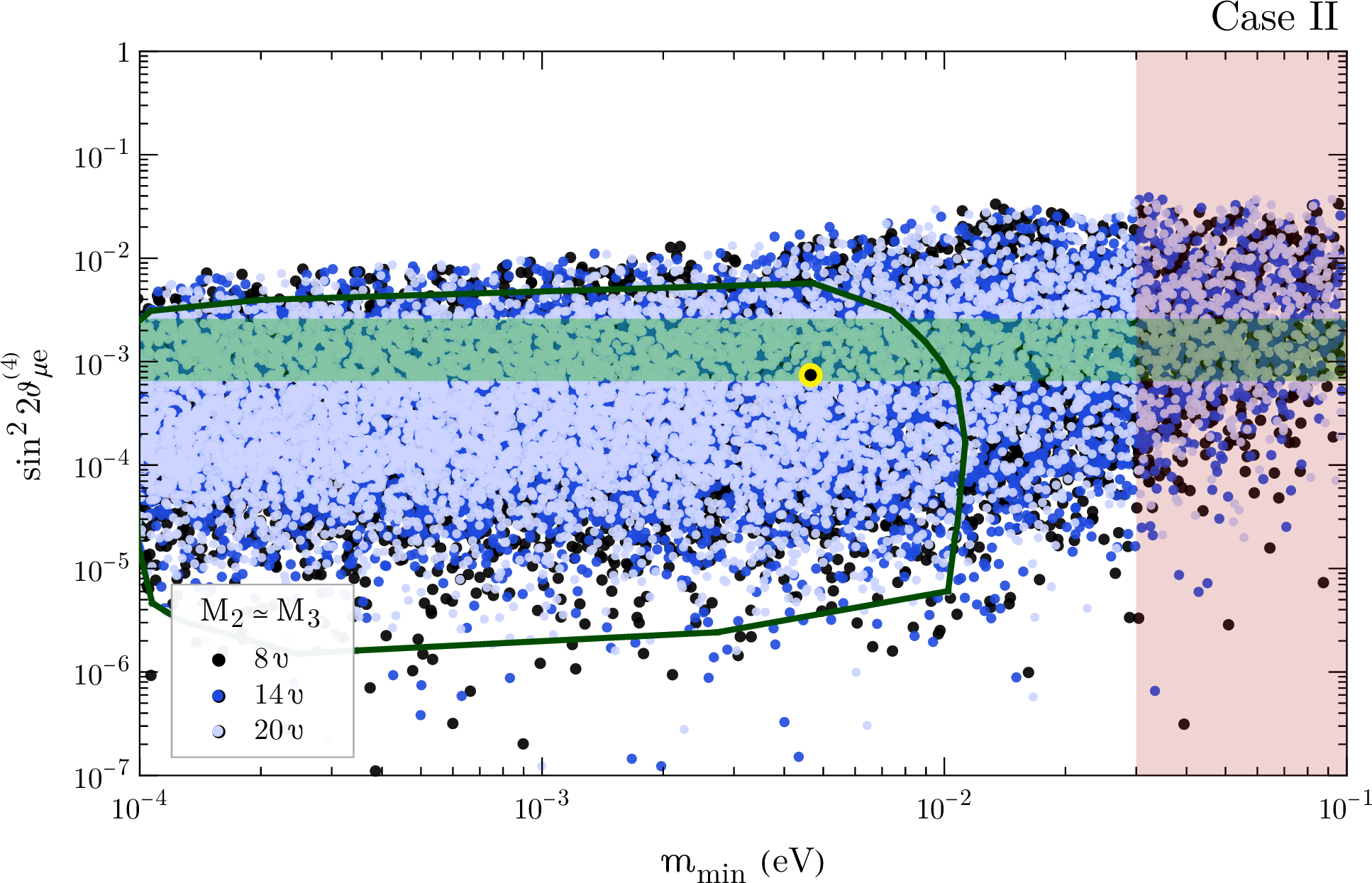}
\caption[Active-sterile mixing measure $\sin^2 2\vartheta_{\mu e}^{(4)}$ 
{\it versus} the lightest neutrino mass $m_\text{min}$ from a scan of the case-\textbf{II} parameter space,
with NO ($m_\text{min} = m_1$). The heavy spectrum at tree level has
$M_1 = 1$ eV, while three values of the heaviest quasi-degenerate masses are considered,
$M_2 \simeq M_3 = 8\,v\, (14\,v)\, \{20\,v\}$,
corresponding to the black (dark blue) \{light blue\} points in the scatter plot.
Here, $v \simeq 174$ GeV is the Higgs VEV.
The horizontal green band shows the $99.7\%$ CL interval,
and the vertical red band corresponds to the cosmological constraint, as in Figure~\ref{fig:Ia}.
The dark green contour delimits the region inside which loop-stable points have been found,
while the benchmark of Table~\ref{tab:II} is marked in yellow.]{
Active-sterile mixing measure $\sin^2 2\vartheta_{\mu e}^{(4)}$ 
{\it versus} the lightest neutrino mass $m_\text{min}$ from a scan of the case-\textbf{II} parameter space,
with NO ($m_\text{min} = m_1$). The heavy spectrum at tree level has
$M_1 = 1$ eV, while three values of the heaviest quasi-degenerate masses are considered,
$M_2 \simeq M_3 = 8\,v\, (14\,v)\, [20\,v]$,
corresponding to the black (dark blue) [light blue] points in the scatter plot.
Here, $v \simeq 174$ GeV is the Higgs VEV.
The horizontal green band shows the $99.7\%$ CL interval of Ref.~\cite{Gariazzo:2015rra},
and the vertical red band corresponds to the cosmological constraint, as in Figure~\ref{fig:Ia}.
The dark green contour delimits the region inside which loop-stable points have been found,
while the benchmark of Table~\ref{tab:II} is marked in yellow.}
\label{fig:II}
\end{figure}

To further explore the parameter space of case \textbf{II}, numerical seesaw structures were generated resembling the benchmark in Table~\ref{tab:II}, by adopting a method similar to case \textbf{Ia}. It was set $M_1 = 1$ eV at tree level and three different values for the second heaviest neutrino mass, $M_2, (\simeq M_3) \in {8, 14, 20}~,~v$, where $v\simeq 174$ GeV represents the Higgs VEV. The mass splitting $M_3 - M_2$ was varied within the range $[0.02,200]$ eV. Figure~\ref{fig:II} displays the values of $\sin^2 2\vartheta_{\mu e}^{(4)}$ in Eq.~\eqref{eq:sinmue} plotted against the lightest neutrino mass for the numerical examples in case \textbf{II}. Only points for which $\textrm{Tr}\left[m\,m^\dagger\right] \in [0.001,\,1]\,v^2$ are kept.
Similar to before, the green band across the horizontal axis represents the range of $\sin^2 2\vartheta_{\mu e}^{(4)}$ preferred by the global fit described in Ref.\cite{Gariazzo:2015rra}. The dark green contour outlines the area where relatively loop-stable points are present.
The approximations used in deriving the oscillation formulae of Eqs.~\eqref{eq:LBL2II} and~\eqref{eq:SBL2II}
are valid for all the plotted points.

For the numerical examples of case \textbf{II} that remain stable under loop corrections, the values of $BR(\mu \rightarrow e\gamma)$ can approach the upper bound of $4.2 \times 10^{-13}$ set by MEG. Our scan eliminated points with higher branching ratios. The benchmark in Table~\ref{tab:II} yields $BR(\mu \rightarrow e\gamma) \simeq 2.0 \times 10^{-13}$. These effects are detectable with the MEG II update~\cite{Cattaneo:2017psr}, which is expected to improve the current sensitivity of MEG by an order of magnitude.
For the loop-stable numerical examples in this case, the following limits are obtained: $m_\beta < 15$ meV, $|m_{\beta\beta}| < 27$ meV, and $|(KX^\dag){e4}|^2 \lesssim 0.02$. While KATRIN aims to improve the current limit on $m\beta$ to $0.2$ eV, the next generation of $(\beta\beta){0\nu}$-decay experiments~\cite{Vergados:2016hso} may be able to probe values of $|m{\beta\beta}| \gtrsim 10^{-2}$ eV.

Concerning the prospect of detecting the heavy neutrino pair in future collider searches, the review~\cite{Antusch:2016ejd} is recommended. If the heavy neutrino pair has a mass in the $1-100$ GeV range and is sufficiently long-lived, it could potentially produce displaced vertex signatures~\cite{Graesser:2007pc,Helo:2013esa,Izaguirre:2015pga,Gago:2015vma,Antusch:2016vyf,Antusch:2017hhu,Cottin:2018kmq,Cvetic:2018elt,Cottin:2018nms,Abada:2018sfh,Drewes:2018xma,Drewes:2019fou} and resolvable neutrino-antineutrino oscillations at colliders~\cite{Antusch:2017ebe}.
Finally, the pseudo-Dirac pair of case \textbf{II} might play a role
in explaining the baryon asymmetry of the Universe
through resonant leptogenesis~\cite{Pilaftsis:2003gt,DeSimone:2007edo}.

However, one should take into account the washout from the interactions of the lighter sterile neutrino
species and to reconcile the light sterile neutrino paradigm with cosmology, it may be necessary to consider non-standard interactions.

\vskip 2mm
The numerical examples presented are only for illustration purposes, but they support the claim that models with an approximate lepton number symmetry and an eV-scale sterile neutrino mass are viable and may contribute to the explanation of SBL anomalies.

\section{General Remarks}

 \subsection{On CP Violating Phases}
\label{sec:leptogenesis}
In table \ref{tab:parameters}, the physical parameters of the 3$\nu$SM are summarised. It is important to parametrise all 6 phases.
 A general parameterisation for a $3\times 3$ complex orthogonal $O_c$ is given by \cite{Chrzaszcz:2019inj,Davidson:2008bu}
\begin{equation}
    O_c=\pm~O_c^{23}O_c^{13} O_c^{12} ~,
\end{equation}
where, for instance,
\begin{equation}
    O_c^{23} = \begin{pmatrix}
    1&0&0\\
     0 & \cos{(r_{23}e^{i f_{23}})} &   \sin{(r_{23}e^{i f_{23}})}\\
     0 & -\sin{(r_{23}e^{i f_{23}})} &\cos{(r_{23}e^{i f_{23}})}
    \end{pmatrix}
\end{equation}
and $O_c^{12}$, $O_c^{13}$ have an obviously similar definition.
$U_K$ has a well-known definition \cite{pdg}, as the $3 \times 3$ unitary part of the leptonic mixing matrix, equivalent to the parameterisation of $U_{\textrm{PMNS}}$ in eq. \eqref{eq:Vstd}.
It is important to note that, eq. \eqref{eq:khu} can be rewritten as
\begin{equation}
    K=(\id_3 -\eta) U_K = U_K~H_R ~,
    \label{eq:kh}
\end{equation}
the unitary part on both decompositions is equal and the hermitian parts are related by the equation
\begin{equation}
    H_R= (\id_3+X^\dag X)^{-1/2} =U_K^\dag~ (\id_3 -\eta) U_K ~,
    \label{eq:kh2}
\end{equation}
by the polar decomposition theorem. Due to the rephasing of the charged leptons, 3 phases can be removed from the left of $K$. The phases of $U_K$ are named the \textbf{unitary phases} of the mixing matrix: $\delta$ and the Majorana phases $\alpha_1$, $\alpha_2$, defined in eq. \eqref{eq:Vstd}.
The phases of $H_R$, $p_{ij}$, are named the \textbf{hermitian phases} of the mixing matrix, or the phases of the deviations from unitarity.
\begin{equation}
    H_R= \begin{pmatrix}
h_{11} & h_{12} e^{i p_{12}} & h_{13} e^{i p_{13}}\\ 
 h_{12} e^{-i p_{12}} & h_{22}  & h_{23} e^{i p_{23}}\\ 
h_{13} e^{-i p_{13}} & h_{23} e^{-i p_{23}} & h_{33} \\ 
\end{pmatrix} ~,
\label{eq:hermitian_part}
\end{equation}
 where $p_{ij}$ are linear functions of the $O_c$ phases $f_{ij}$. This means that if all $f_{ij}$ can be written as $ k \pi$ then the same can be done for $p_{ij}$, with $k \in \N$. In the literature, the unitary phases are also known as low-energy phases and the hermitian phases as high-energy phases \cite{Molinaro:2009lud,Branco:2011zb,Pascoli:2006ci,Branco:2002ws}. This is because if deviations from $3 \times 3$ unitarity are negligible the hermitian phases are essentially impossible to measure at low energy, and only have effects at high energy, possibly being the source of leptogenesis.\\
Thus, the 6 physical phases in the $3\nu SM$ are controlled by $\delta$, $\alpha_1$, $\alpha_2$ - the phases of $U_K$ and $f_{12}$, $f_{13}$ and $f_{23}$ - the phases of $O_c$ ( $p_{12}$, $p_{13}$ and $p_{23}$ - the phases of $H_R$).

 \subsection{On the Parameter Space}
 \label{sec:gen_neutrinos}

 The parameterisation used in this study was developed in Ref.~\cite{Agostinho:2017wfs} for heavy neutrinos with TeV-scale masses, in the {\it quasi-decoupled} regime. This is distinct from the {\it decoupled/seesaw} regime, which is more commonly studied in the literature and where heavy neutrinos are entirely decoupled from light ones, and deviations from $3 \times 3$ unitarity are negligible. This work demonstrates that this parameterisation can also be applied to other spectra of heavy neutrinos, including {\it non-decoupled} regimes with heavy almost-sterile neutrinos with masses close to the light ones. Unlike other parameterisations used in the literature, this one can be used for this purpose without significant numerical deviations from the exact result. In this discussion, $m_i$ refers to the masses of light neutrinos and $M_j$ refers to the masses of heavy neutrinos.\par

Regarding its flexibility, in section \ref{sec:quasi} it was discussed the quasi-decoupled case where $m_i/M_j\ll 1$, and from Eq.\eqref{eq:Xdef}, it is clear that $X$ can only have significant entries and deviations from unitarity if $O_c$ also has considerable ($\gg 1$) entries. On the other hand, for the Dirac matrix given in Eq.\eqref{eq:mdiracWB}, and considering the perturbativity condition $m_{ij} \lesssim m_t$, the size of the product $O_c^\dag \sqrt{D}$ is limited, which can counterbalance the suppression from $\sqrt{d}$.\par
Using the same logic for the {\it decoupled/seesaw} regime, where heavy neutrino masses are near the GUT scale, one can observe that $X$ can only have significant entries if $O_c$ also has significant entries. However, achieving an $X$ with substantial entries in this regime would require entries in $O_c$ that are not compatible with perturbative entries in $m$ generated by electroweak symmetry breaking (EWSB). As a result, $O_c$ must have entries of $\mathcal{O}(1)$, which results in small entries for $X$ and small deviations from unitarity.
The non-decoupled regimes discussed in section \ref{sec:non-dec} present similarities to both the aforementioned regimes. 

In those cases, when the complex orthogonal matrix $O_c$ has entries of order $\mathcal{O}(1)$ (which is "small" for such a matrix), it allows for $X$ to have significant entries, leading to connections with short-baseline anomalies. This is because some of the ${m_i}/{M_j}$ can be around $\mathcal{O}(10^{-2}\text{--}10^{-3})$ when $M_{1(,2)} \sim \text{eV}$, as shown in Eq.~\eqref{eq:Xdef}.
However, embedding this eV neutrino in a type-I seesaw with 3 sterile neutrinos and with reasonable Yukawa couplings $m_{ij}$ of EWSB origin, $m_{ij} \sim \mathcal{O}(1-0.01)\, m_t$, while keeping keeping $X$ sizeable, leads to non-trivial constraints on $O_c$.
In particular, 
spectra where the highest heavy neutrino(s) mass(es) is (are) not extremely large (e.g.~at the GeV\,--\,TeV scale) will need an $O_c$ with some sizeable entries like in the quasi-decoupled regime. On the other hand, spectra where one of the $M_j$ is near the GUT scale require an $O_c$ with $\mathcal{O}(1)$ entries, like in the decoupled regime. These requirements on the size of some entries of $O_c$ are necessary to achieve reasonable Yukawa couplings. \par 

Other parameterisations in the literature, such as those in \cite{Chrzaszcz:2019inj,Fernandez-Martinez:2016lgt}, also begin with an exact parameterisation of $\mathcal{V}$. However, these approaches define the mixing matrix using an infinite series in a matrix that controls the deviations from unitarity. To perform explicit calculations, a truncated series is necessary, leading to a good approximation only for $M_j>100$ GeV, as discussed in section \ref{sec:quasi}, and resulting in small deviations from unitarity. This parameterisation is not suitable for scenarios in which deviations from unitarity are significant, as in the case of a non-decoupled sterile neutrino sector. The parameterisation used in this work, however, is appropriate for such scenarios and may provide a direct connection to symmetries in an underlying model.\par

\begin{figure}[h!]
\centering
\includegraphics[width=1.2\linewidth]{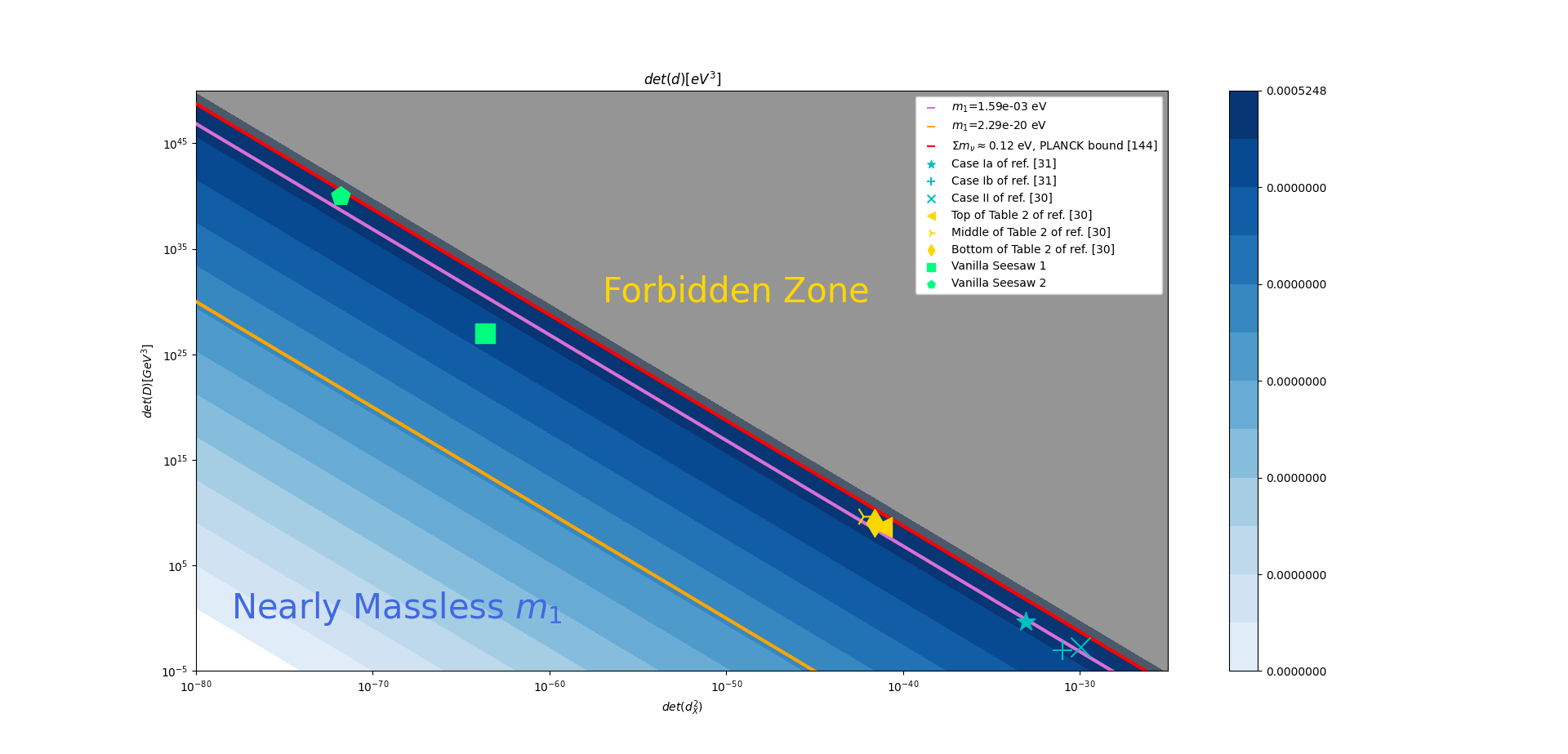}
\caption[Projection of the parameter space of the $n_R\nu$SM in the deviations from unitarity - heavy neutrino masses space. The pink and orange lines are just placeholders to depict the log scale used and to show how fast $m_1$ approaches zero for lighter tones of blue.]{Projection of the parameter space of the $n_R\nu$SM in the deviations from unitarity - heavy neutrino masses space. The pink and orange lines are just placeholders to depict the log scale used and to show how fast $m_1$ approaches zero for lighter tones of blue.}
\label{fig:generalfig}
\end{figure}

Figure \ref{fig:generalfig} is an illustration of the whole parameter space of the $n_R\nu$SM. It is a density plot of the determinant of $d$, hence the darkest the blue the higher the value of
\begin{equation}
    det (d) = m_1 m_2 m_3 ~.
\end{equation}
In the $x$-axis one can find
\begin{equation}
    det (d_X^2) = (d_X^1d_X^2d_X^3)^2 ~,
\end{equation}
where $(d_X^i)^2$ are the eigenvalues of $X^\dag X$ and $d_X^2$ a diagonal matrix with them in the diagonal in increasing order. Hence, $det (d_X^2)$ is a good theoretical measure of the size of deviations from $3 \times 3$ unitarity. Note that due to the properties of orthogonal complex matrices, see. eq. \eqref{eq:Xdef}, one will always have $(d_X^3)^2 \gg (d_X^2)^2 > (d_X^1)^2$.

As for the $y$-axis one has
\begin{equation}
    det (D) = M_1 M_2 M_3 ~.
\end{equation}

 It is easy to distinguish the vanilla seesaw region - green symbols - at high $det(D)$ and low $det(d_X^2)$, the quasi-decoupled region, in the middle range of $det(D)$ and $det(d_X^2)$ and the non-decoupled region - at low $det(D)$ and high $det(d_X^2)$.

 The red line corresponds to the value of $det(d)$ for which the sum of all light neutrino masses matches the bound from Planck \cite{Planck:2018vyg}, one of the most robust bounds \cite{pdg}. This can be deducted because one can write the light neutrino masses as \cite{pdg}

 \begin{equation}
     m_1 = \sqrt{\Delta m_{31}^2 z} ~,~ m_2 = \sqrt{\Delta m_{31}^2 (z+a)} ~,~ m_3 = \sqrt{\Delta m_{31}^2 (z+1)}
 \end{equation}
where 
\begin{equation}
   \Delta m_{ij}^2= m_i^2 - m_j ^2 ~,~ a = \frac{\Delta m_{21}^2}{\Delta m_{31}^2} ~,
\end{equation}
with $z$ being a free parameter. The current best-fit values for normal ordering are
\begin{equation}
     \Delta m_{21}^2 = 7.50 \times 10^{-5} ~ \textrm{eV}^2 ~,~ \Delta m_{31}^2 = 2.55 \times 10^{-3}  \textrm{eV}^2 ~,~ a \approx 0.02941 ~.
\end{equation}

Thus, using these experimental values and the bound from PLANCK \cite{Planck:2018vyg}, one can solve numerically for $z$ and obtain a value for $det(d)$, which is only a function of $z$, as the $m_i$, when the $\Delta m_{ij}^2$ are fixed

\begin{equation}
    m_1 (z) + m_2 (z) + m_3 (z) = 0.12 ~\textrm{eV} \implies z \approx 0.355 \implies det(d) \approx 5.54 \times 10^{-5} ~\textrm{eV}^3 ~.
\end{equation}

Hence, values of $z$ larger than $0.355$, which translates into
\begin{equation}
    m_1 > 0.0301 ~\textrm{eV} ~,~ m_2 > 0.0313 ~\textrm{eV}  ~,~ m_3 > 0.0588 ~\textrm{eV}~,
\end{equation}
 are ruled out.
It's interesting to note that most of this ruled out region, above the red line, would be ruled out due to perturbativity. This is because most of the area of the forbidden region is for high values of $det(D)$ and high values of $det(d_X^2)$. From eq. \eqref{eq:mneutrino} it is easy to see that 
\begin{equation}
    det(m) \sim det(X^\dag)~det (D) ~,
\end{equation}
and too large values of $det(D)$ and $det(d_X^2)$ can easily yield non-perturbative Yukawa couplings.
Before proceeding to another section, it is relevant to discuss how models with very light neutrino masses, $z \ll 1$, are still allowed, although all the points drawn in the plot, taken from published papers, are for higher values of $m_1$.
From eq. \eqref{eq:Xdef} one can understand that very small light neutrino masses ($m_1 \rightarrow 0$) introduces an extra suppression factor on $X$, beyond the usual $D^{-1/2}$. Nonetheless, although more challenging, it is still possible to have sizeable deviations from unitarity, by choosing an appropriate $O_c$, with large enough eigenvalues.

 \chapter{The Standard Model + $n_u$ up-type vector-like quarks + $n_d$ down type vector-like quarks ($n_u$SM$n_d$)}
\label{chapter:framework_quarks}

Vector-like quarks are quarks where the right-handed and the left-handed components transform
in the same way under the gauge group. As a consequence of this, all their interactions are vector-like or, in other words, not “chiral”. That is, wherever the left-handed component of the VLQ can be introduced in the Lagrangian, the same can be done for the right-handed component. This does not happen for the SM fields, since left-handed components are part of a $SU(2)$ doublet and right-handed components are $SU(2)$ singlets. This particularity of VLQs allows the existence of bare mass-terms, meaning that the VLQ mass might have an origin beyond the Higgs mechanism. Hence, their masses can be much larger than the Higgs vev, while still leaving considerable traces at low-energy.

There are only seven different types of VLQ multiplets which may mix with the SM quarks through Yukawa couplings with the Higgs doublet~\cite{delAguila:2000aa,delAguila:2000rc}.
They consist of isosinglets, doublets or triplets, all coloured in the fundamental representation. The complete quantum numbers are listed in Table~\ref{tab:VLQ:irreps}.\par
%
%
\begin{table}[ht!]
  \centering
  \begin{tabular}{cccccccc}
    \toprule    
    Multiplet
 & $U$ & $D$ & $\mtrx{U\cr D}$ &  $\mtrx{X\cr U}$ &  $\mtrx{D\cr Y}$ &  $\mtrx{X\cr U\cr D}$ &  $\mtrx{U \cr D \cr Y}$
\\ \midrule
$SU(2)_L$ & $\mathbf{1}$ & $\mathbf{1}$ & $\mathbf{2}$ & $\mathbf{2}$ & $\mathbf{2}$ & $\mathbf{3}$ & $\mathbf{3}$
\\
$U(1)_Y$ & $2/3$ & $-1/3$ & $1/6$ & $7/6$ & $-5/6$ & $2/3$ & $-1/3$
\\ \bottomrule
  \end{tabular}
  \caption{
  The only VLQ multiplets that couple to the SM quarks through Yukawa interactions ($Q=I_3+Y$).}
  \label{tab:VLQ:irreps}
\end{table}
%
This thesis will only cover extensions of the quark sector with vector-like singlet quarks $U$ and $D$, often denoted in the literature as $T$ and $B$, respectively. Thus, in the context of this thesis, the names isosinglet vector-like quarks and vector-like quarks will be used interchangeably.\par

In section \ref{sec:quarksgen} of this chapter a model where $n_u$ isosinglet vector-like up-type ($Q=+2/3$) quarks and $n_d$ isosinglet vector-like down-type ($Q=-1/3$) quarks are added to the SM, without imposing any symmetry, is discussed. In section \ref{sec:quarkup} (\ref{sec:quarkdown}) a model where only $n_u$ ($n_d$) isosinglet vector-like up-type (down-type) quarks  are added to the SM, without imposing any symmetry, will be covered.
These extension of the SM will be treated exactly using the parameterisation first devised in \cite{Agostinho:2017wfs}, in the more succinct form of \cite{Pereira:2022pqu} and \cite{Alves:2023ufm}.\\

\section{The Most General Case ($n_u$SM$n_d$)}
\label{sec:quarksgen}

To begin, it is important to establish the notation for the most general case - the scenario involving $n_u$ up ($U^0_{Lr}$, $U^0_{Rr}$) and $n_d$ down ($D^0_{Ls}$, $D^0_{Rs}$) vector-like isosinglet quarks. Here, the flavour indices $r$ and $s$ range from 1 to $n_u$ and $n_d$ respectively, and the superscript "0" denotes the flavour basis.
Once these vector-like quarks are introduced, the Yukawa interactions in this basis are given by
\begin{equation}
\label{eq:lagrangian}
  \begin{split}
  \mathcal{L}_{\text{Y}} \,=\, 
  -\,\Big[
  &\,\,\overline Q_{Li}^{0} \, \tilde\Phi \, \left(Y_u\right)_{ij}  \, u^0_{Rj}
  \,+\, \overline Q_{Li}^{0} \, \tilde\Phi \, \left(\oYu\right)_{is} \, U^0_{Rs} 
  \\ +
  &\,\,\overline Q_{Li}^{0} \, \Phi       \, \left(Y_d\right)_{ij}  \, d^0_{Rj} 
  \,+\,\overline Q_{Li}^{0} \, \Phi       \, \left(\oYd\right)_{ir} \, D^0_{Rr}
  \,\Big] + \text{h.c.}\,,
  \end{split}
\end{equation}
where the indices $i$ and $j$ run from 1 to 3 as in the SM and $Y_d$, $Y_u$ are the SM Yukawa couplings. Here, $Q_L = ( u_L  \,\,\, d_L)^T$ are the SM quark doublets, $u_R$ and $d_R$ are the SM quark singlets, $\Phi$ is the Higgs doublet and $\tilde\Phi = i \tau_2 \Phi^*$ is its $SU(2)_L$ conjugate. The matrices ${\oYd}$ and ${\oYu}$ denote the new Yukawa couplings to the extra right-handed fields.

Since the VLQs are $SU(2)$ singlets, the ensuing mass terms are, in general, allowed and must therefore be included
\begin{equation}
\label{eq:baremasses}
\begin{split}
  \mathcal{L}_\text{b.m.} \,=\, 
  -\,\Big[
  &\,\, \overline U^0_{Lr} \, \left(\,\oM_u\right)_{r i} \, u^0_{Ri}
  \,+\, \overline U^0_{Lr} \, \left(M_u\right)_{r r'}   \, U^0_{Rr'}
  \\+
  &\,\, \overline D^0_{Ls} \, \left(\,\oM_d\right)_{s i} \, d^0_{Ri}
  \,+\, \overline D^0_{Ls} \, \left(M_d\right)_{s s'}   \, D^0_{Rs'}
  \,\Big] + \text{h.c.}\,,
\end{split}
\end{equation}
where $\,\oM_q$ and $M_q$ ($q=u,d$) are, respectively, $n_q\times 3$ and $n_q \times n_q$ general complex matrices. At this stage, these terms are interpreted as bare mass (b.m.) terms. In certain models, they may arise from the vacuum expectation value (VEV) of a complex scalar singlet, for instance.

After SSB, one has
$\Phi = \big( 0  \,\,\,  \frac{v+h}{\sqrt{2}}\big)^T$
in the unitary gauge, where $h$ is the Higgs field and $v \simeq 246\;\unit{GeV}$.
Additional mass terms are generated from $\mathcal{L}_\text{Y}$ of~\eqref{eq:lagrangian}. All mass terms can be collected in a compact form,
\begin{equation}
\mathcal{L}_\text{M} \,=\, 
- \begin{pmatrix}
    \overline{d}^0_L &  \overline{D}^0_L
  \end{pmatrix}
  \,\mathcal{M}_d\,
  \begin{pmatrix}
    d^0_{R} \\[1mm] D^0_{R} 
  \end{pmatrix}
- \begin{pmatrix}
    \overline{u}^0_L &  \overline{U}^0_L
  \end{pmatrix}
  \,\mathcal{M}_u\,
  \begin{pmatrix}
    u^0_{R} \\[1mm] U^0_{R} 
  \end{pmatrix}
+ \text{h.c.}\,,
\end{equation}
where
\begin{align}
\label{eq:genmass}
\renewcommand{\arraystretch}{1}
\setlength{\extrarowheight}{6pt}
\mathcal{M}_q \,=\,
\left(\begin{array}{c;{2pt/2pt}c}
\,m_q\,\, & \,\om_q\,
\\[1.5mm] \hdashline[2pt/2pt]
\!{\,\,\,\oM_q\,\,\,}\! &
\!{\,\,\,M_q\,\,\,}\!\\[1mm]
\end{array}\right) \,, 
\end{align}
with $m_q = \frac{v}{\sqrt{2}} Y_q$, $\om_q = \frac{v}{\sqrt{2}} \overline{Y}\!_q$ and $\oM_q$ and $M_q$ defined in eq. \eqref{eq:baremasses}.
It is important to note that, generally speaking, the matrices $\mathcal{M}_q$ aren't symmetric or Hermitian. Due to their different origin, a hierarchy $\oM_q \sim M_q \gg \om_q \sim m_q$ is expected.
Whenever there's no potential for confusion, the indices $q=u,d$ will be omitted.
The mass matrices $\mathcal{M}_q$ can be diagonalized by biunitary transformations --- their singular value decompositions --- of the form
\begin{align}
\renewcommand{\arraystretch}{1}
\setlength{\extrarowheight}{6pt}
\allowdisplaybreaks[0]
    {\mathcal{V}^q_L}^\dagger
    \,\,\mathcal{M}_q\,\,
    \mathcal{V}_R^q
    \,=\, \mathcal{D}_q\,,
    \quad \text{with } \mathcal{D}_q\,=\,
    \left(\begin{array}{c;{2pt/2pt}c}
      \,\,\,d_q\,\,\,\,\, & \,\,\,0\,
      \\[1.5mm] \hdashline[2pt/2pt]
      \!\myunder{\scriptstyle 3}{\,\,\,\,0\,\,\,\,\, }\! &
      \myunder{\scriptstyle \,\,\,n_q}{\,\,\,\,D_q\,\,\,\,}\!\\[1mm]
    \end{array}\right) \!\!\!
    \begin{array}{l}
      \myrightbrace{0}{\scriptstyle \,3}\\[0mm]
      \myrightbrace{D}{\scriptstyle \, n_q}
    \end{array}
\,,
\label{eq:diagonalization}
\\[-1mm] \nonumber
\end{align}
where the unitary rotations $\mathcal{V}_{L,R}^q$ connect the flavour and physical bases by using four rotations (two for each sector).
The diagonal matrices $d_q$ and $D_q$ contain the light  $(d_q)_i \geq 0$ ($i = 1,2,3$)  and heavy $(D_q)_r > 0$ ($r = 1,\ldots,n_q$) quark masses. It is important to note that $D_q$ is a matrix, not a down-type quark field. The notations $d_u = \diag(m_u,m_c,m_t)$, $d_d = \diag(m_d,m_s,m_b)$, $D_u = \diag(m_{U1},\ldots,m_{Un_u})$, and $D_d = \diag(m_{D1},\ldots,m_{Dn_d})$ will be used whenever there is no potential for confusion.
Regardless of how one parameterises the four diagonalization matrices $\ensuremath{\mathcal{V}^q_\chi}$ ($q=u,d$; $\chi=L,R$), it is useful to divide them into two blocks,

\begin{equation}
\cV^q_\chi \,=\,
\setlength{\extrarowheight}{1.2pt}
    \left(\begin{array}{c}
     { }\\[-4mm]
      \qquad \ensuremath{A^q_\chi} \qquad 
      \\[2mm] \hdashline[2pt/2pt]
       { }\\[-4mm]
      \qquad \ensuremath{B^q_\chi}\qquad 
      \\[2mm]
    \end{array}\right) 
    \setlength{\extrarowheight}{6pt}
        \begin{array}{l}
      \myrightbrace{A}{\scriptstyle \,3}\\[0mm]
      \myrightbrace{B}{\scriptstyle \, n_q}
    \end{array}
\,,
\label{def:V:AB}
\end{equation}
where $A^q_\chi$ is a $3 \times (3 + n_q)$ matrix and $B^q_\chi$ is an $n_q \times (3+ n_q)$ matrix.
The flavour basis and the physical basis (no `0' superscript) are thus connected via
\begin{equation}
\begin{pmatrix}
\,u^0_\chi\, \\[1mm] 
U^0_\chi
\end{pmatrix} 
\,=\,
\begin{pmatrix}
\,A^u_\chi\, \\[1mm] 
B^u_\chi
\end{pmatrix} 
\begin{pmatrix}
\,u_\chi\, \\[1mm] 
U_\chi
\end{pmatrix} \,,
\qquad
\begin{pmatrix}
\,d^0_\chi\, \\[1mm] 
D^0_\chi
\end{pmatrix} 
\,=\,
\begin{pmatrix}
\,A^d_\chi\, \\[1mm] 
B^d_\chi
\end{pmatrix} 
\begin{pmatrix}
\,d_\chi\, \\[1mm] 
D_\chi
\end{pmatrix} \,,
\label{eq:connect}
\end{equation}
with $\chi = L,R$.

Within each sector, using the diagonalization relation eq.~\eqref{eq:diagonalization}, one can write the mass matrix blocks of~eq. \eqref{eq:genmass} in terms of the physical masses as
\begin{equation}
\label{eq:mdefs}
\begin{aligned}
  m_q   &= A^q_{L} \,\mathcal{D}_q\, {A_{R}^q}^\dagger\,, \quad 
  \om_q = A^q_{L} \,\mathcal{D}_q\, {B_{R}^q}^\dagger\,, \\[2mm]
  \oM_q &= B^q_{L} \,\mathcal{D}_q\, {A_{R}^q}^\dagger\,, \quad
  M_q   = B^q_{L} \,\mathcal{D}_q\, {B_{R}^q}^\dagger\,.
\end{aligned}
\end{equation}
Finally, unitarity of the $\cV^q_\chi$ implies
\begin{subequations}\begin{align}
\cV^q_\chi ~ \cV^{q \dag}_\chi = \begin{pmatrix}
 \,A^q_\chi\, \\[2mm] 
B^q_\chi
\end{pmatrix} \begin{pmatrix}
\,{A^q_\chi}^\dagger & {B^q_\chi}^\dagger\,
\end{pmatrix} 
\,&=\,
\begin{pmatrix}
\,A^q_\chi\,{A^q_\chi}^\dagger & A^q_\chi\,{B^q_\chi}^\dagger\,\\[2mm] 
\,B^q_\chi\,{A^q_\chi}^\dagger & B^q_\chi\,{B^q_\chi}^\dagger\,
\end{pmatrix} 
\,=\,
\begin{pmatrix}
\,\id_{3}\, & \,0\, \\[2mm] 
\,0\, & \,\id_{n_q}\,
\end{pmatrix}\,, 
 \label{eq:unitAB02}
\\[2mm]
\cV^{q \dag}_\chi ~  \cV^q_\chi = \begin{pmatrix}
\,{A^q_\chi}^\dagger & {B^q_\chi}^\dagger\,
\end{pmatrix} \begin{pmatrix}
\,A^q_\chi\, \\[2mm] 
B^q_\chi
\end{pmatrix}
\,&=\,
\,
{A^q_\chi}^\dagger\,A^q_\chi+ 
{B^q_\chi}^\dagger\,B^q_\chi
\,
\,=\,
\,
\id_{3+n_q}
\,,
 \label{eq:unitAB2}
\end{align}\end{subequations}
for each $q=u,d$ and $\chi = L,R$.

Electromagnetic interactions of quarks, including the VLQ isosinglets, are described by the Lagrangian 
\begin{equation}
    \mathcal{L}_A = - e\, J^\mu_\text{em} A_\mu\,,
\end{equation}
with
\begin{equation}
\begin{aligned}
J^\mu_\text{em} &\,\equiv\,
\frac{2}{3}  
\Big( \overline{u_{i}^0} \gamma^\mu  {u_{i}^0}
+ \overline{U_{r}^0} \gamma^\mu {U_{r}^0}  \Big)  
- \frac{1}{3} 
\Big(\overline{d_{i}^0} \gamma^\mu  {d_{i}^0}
+ \overline{D_{s}^0} \gamma^\mu {D_{s}^0} \Big)\\[2mm]
 &\,=\, 
\frac{2}{3}  
\Big( \overline{u_{i}} \gamma^\mu  {u_{i}}
+ \overline{U_{r}} \gamma^\mu {U_{r}}  \Big)  
- \frac{1}{3} 
\Big(\overline{d_{i}} \gamma^\mu  {d_{i}}
+ \overline{D_{s}} \gamma^\mu {D_{s}} \Big)
\,,
\end{aligned}
\end{equation}
where  $\psi \equiv \psi_L + \psi_R$ for $\psi\in\{u_i^{(0)},d_i^{(0)},U_r^{(0)},D_s^{(0)}\}$. As before, the index $i$ runs from 1 to 3 as in the SM, while the indices $r,s$ run from 1 to $n_{u,d}$. Note that the structure of the electromagnetic current $J^\mu_\text{em}$ is unchanged in going from the flavour to the mass basis.

The charged current Lagrangian for the quarks in the mass basis is given by
\begin{equation}
   \mathcal{L}_W^q=-\frac{g}{\sqrt{2}}
 \begin{pmatrix}
\overline{u}_L & \overline{U}_L 
\end{pmatrix}\,
V\,    \gamma^\mu
\begin{pmatrix}
d_L \\[2mm] D_L 
\end{pmatrix} 
    W_\mu^+
    \,+\,\text{h.c.} ~,
 \end{equation}
where
       \begin{equation}
       \label{eq:mixingmatrixquark}
           V=A_L^{u^\dag} A_L^d ~,
       \end{equation}
is the $(3+n_u)\times (3+n_d)$ non-unitary mixing matrix. The mixing matrix $V$, which may have a rectangular shape when $n_u \neq n_d$, is typically not unitary. The CKM quark mixing matrix, $V_{\textrm{CKM}}$, is the upper-left $3 \times 3$ block of $V$, also not unitary in general.

The weak neutral current Lagrangian in the mass basis is given by

\begin{equation}
\begin{split}
    \mathcal{L}_Z^q&=-\frac{g}{2 \cos{\theta_W}}  Z_\mu \left[
 \begin{pmatrix}\overline{u}_L & \overline{U}_L \end{pmatrix}
\,F^u\, \gamma^\mu
\begin{pmatrix} u_L \\[2mm] U_L \end{pmatrix}
-
 \begin{pmatrix}\overline{d}_L & \overline{D}_L \end{pmatrix}
\,F^d\, \gamma^\mu
\begin{pmatrix} d_L \\[2mm] D_L \end{pmatrix}
- 2 \sin^2{\theta_W} J^{\mu~q}_\text{em} \right] \,,
\end{split}
\end{equation}
while the part of the Lagrangian that contains the Higgs interactions with the quarks in the mass basis is given by
\begin{equation}
 \mathcal{L}_H^q \,=\,
 -\frac{h}{v} \bigg[\begin{pmatrix} \overline{u}_L & \overline{U}_L \end{pmatrix}\,
    F^u\, \mathcal{D}_u\begin{pmatrix} u_R \\[2mm] U_R \end{pmatrix}  + \begin{pmatrix} \overline{d}_L & \overline{D}_L \end{pmatrix}\,
    F^d\, \mathcal{D}_d\begin{pmatrix} d_R \\[2mm] D_R \end{pmatrix}  \bigg]+\,\text{h.c.}
    \,~,
\end{equation}
where
       \begin{equation}
       \label{eq:Fquark}
           F^q=A_{L}^{q^\dag} A_{L}^{q} ~.
       \end{equation}
Note that the $F^q$ are present in interactions with the Higgs and the Z boson and are the matrices that controls Flavour Changing Neutral Couplings (FCNC). This might become clearer if these are written as

\begin{equation}
\begin{aligned}
F^u \,&\equiv\, {A_L^u}^\dagger A_L^u \,=\,
\id - {B_L^u}^\dagger B_L^u
\,=\, VV^\dagger \,,\\
F^d \,&\equiv\, {A_L^d}^\dagger A_L^d  \,=\,
\id - {B_L^d}^\dagger B_L^d
\,=\, V^\dagger V\,.
\end{aligned}
\label{eq:Fud}
\end{equation}
The strength of FCNC is proportional to the off-diagonal elements of $F^{u,d}$, and eq. \eqref{eq:Fud} shows that these come from the deviations from the identity matrix of $F^q$. This also shows that there is a strong connection between the strength of FCNC and deviations of $V$ from unitarity.\\
It should be noted that $F^{u,d}$ matrices are in general not equal to the identity matrix, which leads to the occurrence of FCNC processes mediated by both the Higgs boson and the $Z$ boson. The strength of Higgs-mediated FCNC processes is determined by the off-diagonal elements of $F^{u,d}$ matrices as well as the ratios of the diagonal elements of $\mathcal{D}_{u,d}$ to the vacuum expectation value $v$. This is in contrast to the case of $Z$-mediated FCNC processes. Because of this, in the Higgs sector, for transitions involving only the lighter quarks $u$ and $c$, there is a suppression factor of $m_u/v \sim 10^{-5}$ or $m_c/v\sim 10^{-2}$. \\

The same steps performed in the beginning of chapter \ref{chapter:framework_neutrinos} can be done for the quark sector, allowing one to write the quark form of eq. \eqref{eq:calVfull} as
\begin{align}
\label{eq:uparam-exact}
\renewcommand{\arraystretch}{1.2}
\mathcal{V}_\chi^q \,=\,
\renewcommand{\arraystretch}{1}
\setlength{\extrarowheight}{6pt}
\left(\begin{array}{c;{2pt/2pt}c}
\,\,\,K_\chi^q\,\,\,\,\, & \,\,\,K_\chi^q\,{X_\chi^q}^\dag\,
\\[1.5mm] \hdashline[2pt/2pt]
{- \oK_{\!\chi}^{\!q}\,X_\chi^q\,\, }\! &
{\quad \,\,\oK_{\!\chi}^{\!q}\quad\,\,}\!\\[1mm]
\end{array}\right) ~.
\end{align}
Eq. \eqref{eq:uparam-exact} gives that the $(3+n_q)\times(3+n_q)$ unitary matrices $\cV^q_\chi$ introduced in eq. \eqref{eq:diagonalization} and eq. \eqref{def:V:AB} can be written, without loss of generality, using a non-singular $3 \times 3$  general complex matrix, $K^q_\chi$, a non-singular $n_q \times n_q$  general complex matrix ~$\oK^q_\chi$ and a $n_q \times 3$ matrix, $X^q_\chi$. As in chapter \ref{chapter:framework_neutrinos}, from eq. \eqref{def:V:AB} one obtains
\begin{equation}
    A^q_\chi = ( K^q_\chi~~K^q_\chi {X^q_\chi}^\dag ) ~,~ B^q_\chi = ( -\oK^q_\chi X^q_\chi~~~\oK^q_\chi ) ~,
    \label{eq:ABdefquark}
\end{equation}

Using eq. \eqref{eq:ABdefquark} and the unitarity relations in eqs. \eqref{eq:unitAB02} and \eqref{eq:unitAB2}, eq. \eqref{eq:mdefs} becomes

\begin{equation}
\begin{split}
&m_q \,=\,K_L^q \left( d_q + {X_L^q}^\dag\,D_q\,X_R^q \right ) {K_R^q}^\dagger, \\
 & \om_q \,=\, K_L^q \left( {X_L^q}^\dag\,D_q - d_q\,{X_R^q}^\dag \right ) {\oK_R^q}^\dagger,\\
  & \oM_q \,=\,\oK_L^q \left( D_q\,X_R^q - X_L^q\,d_q \right ) {K_R^q}^\dagger , \\
  & M_q \,=\,\oK_L^q \left( D_q + X_L^q\,d_q\,{X_R^q}^\dag \right ) {\oK_R^q}^\dagger,
\end{split}
\label{eq:mquark}
\end{equation}
where $m_q$ is the $3 \times 3$ Dirac mass matrix for the quarks and the entries of the $3 \times n_q$ matrix, $\om_q$, are proportional to the Higgs vacuum expectation value, as well. The $n_q \times 3$ mass matrix, $\oM_q$, and the $n_q \times n_q$ mass matrix, $M_q$, are bare mass terms involving only singlet quark fields.\\
It is always possible to go to a WB where $\om_q$ is $\textbf{0}$, also for $\oM_q$, see section 3.2.1 of ref. \cite{Alves:2023ufm}. In the vanishing $\om_q$ WB, one can proceed like in the neutrino case to obtain a formula for $X_\chi^q$,
\begin{eqnarray}
\begin{aligned}
\label{eq:Xdefquark}
&X_L^q =  \sqrt{D_q^{-1}}    P^q   \sqrt{d_q} ~,\\
&X_R^q = \sqrt{D_q} P^q \sqrt{d_q^{-1}} ~,
\end{aligned}
\end{eqnarray}
where $P^q$ is a general complex matrix and $d_q$ ($D_q$) is a diagonal matrix with the masses of the light (heavy) quarks in the diagonal.
A keen reader will notice that eq. \eqref{eq:Xdefquark} amounts to nothing, since $P^q$ are general complex matrices, so will $X_L^q$ and $X_R^q$ be. $X_L^q$ and $X_R^q$ are completely general complex matrices which only need to satisfy 
\begin{align}
    d_q X_L^{q \dagger} &= X_R^{q \dagger } D_q \quad\text{in the vanishing } \oM_q \text{ WB}\,,  \\[1mm]
  \text{and}\quad   d_q X_R^{q \dagger} &= X_L^{q \dagger} D_q \quad\text{in the vanishing } \om_q\text{ WB}\,.
\end{align}  
$X_L^q$ will, as in the neutrino sector, control the size of the deviations from unitarity of the $(3+n_u)\times (3+n_d)$ non-unitary mixing matrix $V$.
Naturally, its biggest constraint will be how large one wants these deviations to be. A very important constraint will also be the perturbativity of the theory. More on that in section \ref{sec:perturb}.
The following definitions can also be obtained from the unitarity relations in eqs. \eqref{eq:unitAB02} and \eqref{eq:unitAB2},
\begin{equation}
\begin{split}\label{eq:uparam-exactK}
&K^q_\chi=U^{q}_{K\chi}(\id_3+{X_\chi^q}^\dag X_\chi^q)^{-1/2}\,,\\
&\oK^q_\chi=U^{q}_{\oK\chi}
(\id_{n_q}+ X_\chi^q {X_\chi^q}^\dag)^{-1/2}~,
\end{split}
\end{equation}
where $U^{q}_{K \chi}$ and $U^{q}_{\oK \chi}$ are unitary matrices, since the unitarity relations only define $K$
 and $\oK$ up to a unitary matrix on the left. The combination ${K_L^u}^\dag K_L^d$ will play the role of the CKM mixing matrix and is only unitary when $X_L^u, X_L^d \xrightarrow[]{} \bzero$,

\begin{equation}
    V_{\textrm{CKM}}= {K_L^u}^\dag K_L^d ~.
\end{equation}

The $(3+n_u)\times (3+n_d)$ non-unitary mixing matrix $V$, first defined in eq. \eqref{eq:mixingmatrixquark}, can now be written as
\begin{equation}\label{eq:uparam-exactV}
    V=A^{u\dagger}_LA^d_L=\renewcommand{\arraystretch}{1}\setlength{\extrarowheight}{6pt}
    \left(\begin{array}{c;{2pt/2pt}c}
      \,\,\,V_\text{CKM}\,\,\,\,\, & \,\,\, V_\text{CKM}\,{X_L^{d \dagger}}
      \\[1.5mm] \hdashline[2pt/2pt]
      \!\!\myunder{\scriptstyle 3}{\,\,\,\,{X_L^u} \,V_\text{CKM}\,\,\,\,\, }\!\!\! &
      \!\myunder{\scriptstyle \,\,\,n_d}{\,\,\,{X_L^u} \,V_\text{CKM}\,{X_L^d}^\dag\,\,\,\,}\!\!\!\\[1mm]
    \end{array}\right)\!\!\!
    \begin{array}{l}%
      \myrightbrace{V_\text{CKM}\,{X_L^d}^\dagger}{\scriptstyle \,3}\\[0mm]
      \myrightbrace{V_\text{CKM}\,{X_L^d}^\dagger}{\scriptstyle \, n_u}
    \end{array}.\\[4mm]
\end{equation}

The matrices relevant for FCNC, first defined in eq. \eqref{eq:Fquark}, can also be rewritten as
 \begin{equation}
 \label{eq:quarkF}
F^q= \begin{pmatrix}
(\id_3+{X_L^q}^\dag X_L^q)^{-1} &(\id_3+{X_L^q}^\dag {X_L^q})^{-1}{X_L^q}^\dag\\ 
{X_L^q} (\id_3+{X_L^q}^\dag {X_L^q})^{-1} &{X_L^q} (\id_3+{X_L^q}^\dag {X_L^q})^{-1} {X_L^q}^\dag\\ 
\end{pmatrix} ~,
\end{equation}
Finally, for completeness, the exact formula for the $(3+n_q) \times (3+n_q)$ unitary matrix $\mathcal{V}^q_\chi$ assumes the form
  \begin{equation}
  \mathcal{V}^q_\chi =       \begin{pmatrix}

  U^{q}_{K\chi}(\id_3+{X_\chi^q}^\dag X_\chi^q)^{-1/2} &U^{q}_{K\chi}(\id_3+{X_\chi^q}^\dag X_\chi^q)^{-1/2}X_\chi^{q\dag}\\ 
- U^{q}_{\oK\chi}
(\id_{n_q}+ X_\chi^q {X_\chi^q}^\dag)^{-1/2}X^q_\chi &U^{q}_{\oK\chi}
(\id_{n_q}+ X_\chi^q {X_\chi^q}^\dag)^{-1/2}\\
\end{pmatrix} ~.
\end{equation}

\subsection{Weak Bases and Physical Parameters in the presence of VLQs}
\label{sec:WBphy}
Given that in the context of this thesis no symmetry is being assumed when adding VLQs to the SM Lagrangian, many new parameters will emerge. It is important to count them properly and to disclose which of them are physical. This can be done in a general treatment using Weak-basis (WB) transformations. Even when one is working under a model which restricts the parameter space via symmetries, the procedure and results that follows will be valid. For this discussion for the quark sector of the SM see section 3.1 of ref. \cite{Alves:2023ufm}.\\
As established in eqs. \eqref{eq:lagrangian} \eqref{eq:baremasses}, the flavour sector of the $n_u$SM$n_d$ looks like
\begin{equation}
\label{eq:LF}
  \begin{split}
  \mathcal{F} \,=\, 
  -\,\Big[
  &\,\, \overline Q_{L}^{0} \, \tilde\Phi \, Y_u  \, u^0_{R}
  \,+\, \overline Q_{L}^{0} \, \tilde\Phi \, \oYu \, U^0_{R} 
  \,+\, \overline U^0_{L}   \, \,\oM_u    \, u^0_{R}
  \,+\, \overline U^0_{L}   \, M_u        \, U^0_{R}
  \\ +
  &\,\, \overline Q_{L}^{0} \, \Phi       \, Y_d  \, d^0_{R} 
  \,+\, \overline Q_{L}^{0} \, \Phi       \, \oYd \, D^0_{R}
  \,+\, \overline D^0_{L}   \, \,\oM_d    \, d^0_{R}
  \,+\, \overline D^0_{L}   \, M_d        \, D^0_{R}
  \,\Big] + \text{h.c.}\,,
  \end{split}
\end{equation}
where all matrices are complex.
The most general WB transformation in this context takes the form
\begin{equation}
\label{eq:VLQWB}
\begin{array}{l@{\qquad}l}
  u^0_L \,\to\, W_L  \, u^0_L
  \,,
  &d^0_L \,\to\, W_L  \, d^0_L
  \,,\\[2mm]
  U^0_L \,\to\, W_L^U  \, U^0_L
  \,,
  &D^0_L \,\to\, W_L^D  \, D^0_L
  \,,\\[3mm]
  \begin{pmatrix}
     u^0_R \\[1mm] U^0_R 
  \end{pmatrix}
  \,\to\, \mathcal{W}_R^u 
  \begin{pmatrix}
     u^0_R \\[1mm] U^0_R 
  \end{pmatrix}
  \,,
  &\begin{pmatrix}
     d^0_R \\[1mm] D^0_R 
  \end{pmatrix}
  \,\to\, \mathcal{W}_R^d 
  \begin{pmatrix}
     d^0_R \\[1mm] D^0_R 
  \end{pmatrix}
\,,
\end{array}
\end{equation}
where $W_L \sim 3\times 3$, $W_L^{U,D} \sim n_{u,d}\times n_{u,d}$ and $\mathcal{W}^{u,d}_R \sim (3+n_{u,d})\times (3+n_{u,d})$ are unitary matrices in flavour space.
It should be noted that, in a WB transformation, changes to $u_L^0$ and $d_L^0$ are interdependent, as in the SM case. However, this interdependence is not necessary for the transformations of the isosinglet fields $U_L^0$ and $D_L^0$. Furthermore, the transformation matrices for the right-handed isosinglet fields $\mathcal{W}^{u,d}_R$ are more extensive than those in the Standard Model. They enable mixing between $u_R^0$ and $U_R^0$ (and similarly for $d_R^0$ and $U_R^0$), as the $3+n_q$ right-handed quarks in each sector ($q=u,d$) carry the same quantum numbers.

While the statements above are presented for the case where $n_f=3$ (the number of families of SM quarks), they can be readily generalized and applied to the case where $n_f\neq3$.

\subsubsection{Useful weak bases}
\label{sec:wbzero}
Different WB choices allow for particular forms of the extended quark mass matrices $\mathcal{M}_{q}$, defined in eq.~\eqref{eq:genmass}. Since the mass matrices contain contributions from both Yukawa and bare mass terms, see~eqs. \eqref{eq:lagrangian} \eqref{eq:baremasses}, it is easier to work directly with the mass matrices in this context.
For generic $n_u$ and $n_d$, under the WB change of eq. \eqref{eq:VLQWB}, the mass matrices are transformed as
\begin{equation}
\begin{aligned}
\mathcal{M}_u \,=\,
\begin{pmatrix}
m_u & \om_u  \\[2mm]
\oM_u & M_u
\end{pmatrix}  \quad &\to \quad \mathcal{M}'_u \,=\, 
\begin{pmatrix}
W_L^\dagger \, m_u & W_L^\dagger \, \om_u  \\[2mm]
{W_L^U}^\dagger \, \oM_u & {W_L^U}^\dagger \, M_u
\end{pmatrix}
\mathcal{W}_R^u
\,,  \\[2mm]
\mathcal{M}_d \,=\,
\begin{pmatrix}
m_d & \om_d  \\[2mm]
\oM_d & M_d
\end{pmatrix}  \quad &\to \quad \mathcal{M}'_d \,=\, 
\begin{pmatrix}
W_L^\dagger \, m_d & W_L^\dagger \, \om_d  \\[2mm]
{W_L^D}^\dagger \, \oM_d & {W_L^D}^\dagger \, M_d
\end{pmatrix}
\mathcal{W}_R^d
\,.
\end{aligned}
\end{equation}

The unitary rotations $\mathcal{W}^q_R$ that appear on the right-hand side of these transformations are useful in shaping the mass matrices $\mathcal{M}_q$. Under specific conditions, they can be employed to eliminate some portion of the matrix and produce a triangular matrix.
In particular, it is worth noting that any complex square matrix $\mathcal{M}$ can be expressed as the product of a lower-triangular matrix $L$ and a unitary matrix $W$ to its right, $\mathcal{M} = LW$. Alternatively, it can be written as the product of an upper-triangular matrix $R$ and a different unitary matrix $W'$ to its right, i.e., $\mathcal{M} = RW'$. This follows from the well-known QR decomposition, which states that any complex square matrix $A$ can be written as $A=QR$, with $Q$ being a unitary matrix and $R$ being an upper (or right) triangular matrix. By QR-decomposing $\mathcal{M}^T$, one finds the decomposition $\mathcal{M} = LQ'$, where $Q'$ is unitary and $L$ is lower (or left) triangular. Similarly, by QR-decomposing $\mathcal{M}^T \mathcal{P}$, where $\mathcal{P}$ is the anti-diagonal permutation matrix, one can demonstrate that it is possible to write $\mathcal{M} = R'Q''$, where $Q''$ is unitary and $R'$ is upper triangular.

The matrices $\mathcal{W}^u_R$ and $\mathcal{W}^d_R$ can then be used to eliminate the unitary factors $W^{(\prime)}$ and render the mass matrices $\mathcal{M}_u$ and $\mathcal{M}_d$ lower or upper triangular, as desired. Additionally, they can be utilized to permute the columns of these matrices.
As a result:
\begin{itemize}
\item A WB where $\om_q=0$ exists for all $n_q$ (\it vanishing $\,\om$ WB).
\item A WB where $\oM_q=0$ exists for all $n_q$ (\it vanishing $\,\oM$ WB).
\item A WB where $m_q=0$ is guaranteed to exist only if $n_q \geq n_f = 3$.
\item A WB where $M_q=0$ is guaranteed to exist only if $n_q \leq n_f = 3$.
\end{itemize}
Typically, only one of the sub-matrices $m$, $\om$, $\oM$, or $M$ can be made to vanish at a time through a WB transformation. This means that  one can impose at most  one of the above conditions at a time. However, the vanishing conditions can be enforced separately in the up and down sectors. Therefore, in each sector, it is always possible to choose a weak basis where the mass matrix has the following form:

\begin{equation}
\label{eq:WBzeros}
\mathcal{M} \,=\,
\begin{pmatrix}
m & \om  \\[2mm]
0 & M
\end{pmatrix} \!\!\qquad\text{or}\qquad
\mathcal{M} \,=\,
\begin{pmatrix}
m & 0 \\[2mm]
\oM & M
\end{pmatrix} \,.
\end{equation}
Furthermore, in each sector, the matrix blocks located in the lower-right position, denoted as $M$, can always be transformed into a diagonal form, denoted as $D'$, with real and non-negative entries using a secondary WB transformation. This transformation involves rotating the left and right unitary matrices in the singular value decomposition of $M$ by employing suitable choices of $W^{U,D}_L$ and $\mathcal{W}^{u,d}_R$ (which are now block diagonal). This process may impact the matrix $\oM$ if it is present, which is currently a general complex matrix.

On the other hand, the matrix blocks $m$ located in the upper-left position in each sector cannot be simultaneously converted into a diagonal form using the same procedure. This limitation arises due to the use of a single $W_L$ matrix, which is common to both the up and down sectors.

Consequently, if one begins with upper-triangular mass matrices $\mathcal{M}_u$ and $\mathcal{M}_d$, the most that can be achieved through the WB transformation is a configuration where a unitary rotation remains in one of the sectors (for example, the down sector).
\begin{equation}
\label{eq:WBMorozumi}
\mathcal{M}_u \,=\,
\begin{pmatrix}
d_u' & \om_u  \\[2mm]
0 & D_u'
\end{pmatrix} \,, \qquad
\mathcal{M}_d \,=\,
\begin{pmatrix}
V' d_d' & \om_d  \\[2mm]
0 & D_d'
\end{pmatrix} \,.
\end{equation}
Here ({\it minimal vanishing $\,\oM$ WB}), $d_q'$ and $D_q'$ are diagonal matrices with real non-negative entries, $V'$ is the anticipated residual unitary rotation, and $\om_q$ are general complex matrices.

A similar choice of the WB (Wolfenstein parametrization) has been made in reference \cite{Branco:1992wr}, where the matrix $m_d$ is allowed to be a general complex matrix.
It is important to note that the primed diagonal matrices, denoted as $d_q'$ and $D_q'$, differ from the matrices containing the physical masses as defined in eq. \eqref{eq:diagonalization}.
However, it can be verified that the determinant of $d_q'$ multiplied by the determinant of $D_q'$ is equal to the absolute value of the determinant of $d_q$ multiplied by the determinant of $D_q$, which represents the product of all quark masses of type $q$.
Similarly, it is possible to choose a different WB parametrization where
\begin{equation}
\label{eq:WBlower}
\mathcal{M}_u \,=\,
\begin{pmatrix}
d_u' & 0  \\[2mm]
\oM_u & D_u'
\end{pmatrix} \,, \qquad
\mathcal{M}_d \,=\,
\begin{pmatrix}
V' d_d' & 0  \\[2mm]
\oM_d & D_d'
\end{pmatrix}
\end{equation}
({\it minimal vanishing $\,\om$ WB}),
with $d_q'$ and $D_q'$ diagonal, $V'$ unitary and now $\oM_q$ general complex.
Finally, by the above procedure, it further follows that one may easily arrange for weak bases like
\begin{equation}
\label{eq:WBLavoura}
\mathcal{M}_u \,=\,
\begin{pmatrix}
d_u' & 0  \\[2mm]
\oM_u & D_u'
\end{pmatrix}  \,, \qquad
\mathcal{M}_d \,=\,
\begin{pmatrix}
V' d_d' & \om_d  \\[2mm]
0 & D_d'
\end{pmatrix}
\end{equation}
({\it mixed minimal WB}), where both $\om_u$ and $\oM_d$ are made to vanish.
A closely related WB choice has been made in Ref.~\cite{Branco:1986my},
where again $m_d$ has been kept general complex.

By examining eq. \eqref{eq:WBMorozumi}, it becomes apparent that for arbitrary values of $n_u$ and $n_d$, it is generally not possible to transform either of the matrices $\mathcal{M}_u$ or $\mathcal{M}_d$ into a diagonal form through a WB transformation.

In other words, a diagonal mass matrix in the flavour basis does not typically correspond to a straightforward choice in the WB parametrization. 

However, in the presence of vector-like quarks (VLQs), exceptions occur in certain limit cases where either $n_u$ or $n_d$ equals zero. In these cases, it is indeed possible to diagonalize one of the mass matrices, either $\mathcal{M}_u$ or $\mathcal{M}_d$, respectively. As an illustration, reference \cite{Bento:1991ez} considers the specific scenario with $n_u = 0$ and $n_d = 1$, and a WB parametrization is chosen such that $\mathcal{M}_u$ becomes a diagonal matrix represented by $d_u$, while $\mathcal{M}_d$ takes the form described in eq. \eqref{eq:WBlower}.

\subsubsection{Parameter counting}
\label{sec:par_counting}
In the WB choices of eqs.~\eqref{eq:WBMorozumi} \eqref{eq:WBlower}, \eqref{eq:WBLavoura}, the physical parameter count becomes apparent. Looking at e.g. eq.~\eqref{eq:WBMorozumi}, one finds
\begin{itemize}
\item $n_f + n_q = 3+n_q$ real parameters (moduli) in the matrices $d_q'$ and $D_q'$ of each sector,
\item $n_fn_q = 3 n_q$ complex parameters in each $\om_q$ matrix, i.e.~$3n_q$ moduli and $3n_q$ phases in each sector, and
\item $n_f^2 = 9$ parameters in the unitary matrix $V'$, corresponding to $\frac{1}{2}n_f(n_f-1) = 3$ moduli and $\frac{1}{2}n_f(n_f+1) = 6$ phases.
\end{itemize}
However, it is important to note that, similar to the Standard Model case, there is still some rephasing freedom in the WB parametrization, and not all of the phases mentioned above are physically meaningful.

In order to obtain a WB representation with the minimum number of parameters, one can consider rephasings of the WB where the unitary matrices $W_L$, $W_L^{U,D}$, and $\mathcal{W}_R^{u,d}$ are diagonal matrices consisting of phases. By doing so, one can
\begin{itemize}
\item remove $2n_f-1 = 5$ phases from $V'$ by use of the first $n_f = 3$ phases of $\mathcal{W}_R^d$ (commuting with $d_d'$) and of $W_L$, whose effect on $d_u'$ may be cancelled by $\mathcal{W}_R^u$, 
\item remove $n_q$ phases from $\om_q$ in each sector via the last $n_q$ phases of $\mathcal{W}_R^q$ (see also~\cite{Branco:1992wr}), whose effects on $D_q'$ may be cancelled by the corresponding $W_L^{U,D}$.
\end{itemize}
Therefore, in this {\it minimal WB} the mass matrices display a total of $N_\text{phys} = 10 + 6(n_u+n_d)$ parameters, while keeping the form of eq.~\eqref{eq:WBMorozumi}. Their breakdown into moduli and phases is summarized in table~\ref{tab:params}.\\
%
\begin{table}[h!]
  \centering
  \begin{tabular}{lccc}
    \toprule
    & Moduli & Phases & Total \\
    \midrule
$\mathcal{M}_u$ (minimal WB) & $3+4 n_u $ & $2n_u$ & $3+6n_u$ \\
$\mathcal{M}_d$ (minimal WB) & $6+4 n_d$ & $1+2n_d$ & $7+6n_d$ \\
    \midrule
Total & $9+4(n_u+n_d)$ & $1+2(n_u+n_d)$ & $10+6(n_u+n_d)$ \\
    \bottomrule
  \end{tabular}
  \caption{Physical parameter count in the presence of isosinglet VLQs ($n_f = 3$).}
  \label{tab:params}
\end{table}
%
\vskip 2mm
Not only does one recover the SM result when $n_d=n_u=0$, but one also expects each singlet vector-like quark to bring about 6 new physical parameters: 4 moduli (1 mass and 3 mixing angles) and 2 physical phases.

\subsection{Deviations from Unitarity, FCNC and Perturbativity}
\label{sec:perturb}
Unlike the neutrino case, very light VLQs are very difficult (if not impossible) to accommodate with current phenomenological data. Hence, deviations from unitary in the quark sector can never be too large.
This statement holds true because, similar to the example of neutrinos - see chapter \ref{chapter:framework_neutrinos} - there is a seesaw between deviations from unitarity and the new particles' masses, in order to have sensible Yukawa couplings. This is why one should worry about perturbativity, since for large masses deviations from unitarity are very strongly upper bounded.
 The definitions in section \ref{sec:quarksgen} are useful in theory, but due to the aforementioned fact, more useful formulas can be derived via an approximation valid up to $\mathcal{O}(X^3)$.

To derive these approximations, it is convenient to simplify eqs.~\eqref{eq:uparam-exact} \eqref{eq:uparam-exactK} and consider a specific WB. The following WB transformation
\begin{equation}\label{eq:uparam-wbt}
\begin{array}{l@{\qquad}l}
  u^0_L \,\to\, V_{K^u_L}  \, u^0_L
  \,,
  &d^0_L \,\to\, V_{K^u_L}  \, d^0_L
  \,,\\[2mm]
  U^0_L \,\to\, V_{\oK\!^u_L}  \, U^0_L
  \,,
  &D^0_L \,\to\, V_{\oK\!^d_L}  \, D^0_L
  \,,
\end{array}
\end{equation}
removes the matrices $V_{K^u_L}$, $V_{\oK\!^d_L}$ and $V_{\oK\!^u_L}$ from eq.~\eqref{eq:uparam-exactK} without modifying the charged currents.
Now, doing an approximation valid for small $X$,
\begin{equation}\label{eq:uparam-approx}
    \big(\id_3+X^\dagger X\big)^{-1/2}\simeq\id_3-\frac{1}{2}X^\dagger X\,,\quad            \big(\id_n+X X^\dagger \big)^{-1/2}\simeq\id_n-\frac{1}{2} X X^\dagger \,,
\end{equation}
valid up to $\mathcal{O}(X^3)$, which is the same as saying that terms of higher order are negligible.
It is now possible to write eq.~\eqref{eq:uparam-exact} in the following way
\begin{equation}\label{eq:uparam-approxVX}
\begin{split}
    \cV^d_L&\simeq
    \begin{pmatrix}V'&0\\0&\id_n\end{pmatrix}
    \begin{pmatrix}\id_3-\frac{1}{2}\Theta^\dagger_d \Theta_d &\Theta_d^\dagger\\ -\Theta_d&\id_{n_d}-\frac{1}{2}\Theta_d\Theta^\dagger_d\end{pmatrix}\,,\\[1mm]
    \cV^u_L&\simeq
    \begin{pmatrix}\id_3-\frac{1}{2}\Theta^\dagger_u \Theta_u&\Theta_u^\dagger\\                      -\Theta_u&\id_{n_u}-\frac{1}{2}\Theta_u\Theta^\dagger_u\end{pmatrix}\,,
\end{split}\end{equation}
where $V' = V_{K^u_L}^\dagger V_{K^d_L}$. Furthermore, $X^q_L$ was replaced by $\Theta_q$, to emphasize that an approximation has been employed, this convention will be adopted throughout this thesis.
In this notation, the quark mixing matrix is given by
\begin{equation}\label{eq:uparam-approxV}
    V\simeq\begin{pmatrix}                                                                     V'-\frac{1}{2}\Theta^\dagger_u \Theta_uV'-\frac{1}{2}V'\Theta^\dagger_d \Theta_d&        V'\Theta_d^\dagger\\\Theta_uV'&\Theta_uV'\Theta_d^\dagger\end{pmatrix}\,,
\end{equation}
where the $3\times 3$ unitary matrix $V'$ is related to the CKM matrix $V_\text{CKM}$ through
\begin{equation}
\label{eq:approxCKM}
        V_\text{CKM}\simeq                                       \bigg(\id_3-\frac{1}{2}\Theta^\dagger_u \Theta_u\bigg)\,V' \,           \bigg(\id_3-\frac{1}{2}\Theta^\dagger_d \Theta_d\bigg)\,,
\end{equation}
an expression valid up to terms of order $\Theta^3_q$, as expected. Note that the deviation from unitarity is evident for nonzero $\Theta_u$ or $\Theta_d$.

It is now appropriate to discuss the standard quark FCNC and their relationship to the deviations from unitarity of the CKM matrix, taking advantage of the general and exact parameterization introduced before.

As demonstrated in section \ref{sec:quarksgen}, in models involving singlet VLQs, the FCNC affecting the standard quarks in each sector can be entirely determined by the matrices $F^d = V^\dagger V$ and $F^u = VV^\dagger$.

The standard quark FCNC are controlled by the upper-left $3\times 3$ block of each $F_q$ matrix, and these are equal to the identity matrix, $\id_3$, when the CKM matrix $V_{\textrm{CKM}}$, which represents the upper-left $3\times 3$ block of the full quark mixing matrix, $V$, is unitary.

Hence, FCNC are closely linked to the aforementioned deviations from unitarity. This connection becomes explicit when considering the precise parameterization given by equation~\eqref{eq:uparam-exact}, used to derive eq. \eqref{eq:quarkF}. When the deviations from unitarity are small, this relation may be approximated by
\begin{equation}\label{eq:duFCNC-approx}
    F^q\simeq                                                                              \begin{pmatrix}\id_3-\Theta_q^\dagger\Theta_q&\Theta_q^\dagger\\                                  \Theta_q&\Theta_q\Theta^\dagger_q\end{pmatrix}\,,
\end{equation}
with the replacement $X_L^q \to \Theta_q$, an expression valid up to terms of order $\Theta^3_q$. 
Note that in the limit $\Theta_q \to 0$, $F^q$ approaches the diagonal matrix $\diag(1,1,1,0,\ldots,0)$.

\vskip 2mm

In order to better illustrate the relation between FCNC and deviations from unitarity, the following useful parameters are introduced
\begin{equation}\label{eq:duFCNC-Delta}
    \Delta^d_n\equiv 1-\sum^{3+n_u}_{j=1}\big|V_{jn}\big|^2\,,
    \qquad                             
    \Delta^u_n\equiv 1-\sum^{3+n_d}_{j=1}\big|V_{nj}\big|^2\,.
\end{equation}
Here, $\Delta^d_n$ is a test of the unitarity of the $n$-th column of $V$, with $n=1,...,3+n_d$, and $\Delta^u_n$ is a test of the unitarity of the $n$-th row of $V$, with $n=1,...,3+n_u$. Since $F^d=V^\dagger V$ and $F^u=VV^\dagger$, they are also given by
\begin{equation}
    \Delta^q_n=1-F^q_{nn}\,.
\end{equation}
Combining this equation with eqs. \eqref{eq:quarkF} \eqref{eq:duFCNC-approx} gives
\begin{equation}
    \Delta^q_n=1-\Big(\id_3+X^{q ^\dagger}_LX^{q}_L\Big)^{-1}_{nn}\simeq                    \big(\Theta_q^\dagger \Theta_q\big)_{nn}=\sum^{n_q}_{j=1}
    \big|(\Theta_q)_{nj}\big|^2\,.
\end{equation}
Although these definitions are useful in theory, these parameters are not easily measurable, since one, currently, only has access to the CKM elements of $V$.
Hence, experimentalists test the unitarity of $V_\text{CKM}$ directly (as presented in the beginning of section~\ref{sec:quarkup}).
Taking this into account, definitions for parameters that test the unitarity of the CKM matrix follow. To probe the unitarity of the columns,
\begin{equation}
    \delta^d_n\equiv 1-\sum^3_{j=1}\big|V_{jn}\big|^2=1-|V_{un}|^2-|V_{cn}|^2-|V_{tn}|^2
\end{equation}
and that of its rows,
\begin{equation}
    \delta^u_n\equiv 1-\sum^3_{j=1}\big|V_{nj}\big|^2=1-|V_{nd}|^2-|V_{ns}|^2-|V_{nb}|^2\,,
    \label{eq:rowunitar}
\end{equation}
with $n=1,2,3$ here.
By comparing these expressions with eq.~\eqref{eq:duFCNC-Delta}, a connection between the parameters $\Delta^q_n$ and $\delta^q_n$ becomes apparent
\begin{equation}
    \delta^d_n-\Delta^d_n=\sum^{n_u}_{j=1}\big|V_{j+3,n}\big|^2\,,\qquad                       \delta^u_n-\Delta^u_n=\sum^{n_d}_{j=1}\big|V_{n,j+3}\big|^2\,.
\end{equation}
By using eq.~\eqref{eq:uparam-exactV} and eq. \eqref{eq:uparam-approxVX}, these relations become
\begin{equation}\begin{split}
    \delta^d_n-\Delta^d_n&=                                                          \sum^{n_u}_{j=1}\big|(X^{u}_LV_\text{CKM})_{jn}\big|^2\simeq                \sum^{n_u}_{j=1}\big|(\Theta_uV')_{jn}\big|^2\,,\\
    \delta^u_n-\Delta^u_n&=                                                          \sum^{n_d}_{j=1}\big|(V_\text{CKM}X^{d \dagger}_L)_{nj}\big|^2 \simeq                         \sum^{n_d}_{j=1}\big|(V'\Theta^\dagger_d)_{nj}\big|^2\,.
\end{split}\end{equation}
Finally, 
it is instructive to consider 
the following limit scenarios:
\begin{itemize}
    \item 
    When there are only down-type VLQs in the theory ($n_u=0$, $n_d \neq 0$), the rows of $V$ are orthonormal, which gives $F^u = \id_3$ and $\Delta^u_n=0$.
    However, since these rows have more than three entries, the rows of the $3 \times 3$ $V_\text{CKM}$ generically deviate from unitarity, i.e.~they are not orthonormal, $\delta^u_n=\sum^{n_d}_{j=1}|V_{n,j+3}|^2\neq0$. 
    Meanwhile, since in these models $V$ is a $3\times (3+n_d)$ matrix,
    $\delta^d_n=\Delta^d_n$, and the columns of $V_\text{CKM}$ are also not expected to be orthonormal.
    \item
    When there are only up-type VLQs in the theory ($n_u \neq 0$, $n_d=0$), the columns of $V$ are orthonormal, i.e.~$F^d = \id_3$ and $\Delta^d_n=0$.
    Nevertheless, $\delta^d_n=\sum^{n_u}_{j=1}|V_{j+3,n}|^2\neq0$ generically.
    Meanwhile, since in these models $V$ is a $(3+n_u)\times3$ matrix, $\delta^u_n=\Delta^u_n$. As in the previous case, neither the rows nor the columns of the $3\times 3$ $V_\text{CKM}$ are expected to be orthonormal in general.
\end{itemize}
In the following sections, section \ref{sec:quarkup} and section \ref{sec:quarkdown}, useful parameterizations for $V$ and $F^{u,d}$ in such limit cases, where only one VLQ isosinglet is added to the SM, are discussed.

As stated in section \ref{sec:quarksgen}, the perturbativity of the Yukawa couplings of SM quarks to the Higgs boson is one of the most important constraints on the mixing matrix, $V$, and the matrices that control its deviations from unitarity $X_L^u$ and $X_L^d$.
To make the interplay between deviations from unitarity, mass scales and perturbativity explicit, consider the upper-left block on both sides of the relation $\cM_q\cM^\dagger_q=\cV^q_L\mathcal{D}^2_q\cV^{q\dagger}_L$
\begin{equation}\label{eq:perturb-bidiagonalization}
    m_qm^\dagger_q+\om_q\om^\dagger_q=K^q_L\big(d^2_q+X^{q \dagger}_LD^2_qX^{q}_L\big)K^{q\dagger}_L\,,
\end{equation}
combining it with eq.~\eqref{eq:uparam-exactK}, this relation implies
\begin{equation}
\label{eq:perturb-trace}
\begin{aligned}    
     \textrm{Tr}\big(m_qm^\dagger_q+\om_q\om^\dagger_q\big)&=
     \textrm{Tr} \Big[\big(d^2_q+X^{q\dagger}_LD^2_qX^{q}_L\big)\big(1+X^{q\dagger}_LX^{q}_L\big)^{-1}\Big]\\
     &\simeq    
   \sum^3_{i=1}m^2_{q_i}+\sum^3_{i=1}\sum^{n_q}_{j=1}\big|(\Theta_q)_{ij}\big|^2M^2_{q_j}
    \,,
\end{aligned}
\end{equation}
where in the second line an approximation valid for small deviations from unitarity was considered. Here, $m_{q_i}$ and $M_{q_j}$ refer, respectively, to the the light and heavy quark masses in a given sector.
At this stage, the link to perturbativity is disclosed via the definitions $m_q=\frac{v}{\sqrt{2}}Y_q$ and $\om_q=\frac{v}{\sqrt{2}}\overline{Y}\!_q$ 
and the requirement that
\begin{equation}
    \textrm{Tr}\big(m_qm^\dagger_q+\om_q\om^\dagger_q\big)<\rho\, m^2_t\,,
\end{equation}
for some $\mathcal{O}(1-10)$ numerical factor $\rho$. This requirements reflects the fact that the individual Yukawa couplings should be at most $ \sim \mathcal{O}(1)$, or the theory may not be perturbative.
Combining this bound with eq.~\eqref{eq:perturb-trace}, one finds
\begin{equation}\label{eq:perturb-bound}
 \sum_{i,j}\big|(\Theta_q)_{ij}\big|^2 M^2_{q_j}<\rho\,m^2_t
    -\sum_i
    m^2_{q_i}\,,
\end{equation}
showing how the deviations from unitarity are limited by the ratio between the electroweak and VLQ scales in perturbative models.

For illustrative purposes, a model
with only a single up-type VLQ $T$, has
\begin{equation}\label{eq:perturb-traceU}
\sum_{i,j}\big|(\Theta_u)_{ij}\big|^2 M^2_{u_j}
= \left(|\Theta_u|^2+|\Theta_c|^2+|\Theta_t|^2\right)m^2_T ~,
\end{equation}
where the $3\times1$ matrix, $\Theta_u$, here is implicitly defined as $(\Theta_u,\Theta_c,\Theta_t)^T$.
In such a model, eq.~\eqref{eq:perturb-bound} becomes
\begin{equation}
    \sqrt{|\Theta_u|^2+|\Theta_c|^2+|\Theta_t|^2}
    \lesssim\frac{m_t}{m_T}\,,
\end{equation}
which shows that all deviations from unitarity must be smaller than $\mathcal{O}(m_t/m_T)$, as expected.

\subsection{CP Violation in  the Presence of VLQs}
\label{sec:CP_VLQ}

CP violation arises from a conflict between the CP transformations demanded by the gauge interactions and those required by the Yukawa interactions. The most general CP transformation is defined as the one that preserves the gauge interactions~\cite{Grimus:1995zi}. Subsequently, one may focus on the restrictions that CP invariance imposes on the Yukawa couplings. In the framework of the SM, it can be readily demonstrated that CP invariance implies~\cite{Bernabeu:1986fc}
\begin{equation}
I_\textrm{SM} \,\equiv\, \textrm{Tr}\left[
Y_u Y_u^\dagger ,\, Y_d Y_d^\dagger
\right]^3 \,=\, 0\,,
\label{eq:ISM}
\end{equation}
where $Y_{u,d}$ are the Yukawa couplings, i.e.~$I_\textrm{SM} = 0$ is a necessary condition for CP invariance. The quantity $I_\textrm{SM}$ is a weak-basis invariant, meaning it remains the same in any weak-basis representation. For a single generation, this condition trivially vanishes. For two generations, it can be shown that $I_\textrm{SM} = 0$ follows from the Hermiticity of $Y_q Y_q^\dagger$ ($q=u,d$). However, with three generations, equation \eqref{eq:ISM} imposes non-trivial constraints on the Yukawa couplings. As $I_\textrm{SM}$ can be evaluated in any weak-basis representation, one may consider the basis where $Y_u$ is diagonal and real. In this case, the following relationship holds
\begin{equation}
\begin{aligned}
I_\textrm{SM} &\,=\, 6i\,(y_1^2-y_2^2)(y_2^2-y_3^2)(y_3^2-y_1^2)\,\im\left(H_{12}H_{23}H_{31}\right)
\\[2mm]
&\,\xrightarrow{\textrm{\tiny EWSB}}\, \frac{6i}{v^{12}} \,(m_c^2 - m_u^2)(m_t^2 - m_c^2)(m_t^2- m_u^2) \\
& \qquad\qquad\quad
(m_b^2 - m_s^2)(m_b^2 - m_d^2)(m_s^2- m_d^2) \,\im\left( V_{us} V_{cb} V_{ub}^* V_{cs}^*\right)
\,,
\end{aligned}
\label{eq:ISM2}
\end{equation}
where $Y_u = \diag(y_1,y_2,y_3)$ and $H \equiv Y_d Y_d^\dagger$. Moreover, in this case eq. \eqref{eq:ISM} is both a necessary and sufficient condition to have CP invariance in the SM. It should be noted that in the case of three generations, the condition for CP invariance is equivalent to $\det [M_uM_u^\dagger, M_dM_d^\dagger] = 0$~\cite{Jarlskog:1985ht, Jarlskog:1985cw}, where this determinant is proportional to $I_\textrm{SM}$. For more than three generations, equation \eqref{eq:ISM} remains a necessary condition for CP invariance, but it is no longer sufficient (cf. ref. ~\cite{Branco:1999fs}).\par
In the first line of equation \eqref{eq:ISM2}, the invariant $I_\textrm{SM}$ is expressed solely in terms of Yukawa couplings, and the analysis is conducted at the Lagrangian level, prior to Electroweak Symmetry Breaking (EWSB). However, it is also possible to perform this analysis after EWSB, as shown in the second line of equation \eqref{eq:ISM2}, where the invariant is written in terms of quark masses and mixing. It is worth emphasizing that the CP conditions can be investigated at the Lagrangian level without necessarily relying on quark masses and mixing. The pioneering result by Kobayashi and Maskawa~\cite{Kobayashi:1973fv} could have been derived without much effort by considering a general CP transformation without invoking the matrix $V_\textrm{CKM}$. \par
CP violation is present in the quark sector of the SM if and only if any of the rephasing-invariant functions of the CKM matrix is not real~\cite{Branco:1999fs}. Examples of such functions include \emph{quartets} like $Q_{uscb} \equiv V_{us} V_{cb} V_{ub}^* V_{cs}^*$. In the SM quark sector, there is a single physical phase parameter, as demonstrated in section \ref{sec:par_counting} with $n_u=n_d=0$, which implies the equality of the imaginary parts of all quartets up to a sign \cite{Jarlskog:1985ht, Jarlskog:1985cw, Dunietz:1985uy}. By defining $I \equiv \im Q_{uscb}$, it becomes evident that the condition $I=0$ is not only consistent with CP invariance but also implies it, as evident from eq. \eqref{eq:ISM2}. Moreover, the quantity $|I|$ has a geometrical interpretation, as it corresponds to twice the area of any unitarity triangle. Taking into consideration the magnitudes of the moduli of $V_\text{CKM}$ elements, the standard and most useful unitarity triangle is defined by the unitarity relation:
\begin{equation}
V_{ud}V_{ub}^* + V_{cd} V_{cb}^* + V_{td} V_{tb}^* = 0\,,
\label{eq:triangle}
\end{equation}
with each term in the sum corresponding to a side of the triangle in the complex plane.\par
The experimental determination of the physical CP-violating phases present in the quark mixing matrix serves as a rigorous test of the Standard Model (SM). Without enforcing unitarity, the $3\times 3$ matrix $V_\text{CKM}$ contains four independent rephasing-invariant phases, as five out of the initial nine phases can be eliminated through the rephasing of the standard quark fields, as discussed earlier. It is crucial to emphasize that this result is entirely general and holds true \textbf{regardless of the presence of VLQs} and without relying on the existence of extra generations
These 4 phases can taken to be~\cite{Aleksan:1994if,Botella:2002fr}:
\begin{equation}
\begin{aligned}
\beta &\,\equiv\, \arg\left(-V_{cd}V_{tb}V_{cb}^*V_{td}^*\right)\,,\\
\gamma &\,\equiv\, \arg\left(-V_{ud}V_{cb}V_{ub}^*V_{cd}^*\right)\,,\\
\chi \text{ or } \beta_s &\,\equiv\, \arg\left(-V_{cb}V_{ts}V_{cs}^*V_{tb}^*\right)\,,\\
\chi' \text{ or } \beta_K &\,\equiv\, \arg\left(-V_{us}V_{cd}V_{ud}^*V_{cs}^*\right)\,.
\end{aligned}
\label{eq:CKMphases}
\end{equation}
Another commonly used definition is $\alpha \equiv \arg\left(-V_{td}V_{ub}V_{tb}^* V_{ud}^* \right)$, however, it is not an independent phase since it is related to the phases mentioned earlier by $\alpha + \beta + \gamma = \pi$ (mod $2\pi$) by definition. \par
The matrix $V_\text{CKM}$, which describes the mixing among the known quarks, comprises 13 measurable quantities: 9 moduli and 4 rephasing-invariant phases. In the limit of the Standard Model where $V_\text{CKM}$ is unitary, there are $n_f^2 = 9$ exact and independent relations among these quantities, such as the one given in equation \eqref{eq:triangle}, resulting in a total of 4 independent parameters. These 4 parameters are typically chosen as the standard 3 mixing angles and 1 CP-violation phase. However, they may also be taken to be i) the phases in eq. \eqref{eq:CKMphases}, or ii) four independent moduli. In fact, ref.\cite{Aleksan:1994if} demonstrated that assuming $3\times 3$ unitarity of $V_\text{CKM}$ allows the rephasing-invariant phases to be used for reconstructing the entire CKM matrix. In the SM limit, the phases $\alpha$, $\beta$, and $\gamma$ also correspond to the inner angles of the standard unitarity triangle mentioned earlier. Alternatively, the $3\times 3$ unitary $V_\text{CKM}$, up to a two-fold ambiguity and including the strength of CP violation measured by $|I|$, can be reconstructed by taking 4 independent moduli $|V_{ij}|$ as input\cite{Botella:1985gb, Branco:1987mj}. An additional sign is needed for the complete reconstruction since the moduli do not determine the sign of CP violation. Finally, it is important to note that physical processes can only depend on WB invariant quantities. It has been shown~\cite{Branco:1987mj} that in the SM, any four independent moduli of $V_\text{CKM}$ can be computed from the four WB invariants
\begin{equation}
\textrm{Tr} (H_u H_d)\,, \quad  \textrm{Tr} (H_u H^2_d)\,, \quad \textrm{Tr} (H^2_u H_d)\,, \quad \textrm{Tr} (H^2_u H^2_d)\,,
\label{eq:SMWBinv}
\end{equation}
and the six known quark masses, with $H_q \equiv Y_q Y_q^\dagger$ ($q=u,d$).\par
\vskip 2mm
The presence of VLQs introduces new sources of CP violation, as demonstrated in section \ref{sec:par_counting}. To illustrate this, one may consider the case where a single up-type VLQ is added to the SM ($n_u = 1$, $n_d = 0$), while taking the extreme chiral limit $m_u = m_c = m_d = m_s = 0$. In this chiral limit, $I_\text{SM} = 0$, meaning there is no CP violation in the SM since the masses of same-charge quarks are degenerate.
However, the introduction of VLQ interactions allows for the presence of CP-violating phases even in this chiral limit \cite{delAguila:1997vn}. Specifically, for $n_u = 1$ and $n_d = 0$, it is possible to choose a WB, see section \ref{sec:WBphy}, in which $\mathcal{M}_d$ is diagonal through appropriate choices of $W_L$ and $\mathcal{W}^d_R$, while $\mathcal{M}_u = \mathcal{V}^u_L, \mathcal{D}_u$ with $\mathcal{V}_R^u$ being cancelable by a suitable $\mathcal{W}^u_R$. Consequently, the up-quark mass matrix has two vanishing columns in the chiral limit. By using the remaining rotation and rephasing freedoms, it is possible to arrive at a WB where
\begin{equation}
\mathcal{M}_u \sim \begin{pmatrix} 
0 & 0 & \times & 0 \\
0 & 0 & 0 & \times \\
0 & 0 & \times & \ast\\
0 & 0 & \times & \times
\end{pmatrix} \,, \qquad
\mathcal{M}_d \sim \begin{pmatrix} 
0 & 0 & 0 \\
0 & 0 & 0 \\
0 & 0 & m_b
\end{pmatrix} \,.
\end{equation}
In this WB, all entries in the up-quark mass matrix are real except for the element denoted by ``$\ast$'', which is complex and cannot be removed in general. The presence of this complex element implies the non-vanishing of a WB invariant \cite{Branco:1986my,Albergaria:2022zaq} that is proportional to both $\im(\ast)$ and $m_b^2$, indicating a violation of CP symmetry beyond the SM. The chiral limit is not only physically natural for studying CP violation beyond the SM but also phenomenologically relevant in practice. In high-energy collisions, such as those at the TeV scale, the masses of light fermions are negligible, and distinguishing light quark jets becomes impractical. Therefore, significant CP violation at high energy is expected to originate from beyond the SM physics.

\vskip 2mm

In the presence of VLQs, the extraction of physical phases from data requires reexamination. The mixing matrix $V$ is now of size $(3+n_u) \times (3+n_d)$ and generally non-unitary. As discussed earlier, its upper-left $3\times 3$ block $V_\text{CKM}$ contains 4 independent physical phases, as shown in eq. \eqref{eq:CKMphases}. Without loss of generality, one can adopt the following phase convention~\cite{Branco:1999fs}
\vskip -3mm
\begin{equation}
V = 
  \tikz[baseline=(M.west)]{
    \node[matrix of math nodes,matrix anchor=west,left delimiter=(,right delimiter=),ampersand replacement=\&] (M) {
{}|V_{ud}{}| \& {}|V_{us}{}|\,e^{i\chi'} \& {}|V_{ub}{}|\,e^{- i \gamma} \& \,\cdots\, \\
-{}|V_{cd}{}| \& {}|V_{cs}{}| \& {}|V_{cb}{}| \& 	\cdots \\
{}|V_{td}{}|\,e^{-i \beta} \& -{}|V_{ts}{}|\,e^{i \chi} \& {}|V_{tb}{}| \& \cdots \\
\vdots \& \vdots \& \vdots \& \ddots\\
    };
    \node[draw,fit=(M-3-1)(M-1-3),inner sep = -1pt,label={\scriptsize $V_\text{CKM}$}] {};
  }\,.
\end{equation}

In the presence of NP effects, the extraction of the CKM phases from experimental data can be affected. In models with VLQs, where NP contributions are negligible in weak processes where the SM dominates at tree level, the determination of $\gamma$ and the moduli from the first two rows of $V_\text{CKM}$ remains unaffected. However, the extraction of $\beta$ and $\chi$ from $B$ meson decays may be contaminated by NP effects. Experiments may instead be sensitive to the combinations $\bar\beta \equiv \beta - \phi_d$ and $\bar\chi \equiv \chi + \phi_s$, where $\phi_q$ ($q=d,s$) are phases parameterising NP-induced deviations from the SM in the neutral meson systems $B^0_q$--$\overline{B^0_q}$. For example, the decay $B_d^0 \to J/\psi , K$ is sensitive to $\sin 2(\bar\beta - \chi') \simeq \sin 2\bar\beta$, while the decay $B_s^0 \to J/\psi , \phi$ measures $\sin 2\bar\chi$.\par
In the context of the SM or NP scenarios where $V_\text{CKM}$ is unitary, the remaining phase $\chi'$ is expected to be extremely small, with a value of $\chi' \sim 6 \times 10^{-4}$\cite{Aguilar-Saavedra:2004roc}. Even if $V_\text{CKM}$ is not strictly unitary but can still be described as the upper-left block of an auxiliary $(3+n_u+n_d) \times (3+n_u+n_d)$ unitary matrix \cite{Branco:1992wr}, a bound can still be placed on $\chi'$\cite{Kurimoto:1997ex, Branco:1999fs}. Considering the CKM moduli within their $3\sigma$ ranges~\cite{ParticleDataGroup:2020ssz}, it has been found that $|\chi'|$ is constrained to be less than or around $0.06$, which is approximately equivalent to $3^\circ$. Consequently, in most cases of interest, this phase can be safely neglected. Similarly, the phase $\chi$ can be significantly larger than the SM expectation $\chi \sim 0.02$ in models involving up-type VLQs, but not in models solely comprising down-type VLQs~\cite{Aguilar-Saavedra:2004roc}. Up-type VLQs have been considered to address tensions in the data by introducing a non-zero $\phi_d$\cite{Botella:2012ju}. \par 
Furthermore, both up-type and down-type VLQs have been explored to potentially explain apparent deviations from unitarity in the first row of $V_\text{CKM}$\cite{Belfatto:2019swo, Belfatto:2021jhf, Branco:2021vhs}. It is important to note that in the presence of possible deviations from unitarity, the familiar SM unitary triangle may be modified. For instance, in extensions involving only one VLQ, the unitary triangle may be replaced by various quadrangles~\cite{delAguila:1997vn}. As a result, instead of having only one CP-odd invariant, proportional to $\im\left( V_{us} V_{cb} V_{ub}^* V_{cs}^*\right)$ as in the SM, one may obtain several rephasing-invariants arising from other quartets. These new invariants are proportional to $\im\left( V_{ij} V_{km} V_{im}^* V_{kj}^*\right)$ and, in general, are not equal but may be partly related to each other.\par

\section{Models with $n_d=0$ or $n_u=0$}\
\label{sec:quarkmodels}
To illustrate some features of VLQ models, this section will discuss models where only up-like (down-like) VLQs are added to the SM, hence $n_d=0$ ($n_u=0$). These models are relevant in light of the recent CKM unitarity problem \cite{Czarnecki:2019mwq,Seng:2020wjq,Hayen:2020cxh,Shiells:2020fqp,Belfatto:2019swo,Belfatto:2021jhf}, as they are a natural and simple solution for it. Imposing unitarity of the CKM matrix in the first row gives
\begin{equation}
    |V_{ud}|^{2}+|V_{us}|^{2}+|V_{ub}|^{2}=1\,.  
\end{equation}
However, considering the fact that $|V_{ub}|^2$ is much smaller than the other two entries, with $|V_{ub}|^2 \simeq 1.6\times 10^{-5}$, the following approximate relation
\begin{equation}\label{eq:unitary}
    |V_{ud}|^{2}+|V_{us}|^{2} \simeq 1\,,
\end{equation}
is valid up to $\mathcal{O}\left(|V_{ub}|^2\right)$.
Through precise experimental data and improved control over theoretical uncertainties, the values of $|V_{ud}|$, $|V_{us}|$, and $|V_{ud}|/|V_{us}|$ have been determined with high accuracy. These results, along with the approximate relation stated in equation \eqref{eq:unitary}, provide stringent tests for the unitarity of the Cabibbo-Kobayashi-Maskawa (CKM) matrix.

At energy scales much lower than the mass of the $W$ boson ($M_W$), the matrix elements $|V_{ud}|$ and $|V_{us}|$ play a crucial role in describing the strength of the effective charged-current interaction responsible for the leptonic decays of hadrons involving the valence quarks $u$, $d$, and $s$. Experimental data from neutron decays~\cite{UCNt:2021pcg} primarily contribute to the determination of $|V_{ud}|$, while the entry $|V_{us}|$ is mainly obtained from kaon decay data. The ratio $\left| V_{us}/V_{ud}\right|$ can be independently determined by comparing the decay rates of specific radiative pion and kaon decays.

However, recent advancements in the determination of form factors and inclusion of radiative corrections for relevant meson and neutron decay processes have resulted in a downward shift in the central value of $|V_{ud}|$. This has led to tension with the constraint imposed by the unitarity of the first row of the CKM matrix~\cite{Seng:2018yzq,Seng:2018qru,Czarnecki:2019mwq,Seng:2020wjq,Hayen:2020cxh,Shiells:2020fqp}.
Taking into account these new theory calculations with reduced hadronic uncertainties, the authors of Ref.~\cite{Belfatto:2019swo} find:
\begin{equation}\label{eq:cabibbo2}
    \begin{array}{ccc}
         |V_{us}|=0.22333(60)\,,\quad  & |V_{ud}|=0.97370(14)\,,\quad  & \left|V_{us}/V_{ud}\right| =0.23130(50)\,.
    \end{array}
\end{equation}
These results deviate from the unitarity condition in eq.~\eqref{eq:unitary} by
more than $4\sigma $, disfavouring the SM CKM unitarity at this level.
In fact, the values in eq.~\eqref{eq:cabibbo2} are compatible with
\begin{equation}\label{eq:unitary2}
    |V_{ud}|^{2}+|V_{us}|^{2}=1-\delta\,,
\end{equation}
where 
\begin{equation}\label{eq:delta}
    \sqrt{\delta} =0.04\pm 0.01\, \quad(95\%\text{ C.L.})\,.
\end{equation}
Note that 
\begin{equation}
    \delta \equiv\delta_1^u
\end{equation}
was implicitly defined, from the definition of eq. \eqref{eq:rowunitar}.
Consequently, it is observed that the parameter $\delta$ is of the order of $\mathcal{O}(|V_{cb}|^2)$. Recent lattice calculations performed by the FLAG collaboration~\cite{FlavourLatticeAveragingGroupFLAG:2021npn} also support these findings. This deviation between the predictions of the Standard Model and experimental data is commonly known as the CKM unitarity problem or Cabibbo anomaly.

To address this discrepancy, new physics beyond the Standard Model is required if it indeed persists. One possible solution involves introducing an additional mixing channel. One approach to achieve this is by introducing a singlet vector-like quark (VLQ) into the framework of the Standard Model. The inclusion of an extra VLQ would result in an enlarged mixing matrix, where the effective CKM mixing is described by a non-unitary $3\times 3$ sub-matrix involving the standard quarks. In the literature, models considering both one additional down-type VLQ and one additional up-type VLQ have been explored within this context~\cite{Belfatto:2019swo,Belfatto:2021jhf,Branco:2021vhs,Botella:2021uxz}.

In theories in which only up-type VLQs are introduced, it is useful to work in a WB where the down quarks mass matrix $\mathcal{M}_d$, in this case a $3 \times 3$ matrix, is diagonal, $d_d$. This WB exists when $n_d=0$, as stated in section \ref{sec:WBphy}. In this case, from eq. \eqref{def:V:AB} one obtains
\begin{equation}
\label{eq:VLu}
\mathcal{V}_L^{u \dagger} = \begin{pmatrix} V & B_L^{u \dagger} \end{pmatrix}\,,
\end{equation}
where eq. \eqref{eq:mixingmatrixquark} was used in combination with
\begin{equation}
 A_L^d = \id_3  ~,
\end{equation}
which is valid for the WB in use. $\mathcal{V}_L^{u \dagger}$ is a $(3+n_u) \times (3+n_u)$ unitary matrix, convenient to parameterise because the parameterisations for unitary matrices are known and the physical $(3+n_u) \times 3$ mixing matrix, $V$, can be read off directly from it -  \emph{its first three columns}. Hence, when only one sector has extra mixing channels, this procedure might be more useful and easier to grasp, as one can use parameterisations a la PDG, with extra mixing angles and phases.\\

Again, for models in which only down-type VLQs are introduced, it is useful to work in a WB where the up quarks mass matrix $\mathcal{M}_u$, in this case a $3 \times 3$ matrix, is diagonal, $d_u$. This WB exists when $n_u=0$, as stated in section \ref{sec:WBphy}. In this case, from eq. \eqref{def:V:AB} one obtains
\begin{equation}
\label{eq:VLd}
\mathcal{V}_L^{d} = \begin{pmatrix} V \\ B_L^d \end{pmatrix}\,,
\end{equation}
where eq. \eqref{eq:mixingmatrixquark} was used in combination with
\begin{equation}
 A_L^{u \dagger} = \id_3  ~,
\end{equation}
which is valid for the WB in use. $\mathcal{V}_L^{d}$ is a $(3+n_d) \times (3+n_d)$ unitary matrix. Again, the physical $3 \times (3+n_d)$ mixing matrix, $V$, can be read off directly from it -  \emph{its first three rows}.

\subsection{Model with $n_u=1$, $n_d=0$}
\label{sec:quarkup}
As an example, consider the case $n_u=1$, which is the case of ref. \cite{Branco:2021vhs}. This reference intended to provide a simple and minimal answer to the CKM unitarity problem.
Consider the following parameterisation for the matrix in eq. \eqref{eq:VLu} in terms of 6 mixing angles and 3 phases, first used in ref. \cite{Botella:1985gb},
\begin{equation}\label{eq:SVLQ-Uup}
    \mathcal{V}_L^{u \dagger}=\mkern-4mu
    \begingroup
    \setlength\arraycolsep{2pt}
    \begin{pmatrix}
    1\\&1\\&&c_{34}&s_{34}\\&&-s_{34}&c_{34}
    \end{pmatrix}
    \mkern-8mu
    \begin{pmatrix}
    1\\&c_{24}&&s_{24}e^{-i\delta_{24}}\\&&1\\&-s_{24}e^{i\delta_{24}}&&c_{24}
    \end{pmatrix}
    \mkern-8mu
    \begin{pmatrix}
    c_{14}&&&s_{14}e^{-i\delta_{14}}\\&1\\&&1\\-s_{14}e^{i\delta_{14}}&&&c_{14}
    \end{pmatrix}
    \mkern-8mu
    \begin{pmatrix}
    &&&0\\&V_\text{PDG}&&0\\&&&0\\0&0&0&1
    \end{pmatrix}
    \,,
    \endgroup
\end{equation}
where $V_\text{PDG}$ is the standard PDG parameterization of a $3\times3$ unitary matrix~\cite{Chau:1984fp,pdg},
\begin{equation}\label{eq:SVLQ-PDG}
    V_\text{PDG}=\mkern-4mu
    \begin{pmatrix}
    c_{12}c_{13}&s_{12}c_{13}&s_{13}e^{-i\delta}\\
    -s_{12}c_{23}-c_{12}s_{23}s_{13}e^{i\delta}&c_{12}c_{23}-s_{12}s_{23}s_{13}e^{i\delta}&s_{23}c_{13}\\
    s_{12}s_{23}-c_{12}c_{23}s_{13}e^{i\delta}&-c_{12}s_{23}-s_{12}c_{23}s_{13}e^{i\delta}&c_{23}c_{13}
    \end{pmatrix}\,,
\end{equation}
and $c_{ij} = \cos \theta_{ij}$ and $s_{ij} = \sin \theta_{ij}$, with $\theta_{ij} \in [0,\pi/2]$, $\delta_{ij} \in [0,2\pi]$. Note that $V_\text{CKM}$ is described by the traditional PDG picture when all $\theta_{i4}\rightarrow 0$.
With this parameterization, the matrices which control the tree-level FCNC of these models are $F^d=\id_3$ and
\begin{equation}\label{eq:SVLQ-FU}
    F^u=\mkern-4mu
    \begin{pmatrix}
    1-s^2_{14}&-s_{14}c_{14}s_{24}e^{-i(\delta_{14}-\delta_{24})}&-s_{14}c_{14}c_{24}s_{34}e^{-i\delta_{14}}&-s_{14}c_{14}c_{24}c_{34}e^{-i\delta_{14}}\\
    -s_{14}c_{14}s_{24}e^{i(\delta_{14}-\delta_{24})}&1-c^2_{14}s^2_{24}&-c^2_{14}s_{24}c_{24}s_{34}e^{-i\delta_{24}}&-c^2_{14}s_{24}c_{24}c_{34}e^{-i\delta_{24}}\\
    -s_{14}c_{14}c_{24}s_{34}e^{i\delta_{14}}&-c^2_{14}s_{24}c_{24}s_{34}e^{i\delta_{24}}&1-c^2_{14}c^2_{24}s^2_{34}&-c^2_{14}c^2_{24}s_{34}c_{34}\\
    -s_{14}c_{14}c_{24}c_{34}e^{i\delta_{14}}&-c^2_{14}s_{24}c_{24}c_{34}e^{i\delta_{24}}&-c^2_{14}c^2_{24}s_{34}c_{34}&1-c^2_{14}c^2_{24}c^2_{34}
    \end{pmatrix}
    \,.
\end{equation}
For small angles and neglecting the $\mathcal{O}(\theta^3_{i4})$ terms, one can express the latter as
\begin{equation}\label{eq:SVLQ-FU:exp}
    F^u\simeq
    \begin{pmatrix}
    1-\theta^2_{14}&-\theta_{14}\theta_{24}e^{-i(\delta_{14}-\delta_{24})}&-\theta_{14}\theta_{34}e^{-i\delta_{14}}&-\theta_{14}e^{-i\delta_{14}}\\
    -\theta_{14}\theta_{24}e^{i(\delta_{14}-\delta_{24})}&1-\theta^2_{24}&-\theta_{24}\theta_{34}e^{-i\delta_{24}}&-\theta_{24}e^{-i\delta_{24}}\\
    -\theta_{14}\theta_{34}e^{i\delta_{14}}&-\theta_{24}\theta_{34}e^{i\delta_{24}}&1-\theta^2_{34}&-\theta_{34}\\
    -\theta_{14}e^{i\delta_{14}}&-\theta_{24}e^{i\delta_{24}}&-\theta_{34}&\theta^2_{14}+\theta^2_{24}+\theta^2_{34}
    \end{pmatrix}
    \,.
\end{equation}

\vskip 2mm

Comparing eq. \eqref{eq:duFCNC-approx} and eq. \eqref{eq:SVLQ-FU:exp}, one concludes that
\begin{equation}\label{eq:SVLQ-Tup}
    \Theta_u=-\theta_{14}e^{-i\delta_{14}}\,,\quad\Theta_c=-\theta_{24}e^{-i\delta_{24}}\,,\quad\Theta_t=-\theta_{34}\,,
\end{equation}
where the angles and phases now refer to the parameters introduced in~\ref{eq:SVLQ-Uup}.
Like before, the vector $\Theta_u$ is implicitly defined as $(\Theta_u,\Theta_c,\Theta_t)^T$.

Using the definitions in eq. \eqref{eq:rowunitar} one can derive for this case

\begin{equation}
\begin{aligned}
\delta \,\equiv\, \delta_1^u \,&=\, 1 - |V_{ud}|^2  - |V_{us}|^2 - |V_{ub}|^2 \,&\!\!\!\!=\, \left|\mathcal{V}_{L_{41}}^*\right|^2 \,&=\, s_{14}^2 \,,\\[1mm]
\delta_2^u \,&=\, 1 - |V_{cd}|^2  - |V_{cs}|^2 - |V_{cb}|^2 \,&\!\!\!\!=\, \left|\mathcal{V}_{L_{42}}^*\right|^2 \,&=\, c_{14}^2 \,s_{24}^2 \,,\\[1mm]
\delta_3^u \,&=\, 1 - |V_{td}|^2  - |V_{ts}|^2 - |V_{tb}|^2 \,&\!\!\!\!=\, \left|\mathcal{V}_{L_{43}}^*\right|^2 \,&=\, c_{14}^2\, c_{24}^2\, s_{34}^2 \,.
\label{eq:defdelta}
\end{aligned}
\end{equation}

hence obtaining

\begin{align}
\label{eq:fu}
\renewcommand{\arraystretch}{1.5}
     F^u \simeq\,
\small
\begin{pmatrix}
 1- \delta_1^u
& - \sqrt{\delta_1^u} \sqrt{\delta_2^u} e^{-i (\delta_{14}-\delta_{24})}
& - \sqrt{\delta_1^u} \sqrt{\delta_3^u} e^{-i \delta_{14}}
& - \sqrt{\delta_1^u} e^{-i \delta_{14}} \\
  - \sqrt{\delta_1^u} \sqrt{\delta_2^u} e^{i (\delta_{14}-\delta_{24})}
& 1 - \delta_2^u
& - \sqrt{\delta_2^u} \sqrt{\delta_3^u} e^{-i \delta_{24}}
& - \sqrt{\delta_2^u} e^{-i \delta_{24}} \\
  - \sqrt{\delta_1^u} \sqrt{\delta_3^u} e^{i \delta_{14}}
& - \sqrt{\delta_2^u} \sqrt{\delta_3^u} e^{i \delta_{24}}
& 1 - \delta_3^u
& - \sqrt{\delta_3^u} \\
  - \sqrt{\delta_1^u} e^{i \delta_{14}} 
& - \sqrt{\delta_2^u} e^{i \delta_{24}}
& - \sqrt{\delta_3^u}
& \delta_1^u + \delta_2^u + \delta_3^u
\end{pmatrix}\,.
\end{align}
Finally, one can conclude that, in approximation,
\begin{equation}
    \delta_1^u = |\Theta_u|^2 ~,~ \delta_2^u = |\Theta_c|^2 ~,~\delta_3^u = |\Theta_t|^2 ~.
    \label{eq:deltaTheta}
\end{equation}
This result shows that there is a simple connection between the deviations from unitarity of the rows of the CKM matrix and the standard quark FCNC. This is a consequence of what was discussed in general in section \ref{sec:perturb}.

Since in the context of the CKM unitarity problem one has $\delta \equiv \delta_1^u \gg  \delta_2^u \sim  \delta_3^u$, using eq. \eqref{eq:deltaTheta} in the perturbativity discussion in section \ref{sec:perturb} gives

\begin{equation}
    m_T \lesssim \frac{m_t}{\sqrt{\delta_1^u+\delta_2^u+\delta_3^u}} \sim \frac{m_t}{\sqrt{\delta}}\,.
    \label{eq:perturbativity}
\end{equation}

Which, within the framework of trying to address the CKM unitarity puzzle, see eq. \eqref{eq:delta}, implies $m_T \lesssim 4.4$ TeV for $\sqrt{\delta}= 0.04$. 
Of course that introducing a VLQ does not come without consequences. The rest of this section follows \cite{Branco:2021vhs}. The phenomenological impacts of a up-type VLQ will be discussed, and will be used to restrict the parameter space of this model. With that, a global fit will be performed, with the goal of obtaining a region of the parameter space that is allowed by experiment and that can explain the CKM unitarity problem, while predicting a up-type VLQ, $T$, with a mass that can be probed in the current and next round of experiments.

\subsubsection{New Physics and Global Fit}
\label{sec:ph-fits}
A variety of measurable quantities can undergo modifications either at tree level or through loop processes when up-type vector-like quarks (VLQs) are present \cite{Cacciapaglia:2011fx,Okada:2012gy,Panizzi:2014dwa,Alok:2015iha}. 
The effects of new physics contributions on neutral meson mixing, specifically in the $D^0$, $K^0$, and $B^0_{d,s}$ neutral-meson systems, as well as the enhanced rare decay of the top quark, $t \to q Z$, are particularly significant. These phenomenological quantities play a crucial role in constraining the parameter space of this model. Furthermore, the existence of a new heavy quark, $T$, could potentially allow its production at current colliders if its mass is sufficiently low.

This quark can be generated either through single production or pair production processes \cite{AguilarSaavedra:2009es, DeSimone:2012fs,Buchkremer:2013bha,Aguilar-Saavedra:2013qpa}. The dominant contributions to pair production cross sections are dependent solely on the mass of $T$, $m_T$. On the other hand, single production mechanisms are controlled by functions related to the elements of the mixing matrix, namely $V_{Td}$, $V_{Ts}$, and $V_{Tb}$, hence more model-dependent.

In scenarios where significant couplings to light quark generations are present, single production processes can dominate pair production by several orders of magnitude within the few TeV energy range. This is primarily due to a larger available phase space, an enhancement from longitudinal gauge bosons, and the involvement of the proton's valence quarks ($u$ and $d$) in the initial state \cite{Atre:2008iu, Atre:2011ae}.
Our current scenario aligns with this description, as deviations from unitarity in the first row of the CKM matrix necessitate relatively large values for the quantities $|V_{Td}| \sim |F^u_{41}| \sim 0.04$ (note the approximation $|\mathcal{V}_{L{41}}^u| \sim |\mathcal{V}_{L{14}}^u|$ under the small mixing angle assumption). These quantities govern the primary contributions to the single VLQ production cross-sections via $W$ and $Z$ boson exchanges.

Regarding the decay of the heavy quark, the decay widths can be determined by generalizing the expressions provided in Appendix B of ref.~\cite{Aguilar-Saavedra:2013qpa} (see e.g.~\cite{Botella:2016ibj,Alves:2023ufm}) and they are also influenced by $|V_{Td}|$ and $|F^u_{41}|$. In the parameter space of this model, as will be discussed later, decays to the lightest generation can be approximately an order of magnitude more probable compared to decays to the second and third generations.

At the time of writing of ref. \cite{Branco:2021vhs}, the most stringent lower limits on the VLQ mass were $m_T > 1.3$ TeV at 95\% confidence level (CL) by the ATLAS collaboration \cite{Aaboud:2018pii}, and $m_T \gtrsim 1.0$ TeV at 95\% CL by the CMS collaboration \cite{Sirunyan:2019sza}, based on searches for pair-produced $T$-quarks.\par
For more details regarding new physics, effects on physical observables, the best experimental constraints and the most stringent bounds for models with VLQs see section \ref{sec:res_quarks}.\par
In order to explore the viability of this model, and to check if it can indeed solve the CKM unitarity process, a numerical scan of the parameter space of the model was performed. Initially, the model is restricted by the magnitudes of the elements in the CKM matrix. The current best-fit values of these elements, without enforcing unitarity, are obtained from~\cite{pdg}.
\begin{equation}
    |V_\text{CKM}| = 
    \begin{pmatrix}
      0.97370 \pm 0.00014  & 0.2245 \pm 0.0008 & (3.82 \pm 0.24) \times 10^{-3} \\
      0.221 \pm 0.004      & 0.987 \pm 0.011   & (41.0 \pm 1.4) \times 10^{-3} \\
      (8.0 \pm 0.3) \times 10^{-3} & (38.8 \pm 1.1) \times 10^{-3} & 1.013 \pm 0.030
    \end{pmatrix}\,.
\end{equation}
In addition, it is assumed that the presence of an up-type VLQ does not affect the value of the phase $\gamma = \arg (-V_{ud}V_{cb}V_{ub}^*V_{cd}^*)$ \cite{Botella:2008qm,Botella:2012ju}, which is obtained from SM tree-level dominated $B$ decays, $\gamma = (72.1^{+4.1}_{-4.5})\degree$~\cite{pdg}.
To assess the goodness of fit, the measure $N\sigma = \sqrt{\chi^2}$ is employed, where $\chi^2$ is approximated as the sum of priors,
\begin{equation}
    \chi^2 = \sum_{ij} \left(\frac{V_{ij}-V_{ij}^c}{\sigma(V_{ij})}\right)^2 + \left(\frac{\gamma-\gamma^c}{\sigma(\gamma)}\right)^2\,,
\end{equation}
with the superscript $c$ denoting central values and $\sigma(\gamma) = 4.5\degree$. Furthermore, it is assumed $m_c(M_Z) = 0.619 \pm 0.084$ GeV, $m_t(M_Z) = 171.7 \pm 3.0$ GeV~\cite{PhysRevD.77.113016} and require $m_T > 1$ TeV, in line with collider bounds.

The values of the new angles and phases compatible with the above criteria are shown as the dashed regions in the correlation plot of Figure~\ref{fig:correlations}. These regions contract to the solid green $2\sigma$ and $3\sigma$ contours after all the phenomenological constraints described before are taken into account. These constraints comprise the bounds on $\Delta m_N$ ($N=D^0, K^0, B^0, B^0_s$), and $|\epsilon_K|$ discussed in section~\ref{sec:newphysics} and the perturbativity bound of eq.~\eqref{eq:perturbativity}.
The data suggests that relatively large values for both $\theta_{14} \simeq \sqrt{\delta}$ and $\theta_{34}$ are favored, indicating a disfavoring of $\theta_{34}=0$ at a significance exceeding $2\sigma$. On the other hand, $\theta_{24}$ is consistent with zero, and the inclination towards smaller values of this angle is influenced by the constraint imposed by $D^0$-$\overline{D}^0$ mixing (refer to section~\ref{sec:D0D0bar}).

\afterpage{%
\begin{figure}[h!]
  \centering
\makebox[0pt]{\begin{minipage}{1.15\textwidth}
  \includegraphics[width=\textwidth]{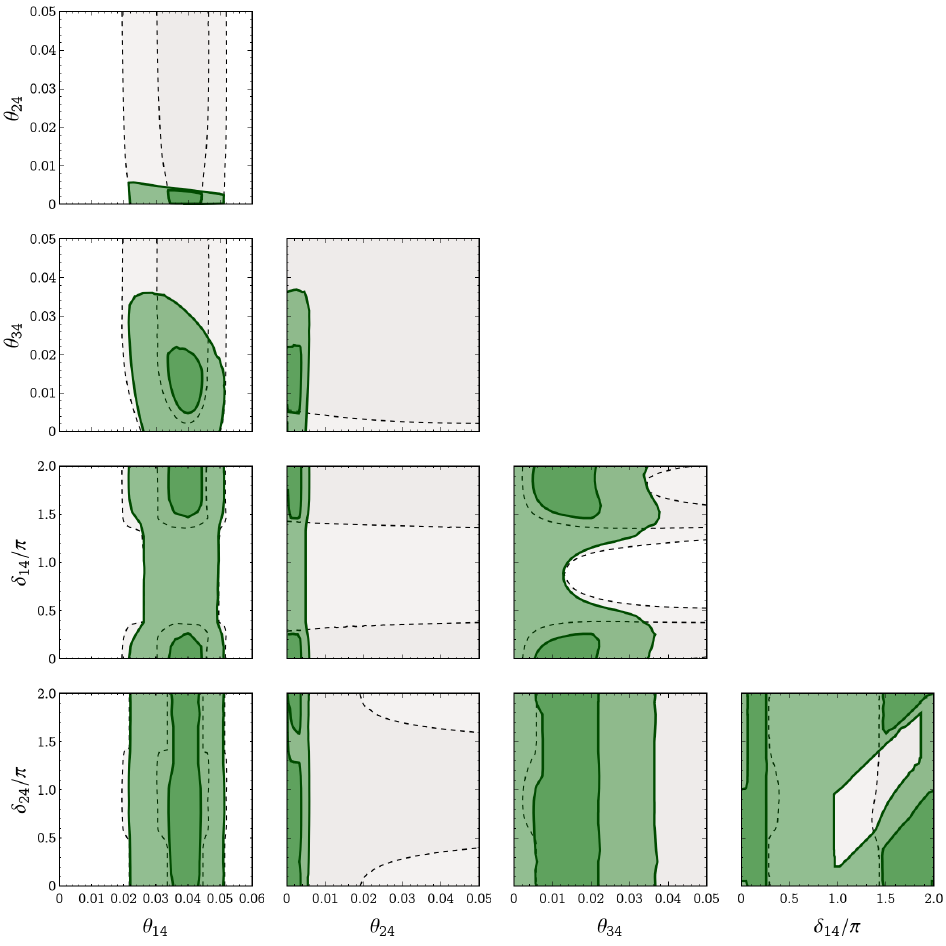}
  \caption{
  Regions of parameter space consistent with the constraints on $|V_{ij}|$ and $\gamma$ (dashed lines) and additionally consistent with the bounds on meson mixing, $|\epsilon_K|$ and the requirement of perturbativity (solid lines). For both sets of constraints, $2\sigma$ ($3\sigma$) contours are shown in darker (lighter) colour. }
  \label{fig:correlations}
\end{minipage}}
\end{figure}
\clearpage
}

\begin{figure}[h!]
  \centering
  \includegraphics[width=0.82\textwidth]{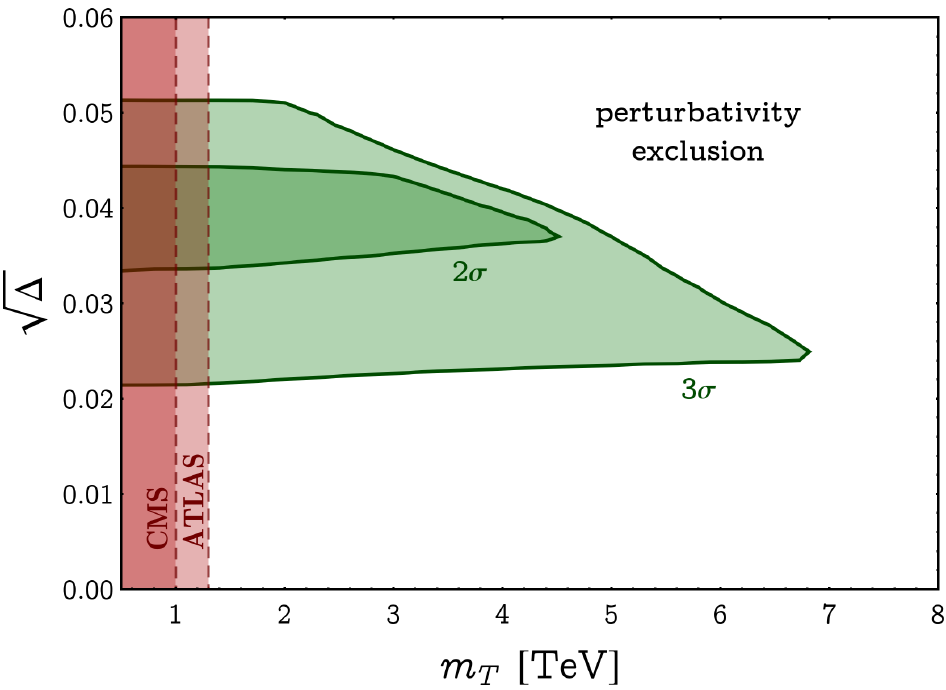}\qquad${}$
  \caption[ Masses $m_T$ of the new heavy quark and first-row deviations from unitarity ($\sqrt{\delta} \simeq \theta_{14}$) compatible with the bounds on $|V_{ij}|$, $\gamma$, meson mixing, $|\epsilon_K|$ and with the requirement of perturbativity. The latter constraint imposes an upper limit on $m_T$, while lower bounds are set by ATLAS and CMS (95\% CL).
   ]{
   Masses $m_T$ of the new heavy quark and first-row deviations from unitarity ($\sqrt{\delta} \simeq \theta_{14}$) compatible with the bounds on $|V_{ij}|$, $\gamma$, meson mixing, $|\epsilon_K|$ and with the requirement of perturbativity. The latter constraint imposes an upper limit on $m_T$, while lower bounds are set by ATLAS~\cite{Aaboud:2018pii} and CMS~\cite{Sirunyan:2019sza} (95\% CL).
   }
  \label{fig:perturbativity}
\end{figure}

The constraint of perturbativity also limits the permissible values of $m_T$, which are depicted in Figure~\ref{fig:perturbativity} against the deviations from unitarity in the first row. As shown by eq. \eqref{eq:perturbativity}, the maximum value of $m_T$ relies on the magnitude of these deviations. Assuming $\sqrt{\delta} = 0.04$, it follows that $m_T \lesssim 5$ TeV. Considering the entire $3\sigma$ region of the fit, this bound extends to $m_T \lesssim 7$ TeV.

\newpage

\begin{figure}[h!]
  \centering
\includegraphics[width=0.5\textwidth]{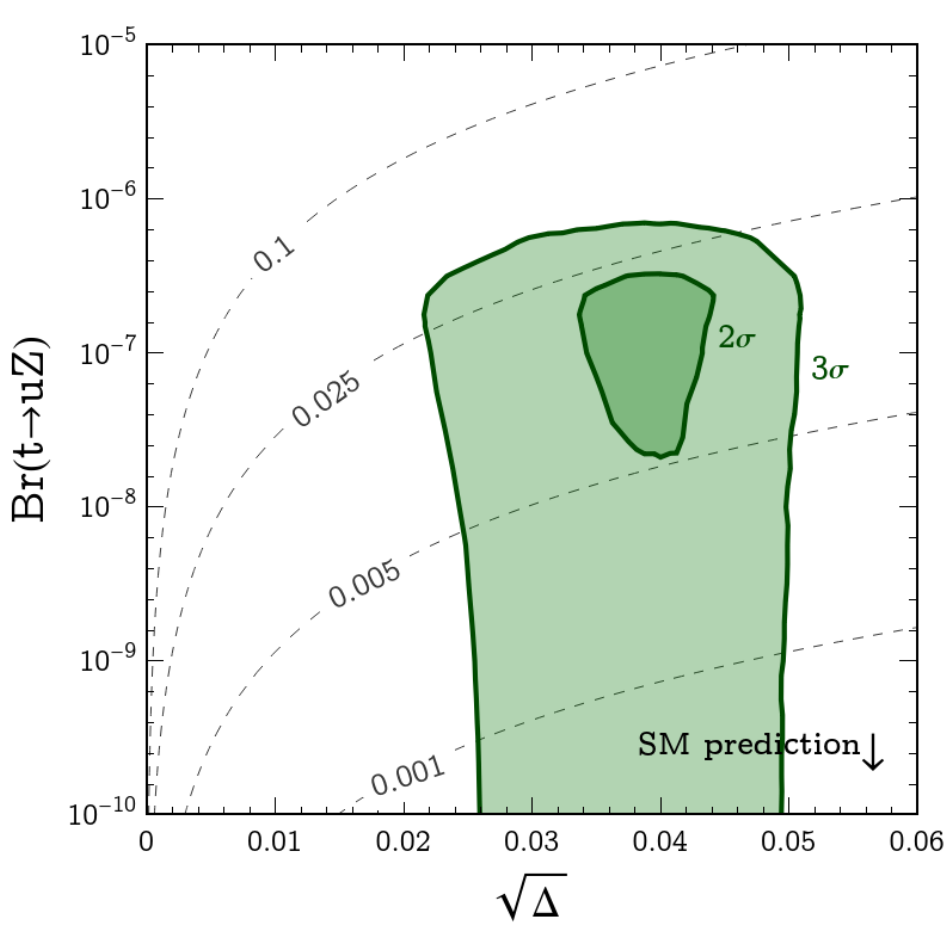}
 \caption[Branching ratios for the rare top decays $t \rightarrow u Z$ as a function of first-row deviations from unitarity $\delta$ ($\sqrt{\delta} \simeq \theta_{14}$). These may considerably exceed the SM expectations Br$(t \to u Z)_\text{SM} \sim 10^{-16}$~(not shown). Dashed contours refer to different values for $\theta_{34}$, in the small angle approximation.]{Branching ratios for the rare top decays $t \rightarrow u Z$ as a function of first-row deviations from unitarity $\delta$ ($\sqrt{\delta} \simeq \theta_{14}$). These may considerably exceed the SM expectations Br$(t \to u Z)_\text{SM} \sim 10^{-16}$ ~\cite{AguilarSaavedra:2004wm} (not shown). Dashed contours refer to different values for $\theta_{34}$, in the small angle approximation.}
  \label{fig:raretop1}
\end{figure}

\begin{figure}[h!]
  \centering
\includegraphics[width=0.5\textwidth]{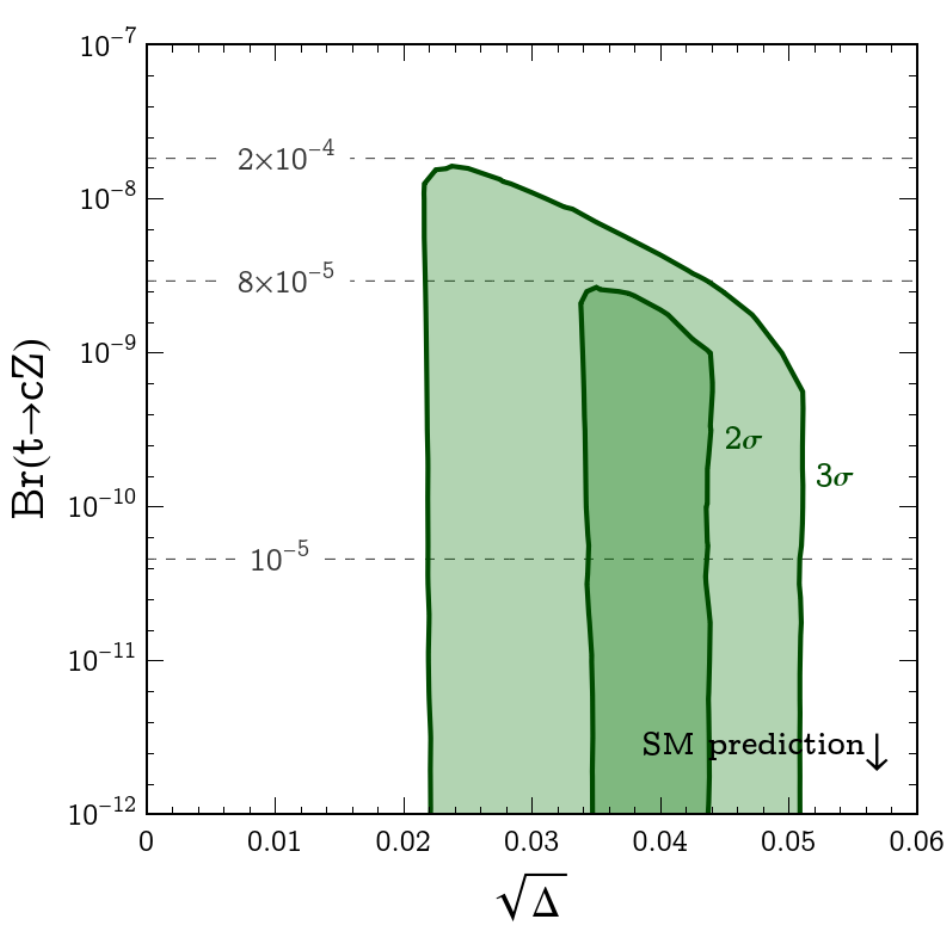}
 \caption[Branching ratios for the rare top decays $t \rightarrow c Z$ as a function of first-row deviations from unitarity $\delta$ ($\sqrt{\delta} \simeq \theta_{14}$). These may considerably exceed the SM expectations  Br$(t \to c Z)_\text{SM} \sim 10^{-14}$ (not shown). Dashed contours refer to different values for the product $\theta_{24}\,\theta_{34}$, in the small angle approximation.]{Branching ratios for the rare top decays $t \rightarrow c Z$ as a function of first-row deviations from unitarity $\delta$ ($\sqrt{\delta} \simeq \theta_{14}$). These may considerably exceed the SM expectations  Br$(t \to c Z)_\text{SM} \sim 10^{-14}$~\cite{AguilarSaavedra:2004wm} (not shown). Dashed contours refer to different values for the product $\theta_{24}\,\theta_{34}$, in the small angle approximation.}
  \label{fig:raretop2}
\end{figure}

Finally, in Figure~\ref{fig:raretop1} and Figure~\ref{fig:raretop2} the predicted values for the branching ratios of rare top decays $t\to uZ$ and $t\to cZ$, respectively, as a function of $\sqrt{\delta}$. These may considerably exceed the SM predictions. One finds $2.0\times 10^{-8} < \text{Br}(t\to uZ) < 3.5 \times 10^{-7}$ at the $2\sigma$ level, which is still 3 orders of magnitude below present bounds (see section~\ref{sec:raretop}). The decay $t\to cZ$ may instead be arbitrarily suppressed, even at the $2 \sigma$ level, by the allowed smallness of $\theta_{24}$.

\newpage
Considering the correlation plot in Figure \ref{fig:correlations}, the $2\sigma$ region of the fit, it is observed that $|V_{41}| > |V_{42}|, |V_{43}|$, indicating that the up-type vector-like quark $T$ exhibits stronger coupling to the up quark compared to the charm and top quarks. This challenges the conventional assumption that the new quark primarily couples to the third generation of Standard Model quarks. Consequently, the VLQ is more inclined to decay into quarks of the first generation, as there are no suppressions associated with the masses of these quarks. The decay widths primarily depend on $m_T$ and $|V_{4j}|^2$ or $|F_{4j}^u|^2$ \cite{Aguilar-Saavedra:2013qpa,Alves:2023ufm}.
A discussion of a low-$\chi^2$ benchmark belonging to this $2\sigma$ region now follows.
The selected point in parameter space corresponds to a mass $m_T = 1.5$ TeV and is described by
\begin{equation}
\begin{aligned}
    &\theta_{12}= 0.2265 ~,~~\theta_{13} =0.003818 ~,~ \theta_{23}=0.03998 ~,~~\\[1mm]
    &\theta_{14}\simeq \sqrt{\Delta_{(1)}} = 0.03951 ~,~~ \theta_{24}\simeq \sqrt{\Delta_2} = 0.002078 ~,~~ \theta_{34}\simeq \sqrt{\Delta_3}=0.01271 ~,~~\\[1mm]
    &\delta_{13} = 0.396\, \pi ~,~~ \delta_{14} = 1.818\, \pi ~,~~ \delta_{24}= 0.728\, \pi\,,
\end{aligned}
\end{equation}
corresponding to $\chi^2 \simeq 3.2$, $m_T \approx \frac{0.36}{\sqrt{\delta}}m_t$ and to the following values for observables:
\begin{equation}
\begin{split}
    &x_D \simeq 0.18\% ~,~~\Delta m_K \simeq 6.3 \times 10^{-13}~\text{MeV} ~,~~
    \Delta m_B \simeq 3.0 \times 10^{-10}~\text{MeV} ~,~~ \\[1mm]
    &\Delta m_{B_s} \simeq 2.7 \times 10^{-10}~\text{MeV} ~,~~
    |\epsilon_K| \simeq 1.9 \times 10^{-4} ~,~~\\[1mm]
    &\text{Br}(t\to uZ) \simeq 1.2 \times 10^{-7}~,~~ 
    \text{Br}(t\to cZ) \simeq 3.2 \times 10^{-10}~,~~\\[1mm]
    &\alpha \simeq 84.92\degree ~,~~ \beta \simeq 23.85\degree ~,~~ \gamma \simeq 71.23\degree\,,
    \end{split}
\end{equation}
where $\alpha = \arg (-V_{td}V_{ub}V_{tb}^*V_{ud}^*)$ and $\beta = \arg (-V_{cd}V_{tb}V_{cb}^*V_{td}^*)$. The absolute values of the diagonalizing matrix in eq.~\eqref{eq:SVLQ-Uup} read
\begin{equation}
    |\mathcal{V}_L^\dagger| \simeq
  \tikz[baseline=(M.west)]{%
    \node[matrix of math nodes,matrix anchor=west,left delimiter=(,right delimiter=),ampersand replacement=\&] (M) {%
 0.9737 \& 0.2243 \& 0.0038 \& \,0.0395\\
 0.2243 \& 0.9737 \& 0.0400 \& \,0.0021 \\
 0.0081 \& 0.0393 \& 0.9991 \& \,0.0127 \\
 0.0390 \& 0.0065 \& 0.0126 \& \,0.9991 \\
    };
    \node[draw,fit=(M-1-1)(M-3-3),inner sep=-1pt] {};
    \node[draw,dashed,fit=(M-1-1)(M-4-3),inner sep=1pt] {};
  }
\,,
\end{equation}
where $|V|$ is given by the first 3 columns of $|\mathcal{V}_L^\dagger|$ (dashed block) and $|V_\text{CKM}|$ is the $3 \times 3$ upper-left block of $|\mathcal{V}_L^\dagger|$ (unbroken line).
The absolute values of the entries of $F^u$ are
\begin{equation}
    |F^u| = \begin{pmatrix}
0.99843969 &0.00008203 &0.00050179 &0.03946672\\
 0.00008203 &0.99999569 &0.00002638 &0.00207496\\
 0.00050179 &0.00002638 &0.99983863 &0.01269224\\
 0.03946672 &0.00207496 &0.01269224 &0.001726  
\end{pmatrix}\,,
\end{equation}
As mentioned in section \ref{sec:perturb}, this matrix encodes the structure of FCNC and deviations from unitarity of $V_{CKM}$.

Lastly, the mass matrix $\mathcal{M}_u$ for this particular scenario can be explicitly expressed in the weak basis, where the down-type quark mass matrix is diagonal, denoted as $\mathcal{M}_d = \mathcal{D}_d = \text{diag}(m_d, m_s, m_b)$. In this basis, the upper-left $2 \times 2$ block of $\mathcal{M}_u$ undergoes a rotation resulting in a zero matrix,
\begin{equation}
    |\mathcal{M}_u|= \left(
\begin{array}{cccc}
 0 & 0 &  1.39 & 58.57 \\
 0 & 0 & 6.776 & 9.775 \\
 0.001658 & 15.38  & 170.9 & 18.92  \\
0.03206 & 1.486 & 2.166 & 1499
 \\
\end{array}
\right)~\text{GeV}\,.
\label{eq:massmatrix}
\end{equation}
The latter rotation can be achieved solely via transformations from the right. 
In this WB, one can directly identify $(\mathcal{M}_u)_{33} \simeq m_t$, $(\mathcal{M}_u)_{44} \simeq m_T$, and ${(\mathcal{M}_u)_{14}}/{(\mathcal{M}_u)_{44}} \simeq \sqrt{\Delta}$. It is evident in this specific instance that $(\mathcal{M}_u)_{14}<(\mathcal{M}_u)_{33}$.

\vskip 2mm
It is worth noting that if in the scenario where a down-type VLQ is added to the SM, the down-type quark mass matrix $\mathcal{M}_d$ would exhibit similar qualitative characteristics to the previously described $\mathcal{M}_u$.
Specifically, in the WB where the up-type quark mass matrix $\mathcal{M}_u$ is diagonal and the upper-left $2 \times 2$ block of the down-type quark mass matrix $\mathcal{M}_d$ is transformed into a zero matrix, the structure of $\mathcal{M}_d$ would need to resemble the form presented in eq.~\eqref{eq:massmatrix} to achieve the necessary deviation from unitarity in the first row of the CKM matrix.
In such a basis, one expects $(\mathcal{M}_d)_{33} \sim m_b$, with $m_b(M_Z) \simeq 2.89$ GeV~\cite{Xing:2007zj}. On the other hand, experimental bounds on the mass $m_B$ of the new heavy quark in this scenario are close to those on $m_T$, i.e.~$m_B \gtrsim 1$ TeV~\cite{Aaboud:2018pii, Sirunyan:2019sza}, implying a large $(\mathcal{M}_d)_{44} \sim m_B$. The requirement of reproducing the observed deviation from unitarity $\sqrt{\delta} \simeq 0.04$ would force $(\mathcal{M}_d)_{14} \gtrsim 50~\text{GeV} \gg (\mathcal{M}_d)_{33}$. Since the terms in the first three rows of the mass matrix share a common origin in electroweak symmetry breaking, this hierarchy between $(\mathcal{M}_d)_{14}$ and $(\mathcal{M}_d)_{33}$ may not be appealing or plausible in a theory of flavour addressing the gap $m_b \ll m_t$ in third-generation quark masses.\\
In conclusion, the addition of one up-type VLQ may explain the CKM unitarity problem, in, possibly, a more appealing way - in terms of flavour -, than adding one down-type VLQ.

\subsection{Model $n_d=1$, $n_u=0$}
\label{sec:quarkdown}

For completeness and comparison to the case of section \ref{sec:quarkup}, consider the case where $n_d=1$, which is briefly discussed in ref. \cite{Alves:2023ufm}. Consider the following parameterisation for the matrix in eq. \eqref{eq:VLd} in terms of 6 mixing angles and 3 phases
\begin{equation}\label{eq:SVLQ-Udown}
  \mathcal{V}_L^{d}=\mkern-4mu
    \begingroup
    \setlength\arraycolsep{2pt}
    \begin{pmatrix}
    &&&0\\&V_\text{PDG}&&0\\&&&0\\0&0&0&1
    \end{pmatrix}
    \mkern-8mu
    \begin{pmatrix}
    c_{14}&&&-s_{14}e^{-i\delta_{14}}\\&1\\&&1\\s_{14}e^{i\delta_{14}}&&&c_{14}
    \end{pmatrix}
    \mkern-8mu
    \begin{pmatrix}
    1\\&c_{24}&&-s_{24}e^{-i\delta_{24}}\\&&1\\&s_{24}e^{i\delta_{24}}&&c_{24}
    \end{pmatrix}
    \mkern-8mu
    \begin{pmatrix}
    1\\&1\\&&c_{34}&-s_{34}\\&&s_{34}&c_{34}
    \end{pmatrix}
    \,.
    \endgroup
\end{equation}

With this choice, the traditional picture of $V_\text{CKM}$ is recovered when all $\theta_{i4}\to0$.
Meanwhile, in these models the matrices which control all tree-level FCNC are given by
\begin{equation}\label{eq:SVLQ-FD}
    F^d=\mkern-4mu
    \begin{pmatrix}
    1-s^2_{14}&-s_{14}c_{14}s_{24}e^{-i(\delta_{14}-\delta_{24})}&-s_{14}c_{14}c_{24}s_{34}e^{-i\delta_{14}}&-s_{14}c_{14}c_{24}c_{34}e^{-i\delta_{14}}\\
    -s_{14}c_{14}s_{24}e^{i(\delta_{14}-\delta_{24})}&1-c^2_{14}s^2_{24}&-c^2_{14}s_{24}c_{24}s_{34}e^{-i\delta_{24}}&-c^2_{14}s_{24}c_{24}c_{34}e^{-i\delta_{24}}\\
    -s_{14}c_{14}c_{24}s_{34}e^{i\delta_{14}}&-c^2_{14}s_{24}c_{24}s_{34}e^{i\delta_{24}}&1-c^2_{14}c^2_{24}s^2_{34}&-c^2_{14}c^2_{24}s_{34}c_{34}\\
    -s_{14}c_{14}c_{24}c_{34}e^{i\delta_{14}}&-c^2_{14}s_{24}c_{24}c_{34}e^{i\delta_{24}}&-c^2_{14}c^2_{24}s_{34}c_{34}&1-c^2_{14}c^2_{24}c^2_{34}
    \end{pmatrix}
\end{equation}
and $F^u=\id_3$.
 For small angles, $F^d$ becomes
 \begin{equation}\label{eq:SVLQ-FD:exp}
    F^d\simeq\mkern-4mu
    \begin{pmatrix}
    1-\theta^2_{14}&-\theta_{14}\theta_{24}e^{-i(\delta_{14}-\delta_{24})}&-\theta_{14}\theta_{34}e^{-i\delta_{14}}&-\theta_{14}e^{-i\delta_{14}}\\
    -\theta_{14}\theta_{24}e^{i(\delta_{14}-\delta_{24})}&1-\theta^2_{24}&-\theta_{24}\theta_{34}e^{-i\delta_{24}}&-\theta_{24}e^{-i\delta_{24}}\\
    -\theta_{14}\theta_{34}e^{i\delta_{14}}&-\theta_{24}\theta_{34}e^{i\delta_{24}}&1-\theta^2_{34}&-\theta_{34}\\
    -\theta_{14}e^{i\delta_{14}}&-\theta_{24}e^{i\delta_{24}}&-\theta_{34}&\theta^2_{14}+\theta^2_{24}+\theta^2_{34}
    \end{pmatrix}\mkern-4mu
\end{equation}
plus terms of order $\mathcal{O}(\theta^3_{i4})$.
Notice that, due to similar choices, eq.~\eqref{eq:SVLQ-FD} and eq. \eqref{eq:SVLQ-FU} are identical. However, this is just a mere coincidence, and fits should prefer different values of the $\theta_{i4}$ in each case, since they represent different things.

Again, comparing eq. \eqref{eq:duFCNC-approx} and eq. \eqref{eq:SVLQ-FD:exp}, one concludes that
\begin{equation}\label{eq:SVLQ-Tdown}
    \Theta_d=-\theta_{14}e^{-i\delta_{14}}\,,\quad\Theta_s=-\theta_{24}e^{-i\delta_{24}}\,,\quad\Theta_b=-\theta_{34}\,,
\end{equation}
where the angles and phases now refer to the parameters introduced in~\ref{eq:SVLQ-Udown}.
Like before, the vector  $\Theta_d$ is implicitly defined as $(\Theta_d,\Theta_s,\Theta_b)^T$.
Fore more details regarding this model and how it can solve the CKM unitarity problem, see section 4 of ref. \cite{Belfatto:2021jhf}.

\vskip 2mm 
\cleardoublepage




\chapter{New Physics and Experimental Signals}
\label{chapter:results}
The smoking gun signals that can reveal the presence of 'heavy'\footnote{Here 'heavy' should be understood as having a larger mass than the active 'light' neutrinos which due to the PLANCK bound \cite{Aghanim:2018eyx} are expected to have a sub eV mass - cf. section \ref{sec:gen_neutrinos} of this thesis.} neutrinos and VLQs primarily hinge on two crucial quantities: their mass and their mixing with their respective SM counterparts.
Notably, the mass of the hypothetical particle is a critical factor, as it determines whether the particle can be produced in experimental setups. In models featuring 'heavy' neutrinos and VLQs, a certain seesaw between their mass and mixing is expected in order to have electroweak perturbativity. This connection is elaborated upon in the final section of \ref{sec:non-dec} and section \ref{sec:gen_neutrinos} for neutrinos, and section \ref{sec:perturb} for VLQs. Consequently, these models share a common characteristic: as the mass of the hypothetical particle increases, its mixing with the active counterparts must decrease. This leads to the existence of a fruitful region in the parameter space of these models, where the particle's mass is not excessively large and its mixing is sufficiently significant to produce measurable effects in observable quantities.\\
The phenomenological analysis presented in refs. \cite{Alves:2023ufm, Branco:2021vhs, Branco:2019avf, Agostinho:2017wfs}, which form the basis of this thesis, precisely focuses on this particular region in the parameter space, yielding results that can be probed in ongoing or forthcoming experimental endeavors.\\
In this chapter, these signals will be reviewed, starting in the neutrino sector and finishing in the quark sector. A comprehensive analysis of each signal will be done, along with an assessment of the existing experimental bounds associated with them. Future prospects and current bounds will be discussed according to the heavy particle's mass. 
For neutrinos, the effects of 'heavy' neutrinos with a mass of some GeV up to a few TeV and with a mass between the MeV and the eV scale will be analysed, as these were discussed in section \ref{sec:quasi} and in ref. \cite{Agostinho:2017wfs} and in section \ref{sec:non-dec} and in ref. \cite{Branco:2019avf}. Finally, for VLQs, only those  with a mass of a few TeV - discussed in section \ref{sec:quarkmodels} and in refs. \cite{Branco:2021vhs,Alves:2023ufm}, will be considered.

\section{Neutrinos}
\label{sec:res_neutrinos}
As implied earlier in this thesis, in terms of experimental viability, only scenarios with $M \lesssim$ TeV hold promise for being tested in the foreseeable future \cite{Deppisch:2015qwa}, as neutrinos within this mass range can potentially leave discernible signatures in ongoing experiments at the energy, cosmic, and intensity frontiers. The next paragraph will list the main signals 'heavy' Majorana neutrinos may leave in current and future experiments. The rest of the section will contain a brief description of each one and their current experimental bounds.\\
Neutrinos, if Majorana particles, can in principle be tested via Lepton number violating processes such as the neutrinoless double beta decay \cite{Rodejohann:2011mu, Bilenky:2014uka,Bolton:2022tds}.
If their mixing with the active neutrinos is sizeable, they can be produced with enough statistics to declare a discovery in the LHC \cite{ATLAS2023tkz}. This mixing with their active counterparts can be inferred from the non-unitarity of $3 \times 3$ leptonic mixing matrix \cite{Antusch:2006vwa, Abada:2007ux, Antusch:2014woa, Blennow:2016jkn, Escrihuela:2015wra} and have impact on low-energy observables related to lepton flavour violation (LFV), lepton flavour non-universality (LFNU) and electroweak precision tests \cite{Fernandez-Martinez:2016lgt, Antusch:2014woa}. Special emphasis will be given to the LFV process $\mu \rightarrow e \gamma$ \cite{MEG:2016leq}, as it was one of the most restrictive to the parameter space of the models studied in ref. \cite{Branco:2019avf}.
Sizeable deviations from $3 \times 3$ unitary may also leave a clear trace in short baseline and long baseline oscillation experiments \cite{Escrihuela:2015wra, Fernandez-Martinez:2016lgt,Dentler:2018sju}.
Beta decay experiments, such as KATRIN, may also play an important role since the existence of a eV almost-sterile 'heavy neutrino' \cite{Riis:2010zm} or a keV one \cite{Mertens:2014nha} changes the shape of the beta decay spectrum. Neutrinos in the discussed mass and mixing ranges may also leave traces of CP Violation at low-energy \cite{Fernandez-Martinez:2007iaa, Branco:2019avf} detectable in future oscillation experiments and, if long-lived, leave displaced vertex \cite{Helo:2013esa, Gago:2015vma, Antusch:2016vyf} and resolvable oscillation \cite{Antusch:2017ebe,Cvetic:2018elt} signals  at colliders.\\
To end section \ref{sec:res_neutrinos}, a discussion on the prospects of future experiments regarding neutrino physics is presented.

\subsection{Neutrinoless double beta decay}
\label{sec:neutrinoless}
Neutrinoless double $\beta$ decay ($0\nu \beta \beta$) is a a Lepton number violating process  \cite{Rodejohann:2011mu,Vergados:2016hso} that consists of
\begin{equation}
    (A,Z) \rightarrow (A,Z+2) + e^{-} + e^{-} ~,
\end{equation}
where $Z$ is the atomic number of a given atom, or, put simply, its number of protons, and $A$ is its atomic mass number, or its number of protons and neutrons, $e^-$ represents an electron.
The decay rate for this process is given by:
\begin{equation}
    \Gamma_{0\nu \beta \beta} =G(Q,Z) |M^0_{nuc}|^{2}| = G(Q,Z) |M_{nuc}|^{2}\cdot | m_{\beta \beta}|^2 ~,
    \label{eq:widthneutrinoless}
\end{equation}
where $G(Q,Z)$ is a phase-space factor that depends on the nucleus and $M^0_{nuc}$ is the total nuclear matrix element. $M^0_{nuc}$ contains the leptonic amplitude $(W^-_{\mu})^* (p_3)+ (W^-_{\nu})^* (p_4) \rightarrow e^-(p_1) + e^-(p_2) $, which has two diagrams (channel t and u) because of the anti-symmetrization of the final state:
\begin{equation}
       M^0_{nuc} \propto (M_t -M_u) ~.
\end{equation}
It is only possible to write the amplitude for this process using a lepton number violating vertex, thus making the process only possible if neutrinos are Majorana \cite{Bilenky:2014uka}. For illustrative purposes, consider the t-channel amplitude, specifically the part that is sandwiched between the spinors, 
\begin{equation}
 iM_t= \sum_j \left[ \overline{u(p_1)}\left(\frac{-ig}{\sqrt{2}} \gamma_{\mu} P_L  (K)_{1j} \right) \frac{i( \gamma^{\delta} p_{\delta} +m_j)} {p^2 - m_j^2} \left(-\frac{-ig}{\sqrt{2}} \gamma_{\nu} P_L  (K)_{1j} \right) v(p_2) \right] ~,
    \label{eq:leptonviolating}
\end{equation}
where $p=p_1-p_3=p_4-p_2$ and $e^-(p_2)=\left(e^+(p_2)\right)^c$ represents the Conjugated Dirac fermion, the vertex connecting momenta $p_2$ and $p_4$ exhibits lepton number violation.

In the computation of $|M^0_{nuc}|^2$, upon evaluating the traces of the leptonic part, it becomes evident that terms proportional to $\gamma^{\delta} p_{\delta}$ vanish due to the presence of an odd number of gamma matrices. Consequently, the only surviving terms are those proportional to $m_j$. Consequently, by neglecting the neutrino mass in the denominator, it becomes possible to factor out a meaningful quantity,
\begin{equation}
    |M^0_{nuc}|^{2}= |M_{nuc}|^{2}\cdot |\sum_j (K_{1j})^2 m_j|^2 ~,
    \label{eq:m0nuc}
\end{equation}
where $K$ is the $3 \times 3$ non-unitary leptonic mixing matrix, defined in eq. \eqref{eq:kdef}. Combining eq. \eqref{eq:m0nuc} with eq. \eqref{eq:widthneutrinoless} yields the definition
\begin{equation}
    m_{\beta \beta} = \sum_j (K_{1j})^2 m_j ~,
    \label{eq:mbetabeta}
\end{equation}
Note that, in general, this quantity is sensitive to all the CP Violating Phases in the leptonic sector of the $n_R \nu$SM. The unitary phases, the Dirac phase and the Majorana phases, and the hermitian phases, the phases of the deviations from unitarity, as described in section \ref{sec:leptogenesis}.
The quantity $ m_{\beta \beta} $ changes if $p$ of the $n_R$ "heavy" states have a mass of the order of the light neutrinos, $\sim eV, keV$, such that it is kinematically allowed, since this is a nuclear process \cite{Gariazzo:2015rra,Giunti:2019aiy,Abada:2018qok}. In that case,
\begin{equation}
    m_{\beta \beta}=  \sum_i^3 (K_{1i})^2 m_i + \sum_j^{p} ((KX^\dag)_{1j})^2 M_j ~.
    \label{eq:mbetabeta_true}
\end{equation}
\par
The current experimental bounds on $m_{\beta \beta}$, considering 3 light/active neutrinos and very heavy almost-sterile neutrinos, are  depicted graphically in fig. \ref{fig:neutrinolessbetabeta}:
\begin{figure}[H]
  \centering
  \includegraphics[width=0.7\textwidth]{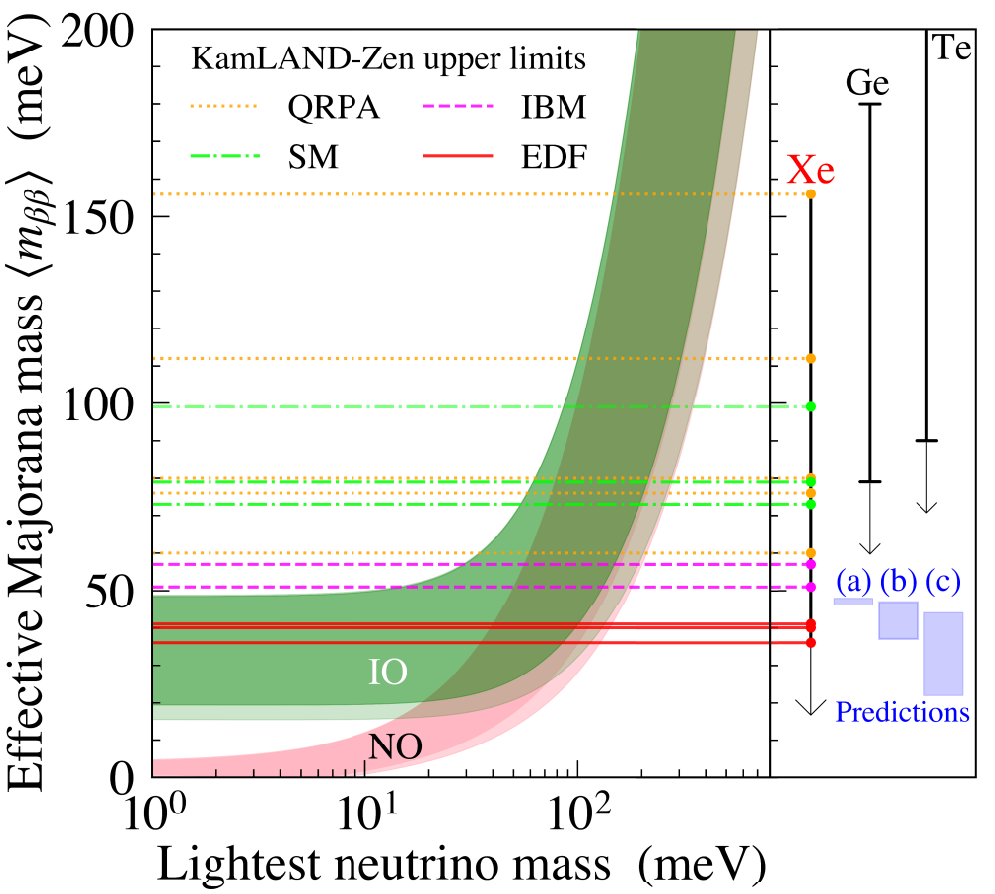} 
  \caption[$m_{\beta \beta}$ as given in eq. \eqref{eq:mbetabeta} as a function of the lightest neutrino mass. The regions below the horizontal lines are allowed at 90\%
C.L. with $\textrm{Xe}^{136}$ from KamLAND-Zen, the side panel shows the corresponding limits for $\textrm{Xe}^{136}$, $\textrm{Ge}^{76}$ , and
$\textrm{Te}^{130}$.]{$m_{\beta \beta}$ as given in eq. \eqref{eq:mbetabeta} as a function of the lightest neutrino mass. The dark shaded regions are
predictions based on best-fit values of neutrino oscillation parameters for the normal ordering (NO) and the inverted ordering (IO), and the light shaded regions indicate the $3\sigma$ ranges calculated from oscillation parameter uncertainties \cite{DellOro:2014ysa,Nufit}.
The regions below the horizontal lines are allowed at 90\%
C.L. with $\textrm{Xe}^{136}$ from KamLAND-Zen \cite{KamLAND-Zen:2022tow} considering an improved phase space factor calculation \cite{Kotila:2012zza,Mirea:2014dza} and
commonly used nuclear matrix element estimates: energy density functional (EDF) theory \cite{LopezVaquero:2013yji,Yao:2014uta,Rodriguez:2010mn} (solid lines), interacting boson model (IBM)\cite{Deppisch:2020ztt,Barea:2015kwa} (dashed lines), shell model
(SM) \cite{Coraggio:2020hwx,Neacsu:2014bia,Menendez:2008jp} (dot-dashed lines), and quasi-particle random phase approximation (QRPA) \cite{Terasaki:2020ndc,Hyvarinen:2015bda, Simkovic:2013qiy, Mustonen:2013zu, PhysRevC.97.045503} (dotted lines). The side
panel shows the corresponding limits for $\textrm{Xe}^{136}$, $\textrm{Ge}^{76}$ \cite{GERDA:2020xhi}, and
$\textrm{Te}^{130}$ \cite{CUORE:2021mvw}, and theoretical model predictions on $m_{\beta \beta}$, (a)
Ref. \cite{PhysRevD.86.013002}, (b) Ref. \cite{ASAKA2020135956}, and (c) Ref. \cite{Asai:2017ryy} (shaded boxes), in the IO region. Image taken from ref. \cite{KamLAND-Zen:2022tow}.}
  \label{fig:neutrinolessbetabeta}
\end{figure}

\subsubsection{$M \gtrsim$ MeV}

Due to the inverted-ordering (IO) hypothesis of light neutrino masses being disfavoured in global fits of leptonic mixing angles \cite{deSalas:2017kay, DeSalas:2018rby, deSalas:2020pgw}, the prospects of measuring $m_{\beta \beta}$ have been affected, since for the normal-ordering (NO) scenario $m_{\beta \beta}$ can be significantly smaller. If the 'heavy' neutrinos all have masses much beyond $\sim 100$ MeV, the nuclear scale of the process, then they can't be produced and contribute to $m_{\beta \beta}$. In that case, the leading experiments in neutrinoless double beta decay, namely KamLAND-Zen \cite{KamLAND-Zen:2022tow}, GERDA \cite{GERDA:2020xhi}, and CUORE \cite{CUORE:2021mvw}, currently lack immediate prospects of observing significant signals, especially with IO being effectively excluded. Figure \ref{fig:neutrinolessbetabeta} presents the current most stringent constraints on $m_{\beta \beta}$ under this framework.\\

\subsubsection{$M \lesssim$ MeV $\sim$ eV, keV}

The layout obviously changes if there are 'heavy' neutrinos with masses closer to the MeV scale \cite{Abdullahi:2022jlv}, since in this case $m_{\beta \beta}$ has new contributions and its value might be  substantially larger \cite{Gariazzo:2015rra,Giunti:2019aiy,Abada:2018qok}.
For instance, the upper bound of the values of $m_{\beta \beta}$, calculated with eq. \eqref{eq:mbetabeta_true}, for the cases of section \ref{sec:non-dec} ranged from $4.6$ meV to $27$ meV, below the current experimental upper bounds. Nonetheless, if heavy neutrinos with these masses and mixings exist, a discovery may be imminent in the next round of experiments \cite{Bolton:2022tds}.

\subsection{Prompt Decay Searches at LHC: ATLAS, CMS and LHCb}
\label{sec:LHC}

The LHC is the most powerful proton-proton collider built by mankind and is host to nine approved experiments. The primary experiments involved in the pursuit of new physics are ATLAS, CMS and LHCb. To date, these collisions have taken place at center-of-mass energies of 7, 8, and 13 TeV. 

Given the limitations of the LHC, ATLAS and CMS searches mainly focused on heavy neutrinos with masses in the GeV/TeV scale \cite{Abdullahi:2022jlv}, as these had the better discovery prospects.
The LHC's array of detector experiments offers comprehensive coverage across a wide range of potential scenarios related to heavy neutrinos, with multiple searches \cite{Deppisch:2015qwa,Cai:2017mow,delAguila:2008cj,Atre:2009rg,Pascoli:2018heg,Tello:2010am} addressing the matter.

Another experimental configuration within the LHC is the LHCb experiment, which primarily focuses on searching for new physics through the decay of heavy-flavour hadrons. Its unique setup has yielded intriguing indications of potential LFNU, as reported by the LHCb experiment \cite{LHCb:2017avl,LHCb:2021lvy}. However, these findings require confirmation and establishment with greater statistical significance and preferably through independent measurements.

To summarize, ATLAS, CMS, and LHCb experiments have conducted searches for heavy neutrinos in different regions of mass, active-sterile neutrino mixing, and lifetimes, encompassing both minimal and non-minimal scenarios. This thesis will only cover the minimal scenario, $n_R\nu$SM, as discussed in chapter \ref{chapter:framework_neutrinos}.

The LHC has the capability to generate heavy neutrinos in the mass range of GeV/TeV by means of decays involving heavy mesons, $\tau$ leptons, $W$ bosons, Higgs bosons, and even top quarks. If there are novel interactions or gauge symmetries present in nature, or if the masses of heavy neutrinos exceed the electro-weak scale, then additional production mechanisms for heavy neutrinos become feasible. In the minimal model, the production of high-mass heavy neutrinos involves processes such as the Drell-Yan process \cite{PhysRevLett.50.1427,PETCOV1984421}, gluon fusion \cite{Willenbrock:1985tj,Hessler:2014ssa,Ruiz:2017yyf}, $W \gamma$ fusion \cite{Dev:2013wba,Datta:1993nm,Alva:2014gxa,Degrande:2016aje} and same-sign WW fusion \cite{Dicus:1991fk,Fuks:2020att}. When neutrinos are Majorana particles, they can decay into both lepton number conserving and lepton number violating channels. However, if they are Dirac particles, only lepton number conserving processes are allowed. However, both Majorana and Dirac neutrinos can have lepton flavour violating processes. Furthermore, depending on the mass and mixing angle of the heavy neutrino, its decays can either be prompt (short lifetime) or displaced (long lifetime) from the production vertex. This section will only cover prompt decays, for displaced vertex decay searches see section \ref{sec:displaced}.

It is important to note that in the experimental searches for heavy neutrinos performed at the LHC, it is common to assume the presence of a single heavy neutrino mixing with a single neutrino flavour. Naturally, the exclusion limits obtained from these searches may not directly apply to the mixing angles in realistic neutrino models that can explain neutrino oscillations. It is evident that many of these searches require reinterpretation\cite{Tastet:2021vwp}.

\subsubsection{$M \sim$ GeV/TeV}

\begin{figure}[H]
  \centering
  \includegraphics[width=1.0\textwidth]{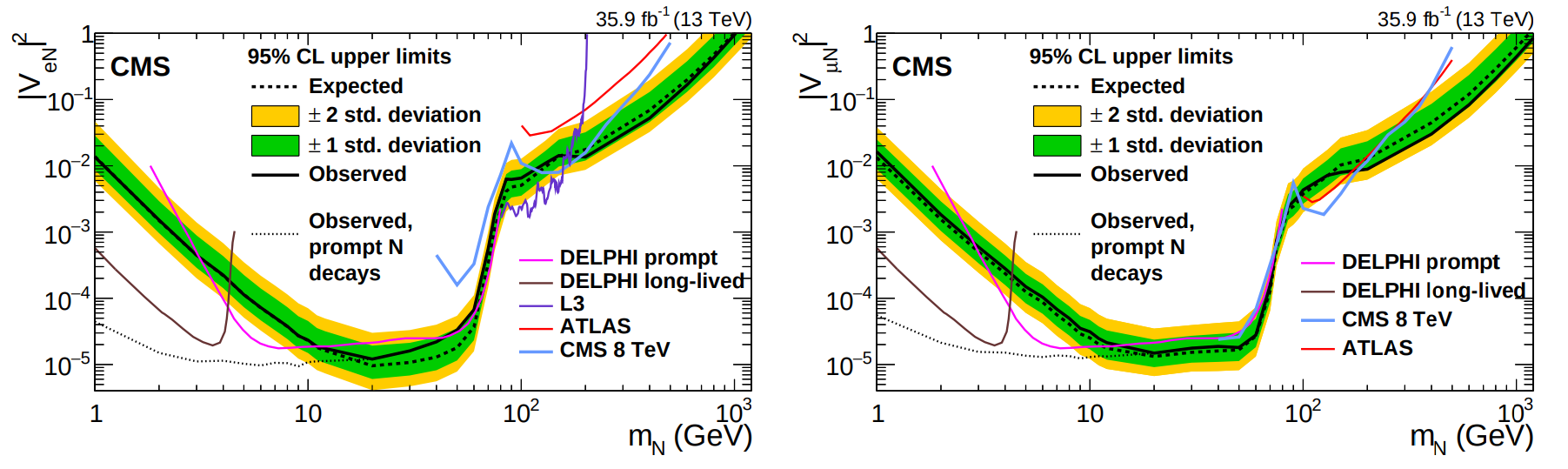} 
  \caption[The CMS experiment has established exclusion regions at a 95\% confidence level (CL) for heavy neutrinos in the $|V_{eN}|^2$ vs. $m_N$ (left) and $|V_{\mu N}|^2$ vs. $m_N$ (right) planes. In these plots, the dashed black curve represents the expected upper limit, accompanied by dark green and light yellow bands denoting one and two standard-deviation uncertainties, respectively. The solid black curve represents the observed upper limit, while the dotted black curve represents the observed limit assuming prompt decays of heavy neutrinos. Additionally, the plots include the best upper limits at a 95\% CL obtained from other collider searches conducted by L3, DELPHI, ATLAS (8 TeV), and CMS (8 TeV).]{The CMS experiment has established exclusion regions at a 95\% confidence level (CL) for heavy neutrinos in the $|V_{eN}|^2$ vs. $m_N$ (left) and $|V_{\mu N}|^2$ vs. $m_N$ (right) planes. In these plots, the dashed black curve represents the expected upper limit, accompanied by dark green and light yellow bands denoting one and two standard-deviation uncertainties, respectively. The solid black curve represents the observed upper limit, while the dotted black curve represents the observed limit assuming prompt decays of heavy neutrinos \cite{CMS2018iaf}. Additionally, the plots include the best upper limits at a 95\% CL obtained from other collider searches conducted by L3, DELPHI, ATLAS (8 TeV), and CMS (8 TeV). Image taken from ref. \cite{CMS2018iaf}.}
  \label{fig:CMS2018iaf}
\end{figure}

A CMS search from 2018 \cite{CMS2018iaf} excluded at 95\% confidence level heavy neutrinos with masses from $1$ GeV up to $1.2$ TeV for mixing $|V_{lN}|^2$ from $1.2 \times 10^{-5}$ up to 1.8, with $l=e,\mu$, at a center-of-mass energy of 13 TeV, with an integrated luminosity of  $35.9 \textrm{fb}^{-1}$. The results are displayed in Figure \ref{fig:CMS2018iaf}.

This search looked for a signal comprised of a heavy neutrino produced via W boson decay $W \rightarrow N l$ with subsequent decay $N \rightarrow Wl$ and the $W$ proceed to decay to $\nu l$, hence the mixing element involved is $V_{lN}$, in the notation of eq. \eqref{eq:mixingmatrix_neutrino} and eq. \eqref{eq:ABdef}.

\begin{figure}[H]
  \centering
  \includegraphics[width=0.7\textwidth]{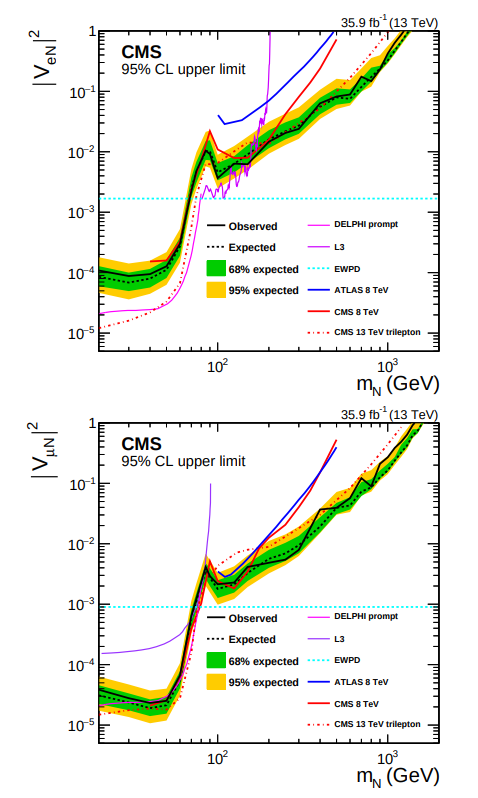} 
  \caption[The CMS experiment has established exclusion regions at a 95\% confidence level (CL) for heavy neutrinos in the $|V_{eN}|^2$ vs. $m_N$ (upper) and $|V_{\mu N}|^2$ vs. $m_N$ (lower) planes. In these plots, the dashed black curve represents the expected upper limit, accompanied by dark green and light yellow bands denoting one and two standard-deviation uncertainties, respectively. The solid black curve represents the observed upper limit and the dashed cyan line shows constraints from EWPD. Also shown are the upper limits from other direct searches: DELPHI, L3, ATLAS, and the upper limits from the CMS $\sqrt{s} = 8$ TeV 2012 data and the trilepton analysis based on the same 2016 data set as used in this analysis.]{The CMS experiment has established exclusion regions at a 95\% confidence level (CL) for heavy neutrinos in the $|V_{eN}|^2$ vs. $m_N$ (left) and $|V_{\mu N}|^2$ vs. $m_N$ (right) planes. In these plots, the dashed black curve represents the expected upper limit, accompanied by dark green and light yellow bands denoting one and two standard-deviation uncertainties, respectively. The solid black curve represents the observed upper limit and the dashed cyan line shows constraints from EWPD. Also shown are the upper limits from other direct searches: DELPHI, L3, ATLAS, and the upper limits from the CMS $\sqrt{s} = 8$ TeV 2012 data and the trilepton analysis based on the same 2016 data set as used in this analysis. Image taken from ref. \cite{CMS2018jxx}.}
  \label{fig:CMS2018jxx}
\end{figure}

Another CMS search from 2018  \cite{CMS2018jxx} excluded, at 95\% C.L., heavy neutrinos with masses from $20$ GeV to $1.6$ TeV for mixing $V_{lN}$ from $2.3 \times 10^{-5}$ up to $1$, with $l=e,\mu$, at a center-of-mass energy of 13 TeV, with an integrated luminosity of  $35.9 \textrm{fb}^{-1}$. The results are displayed in Figure \ref{fig:CMS2018jxx}.

 The search in ref. \cite{CMS2018jxx} probes the decay of a W boson, in which an SM neutrino oscillates into a new heavy state $N$.

\begin{figure}[H]
  \centering
  \includegraphics[width=0.6\textwidth]{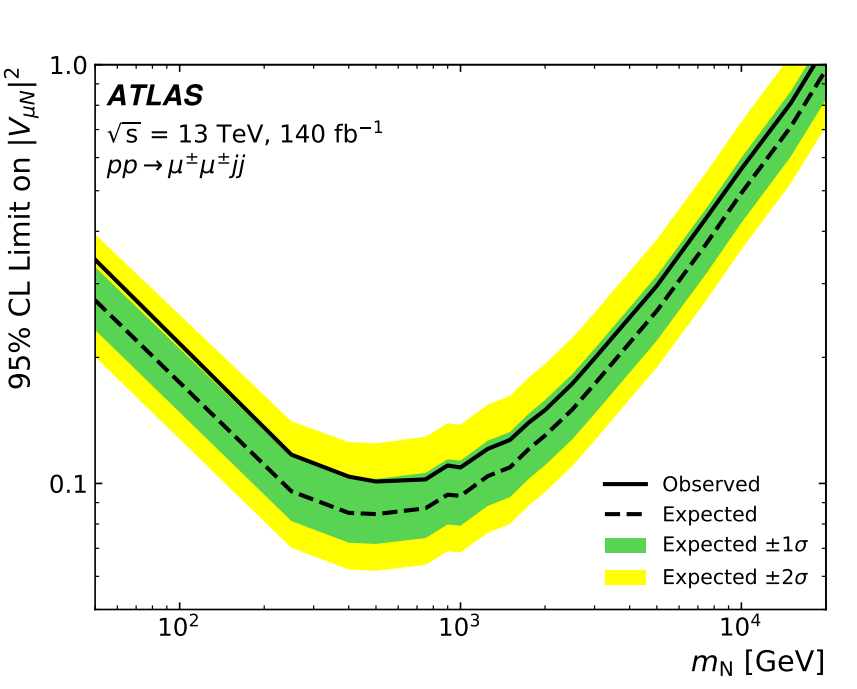} 
  \caption[Observed and expected 95\% CL upper limits on the heavy Majorana neutrino mixing element $|V_{\mu N}|^2$ function of $m_N$ in the Phenomenological Type-I Seesaw model. The one and two standard deviation bands of the expected limit are indicated in green and yellow, respectively.]{Observed and expected 95\% CL upper limits on the heavy Majorana neutrino mixing element $|V_{\mu N}|^2$ function of $m_N$ in the Phenomenological Type-I Seesaw model. The one and two standard deviation bands of the expected limit are indicated in green and yellow, respectively. Image taken from ref. \cite{ATLAS2023tkz}.}
  \label{fig:ATLAS2023tkz}
\end{figure}

An ATLAS search from 2023  \cite{ATLAS2023tkz}  excluded, at 95\% C.L., heavy neutrinos with masses from $50$ GeV to $20$ TeV for mixing $V_{\mu N}$ from $5 \times 10^{-2}$ up to $1$, at a center-of-mass energy of 13 TeV, with an integrated luminosity of  $140 \textrm{fb}^{-1}$. The results are displayed in Figure \ref{fig:ATLAS2023tkz}.

 The search in ref.  \cite{ATLAS2023tkz} probes same-sign $\mu^\pm \mu^\pm$ production in $W^\pm W^\pm$ scattering mediated by a Majorana neutrino $N$.

\begin{figure}[H]
  \centering
  \includegraphics[width=0.7\textwidth]{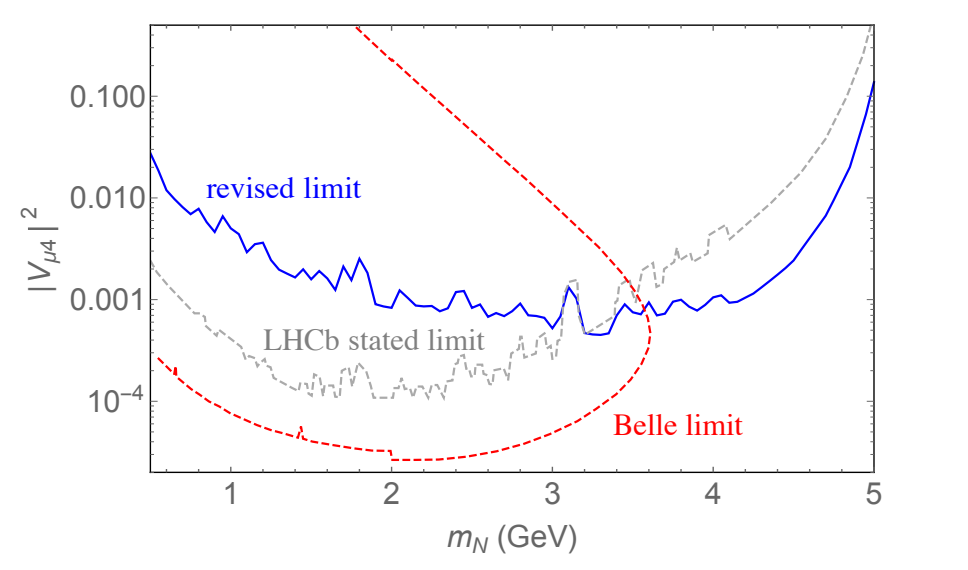} 
  \caption[Upper limit on $|V_{\mu4}|^2$ at 95\% confidence level from the LHCb experiment. The dashed line shows the limit from. The solid line shows the limit that would be extracted using the decay width formulae in this paper. For comparison, the lower dotted line shows the recently revised limit from Belle. All three limit curves are constructed with the assumption $V_{e4} = V_{\tau 4} = 0$.] {Upper limit on $|V_{\mu4}|^2$ at 95\% confidence level from the LHCb experiment. The dashed line shows the limit from \cite{LHCb:2014osd}. The solid line shows the limit that would be extracted using the decay width formulae in this paper. For comparison, the lower dotted line shows the recently revised limit from Belle \cite{Belle:2013ytx}. All three limit curves are constructed with the assumption $V_{e4} = V_{\tau 4} = 0$. Image taken from ref. \cite{Shuve:2016muy}.}
  \label{fig:Shuve2016muy}
\end{figure}

All these experiments exclude neutrinos with a mass between $100$ GeV and $1$ TeV for a $|V_{lN}|^2 \gg 10^{-3}$. The models in section \ref{sec:quasi} and section \ref{sec:non-dec} have neutrinos in these mass ranges with $|V_{lN}|^2 \in [10^{-6},10^{-4}]$, hence these models are not yet ruled out and the next round of experiments may start probing them \cite{Antusch:2016ejd,Antusch:2015mia,Kwok:2023dck,Abdullahi:2022jlv,Das:2015toa,Das:2017nvm,Das:2017zjc}.
For low GeV masses, LHCb provides more stringent bounds \cite{LHCb:2014osd,Shuve:2016muy,LHCb:2020wxx,Abdullahi:2022jlv}.
A search for a heavy Majorana neutrino involved in the $B^- \rightarrow \pi^+ \mu^- \mu^-$ decay mode was performed using $3~\textrm{fb}^{-1}$ of integrated luminosity, at a center-of-mass energies of $7$ TeV and $8$ TeV. Note that the search in ref. \cite{LHCb:2014osd} was revised in ref. \cite{Shuve:2016muy}, its results are summarized in fig. \ref{fig:Shuve2016muy}.

\begin{figure}[H]
  \centering
  \includegraphics[width=0.8\textwidth]{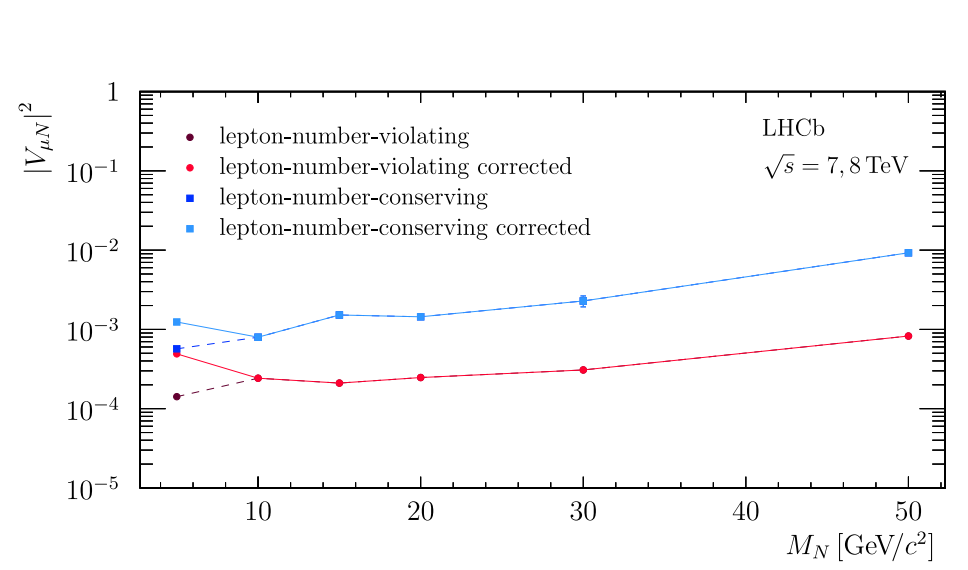} 
  \caption[Observed upper limit on the mixing parameter  $|V_{\mu N}|^2$
between a heavy neutrino and a muon neutrino in the mass range $5-50$ GeV for same-sign and opposite-sign muons in the final states with and without lifetime correction.]{Observed upper limit on the mixing parameter  $|V_{\mu N}|^2$
between a heavy neutrino and a muon neutrino in the mass range $5-50$ GeV for same-sign and opposite-sign muons in the final states with and without lifetime correction. Image taken from ref. \cite{LHCb:2020wxx}.}
  \label{fig:LHCb2020wxx}
\end{figure}

Recently, a search for heavy neutrinos in the decay of a W boson into two muons
and a jet was performed at the LHCb \cite{LHCb:2020wxx}. The data set was collected in the same conditions as the aforementioned search in refs. \cite{LHCb:2014osd,Shuve:2016muy}. Its results are depicted in fig. \ref{fig:LHCb2020wxx}

\subsubsection{$M \lesssim$ GeV}
Bounds for heavy neutrinos with masses below the GeV scale have better future prospects and are more stringent in other experiments that do not work in the energy frontier \cite{Bryman:2019bjg,Drewes:2016upu,Giunti:2015wnd,Giunti:2019aiy,Acero:2022wqg}. For keV neutrinos refer to section \ref{sec:beta_decay}, for eV neutrinos see section \ref{sec:LBL} and section \ref{sec:beta_decay}.


\subsection{LFV, LFNU and Electroweak Precision Observables}
\label{sec:elecweak_observables}
The presence of heavy almost-sterile neutrinos produces effects on electroweak precision observables, specifically lepton number and flavour violating decays \cite{Akhmedov:2013hec,deBlas:2013gla, Basso:2013jka}. Constraints on these typically do not rely on the mass of the heavy neutrinos, but it is important to consider the underlying assumptions made when deriving these constraints \cite{Blennow:2016jkn}. In Section  \ref{sec:devunit_non}, it is explained that if all heavy neutrinos have masses above the electroweak scale (approximately $M \gtrsim m_W$, where $m_W$ represents the mass of the W boson and $M$ the mass scale of all heavy neutrinos), one can solely constrain the hermitian part of the leptonic mixing matrix, as defined in eq. \eqref{eq:kh}. \par
According to the polar decomposition of a square invertible matrix, it is possible to express such a matrix as the product of a unitary matrix with a hermitian matrix. This hermitian matrix can be positioned either on the left or on the right of the unitary matrix \cite{Fernandez-Martinez:2007iaa}. In terms of notation, the hermitian part on the left is commonly written in terms of a matrix $\eta$, cf. eq. \eqref{eq:etadef}. The most stringent constraints on the $\eta$ matrix are provided in equation \eqref{eq:hen}, originating from references \cite{Fernandez-Martinez:2016lgt,Antusch:2014woa,Escrihuela:2015wra}.  In particular, this discussion pertains to scenarios where the mass of the heavy neutrinos is around or above the GeV/TeV scale ($M \gtrsim$ GeV/TeV), while more stringent bounds for lighter heavy neutrinos can be found in sections \ref{sec:neutrinoless}, \ref{sec:LBL} and \ref{sec:beta_decay}. \par
In the $n_R\nu$SM, when all $M \gtrsim$ GeV/TeV, the charged and neutral electroweak current are modified \cite{Antusch:2014woa}
\begin{equation}
    j_\mu^{\pm} = l_\alpha \gamma_\mu K_{\alpha j} \nu_j ~,~ j_\mu^0 = \overline{\nu}_i (K^\dag K)_{ij} \gamma_\mu \nu_j ~,
    \label{eq:currents}
\end{equation}
where $l$ are charged leptons, $\gamma$ are the Gamma matrices, $\nu$ are light active neutrinos and $K$ is the $3 \times 3$ non-unitary leptonic mixing matrix, defined in eqs. \eqref{eq:mixingmatrix_neutrino} and \eqref{eq:ABdef}. Using eq. \eqref{eq:ABdef} and comparing eq. \eqref{eq:currents} with eqs. \eqref{eq:lW}, \eqref{eq:mixingmatrix_neutrino} and eqs. \eqref{eq:lZ}, \eqref{eq:Fdef} one can see indeed that eq. \eqref{eq:currents} is an approximation of the exact result, on which it is assumed that only the active light neutrinos enter in the low-energy electroweak processes one is studying. This is consistent with a effective description approach, where the heavy neutrinos are integrated out, compatible with what is known in the literature as the Minimum Unitarity Violation (MUV) scheme\cite{Antusch:2014woa}. The $n_R\nu$SM along with all $M \gtrsim$ GeV/TeV is equivalent to such approach and the decay rates and cross sections are proportional to $ (KK^\dag)_{\alpha \beta}$ and $\sum_\beta (KK^\dag)_{\alpha \beta}$ for processes involving charged currents and $ \sum_{ij}|(K^\dag K)_{ij}|^2$ for processes involving neutral currents.\par
The $\eta$ matrix is commonly used in the literature because, using eq. \eqref{eq:kh}, there is  a straight forward connection between $KK^{\dagger}$, what appears in the charged currents, and $\eta$
\begin{align}
K K^\dagger \,=\,  \id - 2\eta + \eta^2 \approx \id - 2 \eta~,
\label{eq:KReta}
\end{align}%
where the unitarity relation eq. \eqref{eq:unitAB02n} and eq.~\eqref{eq:khu} were used. \par
Similarly, there is a connection between $K^\dag K$, what appears in neutral currents, and $\eta$
\begin{align}
K^\dagger K \,= U_K^\dag (\id - 2\eta + \eta^2) U_K \approx \id - 2 U_K^\dag \eta U_K~.
\label{eq:KLeta}
\end{align}%
Neglecting the off-diagonal entries of $\eta$ with respect to the diagonal entries, motivated by the Schwarz inequality for square hermitian matrices 
\begin{equation}
    |\eta_{\alpha \beta} | \leq \sqrt{\eta_{\alpha \alpha} \eta_{\beta \beta}} ~,
\end{equation}
and keeping only terms linear in $\eta_{ij}$, one can easily obtain simple expression for the factors that appear in charged currents
\begin{equation}
    \sum_\beta (KK^\dag)_{\alpha \beta} \approx  1 - 2 \eta_{\alpha \alpha} \approx (KK^\dag)_{\alpha \alpha} ~,
    \label{eq:charged1}
\end{equation}
\begin{equation}
    (KK^\dag)_{\alpha \beta} = - 2 \eta_{\alpha \beta} \equiv \epsilon_{\alpha \beta} ~,~ \alpha \neq \beta ~,
    \label{eq:charged2}
\end{equation}
and neutral currents
\begin{equation}
    \sum_{ij}|(K^\dag K)_{ij}|^2 \approx 3-  4 \eta_{11} -  4 \eta_{22} -  4 \eta_{33} ~,
    \label{eq:neutral}
\end{equation}
where the unitarity of $U_K$ was used to obtain the last equation.\par
When $s$ light heavy neutrinos are kinematically accessible, the usual bounds cannot be directly applied because the observables are no longer only proportional to   $(KK^\dag)_{\alpha \beta}$, $\sum_{\beta=1}^3(KK^\dag)_{\alpha \beta}$ and $ \sum_{i,j=1}^3|(K^\dag K)_{ij}|^2$ but to 
\begin{equation*}
     (KK^\dag)_{\alpha \beta}  +(KX^\dag)_{\alpha s}  (X K^\dag)_{s \beta} ~,
\end{equation*}
\begin{equation*}
    \sum_{\beta=1}^3 \Big(  (KK^\dag)_{\alpha \beta}  +(KX^\dag)_{\alpha s}  (X K^\dag)_{s \beta} \Big)
\end{equation*}
and
\begin{equation*}
   \sum_{i,j=1}^{3+s} \Bigg|\begin{pmatrix}
       K^\dag K & K^\dag K X^\dag \\
       X K^\dag K & X K^\dag K X^\dag \\
   \end{pmatrix}_{ij}\Bigg|^2 ~,
\end{equation*}
respectively \cite{Blennow:2016jkn,Branco:2019avf}. An updated version of the study described in refs. \cite{Fernandez-Martinez:2016lgt,Antusch:2014woa,Escrihuela:2015wra} that takes into account such cases would be pertinent but maybe unfeasible in the present due to the expected smallness of the deviations from unitarity of the leptonic mixing matrix, which are proportional to the couplings among light active neutrinos and heavy almost-sterile neutrinos, cf. chapter \ref{chapter:framework_neutrinos} and ref. \cite{Branco:2019avf}.\par
A discussion on the most relevant observables that are sensitive to these effects follows.

\subsubsection{Muon Decay: $G_F$}
The muon lifetime is one of the best measured quantities in particle physics \cite{Crivellin:2021njn}. The best measurement of $G_F$, which is one of most important parameters in the SM, is obtained from it. The decay rate of the muon in the $3n_R\nu$SM with all $M \gtrsim$ GeV/TeV is approximately given by \cite{Fernandez-Martinez:2016lgt,Antusch:2014woa,Escrihuela:2015wra}
\begin{equation}
    \Gamma_\mu = \frac{m_\mu^5G_F^2}{192 \pi^3} \sum_i (KK^\dag)_{2i} \sum_j (KK^\dag)_{1j} \approx  \frac{m_\mu^5G_F^2}{192 \pi^3} (1-2\eta_{11} - 2\eta_{22}) \equiv \frac{m_\mu^5G_\mu^2}{192 \pi^3} ~,
    \label{eq:Gamamu}
\end{equation}
where eq. \eqref{eq:charged1} was used and only terms linear in $\eta$ were kept. In eq. \eqref{eq:Gamamu} it was also implicitly defined
\begin{equation}
    G_\mu = G_F  \sqrt{(1-2\eta_{11} - 2\eta_{22})} \approx G_F  (1-\eta_{11} - \eta_{22})~,
    \label{eq:gmu}
\end{equation}
which is $G_F$ as determined through muon decay. Using the smallness of $\eta_{ii}$ one can also obtain
\begin{equation}
    G_F = G_\mu (1+\eta_{11} + \eta_{22}) ~.
    \label{eq:GF}
\end{equation}
The present value of $G_\mu$ is \cite{pdg}
\begin{equation}
    G_\mu = 1.1663788(6) \times 10^{-5}~\textrm{GeV}^{-2} ~.
\end{equation}

Hence, in this model, it acquires a non-unitary correction that spreads to other observables  \cite{Fernandez-Martinez:2016lgt,Antusch:2014woa,Escrihuela:2015wra}. Thus, it is crucial to compare the measurement of $G_F$ via the muon decay with other independent determinations of it \cite{Crivellin:2021njn}, like via super-allowed $\beta$ decays, where similarly to eq. \eqref{eq:gmu}, one can define $G_F$ determined through it as
\begin{equation}
    G_\beta = G_F  \sqrt{\sum_j (KK^\dag)_{1j}} \approx\sqrt{(1-2\eta_{11})} \approx G_F  (1-\eta_{11})~,
    \label{eq:gbeta}
\end{equation}
where eq. \eqref{eq:charged1} was used. It is important to keep in mind that one may be propagating non-unitary corrections to other measurements when using $G_F$ determined through these weak processes.

\subsubsection{Invisible $Z$ width: $N_\nu$}
An important test to the number of active neutrino species with masses below the Z boson mass, $N_{\nu}$, was the measurement of the Z invisible width at LEP \cite{d.decampetal.alephcollaboration1989}, obtaining $N_{\nu} = 2.984 \pm 0.008$.
In the $3n_R\nu$SM with all $M \gtrsim$ GeV/TeV the Z invisible width is approximately given by \cite{Fernandez-Martinez:2016lgt,Antusch:2014woa,Escrihuela:2015wra}
\begin{equation}
    \Gamma_{\textrm{inv}}= \frac{G_F M_Z^3}{12 \sqrt{2}\pi}  \sum_{ij}|(K^\dag K)_{ij}|^2 \approx  \frac{G_\mu M_Z^3}{12 \sqrt{2}\pi} (3- (4\eta_{33} + \eta_{11} + \eta_{22})) \equiv  \frac{G_\mu M_Z^3}{12 \sqrt{2}\pi} N_\nu ~,
    \label{eq:inv}
\end{equation}
where eqs. \eqref{eq:neutral}, \eqref{eq:gmu} and \eqref{eq:GF} were used. In eq. \eqref{eq:inv} it was also implicitly defined
\begin{equation}
   N_\nu = 3- (4\eta_{33} + \eta_{11} + \eta_{22})~,
    \label{eq:Nnu}
\end{equation}
The present value of $N_\nu$, averaged from several measurements \cite{pdg} is
\begin{equation}
    N_\nu = 2.92 \pm 0.05 ~.
\end{equation}
It is remarkable that the $n_R \nu$SM with all $M \gtrsim$ GeV/TeV predicts this deviation from $3$, caused by the deviations from unitarity expressed in $\eta_{ii}$, cf. eq. \eqref{eq:Nnu}.

\subsubsection{$W$ Decay to leptons: $W \rightarrow l_\alpha \nu_\alpha$}
The decay width of the W boson into a charged lepton-neutrino pair in the $3n_R \nu$SM with all $M \gtrsim$ GeV/TeV is given by \cite{Fernandez-Martinez:2016lgt,Antusch:2014woa,Escrihuela:2015wra}
\begin{equation}
    \Gamma_{W, \alpha} = \sum_i \Gamma (W \rightarrow l_\alpha \nu_i) = \frac{G_F m_W^3}{6 \sqrt{2} \pi}  (KK^\dag)_{\alpha \alpha} F_W(m_{l_\alpha}) \approx  \frac{G_\mu m_W^3}{6 \sqrt{2} \pi}  F_W(m_{l_\alpha}) (1 + \eta_{22} + \eta_{11} - 2\eta_{\alpha \alpha})~,
    \label{eq:Wdecay}
\end{equation}
where eqs.  \eqref{eq:charged1} and \eqref{eq:GF} were used and with
\begin{equation}
  F_W(m_{l_\alpha}) =\Big(1-\frac{m^2_{l_\alpha}}{m^2_W}\Big)\Big( 1+\frac{m^2_{l_\alpha}}{m^2_W}\Big) ~.
\end{equation}
The flavour structure of this decay is common to many weak processes, hence its importance. Nevertheless, the $W$ boson branching ratios to leptons is a well measured quantity \cite{pdg}, making it a great test for non-unitarity.

\subsubsection{LFNU: Lepton Flavour Universality Tests}
From eq. \eqref{eq:Wdecay} one can define the following lepton-universality observables
\begin{equation}
    R_{\alpha \beta}^W = \sqrt{ \frac{\Gamma_{W, \alpha} F_W(m_{l_\beta}) }{\Gamma_{W, \beta} F_W(m_{l_\alpha}}} = \sqrt{\frac{(KK^\dag)_{\alpha \alpha}}{(KK^\dag)_{\beta \beta}}}\approx (1-  \eta_{\alpha \alpha} +  \eta_{\beta \beta}) ~.
    \label{eq:RW}
\end{equation}
These observables allow to directly constrain the ratios of the diagonal elements of $\eta_{ii}$.
The flavour structure of $ \Gamma_{W, \alpha}$, defined in eq. \eqref{eq:Wdecay}, is the same in the width for other weak processes such as $\pi/ K \rightarrow l_\alpha \nu_\alpha$ for instance, thus one can define \cite{Fernandez-Martinez:2016lgt,Antusch:2014woa,Escrihuela:2015wra}
\begin{equation}
    R_{\alpha \beta} = \sqrt{\frac{(KK^\dag)_{\alpha \alpha}}{(KK^\dag)_{\beta \beta}}}\approx (1-  \eta_{\alpha \alpha} +  \eta_{\beta \beta}) ~,
        \label{eq:Rdef}
\end{equation}
for ratios of decay widths of $ W \rightarrow l_\alpha \nu_\alpha$ processes, as in eq. \eqref{eq:Wdecay}, for ratios of decay widths of $l_\alpha \rightarrow l_\beta \nu_\alpha \overline{\nu}_\beta$ processes or for ratios of decay widths of $K/\pi \rightarrow l_\alpha \overline{\nu}_\alpha$ processes to $l_\alpha \rightarrow K/\pi \nu_\alpha$ processes. Note that the following ratios of decay widths
\begin{equation}
    R_{\mu e} = \sqrt{ \frac{ \Gamma (\tau \rightarrow \nu_\tau \mu \overline{\nu}_\mu)} {\Gamma (\tau \rightarrow \nu_\tau e \overline{\nu}_e)}} ~,~  R_{\tau \mu} = \sqrt{ \frac{ \Gamma (\tau \rightarrow \nu_\tau e \overline{\nu}_e)} {\Gamma (\mu \rightarrow \nu_\mu e \overline{\nu}_e)}} ~,~    R_{\mu e} = \sqrt{ \frac{ \Gamma (\pi \rightarrow  \mu \overline{\nu}_\mu)} {\Gamma (\pi \rightarrow e \overline{\nu}_e)}} ~,~  R_{\tau \mu} = \sqrt{ \frac{ \Gamma (\tau \rightarrow K \nu_\tau) } {\Gamma (K \rightarrow \mu \overline{\nu}_\mu)}} ~,
\end{equation}
have the same flavour dependence, expressed in eq. \eqref{eq:Rdef}. It is important to emphasize that in the SM $R_{\alpha \beta}$ is given by a function of the masses of the involved charged leptons, incorporating phase space and chirality flip factors, along with various loop corrections. Since a flavour dependence is induced in the presence of non-unitary mixing, weak interaction lepton flavour universality constraints become powerful probes of light-heavy neutrino mixing.

\subsubsection{Unitarity of the CKM Matrix}
Both within the Standard Model (SM) and in the $n_R\nu$SM, the CKM matrix is unitary. Denoting the theoretical values of the CKM matrix elements with a superscript "th," the unitary condition for the first row can be expressed as follows
\begin{equation}
    |V_{ud}^{\textrm{th}}|^2 + |V_{us}^{\textrm{th}}|^2  +|V_{ub}^{\textrm{th}}|^2  =1 ~.
    \label{eq:CKMunitth}
\end{equation}
As discussed in section \ref{sec:quarkmodels}, presently a tension exists regarding the unitarity of the first row of the CKM matrix. This is known in the literature as the CKM unitarity problem or Cabibbo angle anomaly \cite{Seng:2018yzq,Seng:2018qru,Czarnecki:2019mwq,Seng:2020wjq,Hayen:2020cxh,Shiells:2020fqp}.\par
The current experimental values for the entries of the first row of the CKM matrix are given in eq. \eqref{eq:cabibbo2}. From those values, one can obtain the experimental version of eq. \eqref{eq:CKMunitth} \cite{Belfatto:2019swo},
\begin{equation}
    |V_{ud}^{\textrm{exp}}|^2 + |V_{us}^{\textrm{exp}}|^2  +|V_{ub}^{\textrm{exp}}|^2  \approx 0.9984 \pm 0.0001 ~.
\end{equation}
It is remarkable that in the MUV scheme, equivalent to the $3n_R\nu$SM with all $M \gtrsim$ GeV/TeV, the weak processes involving leptons from which the $V_{ij}^{\textrm{exp}}$ are measured get modified such that \cite{Fernandez-Martinez:2016lgt,Antusch:2014woa,Escrihuela:2015wra}
\begin{equation}
    |V_{ij}^{\textrm{exp}}|^2 = |V_{ij}^{\textrm{th}}|^2 (1+f^{\textrm{process}}(\eta_{\alpha \alpha})) ~,
\end{equation}
where $f^{\textrm{process}}$ refers to a function of $\eta_{\alpha}$ with $\alpha$ the leptonic flavour involved in the process in which the CKM matrix element is measured.
For instance, in the determination of $|V_{ud}|$ via super-allowed $\beta$ decays, one has
\begin{equation}
  |V_{ud}^{\textrm{exp},\beta}|^2  \approx | V_{ud}^{\textrm{th}}|^2 (1 -2 \eta_{11}) (1+ 2\eta_{11} + 2\eta_{22}) \approx |V_{ud}^{\textrm{exp},\beta}| (1+ 2\eta_{22}) ~,
\end{equation}
where eq. \eqref{eq:GF} and the smallness of $\eta_{ii}$ was used, since the process is proportional to $G_F^2$, and the factor $(1 -2 \eta_{11}) $ comes from the electrons in the final state. A similar argument can be made regarding $V_{us}$ and its measurement via  kaon or tau decays \cite{Fernandez-Martinez:2016lgt,Antusch:2014woa,Escrihuela:2015wra}
\begin{equation}
 |V_{us}^{\textrm{exp},K \rightarrow \pi e \overline{\nu}_e}|^2   \approx |V_{us}^{\textrm{th}}|^2 (1+ 2\eta_{22})~,
\end{equation}
\begin{equation}
   |V_{us}^{\textrm{exp},K \rightarrow \pi \mu \overline{\nu}_\mu}|^2   \approx |V_{us}^{\textrm{th}}|^2 (1+ 2\eta_{11}) ~,
\end{equation}
and
\begin{equation}
   |V_{us}^{\textrm{exp},\tau \rightarrow K \nu_\tau}|^2   \approx |V_{us}^{\textrm{th}}|^2 (1+ 2\eta_{11} + 2 \eta_{22} - 2 \eta_{33}) ~.
\end{equation}
\par
As explained in section \ref{sec:quarkmodels}, the present accuracy and smallness of $V_{ub}$ dictates that it can be safely neglected in eq. \eqref{eq:CKMunitth} and in this discussion, hence, when doing these types of studies, one can consider $|V_{ub}^{\textrm{exp}}|^2 \equiv |V_{ub}^{\textrm{th}}|^2$. \par
The corollary of this section is that the measurement of $V_{\textrm{CKM}}$ elements via weak processes involving leptons may propagate leptonic non-unitarity effects to the quark sector, which may alleviate the tension regarding the unitarity of the first row of the CKM matrix \cite{Coutinho:2019aiy,Crivellin:2020lzu}.

\subsubsection{LFV: Rare Charged Lepton Decays, $\mu \rightarrow e \gamma$}

Although they are absent in the SM due to the masslessness of neutrinos, charged lepton decays of the type $l_\sigma \rightarrow l_\beta \gamma$ occur at one loop in the MUV scheme. The decay width for this LFV process is given by
\begin{equation}
    \Gamma_{l_\sigma \rightarrow l_\beta \gamma} = \frac{\sigma G_\mu^2 m_\sigma^5}{2048 \pi} \Big|\sum_j K_{\sigma j} K_{j\beta}^\dag F(x_j)\Big|^2 ~,
\end{equation}
where $\alpha$ is the fine structure constant, $m_\sigma$ is the mass of the charged lepton with flavour $\sigma$ and the terms of order $\mathcal{O}(m_{l_\sigma} / m_{l_\beta})^2)$ were neglected. $F(x_j)$ is a loop function where $x_j=m_j/m_W \approx 0$, where $m_i$ are the masses of the light neutrinos. The smallness but non-zero nature of $F(x_j)$ allows the excellent approximation
\begin{equation}
    \sum_j K_{\sigma j} K_{j\beta}^\dag F(x_j) \approx - 2F(0) \eta_{\sigma \beta}~,~ F(0) = \frac{10}{3} ~.
\end{equation}
Using this approximation one can write simple formulas for the branching ratios of the LFV processes $l_\sigma \rightarrow l_\beta \gamma$ \cite{Fernandez-Martinez:2016lgt,Antusch:2014woa,Escrihuela:2015wra}
\begin{equation}
    \textrm{BR}(l_\sigma \rightarrow l_\beta \gamma) = G_{\sigma \beta} \frac{100 \alpha}{96 \pi} 4 |\eta_{\sigma\beta}|^2 ~,
\end{equation}
where $G_{\sigma \beta}$ is a numerical factor that contains phase space factors
\begin{equation}
    G_{\mu e} = 1 ~,~G_{\tau e} = \frac{1}{5.6} ~,~G_{\tau \mu} = \frac{1}{5.9} ~.
\end{equation}
As pointed out in section \ref{sec:devunit_non}, the experimental bound obtained from the search for the LFV decay $\mu \rightarrow e \gamma$ is one of the most stringent constraints on the parameter space of $n_R\nu$SM models with sizeable deviations from unitarity. Hence, the search by the MEG collaboration \cite{MEG:2016leq} and also for the other rare tau decays $\tau \rightarrow e \gamma$, $\tau \rightarrow \mu \gamma$ \cite{Hayasaka:2011jw} provide a great probe for the non-unitarity of the leptonic mixing matrix and one of the best ways to constrain the off-diagonal terms of $\eta$.\par

\subsection{Signals in Neutrino Oscillation Experiments}
\label{sec:LBL}
Neutrino oscillation experiments, such as SNO \cite{PhysRevLett.89.011301} and Super-Kamiokande \cite{PhysRevLett.81.1158}, played a crucial role in the discovery that neutrinos have mass. These experiments are known as long-baseline (LBL) experiments and are characterized by having a large source-detector distance, often denominated as $L$, allowing for oscillations due to extremely small mass differences to develop, leading to a sufficient phase difference to cause flavour transformations \cite{Diwan:2016gmz}. T2K \cite{T2K:2019bcf} and No$\nu$A \cite{NOvA:2019cyt} are prominent examples.  \\
In contrast, short-baseline (SBL) experiments, as implied by their name, involve significantly shorter source-detector distances, typically on the order of a few meters \cite{Katori:2014vka}, as is the case of MiniBooNE \cite{MiniBooNE:2008paa}. \par
\begin{figure}[H]
  \centering
  \includegraphics[width=0.8\textwidth]{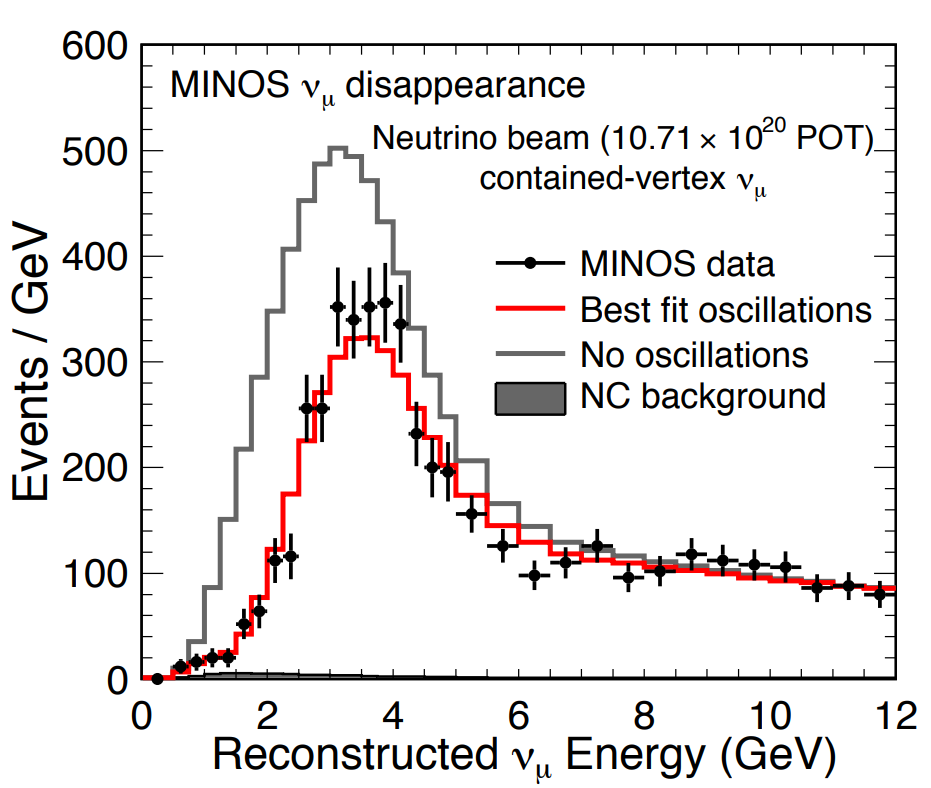} 
  \caption[Distribution of muon neutrino events plotted as a function of reconstructed energy in the MINOS far detector at a distance of $735$ km. The
solid grey line shows the expected spectrum under no oscillation hypothesis. The
depletion of events at $\gtrsim 1.5$ GeV and the shape of the depleted spectrum is
consistent with dominant $\nu_\mu \rightarrow \nu_\tau$ oscillations with maximal mixing.]{Distribution of muon neutrino events plotted as a function of reconstructed energy in the MINOS far detector at a distance of $735$ km. The
solid grey line shows the expected spectrum under no oscillation hypothesis. The
depletion of events at $\gtrsim 1.5$ GeV and the shape of the depleted spectrum is
consistent with dominant $\nu_\mu \rightarrow \nu_\tau$ oscillations with maximal mixing. Image taken from ref. \cite{Diwan:2016gmz}.}
  \label{fig:MINOS}
\end{figure}

Considering a unitary leptonic mixing matrix, $U \equiv U_{\textrm{PMNS}}$, defined in eq. \eqref{eq:Vstd}, the oscillation probability between a (anti-)neutrino of flavour $\alpha$ to a (anti-)neutrino of flavour $\beta$, with $\alpha, \beta = e, \mu, \tau$, can be shown to take the form
\begin{equation}
\begin{aligned}
P_{\stackon[-.7pt]{$\nu$}{\brabar}_\alpha \rightarrow \stackon[-.7pt]{$\nu$}{\brabar}_\beta}(L,E)
\,=\,
\delta_{\alpha\beta}
&-   4 \sum_{i>j}^{3+n}\,\re
\left(U_{\alpha i}^*\,U_{\beta i}\,U_{\alpha j}\,U_{\beta j}^*\right)
\sin^2 \Delta_{ij} \\
&\pm 2 \sum_{i>j}^{3+n}\,\im
\left(U_{\alpha i}^*\,U_{\beta i}\,U_{\alpha j}\,U_{\beta j}^*\right)
\sin 2 \Delta_{ij} \,,
\end{aligned}
\label{eq:probability_unit}
\end{equation}%
where the plus or minus sign in the second line refers to neutrinos or anti-neutrinos, respectively.
Here, again $L$ denotes the source-detector distance, $E$ is the (anti-)neutrino energy,
and one has defined
\begin{equation}
\Delta_{ij} \,\equiv \, \frac{\Delta m^2_{ik}\, L}{4E}
\,\simeq\, 1.27\,\frac{\Delta m_{ij}^{2}[\text{eV}^{2}]\,L[\text{km}] }{ E[\text{GeV}]}\,,
\end{equation}
with mass-squared differences $\Delta m_{ij}^2 \equiv m_i^2 - m_j^2$. In order for experiments to be sensitive to neutrino oscillations, it is necessary for the neutrino energy and the distance to the detector to satisfy the condition $\Delta_{ij} \,\equiv \, \frac{\Delta m^2_{ik}\, L}{4E} \gtrsim 1$. However, it is often impractical to physically move either the detector or the source to cover a wide range of $L/E$ values. Therefore, having a broad energy spectrum proves advantageous as it allows for the detection of oscillations through the distortion of the expected neutrino energy distribution, as depicted in figure \ref{fig:MINOS}, which is a reconstructed $\nu_\mu$ energy spectrum from the MINOS experiment \cite{Evans:2013pka}.

\pagebreak

The sensitivity ranges for different types of neutrino experiments are as follows:
\begin{itemize}
    \item For solar neutrino experiments ($L \approx 10^{11}$ m, $E \approx 15$~MeV), the sensitivity range is $\Delta m_{ij} \approx 10^{-10}~\textrm{eV}^2$.
    \item  For atmospheric neutrino experiments ($20$ km $< L < 10^3$ km, $E \approx 1$ GeV), the sensitivity range is $\Delta m_{ij} \approx 10^{-4}~\textrm{eV}^2$.
    \item For reactor neutrino experiments ($L \approx [10,10^4]$ m, $E \approx 3$ MeV), the sensitivity range is $\Delta m_{ij} \approx 10^{-5}~\textrm{eV}^2$.
    \item For accelerator neutrino experiments ($L \leq 500$ km, $E \approx 1$ GeV), the sensitivity range is $\Delta m_{ij} \approx 10^{-3}~\textrm{eV}^2$.
    \item For a typical SBL experiment ( $L \sim \mathcal{O} (100)$ m, $E \approx \mathcal{O} (100)$ MeV), the sensitivity range is $\Delta m_{ij} \approx \mathcal{O}(1)~\textrm{eV}^2$.
\end{itemize}
One cannot end this summary without mentioning IceCube \cite{Halzen:2022pez}, an experiment that is sensitive to typical LBL $\Delta m_{ij}$ \cite{PhysRevD.91.072004} but also to the usual SBL $\Delta m_{ij}$ \cite{PhysRevLett.117.071801}, as it can detect both cosmic and atmospheric neutrinos, hence being able to acquire data for various values of $L/E$. \par
Comprehensive summaries of current results based on global fits from LBL experiments, considering a unitary leptonic mixing matrix, can be found in references \cite{Esteban:2020cvm,deSalas:2020pgw}. The results from ref. \cite{deSalas:2020pgw} can be found in Table \ref{tab:bestfit2}. These results provide information about the leptonic mixing angles and the Dirac phase $\delta$. However, the models studied in this thesis do not have unitary leptonic mixing matrices, as discussed in chapter \ref{chapter:framework_neutrinos}. Consequently, the equations governing neutrino oscillation probabilities, such as eq. \eqref{eq:probability_unit}, are modified due to non-unitarity effects, cf. section \ref{sec:non-dec} and eq. \eqref{eq:probability}. This modified equation can be simplified for specific cases, such as SBL and LBL experimental setups with the consideration of matter effects in  LBL experiments being mandatory in order to obtain realistic results. \par
Interestingly, experiments with sensitivities of $\Delta m_{ij} \sim \mathcal{O}(1)~\textrm{eV}^2$ are particularly susceptible to non-unitarity effects \cite{Branco:2019avf}, as discussed in section \ref{sec:non-dec}. Notably, SBL experiments, which fall within this sensitivity range \cite{Abazajian:2012ys}, observed an anomaly now known as SBL anomaly, initially referred to as the LSND anomaly \cite{PhysRevD.64.112007,PhysRevLett.75.2650}, based on the experiment that detected it. This anomaly involves an excess of $\overline{\nu}_e$ at the detector from a $\overline{\nu}_\mu$ source. The $L/E$ settings of the detector points towards a large $\Delta m^2$ if interpreted as a two-neutrino oscillation $\overline{\nu}_\mu \xrightarrow{} \overline{\nu}_e$. 
Reactor and radioactive source experiments also reported excess or deficit of $\nu_e$ or $\overline{\nu}_e$ in the detector \cite{Mueller:2011nm,Huber:2011wv,Dentler:2018sju,Bahcall:1994bq}. Later, KARMEN reported results that disagreed with the results from LSND \cite{Armbruster:2002mp}, but the MiniBOONE experiment 
\cite{MiniBooNE:2008paa} confirmed the results obtained by LSND \cite{MiniBooNE:2018esg,MiniBooNE:2020pnu}. MiniBOONE, despite having similar $L/E$ as LSND, differs in terms of energy, beam, detector systematics, event signatures and backgrounds, which makes it an excellent tester of the LSND results. \par
A global fit combining all these results and other constraints \cite{PhysRevD.83.015015,PalomaresRuiz:2005vf} suggests the existence of a heavy sterile neutrino with the data exhibiting a preference for the $(3+1)$ scenario, where there are three light neutrinos and one neutrino with a mass in the eV range and a sizeable coupling with at least one light neutrino\cite{Giunti:2019aiy,deGouvea:2019qre,Cianci:2017okw}. Hence, these anomalies may be explained if the spectrum of heavy neutrinos includes at least one neutrino with an eV-scale mass \cite{Dentler:2018sju,Dentler:2019dhz}, and if there are significant deviations from the unitarity of the $3 \times 3$ mixing matrix (DU) 
\cite{Giunti:2006bj,Branco:2019avf}. \par
However, it is crucial to note that a closer examination of the dataset reveals tension between the appearance and disappearance datasets, as well as the neutrino and anti-neutrino datasets \cite{Acero:2022wqg}. Furthermore, recently, the IceCube experiment also reported results that conflict with the ones from LSND and MiniBOONE \cite{PhysRevLett.117.071801}. 
In conclusion, all of these anomalies and hypotheses are still subject to ongoing debate and discussion.
\par

\subsubsection{Effects in LBL Experiments}
Obtaining direct evidence of heavy neutrinos in LBL oscillation experiments is extremely challenging because these experiments are sensitive to very small mass differences, $\Delta m_{ij}$. One possibility for detection would arise if there exists a pair of quasi-degenerate heavy neutrinos, hence with a small enough mass splitting, with significant mixing and adequate mass in order for it to be produced and/or detectable in the experiment. For the spectra considered in this thesis, which contain a quasi-degenerate pair of heavy neutrinos with TeV masses, cf. case \textbf{II} of \ref{sec:non-dec} , an experiment like IceCube\cite{IceCube:2017roe, Halzen:2022pez} could potentially detect them, although the likelihood of this occurring is extremely low. \par
LBL experiments have a higher probability of detecting non-unitarity effects associated with large mass splittings. This is because terms proportional to $\sin^2(\Delta m~L /4E)$ can be effectively averaged out to approximately 0.5 \cite{Super-Kamiokande:2014ndf}, thus the oscillation probabilities in these cases only depend on the usual mixing among light neutrinos and the mixing between light and heavy neutrinos. The modifications of neutrino oscillation probabilities in vacuum in LBL experiments for the cases studied in ref. \cite{Branco:2019avf} is discussed in section \ref{sec:benc_non-dec}. Using the definitions of eq. \eqref{eq:sinmue} and eq. \eqref{eq:3nu}, the transition probability of muon to electron (anti-)neutrinos for the \textbf{Ia}  case can be approximated by eq. \eqref{eq:LBL2Ia}, for the \textbf{Ib}  case by eq. \eqref{eq:LBL2Ib} and for the \textbf{II} case by eq. \eqref{eq:LBL2II}. \par
Hence, heavy almost-sterile neutrinos may leave traces in LBL experiments. This should come as no surprise since, as stated earlier in this section, in the presence of a non-unitary mixing matrix, and possibly a sizeable mixing with almost-sterile neutrinos, the neutrino oscillation probabilities are modified. This can have noticeable effects in the conversion and survival probabilities of a neutrino species in a LBL experiment\cite{Escrihuela:2015wra}, mostly due to the effect of the hermitian phases, the phases of the deviations from unitarity, first discussed in this thesis in section \ref{sec:leptogenesis}.
A summary of the effect of these phases in CP Violation measurements can also be found in section \ref{sec:cp}. However, the limited sensitivity of active LBL experiments to these effects, see ref. \cite{Super-Kamiokande:2014ndf} and the relatively weak limits obtained, means that perhaps only experiments like DUNE \cite{Kudryavtsev:2016ybl, DUNE:2016evb} have the potential to truly test them.\par
Nevertheless, is interesting to point out that a global-fit from a combined analysis of 2019 and 2020 data from T2K and NO$\nu$A prefers non-unitarity at 2 $\sigma$ C.L. \cite{Miranda:2019ynh}. The tension between NO$\nu$A and T2K 2019 data is also reduced when
both the experiments are analysed with the non-unitary hypothesis. If in the future the two experiments consistently show improved agreement with the non-unitary hypothesis, it could serve as a compelling indication of the existence of new physics in the neutrino sector.

\subsubsection{Effects in SBL Experiments}

\begin{figure}[H]
  \centering
  \includegraphics[width=0.8\textwidth]{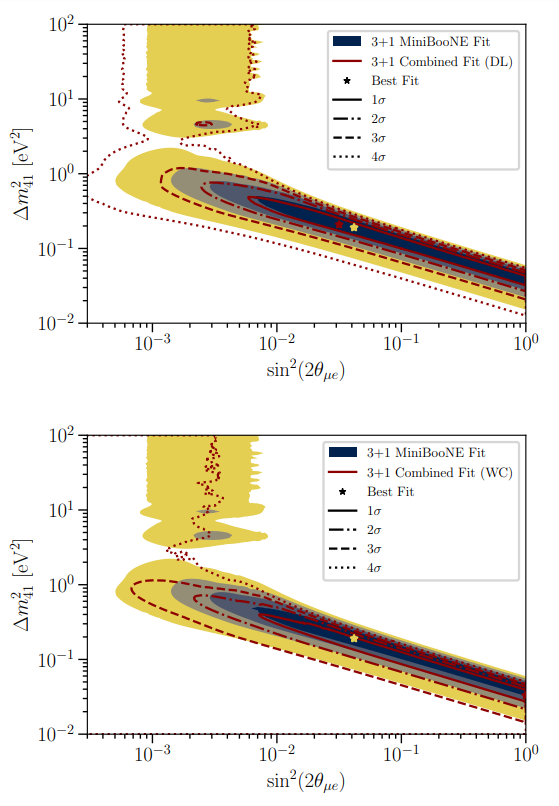} 
  \caption[The results of the MiniBooNE-only and combined
fits with MicroBooNE’s CCQE sample (top) and MicroBooNE’s Inclusive sample (bottom). The likelihood is
obtained by profiling over all parameters $\Delta m^2$ and  $\sin^2(2 \theta_{\mu e}) \equiv \sin^2 2 \vartheta^{(4)}_{\mu e}$. The two best-fit points are shown as appropriately
colored stars, and the contours are obtained by comparing the
profile-likelihood-ratio test-statistic to the asymptotic distribution provided by Wilks’ theorem, and assuming a difference of two degrees of freedom.]{ The results of the MiniBooNE-only and combined
fits with MicroBooNE’s CCQE sample \cite{MicroBooNE:2021pvo} (top) and MicroBooNE’s Inclusive sample \cite{MicroBooNE:2021nxr} (bottom). The likelihood is
obtained by profiling over all parameters $\Delta m^2$ and  $\sin^2(2 \theta_{\mu e}) \equiv \sin^2 2 \vartheta^{(4)}_{\mu e}$. The two best-fit points are shown as appropriately
colored stars, and the contours are obtained by comparing the
profile-likelihood-ratio test-statistic to the asymptotic distribution provided by Wilks’ theorem, and assuming a difference of two degrees of freedom. Image taken from ref. \cite{MiniBooNE:2022emn}.}
  \label{fig:miniboone}
\end{figure}

SBL experiments and LBL experiments have distinct discovery prospects. SBL experiments are capable of detecting oscillations only when the magnitude of $\Delta m_{ij}$ falls within their typical sensitivity range, which is approximately $\Delta m_{ij} \approx \mathcal{O}(1)\textrm{eV}^2$. Consequently, SBL experiments can test the existence of almost-sterile neutrinos with a mass of around $1$ eV, as in case \textbf{Ia} of ref. \cite{Branco:2019avf}, or the presence of a pair of nearly-sterile neutrinos with $\Delta m_{ij} \approx \mathcal{O}(1)\textrm{eV}^2$, as outlined in case \textbf{Ib} of ref. \cite{Branco:2019avf}.
The observability of these oscillations is enhanced when there is a substantial mixing between the heavy and light neutrinos, as demonstrated by the figures and benchmarks presented in section \ref{sec:benc_non-dec} and reference \cite{Branco:2019avf}. In fact, the resolution of the aforementioned SBL anomalies, within the 3+1 scheme, which closely resembles the \textbf{Ia} case, requires $\Delta m^2 = 0.209$ eV$^2$ and $\sin^2 2 \vartheta^{(4)}_{\mu e} =0.0316$ and $|V_{e4}|^2=0.016$, a 4.3 $\sigma$ preference comparing to the no-oscillation scenario, per the most recent work of the MiniBooNE and MicroBooNE collaborations\cite{MiniBooNE:2022emn}, cf. figure \ref{fig:miniboone}. 
MicroBooNE \cite{MicroBooNE:2015bmn} is another SBL experiment built in the same apparatus as MiniBooNE, located in the Booster Neutrino Beam Beamline at Fermilab. MicroBooNE's two main physics goals are to investigate the MiniBooNE low-energy excess and neutrino-argon cross sections.\par
The modifications of neutrino oscillation probabilities in vacuum for the cases examined in ref. \cite{Branco:2019avf} are discussed in section \ref{sec:benc_non-dec}. By utilizing the definitions in equation \eqref{eq:sinmue} and equation \eqref{eq:3nu}, the transition probability from muon to electron (anti-)neutrinos can be approximated as equation \eqref{eq:SBL2Ia} for the \textbf{Ia} case, equation \eqref{eq:SBL2Ib} for the \textbf{Ib} case, and equation \eqref{eq:SBL2II} for the \textbf{II} case.
Hence, SBL experiments provide the most stringent constraints on light heavy neutrinos with eV-scale masses  \cite{Blennow:2016jkn}. In these cases, the observables directly constrain the entries of the mixing matrix  $V_{\alpha k} = (K X^\dag)_{\alpha k}$, where $\alpha=1,2,3$ and $k=1,..,n_R$, defined in eq. \eqref{eq:mixingmatrix_neutrino} and eq. \eqref{eq:ABdef}. To incorporate these constraints from the experiments, the relevant exclusion curves in the $\sin^{2}2\vartheta{\alpha \beta}$--$\Delta m^{2}$ planes are considered and translated into constraints on the elements of the mixing matrix $V_{\alpha k}$.

\begin{figure}[H]
  \centering
  \includegraphics[width=0.8\textwidth]{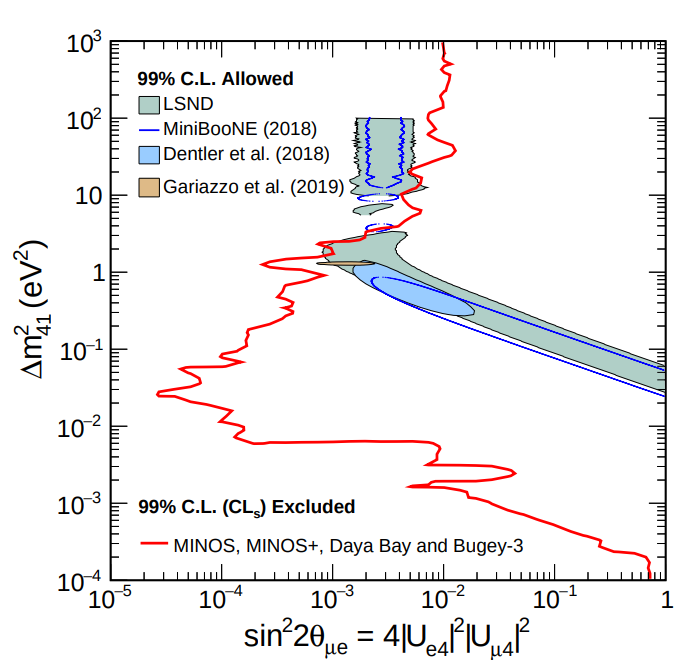} 
  \caption[]{The combined limits at 99\% C.L.(CLs) on $\sin^2(2\theta_{\mu e}) \equiv \sin^{2}2\vartheta{\alpha \beta}$ using the CLs method (red). The exclusion
contours are compared to the LSND and MiniBooNE allowed regions. Also shown are the allowed regions from global fits by Gariazzo et al. \cite{Gariazzo:2015rra} and Dentler et al. \cite{Dentler:2018sju}. Note that  $\sin^2(2 \theta_{\mu e}) \equiv \sin^2 2 \vartheta^{(4)}_{\mu e}$. Image taken from ref. \cite{Hu:2020uvx}.}
  \label{fig:daya}
\end{figure}

For values $\Delta m^2 \in [10^{-2},1] \textrm{eV}^2$ the combined result of MINOS \cite{MINOS:2016viw}, MINOS+ \cite{Evans:2017brt}, Daya Bay \cite{Dohnal:2021rcr} and Bugey-3 \cite{Declais:1994su} gives the best result, shown in fig. \ref{fig:daya}.

Thus, for a heavy neutrino with a $1$ eV mass and $\Delta m^2 \sim 1 \textrm{eV}^2$, one has the bound
\begin{equation}
   \sin^2 2 \vartheta^{(4)}_{\mu e} = 4 |V_{e 4}|^2 |V_{\mu 4}|^2 \lesssim 10^{-3} ~.
\end{equation}
Note that the benchmarks presented in tables \ref{tab:Ia}, \ref{tab:Ib} and \ref{tab:II} satisfy this bound.

\begin{figure}[H]
  \centering
  \includegraphics[width=0.9\textwidth]{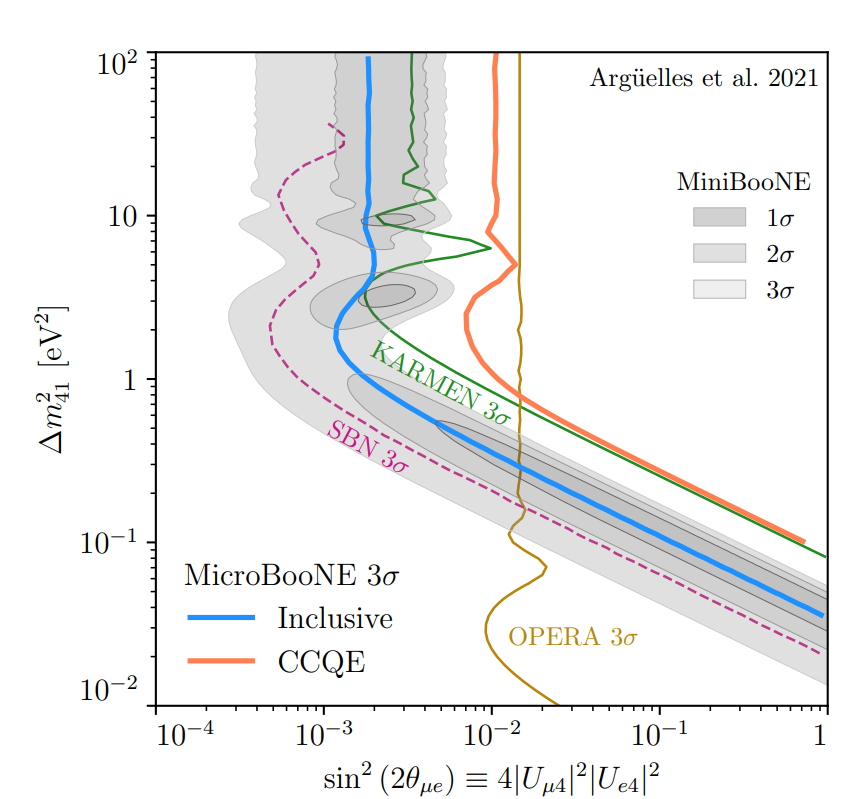} 
  \caption[MicroBooNE constraints on sterile neutrino parameter space at 3$\sigma$ C.L. (blue, Inclusive and orange, CCQE). For reference, the MiniBooNE 1-, 2-, and 3-$\sigma$ preferred regions is shown in shades of grey, the future sensitivity of
the three SBN detectors (pink), and existing constraints
from KARMEN (green) and OPERA (gold).]{MicroBooNE constraints on sterile neutrino parameter space at 3$\sigma$ C.L. (blue, Inclusive and orange, CCQE). For reference, the MiniBooNE 1-, 2-, and 3-$\sigma$ preferred regions is shown in shades of grey \cite{MiniBooNE:2018esg}, the future sensitivity of
the three SBN detectors (pink) \cite{Machado:2019oxb}, and existing constraints
from KARMEN (green) \cite{KARMEN:2002zcm} and OPERA (gold) \cite{OPERA:2018ksq}. Note that  $\sin^2(2 \theta_{\mu e}) \equiv \sin^2 2 \vartheta^{(4)}_{\mu e}$. Image taken from ref.\cite{Arguelles:2021meu}.}
  \label{fig:microboone}
\end{figure}

For larger values of $\Delta m^2$, KARMEN \cite{KARMEN:2002zcm}, MiniBooNE \cite{MiniBooNE:2020pnu} and MicroBooNE \cite{Arguelles:2021meu} give the most stringent limit, depicted in fig. \ref{fig:microboone}.
Note that a large portion of the LSND and MiniBooNE allowed region \cite{Arguelles:2021meu} and the 3+1 scheme best-fit zone \cite{MiniBooNE:2022emn} are excluded by the combined analysis of MINOS, MINOS+, Daya Bay and Bugey-3 \cite{Hu:2020uvx}.  The tension between the indications of electron (anti)neutrino appearance and the absence of corresponding signals in disappearance channels is heightened, further contributing to the ongoing debate surrounding SBL neutrino oscillations.

\subsection{Beta Decay }
\label{sec:beta_decay}
\begin{figure}[H]
  \centering
  \includegraphics[width=0.8\textwidth]{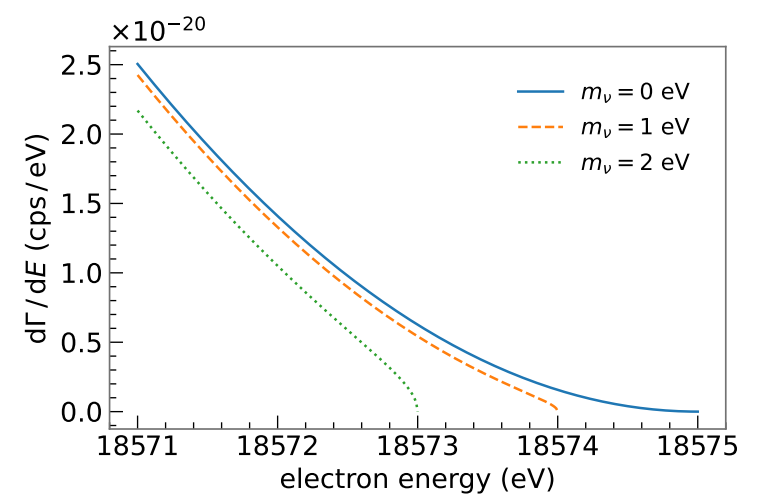} 
  \caption[Illustration of the shape distortion induced by a non-zero $m_\beta$ on the tritium $\beta$-decay spectrum. ]{Illustration of the shape distortion induced by a non-zero $m_\beta$ on the tritium $\beta$-decay
spectrum. Image taken from ref. \cite{Onillon:2023aqz}.}
  \label{fig:mbeta}
\end{figure}

Beta ($\beta$) decay is a nuclear process described by
\begin{equation}
    n \xrightarrow{} p + e^{-} + \overline{\nu_e} ~,
\end{equation}
where $n$ represents a neutron in a given atomic nucleus, $p$ a proton, $e^-$ an electron and $\overline{\nu_e}$ an electron-flavoured anti-neutrino.
The spectra of the emitted electron depends on it energy, $E_e$, and is given by the Kurie function \cite{Gariazzo:2015rra}
\begin{equation}
\begin{split}
 K^2(E_e)&= (Q-E_e) \sum_i |K_{1i}|^2 \sqrt{(Q-E_e)^2 - m^2_{i}} \times \Theta(Q-T_e-m_i) ~. \\
 \end{split}
 \label{eq:kurie}
\end{equation}
$Q$ is the amount of energy released by the reaction, $\Theta$ is the Heaviside step function and $m_i$ is the mass of the emitted anti-neutrino. 
Since there are no prospects of achieving a resolution that may distinguish the light neutrino states, it is customary to perform an approximation that is valid for an experiment in which the energy resolution is such that $m_i << Q-E_e$,
\begin{equation}
     K^2(E_e)  \approx  (Q-E_e)  \sqrt{(Q-E_e)^2 - m^2_{\beta}} \times \Theta(Q-E_e-m_{\beta}) ~,
\end{equation}
where $m_{\beta}$ is the “electron neutrino mass” and the Kurie function depends on it thanks to the phase space factor. It is determined as a real average over the three light neutrino mass eigenstates contributing to the electron neutrino,
\begin{equation}
    m^2_{\beta}= \sum_i^3 |K_{1i}|^2 m^2_i ~,
    \label{eq:mbeta}
\end{equation}
where $K$ is defined in eq. \eqref{eq:kdef} and eq. \eqref{eq:calVfull}. Note that, in general, this quantity is sensitive to all the CP Violating Phases in the leptonic sector of the $n_R \nu$SM. The unitary phases, the Dirac phase and the Majorana phases, and the hermitian phases, the phases of the deviations from unitarity, as described in section \ref{sec:leptogenesis}. Using the full equation for $K$, eq. \eqref{eq:kdef} and eq. \eqref{eq:kh}, with eq. \eqref{eq:Vstd} to parameterise the unitary part of $K$ and eq. \eqref{eq:hermitian_part} to parameterise the hermitian part of $K$, one can calculate $|K_{11}|^2$, to prove the previous statement,
\begin{equation}
\begin{split}
    |K_{11}|^2 =& \Bigl(c_{12}c_{13}h_{11} + h_{12}s_{12}c_{13} \cos(\alpha_{21} - p_{12}) + 
   h_{13} s_{13} \cos(\alpha_{31} - \delta - p_{13})\Bigr)^2 \\
   & + \Bigl(h_{12} s_{12}c_{13} \sin(\alpha_{21} - p_{12}) + h_{13} s_{13} \sin(\alpha_{31} - \delta - p_{13})\Bigr)^2 ~.
\end{split}
\end{equation}
However, in practical terms, the effect of the phases is only measurable if the deviations from unitarity are sizeable, otherwise $h_{11} \sim 1$ and $h_{12} \sim h_{13} \sim 0$, and $m^2_\beta$ only depends on mixing angles.
The measurement of $m_\beta$ consists of looking for a decrease in the observed maximum electron energy (the end-point of the spectrum) and a subtle distortion in the spectral shape, relative to the expected theoretical prediction. Figure \ref{fig:mbeta} illustrates this. Note how a $m_{\beta} \neq 0$ ($m_\beta = m_\nu$ in the figure) affects the shape and the end-point of the spectrum.

\begin{figure}[H]
  \centering
  \includegraphics[width=0.8\textwidth]{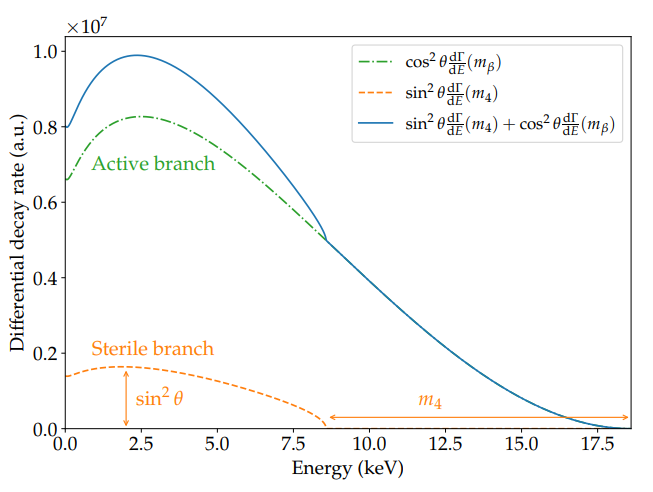} 
  \caption[Illustration of the almost-sterile neutrino signature in a differential tritium $\beta$-decay spectrum. For the sake of clarity, a almost-sterile neutrino with a mass of $10$ keV and an unrealistic large mixing angle is considered.]{Illustration of the almost-sterile neutrino signature in a differential tritium $\beta$-decay spectrum.
For the sake of clarity, a almost-sterile neutrino with a mass of $10$ keV and an unrealistic large mixing angle is considered. Image taken from ref. \cite{Onillon:2023aqz}.}
  \label{fig:kink}
\end{figure}

As discussed in section \ref{sec:non-dec}, the deviations from unitarity, and as a consequence the effect of the phases on $m^2_\beta$, might be sizeable precisely in scenarios where $k$ of the $n_R$ "heavy" states have a mass of the order of the eV or keV. These neutrinos would be kinematically allowed to be produced in a beta decay, hence the electron energy spectrum would be a superposition of the light neutrino spectrum and the "heavy" neutrino spectrum \cite{Gariazzo:2015rra,Giunti:2019aiy}. In this case, eq. \eqref{eq:kurie} becomes
\begin{equation}
\begin{split}
 K^2(E_e) & \approx (Q-E_e)  \sqrt{(Q-E_e)^2 - m^2_{\beta}} \times \Theta(Q-E_e-m_{\beta})  \\
& + (Q-E_e) \sum_{k=4}^{n_R} |V_{ek}|^2  \sqrt{(Q-E_e)^2 - m^2_{k}} \times \Theta(Q-E_e-m_{k}) ~,
\end{split}
\end{equation}
The above expression shows that a "heavy" neutrino
mass, $m_k$ can be measured by observing a kink of the kinetic energy spectrum at $E_e=Q-m_k$, the point where the "heavy" neutrino spectrum ends \cite{Gariazzo:2015rra}.
Figure \ref{fig:kink} illustrates how the existence of a light heavy neutrino that is almost-sterile affects the spectrum with a kink-like signature.

KATRIN \cite{Arenz:2018aly,Arenz:2018bjh,Arenz:2018jpa,Arenz:2018kma,Arenz:2018ymp} is an experiment that aims to measure $m_{\beta}$, from the beta decay of Tritium ($H^3$) with endpoint $Q=18.6$ keV. The experiment is sensitive to $m_{\beta} > 0.2~eV$ \cite{Brunst:2018vka}. With the data acquired during the first two measurement campaigns of 2019, the collaboration obtained the most stringent limit on $m_\beta$ with $m_\beta < 0.8$ eV at 90\% CL. \cite{KATRIN:2022ith,Onillon:2023aqz}. TRISTAN (TRitium Investigation on STerile to Active Neutrino mixing) is a subsequent experiment to KATRIN, building upon its advancements by utilizing the complete KATRIN source. The primary objective of TRISTAN is to explore the existence of heavy neutrinos within the mass range of 1 keV to 18 keV. By mitigating theoretical and experimental systematic uncertainties, TRISTAN has the potential to investigate $|V_{14}|^2$ values as low as $10^{-7}$ \cite{Onillon:2023aqz,Abdullahi:2022jlv}. The planned upgrade for TRISTAN is scheduled to occur following the completion of the KATRIN experiment's $m_\beta$ measurement, which is expected to take place around 2025. At the conclusion of the data collection, it is anticipated that a total of approximately $70 \times 10^6$ electrons will be obtained, aiming to achieve a target sensitivity  $m_\beta < 0.2 - 0.3$ eV (90\% confidence level).
Of particular interest is the observation that the upper bound of the values of $m_{\beta}$, for the cases of section \ref{sec:non-dec} ranged from $0.0093$ eV to $0.015$ eV, considerably lower than this target sensitivity.

\subsubsection{$M \sim$ eV}
\begin{figure}[H]
  \centering
  \includegraphics[width=0.9\textwidth]{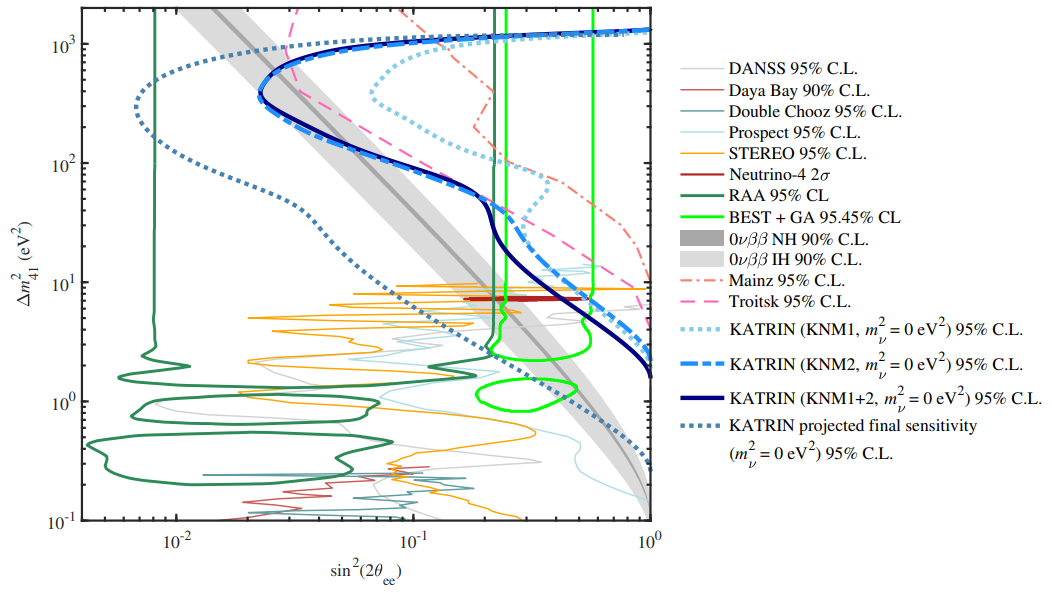} 
  \caption[eV-scale sterile neutrino exclusion contours at 95\% C.L. for the first and second campaign of KATRIN.  Note that $\sin^2(\theta_{ee})$ in the figure refers to  $|V_{14}|^2$ in the text.  The combined results of the two campaigns and the projected final sensitivity are shown. These exclusions are compared to the most recent results from reactor, gallium, beta decay
and double beta decay experiments.]{eV-scale sterile neutrino exclusion contours at 95\% C.L. for the first and second campaign of KATRIN \cite{KATRIN:2022ith}. Note that $\sin^2(\theta_{ee})$ in the figure refers to  $|V_{14}|^2$ in the text. The combined results of the two campaigns as well the projected final sensitivity are also shown. These exclusions are compared to the most recent results from reactor, gallium, beta decay
and double beta decay experiments. References for the different experimental results used in this plot
can be found in ref. \cite{KATRIN:2022ith}. Image taken from ref. \cite{Onillon:2023aqz}.}
  \label{fig:eVKATRIN}
\end{figure}

Before KATRIN started acquiring data it was already speculated that it could probe heavy neutrinos with a eV mass \cite{Riis:2010zm, Giunti:2015wnd, Giunti:2019aiy}. By utilizing the same datasets employed for determining the value of $m_\beta$, the KATRIN experiment conducted an investigation into the possible existence of a fourth neutrino mass eigenstate \cite{KATRIN:2022ith,Onillon:2023aqz}. This analysis possesses sensitivity to the range of $m_4^2 \lesssim 1600~\textrm{eV}^2$ for the fourth neutrino mass eigenstate, as well as active-to-sterile mixing $|V_{14}|^2 \gtrsim 6 \times 10^{-3}$ within the $3\nu + 1$ framework \cite{Giunti:2019aiy}. No significant signal indicating the presence of an eV-scale sterile neutrino was observed, and an improved exclusion limit was established. The outcomes of the analysis are summarized in Figure \ref{fig:eVKATRIN}, which also includes the constraints derived from previous experiments.

The eV neutrinos in the models from section \ref{sec:non-dec} have $|V_{14}|^2 \sim 10^{-4}$, hence they are not yet ruled out by KATRIN, and may start being probed in the next round of results.

\subsubsection{$M\sim$ keV}
\begin{figure}[H]
  \centering
  \includegraphics[width=0.8\textwidth]{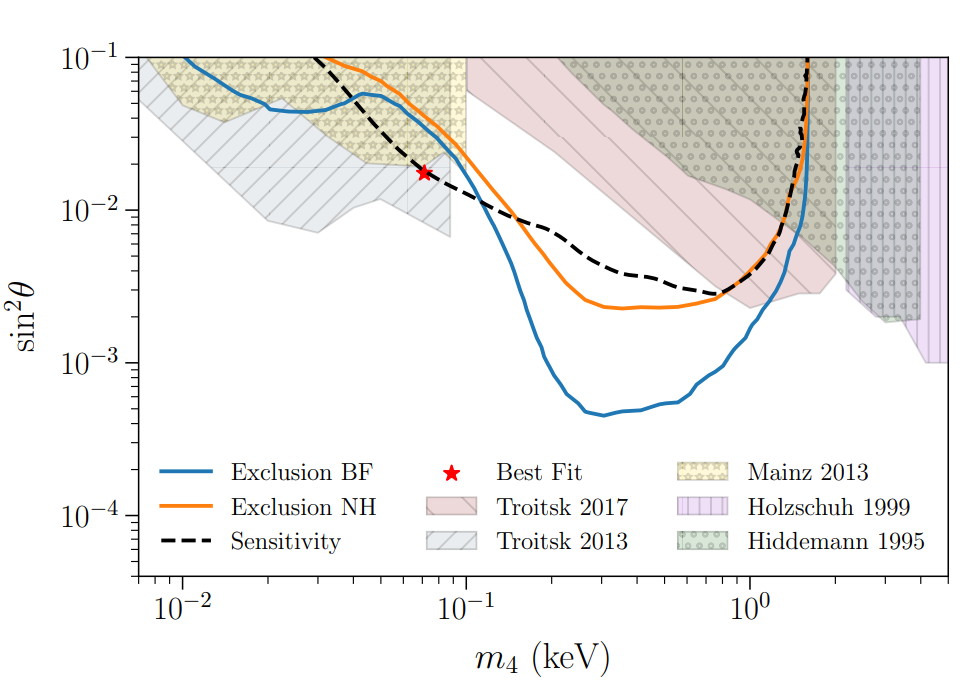} 
  \caption[95 \% C.L. exclusion limit obtained based on the first
tritium data set of KATRIN with respect to the best fit (BF)
(blue). The exclusion limit obtained by comparing the $\chi^2$ values to the null hypothesis (NH) (orange) is in reasonable
agreement with the sensitivity (dashed black). The current laboratory limits were improved (colored shaded areas) on the active-to-sterile mixing amplitude in a mass range of
$0.1$ keV $< m_4 < 1.0$ keV by up to an order of magnitude. The main result is given by the exclusion limit with respect to the best fit (blue line).]{95 \% C.L. exclusion limit obtained based on the first
tritium data set of KATRIN with respect to the best fit (BF)
(blue). The exclusion limit obtained by comparing the $\chi^2$ values to the null hypothesis (NH) (orange) is in reasonable
agreement with the sensitivity (dashed black). The current laboratory limits were improved (colored shaded areas) on the active-to-sterile mixing amplitude in a mass range of
$0.1$ keV $< m_4 < 1.0$ keV by up to an order of magnitude. The main result is given by the exclusion limit with respect to the best fit (blue line). Image taken from ref. \cite{KATRIN:2022spi}.}
  \label{fig:kevKATRIN}
\end{figure}

Prior to the commencement of the operations at KATRIN, preliminary studies highlighted its potential to investigate almost-sterile neutrinos at the keV scale\cite{Mertens:2014nha,Mertens:2014osa,Abada:2018qok,Mertens:2018vuu,Drewes:2016upu,Dolde:2016wnv}.
Hence, in 2022, KATRIN also presented the outcomes of its initial search for keV-scale almost-sterile neutrinos using tritium data obtained during the commissioning run \cite{KATRIN:2022spi}. No significant signal indicating the presence of keV-scale sterile neutrinos was observed, leading to the establishment of exclusion limits. These results significantly enhance the existing laboratory constraints on the amplitude of active-to-sterile mixing within the mass range of $0.1$ keV $< m_4 < 1.0$ keV by up to an order of magnitude. Consequently, an active-sterile mixing amplitude of $\sin^2 \theta < 5 \times 10^{-4}$ is excluded for an almost-sterile neutrino mass of $m_4 = 300$ eV \cite{KATRIN:2022spi,Onillon:2023aqz}. These results are summarised in figure \ref{fig:kevKATRIN}.

The keV neutrinos in the models from section \ref{sec:non-dec} have $\sin^2 \theta  \in [10^{-6},10^{-5}]$, note that in case \textbf{Ia} of section \ref{sec:non-dec} $\sin^2 \theta \equiv |V_{15}|^2$, hence they are not yet ruled out by KATRIN, and may start being probed in the next round of results.

\subsubsection{$M >$ keV}
Masses beyond the keV scale are not kinematically allowed since the process is nuclear and its scale is the MeV. For tests for these mass scales see section \ref{sec:LHC} and section \ref{sec:elecweak_observables}.

\subsection{CP Violation Effects at low-energy}
\label{sec:cp}
The $n_R\nu$SM contains several CP Violating phases. The unitary phases, the Dirac phase and the Majorana phases, and the hermitian phases, the phases of the deviations from unitarity, as described in section \ref{sec:leptogenesis}.
All these phases play a role in neutrino oscillations, since in the presence of deviations from unitarity, neutrino oscillation probabilities are modified  \cite{Antusch:2006vwa,Blennow:2016jkn}. This subject was originally discussed in ref. \cite{Branco:2019avf}, as an extension of ref. \cite{Fernandez-Martinez:2007iaa}.
In order to analyse the CP Violation effects of these phase,
it is instructive to define CP asymmetries $A_{\nu \overline{\nu}}^{\alpha \beta}$
at the level of oscillation probabilities \cite{Gandhi:2015xza}:
\begin{align}
A_{\nu \overline{\nu}}^{\alpha \beta}
\,\equiv\,
\frac
{P_{\nu_\alpha \to \nu_\beta} - P_{\overline{\nu}_\alpha \to \overline{\nu}_\beta}}
{P_{\nu_\alpha \to \nu_\beta} + P_{\overline{\nu}_\alpha \to \overline{\nu}_\beta}}
\,\equiv\,
\frac{\Delta P_{\alpha\beta}}
{P_{\nu_\alpha \to \nu_\beta} + P_{\overline{\nu}_\alpha \to \overline{\nu}_\beta}}
\,.
\end{align}
This analysis focus exclusively on the vacuum scenario, acknowledging that in a more realistic context, the violation of CP and CPT due to the matter-induced asymmetry encountered by neutrinos needs to be considered. The principle of CPT invariance gives rise to the relations $\Delta P_{\alpha\beta} = -\Delta P_{\beta\alpha}$ and $\Delta P_{\alpha \alpha} = 0$. Additionally, the unitarity of the complete mixing matrix imposes the constraints
\begin{align}
\sum_\beta\,\Delta P_{\alpha\beta} = 0\,,
\end{align}
for all indices $\alpha$ and $\beta$ encompassing the complete set, which includes $e$, $\mu$, $\tau$, $s_1$, $s_2$, ..., $s_q$. In a $3\times 3$ unitary framework, these relations indicate the existence of a single independent difference, which can be selected as $\Delta P_{e\mu}$. On the other hand, within a $4\times 4$ unitary framework, as demonstrated in ref. \cite{Gandhi:2015xza}, these relations imply the presence of three independent differences, denoted as $\Delta P_{e\mu}$, $\Delta P_{\mu\tau}$, and $\Delta P_{\tau e}$.

In the case of a $6\times 6$ unitary context, it is determined that there are 10 independent differences $\Delta P_{\alpha \beta}$ (as also mentioned in ref. \cite{Reyimuaji:2019wbn}), but only the three involving solely active neutrinos hold experimental significance. Consequently, it is generally anticipated that different values for $\Delta P_{e\mu}$, $\Delta P_{\mu\tau}$, and $\Delta P_{\tau e}$ will manifest in a given minimal seesaw-type model.
Using Eq.~\eqref{eq:probability}, with $n$ mostly-sterile neutrinos kinematically accessible at an oscillation experiment, one finds:
\begin{align}
\Delta P_{\alpha\beta} \,=\,
\frac{4}{(\Theta\Theta^\dagger)_{\alpha\alpha}(\Theta\Theta^\dagger)_{\beta\beta}}
\, \sum_{i>j}^{3+n}\,\im
\left(\Theta_{\alpha i}^*\,\Theta_{\beta i}\,\Theta_{\alpha j}\,\Theta_{\beta j}^*\right)
\sin 2 \Delta_{ij}\,.
\end{align}
Even in scenarios where none of the new sterile states are accessible, meaning $n=0$, it is still anticipated that $\Delta P_{e \mu}$, $\Delta P_{\mu \tau}$, and $\Delta P_{\tau e}$ will remain independent. This is due to the fact that the relevant $3\times 3$ mixing submatrix $\Theta$, defined in eq. \eqref{eq:wdef}, (which in that case is equivalent to $K$) is not unitary.
This means that it is possible for CP invariance to hold in one oscillation channel, such as 
$\nu_\mu \rightarrow \nu_e$
and yet be violated in another, such as 
$\nu_\mu \rightarrow \nu_\tau$
Indeed, one has:
\begin{align}
\Delta P_{\mu \tau} \,&=\, \Delta P_{e \mu} \,+\,
\frac{4}{\prod_{\alpha = e,\mu,\tau} (\Theta\Theta^\dagger)_{\alpha\alpha}}
\, \sum_{i>j}^{3}\,\im
\Big[\Theta_{\mu i}^*\,\Theta_{\mu j}\Big(
  \Theta_{e i}\,\Theta_{e j}^*      \, (\Theta\Theta^\dagger)_{\tau\tau}
+ \Theta_{\tau i}\,\Theta_{\tau j}^*\, (\Theta\Theta^\dagger)_{ee}
\Big)\Big]
\sin 2 \Delta_{ij}
 \,,\\
\Delta P_{\tau e}   \,&=\, \Delta P_{e \mu} \,-\,
\frac{4}{\prod_{\alpha = e,\mu,\tau} (\Theta\Theta^\dagger)_{\alpha\alpha}}
\, \sum_{i>j}^{3}\,\im
\Big[\Theta_{e i}^*\,\Theta_{e j}\Big(
  \Theta_{\mu i}\,\Theta_{\mu j}^*  \, (\Theta\Theta^\dagger)_{\tau\tau}
+ \Theta_{\tau i}\,\Theta_{\tau j}^*\, (\Theta\Theta^\dagger)_{\mu\mu}
\Big)\Big]
\sin 2 \Delta_{ij}
 \,.
\end{align}
In such cases, it is possible for $\Delta P_{e\mu}$ to be zero while $\Delta P_{\mu \tau}$ and/or $\Delta P_{\tau e}$ are non-zero. It should be noted that if the mixing submatrix $\Theta$ were unitary, one would have $\Delta P_{e \mu} = \Delta P_{\mu \tau} = \Delta P_{\tau e}$. Therefore, deviations from unitarity can serve as a potential source of CP violation. This observation aligns with the fact that the hermitian part of $K$, defined in eq. \eqref{eq:kh} and eq. \eqref{eq:kh2}, contains complex hermitian elements, which can encompass CP-violating physical phases, eq. \eqref{eq:hermitian_part}.

\subsubsection{$M \sim$ eV, keV}
In the cases examined in section~\ref{sec:non-dec}, where $n = 1,2$, we can derive explicit expressions for the CP asymmetries that are relevant in the context of short-baseline oscillations. These expressions can be obtained from the approximate relations given by equations~\eqref{eq:SBL2Ib} and~\eqref{eq:SBL2II} for cases \textbf{Ib} and \textbf{II}, respectively. For case \textbf{Ia}, it can be observed from equation \eqref{eq:SBL2Ia} that the CP asymmetries in that case are negligible in the short-baseline regime.
One has, for case \textbf{Ib}:
\begin{align}
\Delta P_{e\mu}^\text{SBL,\:Ib} \,\simeq\,
4 \,\im\left(\Theta_{\mu 4}^*\,\Theta_{e 4}\,\Theta_{\mu 5}\,\Theta_{e 5}^*\right)\sin 2 \Delta_{54}
\,,
\end{align}
while for case \textbf{II}:
\begin{align}
\Delta P_{e\mu}^\text{SBL,\:II} \,\simeq\,
8 \,\im\left(\Theta_{\mu 4}^*\,\Theta_{e 4}\,(KX^\dag)_{\mu 2}\,(KX^\dag)_{e 2}^*\right)\sin 2 \Delta_{41}
\,.
\end{align}

These asymmetries might be measurable in future short-baseline oscillation experiments under two conditions:
\begin{itemize}
    \item There is a mass-squared difference, $\Delta_{ij}$, among any neutrino (light and heavy) pair of the order of the sensitivity of the experiment. For instance, active SBL experiment MicroBooNE \cite{MicroBooNE:2022sdp} is sensitive to $\Delta_{ij} \sim \mathcal{O}(1)~\textrm{eV}^2$.
    \item The mixing between the neutrinos that comprise this pair and the active light neutrinos, $\Theta_{in}$ with $i=e,\mu,\tau$ and $n=4$ or $n=4,5$, is sizeable enough. That implies sizeable deviations from unitarity, only possible if there are heavy neutrinos with a mass in the eV or keV scale, cf. beginning of chapter \ref{chapter:results}.
\end{itemize}

\subsection{Displaced Vertex Signatures}
\label{sec:displaced}
\begin{figure}[H]
  \centering
  \includegraphics[width=0.8\textwidth]{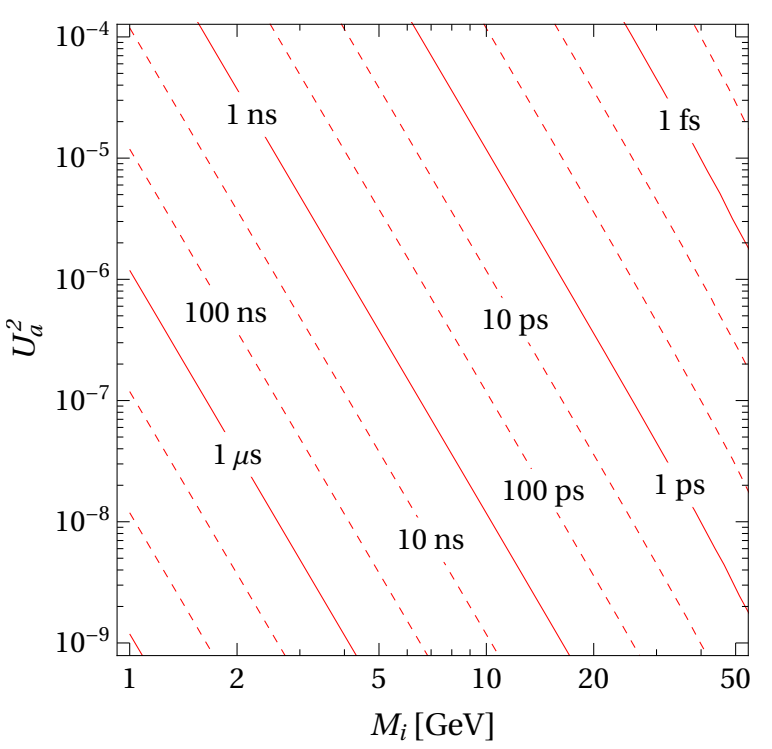} 
  \caption[The heavy neutrino life time as a function of its mass and coupling calculated with MadWidth.]{The heavy neutrino life time as a function of its mass and coupling calculated with MadWidth. Image taken from ref. \cite{Drewes:2019fou}.}
  \label{fig:lifetime}
\end{figure}

In recent years it has been pointed out how in many realistic models these heavy neutrinos may have a large lifetime
 \cite{Helo:2013esa,Abada:2018sfh,Gago:2015vma,Drewes:2019fou}. This means they can leave a noticeable signature in detectors: a displaced vertex decay  \cite{Cottin:2018nms,Antusch:2017hhu,Drewes:2018xma}. This can be searched for using recently developed machine learning techniques \cite{AngLi:2023oax}. The signal is especially traceable if these neutrinos have a mass between $1$ and $50$ GeV \cite{Cottin:2018kmq,Gago:2015vma}, as in realistic models with $\mathcal{O}(1)$ Yukawa couplings, these naturally have smaller mixing with active neutrinos, cf. beginning of chapter \ref{chapter:results}, hence having larger lifetimes. Figure \ref{fig:lifetime} illustrates this.
Recently, the possibility of a new signature - a displaced shower-, was suggested 
\cite{Cottin:2022nwp}. It is claimed heavy neutrinos masses between $1-6$ GeV can can be accessed for mixings as low as $|V_{\tau 4}|^2 \sim 10^{-7}$, enabling the probing of unique areas of the parameter space in the $\tau$ sector.
The concern mentioned in section \ref{sec:LHC} regarding the need to interpret many of these searches \cite{Tastet:2021vwp} also applies to these kinds of searches, as it also customary to to assume the presence of a single heavy neutrino mixing with a single neutrino flavour.
\subsubsection{$M \in [1,50]$ GeV}

A 2020 ATLAS search \cite{ATLAS2019kpx} published results obtained at a center-of-mass energy of 13 TeV, which places constraints on the mixing of heavy neutrinos with muon-neutrinos at the level of $|V_{\mu 4}|^2 \approx 2 \times 10^{-6}$ for heavy neutrino masses ranging from $4$ to $10$ GeV, based on a data sample of $33-36 \textrm{fb}^{-1}$ . These results are illustrated in Figure 5.

 \begin{figure}[H]
  \centering
  \includegraphics[width=1.0\textwidth]{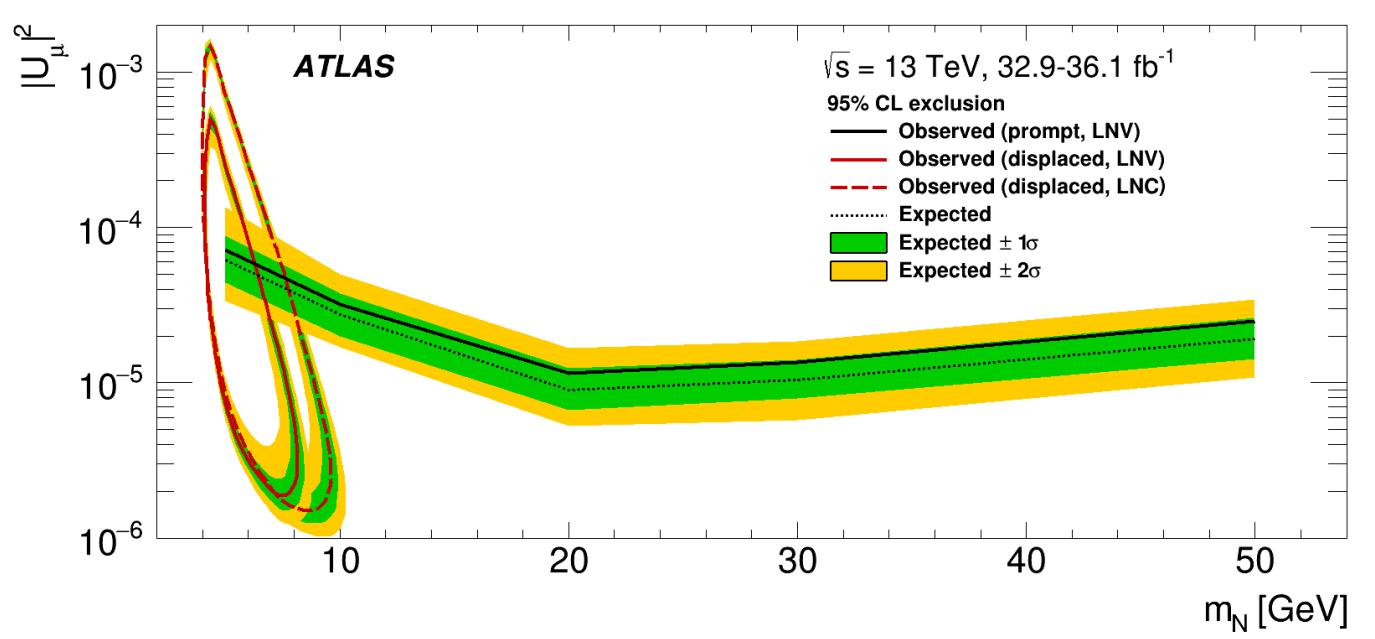} 
  \caption[Observed 95\% confidence-level exclusion in  $|V_{\mu 4}|^2$ versus the heavy neutrino mass for the
prompt signature (the region above the black line is excluded) and the displaced signature (the region enclosed by the
red line is excluded). The solid lines show limits assuming lepton-number violation (LNV) for 50\% of the decays and the long-dashed line shows the limit in the case of lepton-number conservation (LNC). The dotted lines show expected limits and the bands indicate the ranges of expected limits obtained within $1\sigma$ and $2\sigma$ of the median limit, reflecting uncertainties in signal and background yields.]{Observed 95\% confidence-level exclusion in  $|V_{\mu 4}|^2$ versus the heavy neutrino mass for the
prompt signature (the region above the black line is excluded) and the displaced signature (the region enclosed by the
red line is excluded). The solid lines show limits assuming lepton-number violation (LNV) for 50\% of the decays and the long-dashed line shows the limit in the case of lepton-number conservation (LNC). The dotted lines show expected limits and the bands indicate the ranges of expected limits obtained within $1\sigma$ and $2\sigma$ of the median limit, reflecting uncertainties in signal and background yields. Image taken from ref. \cite{ATLAS2019kpx}.}
  \label{fig:ATLAS2019kpx}
\end{figure}

\begin{figure}[H]
  \centering
  \includegraphics[width=1.0\textwidth]{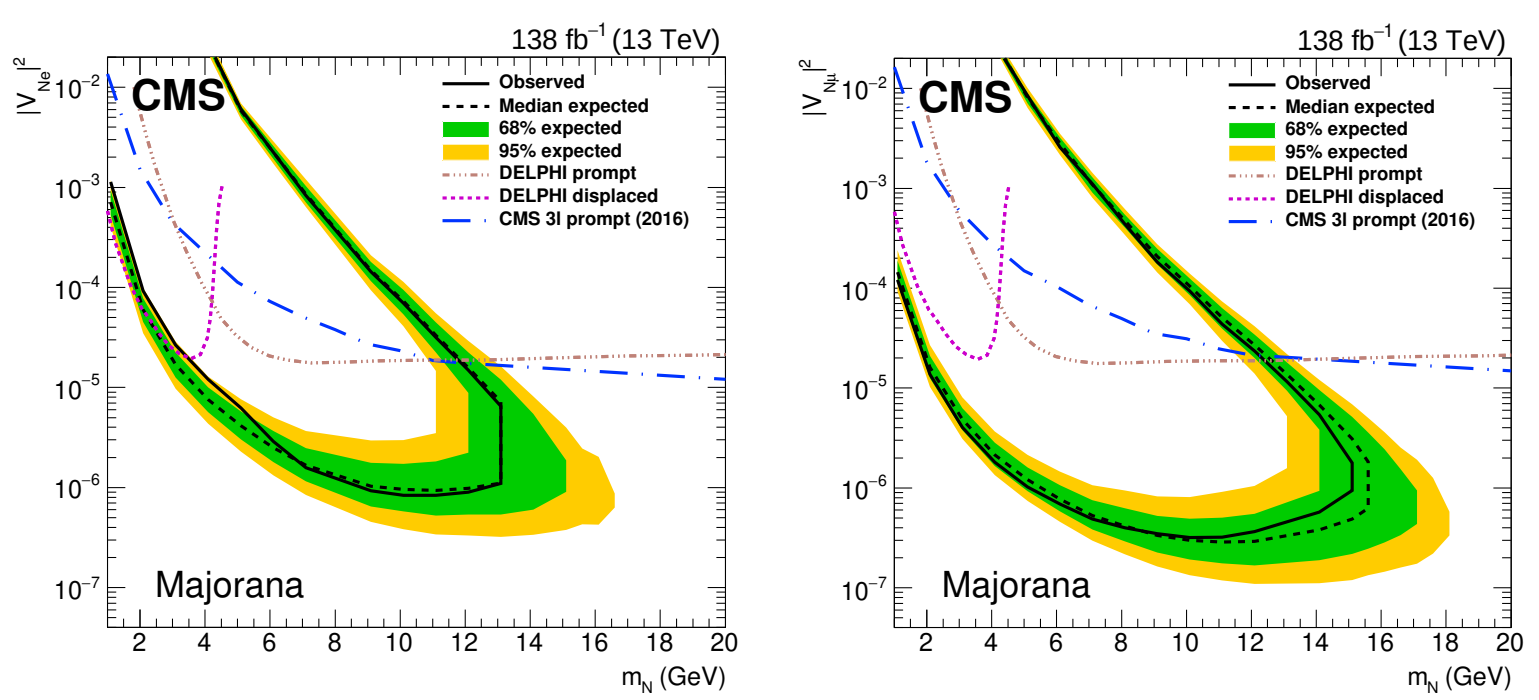} 
  \caption[The 95\% CL limits on $|V_{4e}|^2$ (left) and $|V_{4\mu}|^2$
(right) as functions of $m_N$ for a Majorana heavy neutrino. The area inside the solid (dashed) black curve indicates the observed (expected) exclusion region. Older results from the DELPHI and the CMS Collaborations are shown for reference.]{The 95\% CL limits on $|V_{4e}|^2$ (left) and $|V_{4\mu}|^2$
(right) as functions of $m_N$ for a Majorana heavy neutrino. The area inside the solid (dashed) black curve indicates the observed (expected) exclusion region. Older results from the DELPHI and the CMS Collaborations are shown for reference. Image taken from ref. \cite{CMS2022fut}.}
  \label{fig:CMS2022fut}
\end{figure}

In a recent study by CMS \cite{CMS2022fut}, the focus was on searching for low-mass heavy neutrinos. The analysis made use of the complete dataset from Run 2, corresponding to an integrated luminosity of $138~\textrm{fb}^{-1}$. The investigation specifically targeted the decay of heavy neutrinos into oppositely charged leptons and a neutrino, resulting in a final state characterized by three charged leptons, two of which exhibit displacement. The results of this study are presented in Figure \ref{fig:CMS2022fut}. The obtained results impose bounds on the mixing of heavy neutrinos with muon-neutrinos, with a value of approximately $|V_{\mu 4}|^2 \approx 4 \times 10^{-7}$ for neutrino masses ranging from 8 to 14 GeV. In the case of electron-neutrino channels, the sensitivity is approximately $|V_{e 4}|^2 \approx 4 \times 10^{-6}$ within the same mass range.

The most recent displaced vertex search was done by the ATLAS collaboration \cite{ATLAS2022atq}, and it addressed the concerns explained in ref. \cite{Tastet:2021vwp} regarding the need to reinterpret searches considering multiflavour scenarios. Hence, it is the first study that provides limits for both single-flavour and multiflavour mixing scenarios, which are motivated by the findings of neutrino flavour oscillation studies  \cite{Tastet:2021vwp}. The search utilizes a dataset of $139~\textrm{fb}^{-1}$ of proton-proton collision data at a center-of-mass energy of $13$ TeV. The search is centered around the production of heavy neutrinos through the decay process of $W$ bosons, specifically $W \rightarrow N l_\alpha$, where $\alpha=e,\mu$ represents the flavour of the prompt lepton. The heavy neutrinos subsequently decays into two oppositely charged leptons and a neutrino: $N \rightarrow l_\beta l_\alpha \nu_\gamma$ via an intermediate $W^*$ boson, or $N \rightarrow l_\gamma l_\gamma \nu_\beta$ via a $Z^*$ boson, where $\beta$ and $\gamma$ can be either $e$ or $\mu$. The search specifically focuses on the mixing and mass range (up to $20$ GeV) in which the heavy neutrino exhibits long lifetimes.

\begin{figure}[H]
  \centering
  \includegraphics[width=0.75\textwidth]{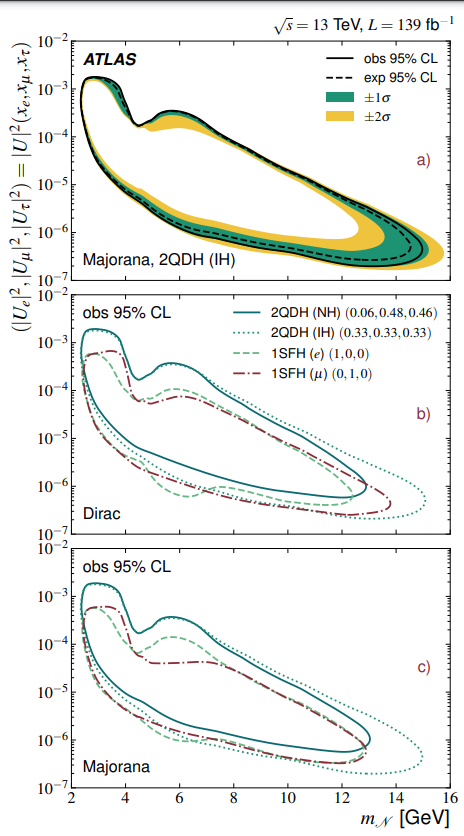} 
  \caption[The observed and expected 95\% CL limits on $|V_{N\alpha}|^2$ vs. $m_N$ in the Majorana-limit case, with green and yellow bands showing the one and two standard deviation spreads for the expected limits. (b,c) The observed limits in the 2QDH scenario with inverted (IH) and normal (NH) mass hierarchy, and in 1SFH scenarios. 1SFH refers to the model where the heavy neutrino only couples with one flavour of active neutrinos. 2QDH refers to the more realistic case of two quasi-degenerate neutrinos with masses in the range between $1$ and $20$ GeV that couple to all active neutrino flavours.]{The observed and expected 95\% CL limits on $|V_{N\alpha}|^2$ vs. $m_N$ in the Majorana-limit case, with green and yellow bands showing the one and two standard deviation spreads for the expected limits. (b,c) The observed limits in the 2QDH scenario with inverted (IH) and normal (NH) mass hierarchy, and in 1SFH scenarios. 1SFH refers to the model where the heavy neutrino only couples with one flavour of active neutrinos. 2QDH refers to the more realistic case of two quasi-degenerate neutrinos with masses in the range $m_N \in [1,20]$ GeV that couple to all active neutrino flavours. Image taken from ref. \cite{ATLAS2022atq}.}
  \label{fig:ATLAS2022atq}
\end{figure}

No excess was observed and limits were set at 95\% CL on the squared mixing coefficient $|V_{N\alpha}|^2$, where $\alpha=e,\mu,\tau$ in different scenarios for neutrino masses
masses in the approximate range $M \in [3,15]$ GeV. The results of the search are summarised in fig. \ref{fig:ATLAS2022atq}.

\subsection{Signals at Future Experiments: FCC, DUNE, HyperK}
Several experiments are scheduled for the upcoming years, including the High-Luminosity Large Hadron Collider (HL-LHC), which will collect data at a significantly higher luminosity than the LHC. This increase in luminosity will greatly enhance the statistical significance of future analyses conducted with the acquired data. It is anticipated that these analyses significantly improve the constraints on the direct detection of heavy neutrinos \cite{Das:2015toa, Das:2017nvm, Das:2017zjc}. \par
The next step in the energy frontier is the  the Future Circular Collider (FCC) \cite{Bernardi:2022hny,FCC:2018byv}. The initial phase of the FCC consists of an $e^+e^-$ collider known as FCC-ee. This collider is designed to serve as a Higgs factory, an electroweak and top factory, operating at the highest luminosities. It will operate at four distinct center-of-mass energies, specifically targeting the $Z$ pole, the $WW$ threshold, the $ZH$ production peak, and the $tt$ threshold. Subsequently, in the second phase, the FCC-ee would be followed by the FCC-hh, a proton collider with a center-of-mass energy of $100$ TeV.\par
During the electron-positron stage of the Future Circular Collider (FCC), there will be a focus on conducting high-precision measurements. However, the FCC-ee also presents significant opportunities for the exploration and discovery of new phenomena. By combining precise measurements and direct searches, both indirect and direct evidence for new physics can be obtained. In particular, the high-luminosity Z pole run at FCC-ee holds exceptional potential for the detection of heavy neutrinos  \cite{Antusch:2015mia,Antusch:2016ejd,Antusch:2016vyf}. The primary production mode for heavy neutrinos at this stage will be through the process $Z \rightarrow \nu N$, where $N$ is the heavy neutrino and $\nu$ a light active neutrino, followed by the $N$ decaying into off-shell $W$ or $Z$ bosons. The FCC-ee is expected to exhibit excellent sensitivity above the charm mass and extending up to the mass of the W boson, for various mixing angles with the known neutrinos. In refs. \cite{Antusch:2015mia,Antusch:2016ejd,Antusch:2016vyf,Blondel:2014bra,Drewes:2022rsk,Blondel:2022qqo}, sensitivity down to heavy-light mixing of $|V_{\nu N}|^2 \sim 10^{-8} - 10^{-12}$ was achieved, covering a broad parameter space for heavy neutrino masses in the GeV scale.\par
As for the FCC-hh, it will have a much higher sensitivity than the FCC-ee to higher masses, both for prompt and displaced searches \cite{Das:2015toa,Das:2017nvm,Das:2017zjc,Abdullahi:2022jlv,Drewes:2022akb}. For instance, refs. \cite{Pascoli:2018heg,Abdullahi:2022jlv} suggest that it could be sensitive to masses up to $10$ TeV ($100$ GeV) to the level of $|V_{\nu N}|^2 \sim 10^{-3}$ ( $|V_{\nu N}|^2 \sim 10^{-4}$) and much smaller mixings can potentially be probed at lower masses \cite{Pascoli:2018heg}.\par
The FCC and other projected colliders such as the International Linear Collider (ILC) or the Compact Linear Collider (CLIC) also hold great promise in detecting heavy neutrino-antineutrino oscillations \cite{Antusch:2017ebe,Cvetic:2018elt,Antusch:2022hhh}, a phenomenon that occurs when there is a pair of quasi-degenerate heavy neutrinos, and also in probing some specific models where leptogenesis explains the BAU \cite{Antusch:2017pkq,Antusch:2018ahh}.\par
Regarding future neutrino oscillation experiments, the most notable ones are JUNO \cite{JUNO:2015zny}, DUNE \cite{DUNE:2016evb,Kudryavtsev:2016ybl} and HyperKamiokande (HyperK) \cite{Hyper-Kamiokande:2022smq}. JUNO is a reactor anti-neutrino experiment in China that intends on performing a $>3 \sigma$ mass ordering determination, and to measure $\sin^2 \theta_{12}$, $\Delta m^2_{21}$ and $|\Delta m^2_{21}|$ with a precision better than $0.6 \%$ \cite{NavasNicolas:2023fza}. DUNE, a long-baseline experiment in the USA, and HyperK, an improved iteration of SuperKamiokande in Japan, have similar goals - measuring $\sin^2 2\theta_{13}$, $\sin^2 \theta_{23}$ and $\Delta m^2_{32}$ with great precision with a special emphasis in the determination of the octant of $\theta_{23}$ and the discovery (or not) of CP violation in the leptonic sector via the precise determination of $\delta_{CP}$.
Furthermore, both DUNE and HyperK will have a nearby detector complex situated close to the neutrino source. These detectors serve multiple purposes, including observing and tracking the neutrino flux generated by the facility. Moreover, the high-intensity beam interactions with the target offer the possibility of generating heavy neutrinos and other novel particles, particularly through the decay of light mesons. Given the energy and intensity of the beams, the near detectors of these experiments provide an intriguing and unexplored search area for heavy neutrinos undergoing decay while in motion, with masses of up to $2$ GeV\cite{Abdullahi:2022jlv}. Despite having similar goals and operational principles, DUNE and HyperK have largely uncorrelated systematic uncertainties. This is due to their differences in neutrino energies, far detector target nuclei, and the effects of calibration, reconstruction, and event selection. As a result, these two experiments complement each other greatly.

\pagebreak
\section{Vector-like Quarks}
\label{sec:res_quarks}

In recent decades, extensive searches for heavy quarks have been conducted at the Tevatron and the LHC. Despite the absence of concrete evidence for the existence of heavy quarks, experimental limits have been progressively tightened. \par

The presence of VLQs has a significant impact on various electroweak precision observables. In the SM, the absence of FCNC at tree level in the quark sector results in the suppression of several quark-related processes, such as $D^0$-$\overline{D}^0$ mixing, as they only occur at the one-loop through a version of the GIM mechanism. However, VLQs introduce FCNC at tree level, and the amplitude of these processes is proportional to $|F^q_{ij}|^2$, which are defined in equations \eqref{eq:Fquark} and \eqref{eq:quarkF}. The matrix $F$ controls FCNC and is proportional to the deviations from unitarity of the quark mixing matrix $V$, as defined in equations \eqref{eq:mixingmatrixquark} and \eqref{eq:uparam-exactV}, cf. eq. \eqref{eq:Fud}. A detailed explanation of the connection among $F$, $V$, FCNC, and the deviations from unitarity of $V_{\textrm{CKM}}$ (the upper-left $3 \times 3$ block of $V$) can be found in section \ref{sec:perturb}. The absence of significant signals in the searches for these rare quark processes can be interpreted in one of two ways: either there are no additional quarks, or the elements of $F^q$ are highly suppressed. The latter scenario could indicate the existence of VLQs with large masses, as elaborated in section \ref{sec:perturb}, as large masses naturally suppress FCNC in VLQ models \cite{Alves:2023ufm}. Consequently, the precision frontier is imposing increasingly stringent constraints on the matrix elements of $F$ and $V$ and, consequently, the deviations from unitarity of $V_{\textrm{CKM}}$. A discussion on the effects of VLQ on these observables and the most recent experimental bounds can be found in section \ref{sec:newphysics}. \par

In terms of direct searches, the bounds on the matrix elements of $V$, particularly those connecting the SM quarks with the new quark states, are becoming increasingly restrictive. Otherwise, these new quark states would have already been detected at the LHC, considering the current center of mass energy it operates at. Moreover, the lower mass bounds on VLQs have been steadily increasing over the last decade and VLQs that decay predominantly to third generation quarks are essentially ruled out for masses below the TeV scale \cite{Alves:2023ufm,
CMS2022fck,ATLAS2022tla,ATLAS2022ozf,CMS2023agg,ATLAS2023uah,ATLAS2023pja,CMS2017gsh}.\par

The majority of model-independent constraints are derived from searches focusing on pair production, as the production mechanism of heavy quarks is attributed to Quantum Chromodynamics (QCD). Typically, the outcome of a pair production search is an exclusion, which establishes, at a specific confidence level (e.g. 95\% C.L.), a lower bound for VLQ masses, indicating that masses below this threshold are ruled out by the search.\par

The single production of VLQs is highly dependent on the specific theoretical model, as the production mechanism involves electroweak interactions. Consequently, the outcomes of single production VLQ searches are typically presented in terms of constraints on cross sections and couplings of the heavy quarks to SM quarks, for different heavy quark masses. In the context of a particular model, it becomes feasible to establish bounds on VLQ masses at specific confidence levels. It is worth noting that if the coupling between light quarks and VLQs is substantial, the single production of VLQs is anticipated to exhibit a higher cross section compared to pair production, particularly in proton colliders like the LHC, as outlined in refs. \cite{Atre:2008iu,Atre:2011ae,Branco:2021vhs}. The results of the most recent direct searches for VLQ are discussed in section \ref{sec:direct_searches_quarks}. \par
Before proceeding, it is crucial to address the customary conventions, assumptions, and terminology found in the relevant literature. The majority of searches are conducted with the assumption that VLQs predominantly decay to third-generation quarks or solely decay to third-generation quarks. However, it is important to acknowledge that such an assumption may not always be justified, particularly in scenarios where VLQs are proposed to address the CKM unitarity problem, cf. ref. \cite{Branco:2019avf} and section \ref{sec:quarkup}. Nevertheless, this hypothesis of only decaying into third-generation quarks, as expressed in eq. \eqref{eq:3rdgen}, has become ingrained in the literature as a form of "conventional wisdom".\par

\begin{figure}[!t]
\centering
\includegraphics[scale=0.42]{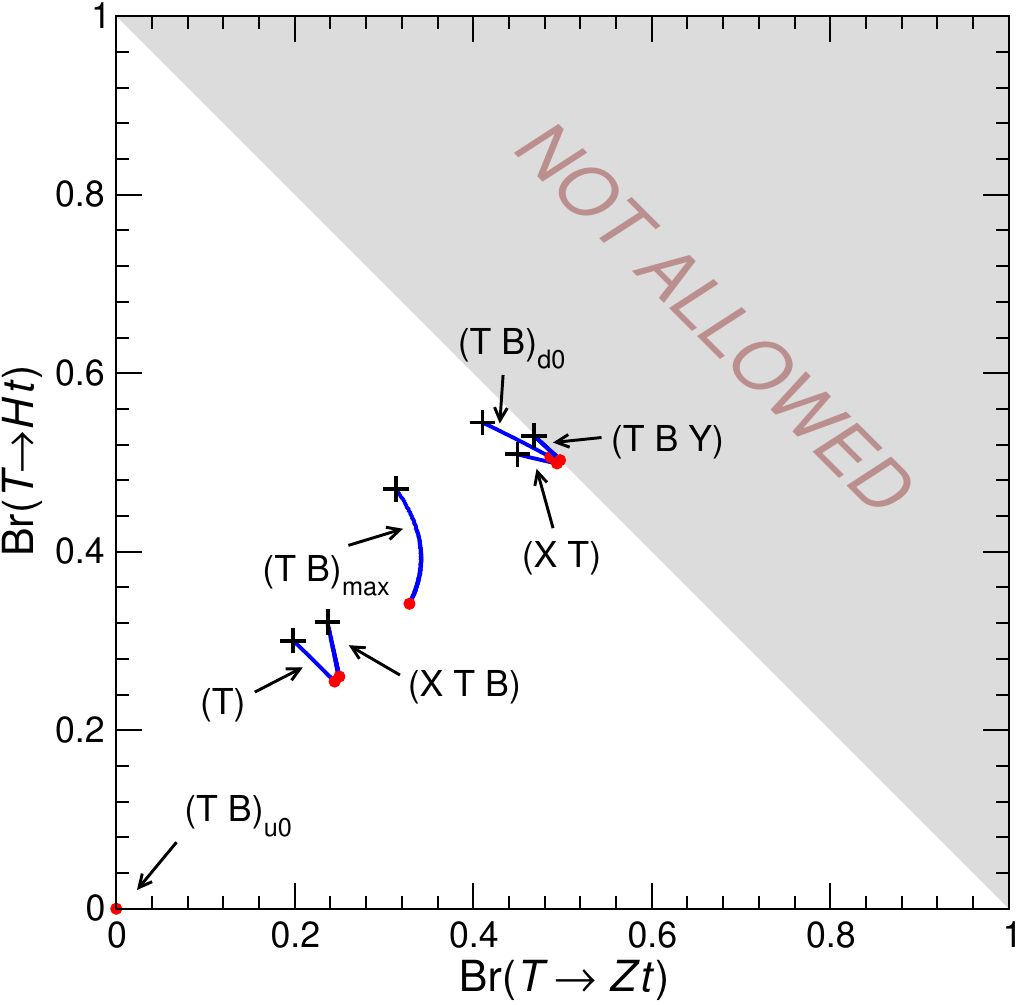}
\includegraphics[scale=0.42]{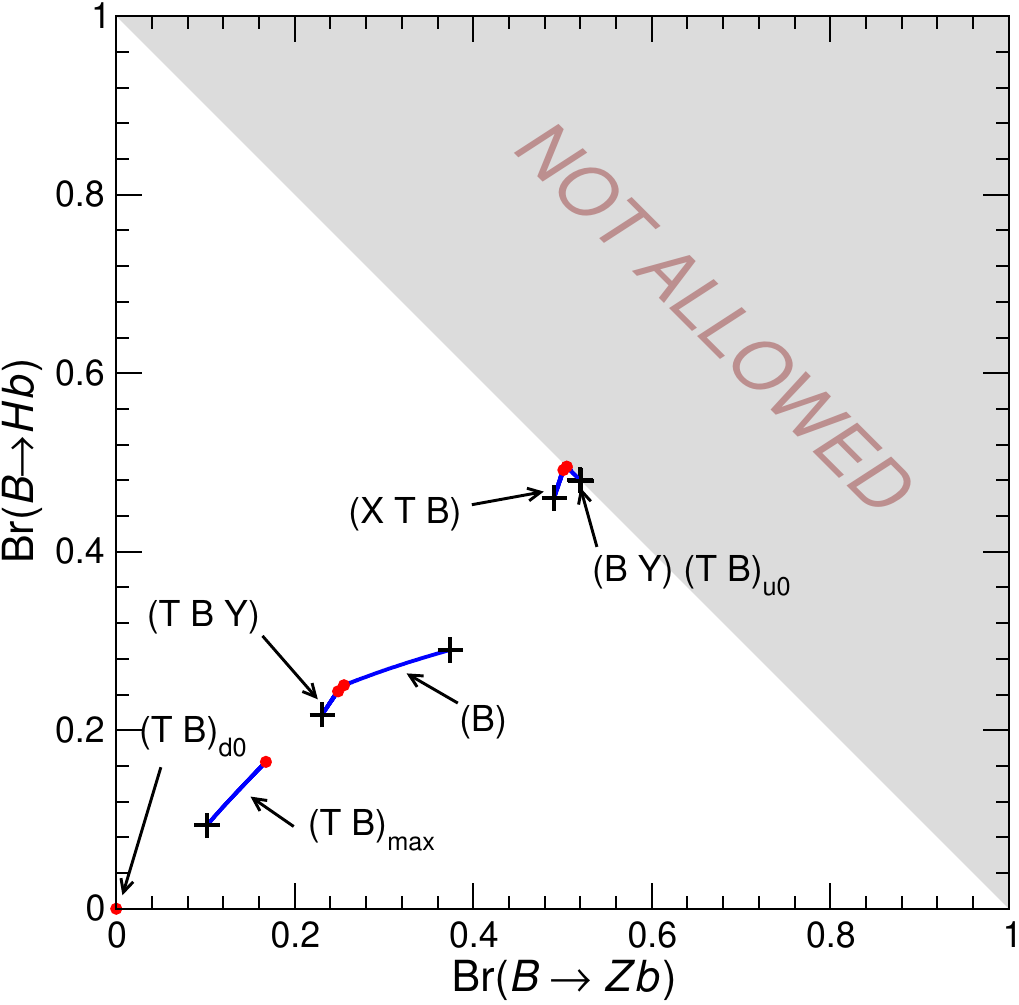}
\caption[Allowed branching ratios for singlet VLQs $T$ (left) and $B$ (right) assuming they only couple with the third family quarks. VLQs in non-trivial $SU(2)_L$ multiplets are also shown. Red dots correspond to $M_{T,B}=2~\unit{TeV}$ and are 
representative of the asymptotic high-mass value.
The crosses indicate $M_T=640~\unit{GeV}$ or $M_B=590~\unit{GeV}$.
The blue lines refer to intermediate masses.]{\label{fig:branching-ratio}
Allowed branching ratios for singlet VLQs $T$ (left) and $B$ (right) assuming they only couple with the third family quarks. VLQs in non-trivial $SU(2)_L$ multiplets are also shown. Red dots correspond to $M_{T,B}=2~\unit{TeV}$ and are 
representative of the asymptotic high-mass value.
The crosses indicate $M_T=640~\unit{GeV}$ or $M_B=590~\unit{GeV}$.
The blue lines refer to intermediate masses.
Taken from ref.~\cite{Aguilar-Saavedra:2013qpa}.
}
\end{figure}

Figure \ref{fig:branching-ratio} depicts the potential BRs of VLQs, based on their gauge group representation, with the constraint that they exclusively decay into third-generation quarks. While these BRs are contingent on the masses of VLQs, it is evident that $SU(2)$ singlet VLQs, limited to decaying into third-generation quarks, exhibit a distinct range of permissible BRs. These values closely align with the specific points where
\begin{equation}
\begin{aligned} \label{eq:SU2BRs}
     \textrm{BR}(T \rightarrow W b) &= 0.5\,,\,\, \textrm{BR}(T \rightarrow Zt)= 0.25\,,\,\, \textrm{BR}(T \rightarrow ht) = 0.25\,, \text{ and}\\[2mm]
     \textrm{BR}(B \rightarrow W t) &= 0.5\,,\,\, \textrm{BR}(B \rightarrow Zb)= 0.25\,,\,\, \textrm{BR}(B \rightarrow hb) = 0.25 \,.
\end{aligned}
\end{equation}
To maintain consistency, the term "$SU(2)$ singlet (VLQ) BRs" will be employed to refer to the BRs in eq. \eqref{eq:SU2BRs}, as it is commonly used in the majority of literature. Nevertheless, it is important to note that these BRs specifically pertain to the allowed branching ratios of $SU(2)$ singlet VLQs under the condition that they exclusively decay into third-generation quarks. Under that same strong condition but allowing for deviations from the values of eq. ~\eqref{eq:SU2BRs}, one has instead the weaker constraints
\begin{equation}
\begin{aligned}\label{eq:3rdgen}
    \textrm{BR}(T \rightarrow W b) + \textrm{BR}(T \rightarrow Zt) + \textrm{BR}(T \rightarrow ht) &= 1\,, \text{ and}\\[2mm]
    \textrm{BR}(B \rightarrow W t) +\textrm{BR}(B \rightarrow Zb) + \textrm{BR}(B \rightarrow hb) &= 1 \,.
\end{aligned}
\end{equation}
It is crucial to bear in mind that these assumptions may be violated in more complex scenarios, where other significant decay channels can arise. Examples of such extended scenarios can be found in references such as~\cite{Aguilar-Saavedra:2017giu,Chala:2017xgc,Kim:2018mks,Alhazmi:2018whk,Kim:2019oyh,Criado:2019mvu,Wang:2020ips,Banerjee:2022xmu,Bhardwaj:2022wfz,Verma:2022nyd,Bardhan:2022sif}. This motivates searches for VLQs which decay predominantly to light quarks $u,d,s$~\cite{ATLAS2015lpr,CMS2017asf}. Because of this, in section \ref{sec:direct_searches_quarks}, where the results of the most recent searches at ATLAS and CMS will be discussed, only mass bounds that stem from a search whose assumptions roughly correspond to an $SU(2)$ singlet VLQ with realistic BRs will be covered. \par 

Future VLQ search prospects are discussed in~section \ref{sec:future}. \par

\subsection{Direct Searches}
\label{sec:direct_searches_quarks}

Pair production and single production are the two primary mechanisms by which heavy VLQs can be generated. The specific production mechanism and subsequent decay modes of VLQs at high energies are distinctive characteristics of different VLQ models. In most experimental analyses, it is assumed that VLQs predominantly decay into or exclusively decay to third-generation quarks. Under this assumption, previous studies have extensively examined pair production \cite{Aguilar-Saavedra:2009xmz} and single production \cite{Aguilar-Saavedra:2013qpa} of VLQs. These studies have emphasized the relationship between decay widths, electroweak observables, and the production modes of VLQs. Various types of VLQs, including singlets, doublets, and triplets of $SU(2)_L$, have been considered in these investigations.
The decay widths of isosinglet VLQs, the focus of this thesis, via the $Z$, $h$, and $W$ bosons can be expressed at the tree-level as follows:
\begin{align}\label{eq:hd-fullwidths}
    \Gamma\big(Q_i\rightarrow Z\psi^q_j\big)&=  \frac{G_F M^3_i}{16\pi\sqrt{2}}\lambda[r_Z,r_j] \Big[\big(1-r^2_j\big)^2+r^2_Z\big(1+r^2_j\big)-2r^4_Z\Big]\big|F^q_{ji}\big|^2\,,\\
    \Gamma\big(Q_i\rightarrow h\psi^q_j\big)&=  \frac{G_F M^3_i}{16\pi\sqrt{2}}\lambda[r_h,r_j] \Big[\big(1+r^2_j\big)^2+4r^2_j-r^2_h\big(1+r^2_j\big)\Big]\big|F^q_{ji}\big|^2\,,\\
    \Gamma\big(U_i\rightarrow W^+\psi^d_j\big)&= \frac{G_F M^3_i}{8\pi\sqrt{2}}\lambda[r_W,r_j] \Big[\big(1-r^2_j\big)^2+r^2_W\big(1+r^2_j\big)-2r^4_W\Big]\big|V_{ij}\big|^2\,,\\
    \Gamma\big(D_i\rightarrow W^-\psi^u_j\big)&= \frac{G_F M^3_i}{8\pi\sqrt{2}}\lambda[r_W,r_j] \Big[\big(1-r^2_j\big)^2+r^2_W\big(1+r^2_j\big)-2r^4_W\Big]\big|V_{ji}\big|^2\,,
\end{align}
where $\psi^q$ ($q=u,d$) contains the SM and heavy quarks, with $\psi^u = (u,c,t,U_1,\ldots,U_{n_u})^T$ and $\psi^d = (d,s,b,D_1,\ldots,D_{n_d})^T$. In the ratio $r_j=m_j/M_i$, the $m_j$ is the mass of the final-state quark and $M_i$ the mass of the VLQ --- the decaying particle. The function $\lambda[r,r']$ is given by
\begin{equation}\label{eq:lambda-function}
    \lambda^2\big[a,b\big]=1-2\big(a^2+b^2\big)+\big(a^2-b^2\big)^2\,,
\end{equation}
and for the Fermi constant $G_F$ one has, at tree level, $G_F=\frac{\sqrt{2}g^2}{8m_W^2}$. 
The quark mixing matrix $V$ was defined in equations \eqref{eq:mixingmatrixquark} and \eqref{eq:uparam-exactV}, while the matrices $F^u = V V^\dagger$ and $F^d = V^\dagger V$, controlling the FCNC in the up sector and down sector, respectively, were defined in eq~\eqref{eq:Fud}.\par
For sufficiently heavy quarks decaying into SM quarks, with all $r \ll 1$, the expressions above reduce to
\begin{align}\label{eq:hd-widths}
    \Gamma\big(Q_i\rightarrow Z q_j\big)\simeq \frac{G_F M^3_i}{16\pi\sqrt{2}}\big|F^q_{ji}\big|^2&\,,\quad            \Gamma\big(U_i\rightarrow W^+ d_j\big)\simeq \frac{G_F M^3_i}{8\pi\sqrt{2}}\big|V_{ij}\big|^2\,,\\
    \Gamma\big(Q_i\rightarrow h q_j\big)\simeq \frac{G_FM^3_i}{16\pi\sqrt{2}}\big|F^q_{ji}\big|^2&\,,\quad            \Gamma\big(D_i\rightarrow W^-u_j\big) \simeq \frac{G_FM^3_i}{8\pi\sqrt{2}}\big|V_{ji}\big|^2\,,
\end{align}
which can be used to derive the following useful relations,
\begin{align}\label{eq:BR-relation:U}
    \frac{\Gamma(U_i\rightarrow Zu_j)}{|F^u_{ji}|^2}&\simeq          \frac{\Gamma(U_i\rightarrow h  u_j)}{|F^u_{ji}|^2}\simeq          \frac{\Gamma(U_i\rightarrow W^+ d_j)}{2|V_{ij}|^2}\,,\\
\label{eq:BR-relation:D}
    \frac{\Gamma(D_i\rightarrow Z d_j)}{|F^d_{ji}|^2}&\simeq          \frac{\Gamma(D_i\rightarrow h d_j)}{|F^d_{ji}|^2}\simeq
    \frac{\Gamma(D_i\rightarrow W^- u_j)}{2|V_{ji}|^2}\,.
\end{align}

The limiting values in eq. \eqref{eq:SU2BRs} can be seen in the relations of eq.~\eqref{eq:BR-relation:U} and eq. \eqref{eq:BR-relation:D}, taking $|F_{ji}|\simeq |V_{ji}|$ or $|F_{ji}|\simeq |V_{ij}|$, valid when the index $i$ refers to the heavy quark with high mass and $j$ refers to the third-generation quark, neglecting CKM mixing, cf.~\eqref{eq:uparam-approxV} and eq. \eqref{eq:duFCNC-approx}. It is important to note the relation between the decay and single production of VLQs. In a single production event, the vertex involving the heavy quark is of the same form as that of one of its decays. Evidently, single production also depends on the flavour couplings between VLQ, SM quarks and $Z$, $W$ and $h$ bosons. These couplings consist of the elements of the matrices $F^q$ and $V$, previously defined. \par

\subsubsection{Searches for Pair-Produced VLQs that Decay Predominantly to $t,b$ Quarks}

\begin{figure}[H]
\centering
\includegraphics[scale=0.5]{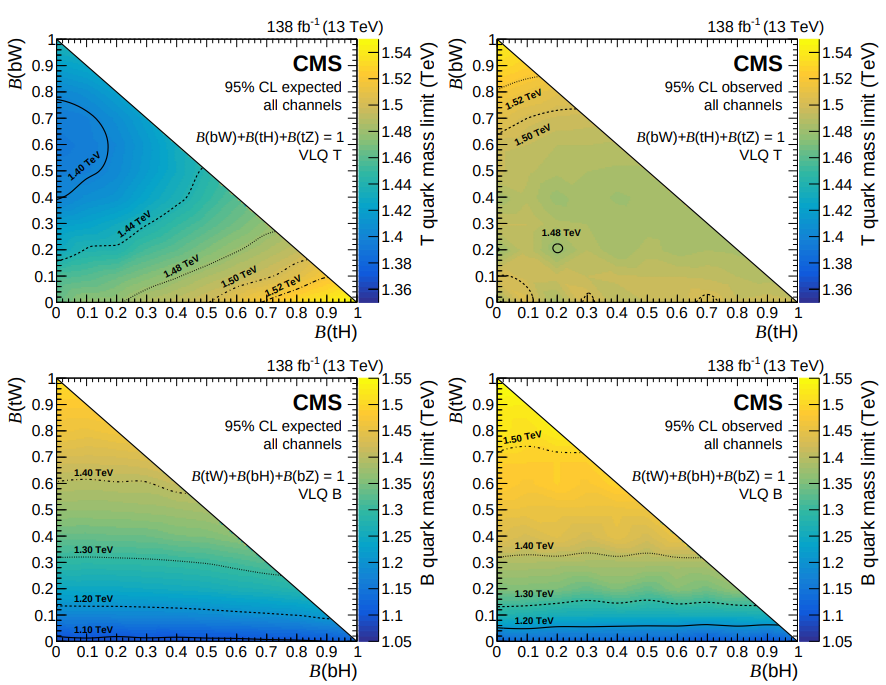}
\caption[The $95\%$ CL expected (left) and observed (right) lower mass limits on pair-produced
$T$ (upper) and $B$ (lower) quark masses, from the combined fit to all channels, as functions of
their branching ratios to H and W bosons. Mass contours are shown with lines of various
styles.]{\label{fig:CMS2022fck}
The $95\%$ CL expected (left) and observed (right) lower mass limits on pair-produced
$T$ (upper) and $B$ (lower) quark masses, from the combined fit to all channels, as functions of
their branching ratios to H and W bosons. Mass contours are shown with lines of various
styles.
Taken from ref.~\cite{CMS2022fck}.
}
\end{figure}

The most recent results of searches for pair-produced singlet VLQs follows. The results of the most recent CMS search \cite{CMS2022fck} are displayed in Fig. \ref{fig:CMS2022fck} resulting in the lower bounds for ``$SU(2)$ singlet BRs''~\cite{CMS2022fck}
\begin{equation*}
m_T > 1480~\unit{GeV}\,,~ m_B >1470~\unit{GeV}\quad {(95\%\text{ C.L.})}\,.
\end{equation*}

\begin{figure}[H]
\centering
\includegraphics[scale=0.4]{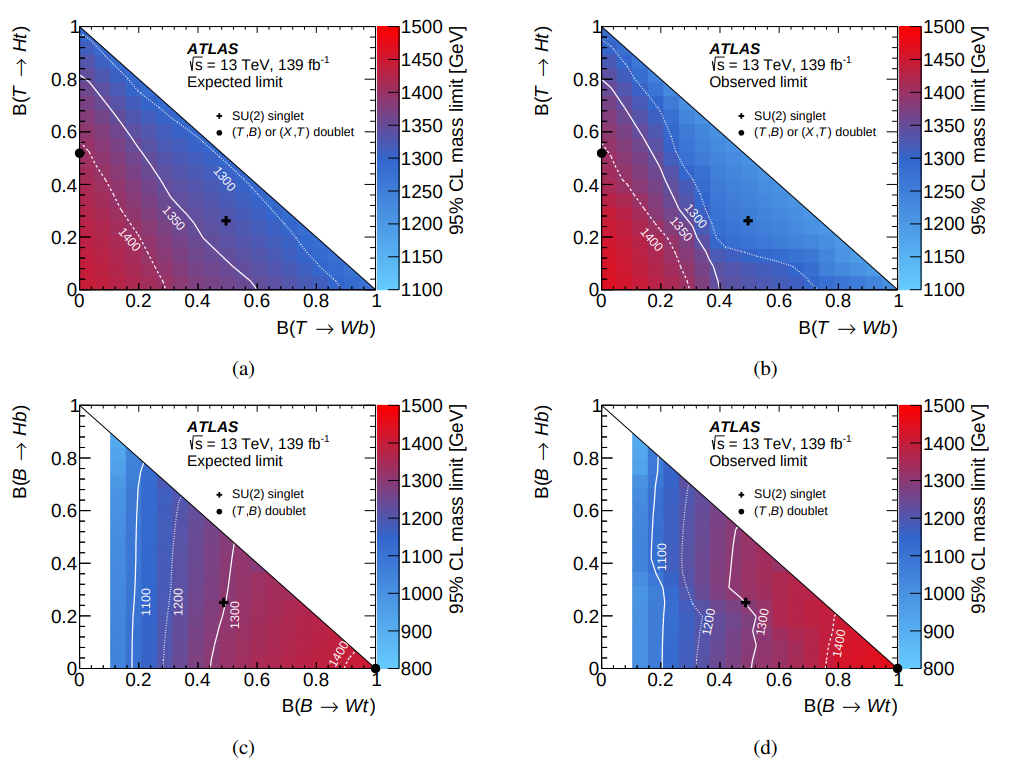}
\caption[Expected (left) and observed (right) mass limits for $T\overline{T}$ (upper row) and $B\overline{B}$ (lower row) production. The mass limit is calculated using the NN giving the most stringent expected limit at each signal mass and branching ratio point. The white lines indicate mass exclusion contours. The black markers indicate the branching ratios for the SU(2) singlet and doublet scenarios for masses above 800 GeV, where they are approximately independent of the VLQ mass.]{\label{fig:ATLAS2022tla}
Expected (left) and observed (right) mass limits for $T\overline{T}$ (upper row) and $B\overline{B}$ (lower row) production. The mass limit is calculated using the NN giving the most stringent expected limit at each signal mass and branching ratio point. The white lines indicate mass exclusion contours. The black markers indicate the branching ratios for the SU(2) singlet and doublet scenarios for masses above 800 GeV, where they are approximately independent of the VLQ mass.
Taken from ref.~\cite{ATLAS2022tla}.
}
\end{figure}

While the results of the most recent ATLAS search for pair-produced VLQs \cite{ATLAS2022tla} are displayed in Fig. \ref{fig:ATLAS2022tla}. It reports the following lower bounds for ``$SU(2)$ singlet BRs''~\cite{ATLAS2022tla}
\begin{equation*}
m_T > 1260~\unit{GeV}\,,~ m_B >1330~\unit{GeV}\quad {(95\%\text{ C.L.})}\,.
\end{equation*}

\subsubsection{Searches for Pair-Produced VLQs that Decay Predominantly to $u,d,s$ Quarks}
\begin{figure}[H]
\centering
\includegraphics[scale=0.3]{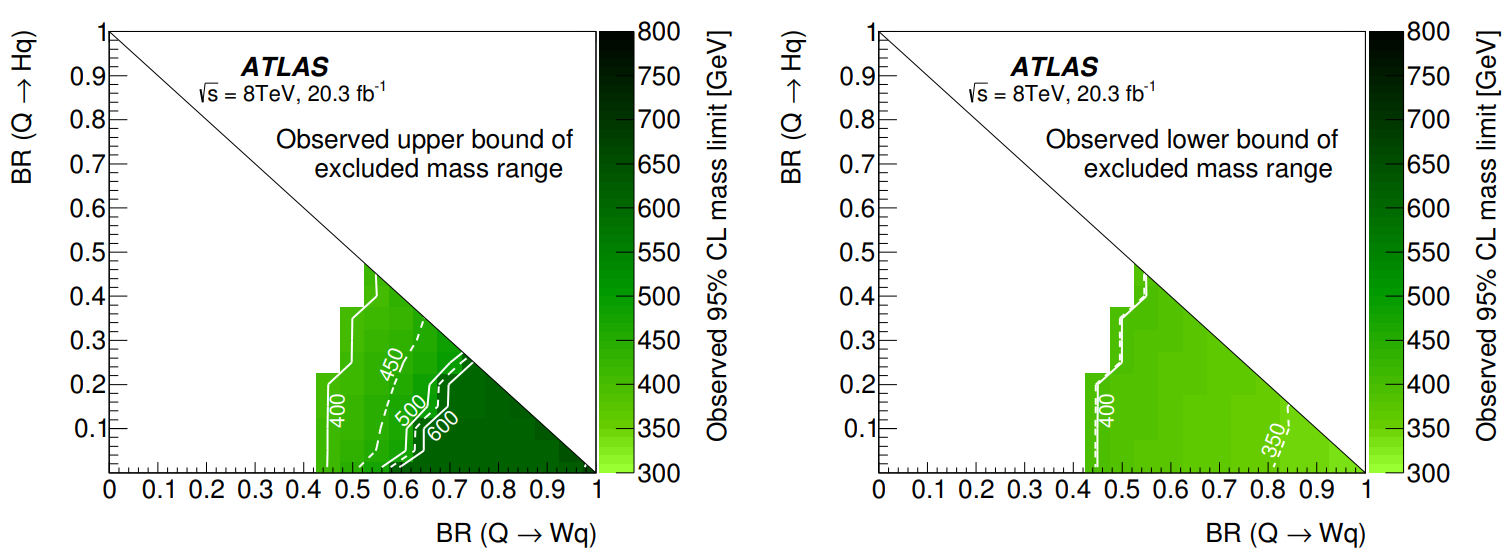}
\caption[The (left) upper and (right) lower bounds on the range of heavy quark masses observed to be excluded
at $95\%$ CL, as a function of the branching ratio of the heavy quark to $Wq$ versus $hq$, with the branching ratio to
$Zq$ fixed by the requirement $\textrm{BR}(Q\rightarrow Zq)  = 1 - \textrm{BR}(Q \rightarrow Wq) - \textrm{BR}(Q \rightarrow hq)$. The region above the diagonal is
forbidden by unitarity.]{\label{fig:ATLAS2015lpr}
The (left) upper and (right) lower bounds on the range of heavy quark masses observed to be excluded
at $95\%$ CL, as a function of the branching ratio of the heavy quark to $Wq$ versus $hq$, with the branching ratio to
$Zq$ fixed by the requirement $\textrm{BR}(Q\rightarrow Zq)  = 1 - \textrm{BR}(Q \rightarrow Wq) - \textrm{BR}(Q \rightarrow hq)$. The region above the diagonal is
forbidden by unitarity. Taken from ref.~\cite{ATLAS2015lpr}.
}
\end{figure}

The latest results of searches for pair-produced VLQs ($Q=T,B$) that mainly decay into light quarks $u, d, s$ were obtained between 2015 and 2017, utilizing data from LHC Run 1 \cite{ATLAS2015lpr,CMS2017asf}. It is of utmost significance to reevaluate these searches with updated datasets. The results are given as a function of the branching ratios $\textrm{BR}(Q \rightarrow Wq)$ versus $\textrm{BR}(Q \rightarrow hq)$, with the branching ratio to $Zq$ fixed by the requirement $\textrm{BR}(Q\rightarrow Zq)= 1 - \textrm{BR}(Q \rightarrow Wq) - \textrm{BR}(Q \rightarrow hq)$. The exclusion region reported by ref. \cite{ATLAS2015lpr}, depending on the different branching ratios, is depicted in Fig. \ref{fig:ATLAS2015lpr}.

As an illustration, let's consider different sets of branching ratios for the VLQs ($Q=T,B$), and examine the corresponding exclusion ranges from the results obtained in the study \cite{ATLAS2015lpr}:
\begin{itemize}
    \item For the branching ratios $\textrm{BR}(Q \rightarrow Wq) = 0.6$, $\textrm{BR}(Q\rightarrow Zq) = 0.4$, and $\textrm{BR}(Q \rightarrow hq) = 0$, the results exclude VLQs with a mass ranging from $370$ GeV to $610$ GeV.
    \item For $\textrm{BR}(Q \rightarrow Wq) = 0.6$, $\textrm{BR}(Q\rightarrow Zq) = 0.3$, and $\textrm{BR}(Q \rightarrow hq) = 0.1$, the exclusion range is from $370$ GeV to $470$ GeV.
    \item If $\textrm{BR}(Q \rightarrow Wq) = 0.5$, $\textrm{BR}(Q\rightarrow Zq) = 0.25$, and $\textrm{BR}(Q \rightarrow hq) = 0.25$, the exclusion range extends from $390$ GeV to $410$ GeV.
\end{itemize}

\begin{figure}[H]
\centering
\includegraphics[scale=0.3]{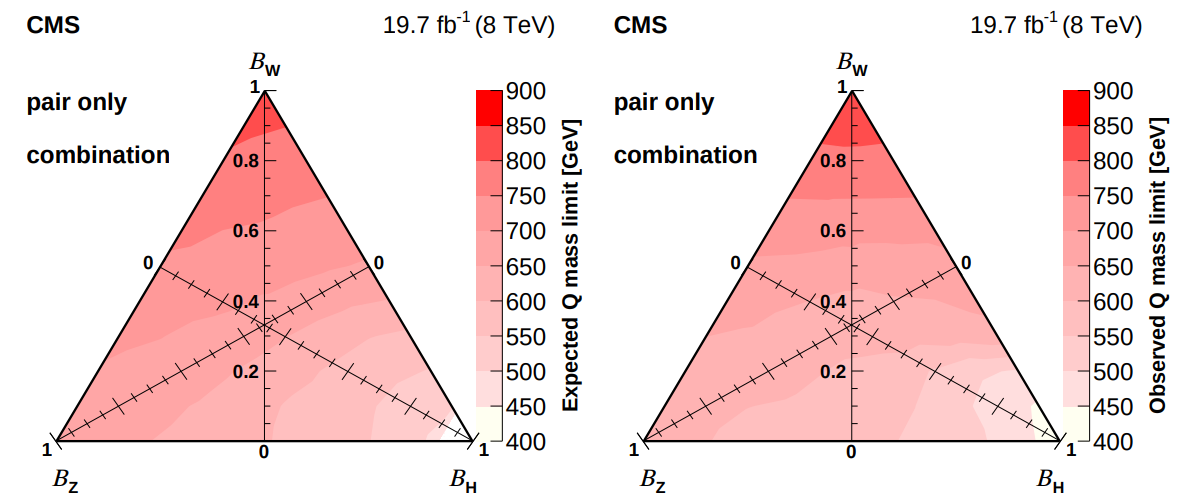}
\caption[ The median expected (left) and observed (right) combined lower mass limits represented in a triangular form, where each point of the triangle corresponds to a given set of
branching fractions for a VLQ decaying into a boson and a first-generation quark. The limit
contours are determined assuming that $\Tilde{k}_W$ and $\Tilde{k}_Z$ are so small that the single-production
modes can be neglected, and therefore that the heavy quarks can only be produced in pairs
via strong interaction. The white area in the triangle with expected limits indicates mass limits
below 400 GeV.]{\label{fig:CMS2017asf}
: The median expected (left) and observed (right) combined lower mass limits represented in a triangular form, where each point of the triangle corresponds to a given set of
branching fractions for a VLQ decaying into a boson and a first-generation quark. The limit
contours are determined assuming that $\Tilde{k}_W$ and $\Tilde{k}_Z$ are so small that the single-production
modes can be neglected, and therefore that the heavy quarks can only be produced in pairs
via strong interaction. The white area in the triangle with expected limits indicates mass limits
below 400 GeV.
Taken from ref.~\cite{CMS2017asf}.
}
\end{figure}

As for ref.~\cite{CMS2017asf}, its results for pair production can be found in Fig. \ref{fig:CMS2017asf}. The authors define a quantity
\begin{equation}
    \Tilde{k} = \frac{\sqrt{2} M_Q V_{qQ}}{v} ~,
    \label{eq:tildek}
\end{equation}
where $v$ is the Higgs field expectation value, $M_Q$ is the mass of the VLQ and $V_{qQ}$ is the entry of the quark mixing matrix $V$, defined in equations \eqref{eq:mixingmatrixquark} and \eqref{eq:uparam-exactV}, that connects the light quarks $q=u,d,s$ and the heavy VLQ $Q$. $\Tilde{k}$ is a measure, that takes phase space into account, of how important is single-production comparing to pair-production. Hence, $\Tilde{k}_W=1$ means that electroweak processes dominate and VLQs are predominantly produced via single-production. For very small values of $\Tilde{k}$, pair-production processes, via QCD, dominate.

Again, as an example, exclusion ranges for some benchmark branching ratios based on the findings from the study \cite{CMS2017asf} are listed next.
\begin{itemize}
    \item For the branching ratios $\textrm{BR}(Q \rightarrow Wq) = 0.6$, $\textrm{BR}(Q\rightarrow Zq) = 0.4$, and $\textrm{BR}(Q \rightarrow hq) = 0$, the results exclude VLQs with a mass smaller than $725$ GeV.
    \item For $\textrm{BR}(Q \rightarrow Wq) = 0.6$, $\textrm{BR}(Q\rightarrow Zq) = 0.2$, and $\textrm{BR}(Q \rightarrow hq) = 0.2$, the exclusion range is for masses below $715$ GeV.
    \item If $\textrm{BR}(Q \rightarrow Wq) = 0.5$, $\textrm{BR}(Q\rightarrow Zq) = 0.25$, and $\textrm{BR}(Q \rightarrow hq) = 0.25$, masses below $685$ GeV are excluded.
\end{itemize}

\subsubsection{Searches for Singly-Produced VLQs that Decay Predominantly to $u,d,s$ Quarks}
\begin{figure}[H]
\centering
\includegraphics[scale=0.3]{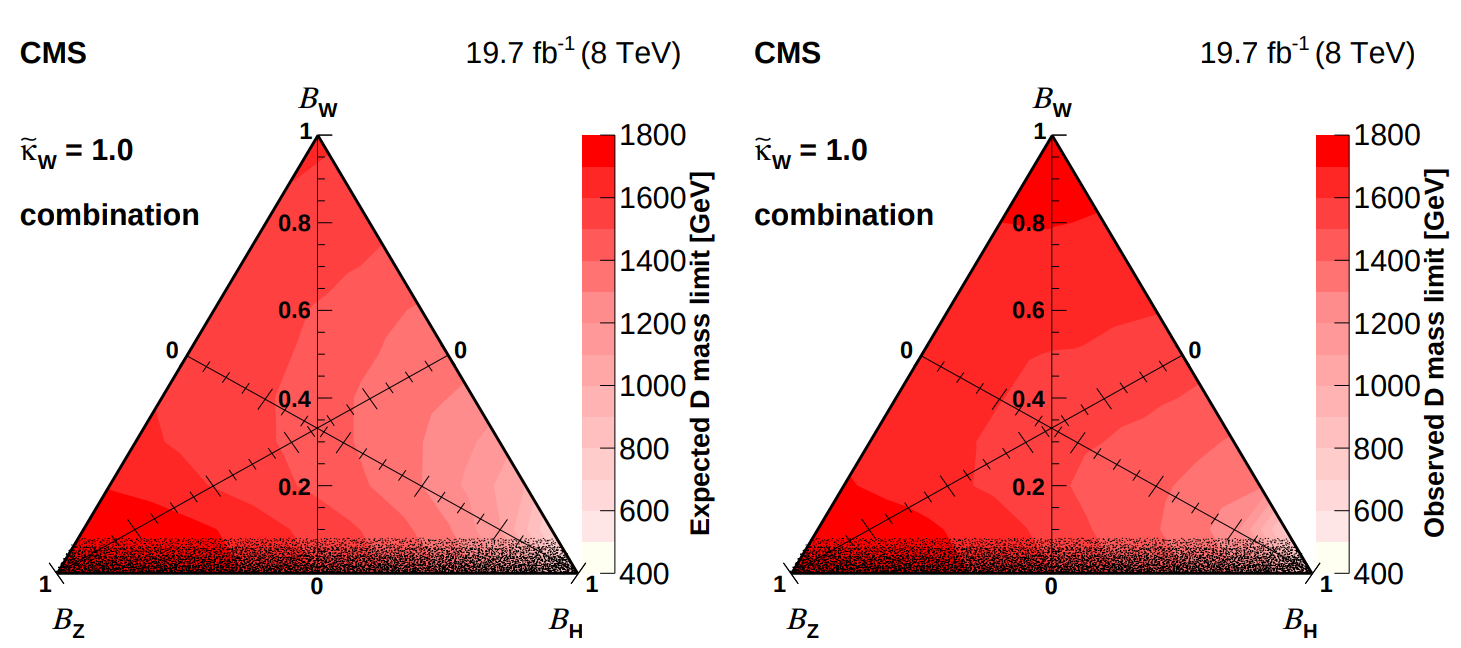}
\caption[ The median expected (left) and observed (right) combined lower mass limits represented in a triangular form, where each point of the triangle corresponds to a given set of
branching fractions for the decay of a VLQ into a boson and a first-generation quark. The limit
contours are determined assuming that $\Tilde{k}_W = 1.0$, which means that the signal is dominated by
electroweak single production. ]{\label{fig:CMS2017asf2}
The median expected (left) and observed (right) combined lower mass limits represented in a triangular form, where each point of the triangle corresponds to a given set of
branching fractions for the decay of a VLQ into a boson and a first-generation quark. The limit
contours are determined assuming that $\Tilde{k}_W = 1.0$, which means that the signal is dominated by
electroweak single production.
Taken from ref.~\cite{CMS2017asf}.
}
\end{figure}

In scenarios where significant couplings to light quark generations are present, single production processes can dominate pair production by several orders of magnitude within the few TeV energy range. This is primarily due to a larger available phase space, an enhancement from longitudinal gauge bosons, and the involvement of the proton's valence quarks ($u$ and $d$) in the initial state \cite{Atre:2008iu, Atre:2011ae}. The study by the CMS collaboration \cite{CMS2017asf}
is the only search for singly-produced VLQs that decay to predominantly or exclusively to light quarks. Its results for when the signal is dominated by
electroweak single production, i.e. for $\Tilde{k}=1$ (defined in eq. \eqref{eq:tildek}), are summarised in Fig. \ref{fig:CMS2017asf2}.

Exclusion ranges for some benchmark branching ratios based on the findings from the study \cite{CMS2017asf} are listed next. It is important to emphasize that this is done for the case where VLQs predominantly or exclusively to light quarks and the production signal is dominated by electroweak single production,  $\Tilde{k}_W = 1.0$.
\begin{itemize}
    \item For the branching ratios $\textrm{BR}(Q \rightarrow Wq) = 0.6$, $\textrm{BR}(Q\rightarrow Zq) = 0.4$, and $\textrm{BR}(Q \rightarrow hq) = 0$, the results exclude VLQs with a mass smaller than $1700$ GeV.
    \item For $\textrm{BR}(Q \rightarrow Wq) = 0.6$, $\textrm{BR}(Q\rightarrow Zq) = 0.2$, and $\textrm{BR}(Q \rightarrow hq) = 0.2$, the exclusion range is for masses below $1645$ GeV.
    \item If $\textrm{BR}(Q \rightarrow Wq) = 0.4$, $\textrm{BR}(Q\rightarrow Zq) = 0.4$, and $\textrm{BR}(Q \rightarrow hq) = 0.2$, masses below $1605$ GeV are excluded.
\end{itemize}

\subsubsection{Searches for Singly-Produced VLQs that Decay Predominantly to $t,b$ Quarks}

\begin{figure}[H]
\centering
\includegraphics[scale=0.3]{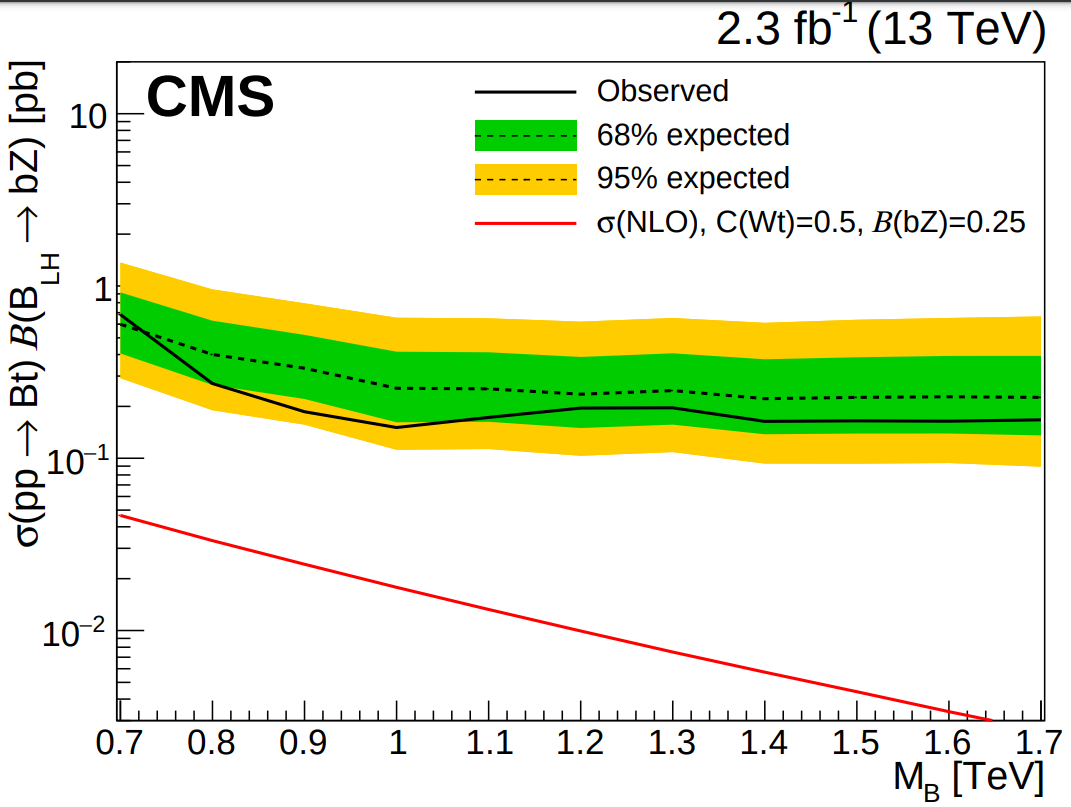}
\caption[ Observed and expected $95\%$ CL upper limit on the product of cross section and
branching fraction for $t W \rightarrow B$ production mode in the singlet LH scenario, with
the $B$ decaying to $bZ$ with "SU(2) singlet BRs". The $68\%$ and $95\%$ expected bands are shown. The branching fraction $\textrm{BR}(Q\rightarrow Zq)$ is $0.25$. ]{\label{fig:CMS2017gsh}
Observed and expected $95\%$ CL upper limit on the product of cross section and
branching fraction for $t W \rightarrow B$ production mode in the singlet LH scenario, with
the $B$ decaying to $bZ$ with "SU(2) singlet BRs". The $68\%$ and $95\%$ expected bands are shown. Theoretical cross sections
as calculated at NLO in Ref. \cite{Matsedonskyi:2014mna} are shown. The branching fraction $\textrm{BR}(Q\rightarrow Zq)$ is $0.25$. 
Taken from ref.~\cite{CMS2017gsh}.
}
\end{figure}

The most recent search for singly-produced down VLQs \cite{CMS2017gsh}, looked for a $B$ VLQ that was produced via $t W \rightarrow B$ and subsequently decays via $B \rightarrow Z b$. The excluded region at 95\% C.L.~is
\begin{equation*}
    m_{B} \in [700,1700]~\unit{GeV}\, ~\text{for}~\sigma(t W \rightarrow B) \in [0.13,1.26]~\unit{pb}\ ~.
\end{equation*}
The excluded region depends on the value of the cross section of the production process, which is model dependent - depends on $V_{tB}$ and $m_B$, for example. The region is significantly distant from the area a $B$ VLQ in a model with $V_{tB}=0.5$ and "SU(2) Singlet BRs" would occupy \cite{CMS2017gsh}. The red line in Fig. \ref{fig:CMS2017gsh} illustrates this theoretical region, with the production cross section via $Wt$, $\sigma(t W \rightarrow B)$, as calculated in ref. \cite{Matsedonskyi:2014mna}.

The most recent search for singly-produced $T$ VLQs\cite{ATLAS2023pja} was performed by the ATLAS collaboration and looks for an up VLQ that is produced via $Wb \rightarrow T$ or $Zt \rightarrow Z T$ and can decay via $T \rightarrow h t$ or via $T \rightarrow Z t$. The authors of ref. \cite{Buchkremer:2013bha} define the universal coupling strength $\kappa$, which for SU(2) singlet VLQs is given by
\begin{equation}
    \kappa \approx \sqrt{2}~V_{Tb} ~,
    \label{eq:kappa}
\end{equation}
where $V_{Tb}$ is the entry of the quark mixing matrix $V$, defined in equations \eqref{eq:mixingmatrixquark} and \eqref{eq:uparam-exactV}, that connects the bottom quark $b$ and the heavy VLQ $T$.

\begin{figure}[H]
\centering
\includegraphics[scale=0.4]{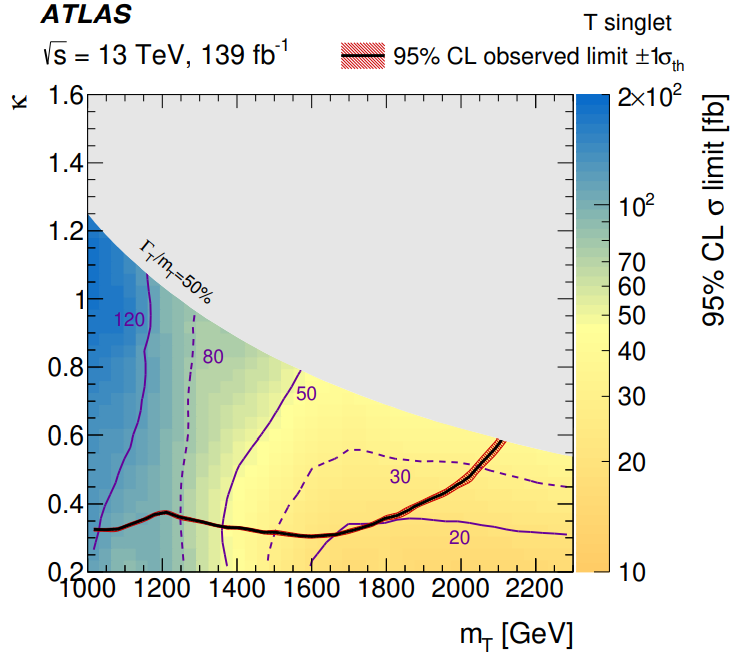}
\caption[ Observed $95\%$ CL exclusion limits on the cross section times branching ratio of
single $T$-quark production as a function of the universal coupling constant $\kappa$, defined in eq. \eqref{eq:kappa} for singlet $T$ VLQs, and the $T$-quark mass in the "SU(2)
singlet scenario". The red hashed area around the observed limit corresponds to the theoretical uncertainty of the
NLO cross-section prediction. All values of $\kappa$ above the black contour line are excluded at each mass point. The
purple contour lines denote exclusion limits of equal cross section times branching ratio in units of fb. ]{\label{fig:ATLAS2023pja}
 Observed $95\%$ CL exclusion limits on the cross section times branching ratio of
single $T$-quark production as a function of the universal coupling constant $\kappa$, defined in eq. \eqref{eq:kappa} for singlet $T$ VLQs, and the $T$-quark mass in the "SU(2)
singlet scenario". The red hashed area around the observed limit corresponds to the theoretical uncertainty of the
NLO cross-section prediction. All values of $\kappa$ above the black contour line are excluded at each mass point. The
purple contour lines denote exclusion limits of equal cross section times branching ratio in units of fb. 
Taken from ref.~\cite{ATLAS2023pja}.
}
\end{figure}
The exclusion limit for singlet $T$ quarks, with SU(2) BRs is

\begin{equation*}
|V_{Tb}| \gtrsim 0.37\,,~\textrm{for}~m_T < 2300~\unit{GeV}\quad {(95\%\text{ C.L.})}\,.
\end{equation*}

One should note that the experimental limits from refs. \cite{CMS2017gsh,ATLAS2023pja} for these values of $V_{Tb}$ and $V_{tB}$ are still far off from reasonable theoretical values, since they would easily contradict several measurements regarding the unitarity of the CKM matrix. Hence, in the short to medium term, the easier ways to detect VLQs may be via pair-production searches or single production of VLQs that decay predominantly to light quarks, since in that case that production mode dominates over pair-production \cite{Alves:2023ufm}.

\subsection{Effects on Observables}
\label{sec:newphysics}

This section provides a comprehensive explanation of observables influenced by the presence of VLQs. The focus will be on electroweak precision quantities and low-energy effects, including neutral meson mixing and meson decays. Additionally, it will cover rare decays of the top quark, oblique parameters, and the $R_{b}$ ratio (associated with the decay of the $Z$ boson).\par
It is crucial to note that some of the quantities discussed here may rely on tree-level contributions, while others involve loop calculations. Each model has its unique set of contributions. Section ~\ref{sec:ph-fits} presents global fits carried out in the presence of one up-type VLQ, clearly highlighting the significance of some observables.\par
Next follows a compilation of the observables commonly used to constrain the parameter space of an extension of the SM with any number of up-type, $n_u$, or down-type, $n_d$, singlet VLQs. 
Throughout this section, the well-known Inami-Lim functions~\cite{Inami:1980fz,Buchalla:1995vs}
\begin{equation}\begin{split} \label{eq:InamiLim}
    S_0(x_i,x_j)&=\frac{x_ix_j}{4}\bigg[-\frac{3}{(1-x_i)(1-x_j)}\\
    &+\frac{4-8x_i+x^2_i}{(1-x_i)^2(x_i-x_j)}\log x_i +\frac{4-8x_j+x^2_j}{(1-x_j)^2(x_j-x_i)}\log x_j\bigg]\,,\\
    S_0(x)&=\frac{x}{4}\bigg[\frac{4-11x+x^2}{(1-x)^2}-\frac{6x^2}{(1-x)^3}\log x\bigg]\,,\\
    X_0(x)&=\frac{x}{8}\bigg[\frac{x+2}{x-1}+\frac{3x-6}{(1-x)^2}\log x\bigg]\,,\\
    Y_0(x)&=\frac{x}{8}\bigg[\frac{x-4}{x-1}+\frac{3x}{(1-x)^2}\log x\bigg]\,,\\
    Z_0(x)&=\frac{x(108-259x+163x^2-18x^3)}{144(1-x)^3}+ \frac{24x^4-6x^3-63x^2+50x-8}{72(1-x)^4}\log x\,,\\
    E_0(x)&=\frac{x(18-11x-x^2)}{12(1-x)^3}-\frac{4-16x+9x^2}{6(1-x)^4}\log x\,,\\
	\end{split}\end{equation}
will be used together with the conventional definitions (save the exception in section ~\ref{sec:D0D0bar})
\begin{equation}
    \lambda^\alpha_{ij}\equiv V^*_{\alpha i}V_{\alpha j}\,,\quad x_i\equiv m^2_i/m^2_W\,,
\end{equation}
and the loop functions
\begin{equation}\label{eq:functionN}
    N\big(x_i,x_j\big)=\frac{x_ix_j}{8}\bigg(\frac{\log x_i-\log x_j}{x_i-x_j}\bigg)\,.
\end{equation}
For $x_i=x_j$, eq. \eqref{eq:functionN} has the following limit,
\begin{equation}
    N\big(x_i\big)\equiv N\big(x_i,x_i\big)=\frac{x_i}{8}\,.
\end{equation}

\subsubsection{Rare Top Decays \texorpdfstring{$t \to q~Z/h$}{t -> q Z/h}}
\label{sec:raretop}
%

In the SM, the decays $t\rightarrow q~Z/h$, where $q=u,c$, are strongly suppressed as they occur only at one-loop level due to a variant of the GIM mechanism~\cite{Eilam:1990zc}.
However, in models with VLQs, both of these suppression factors are bypassed thanks to their tree-level FCNC, which, at the leading order, give rise to the following decay widths
\begin{equation}\label{eq:rtd-width}\begin{split}
    \Gamma\big(t\rightarrow Zq\big)&=\frac{G_Fm^3_t}{16\pi\sqrt{2}}\lambda\big[r_Z,r_q\big] \Big[\big(1-r^2_q\big)^2+r^2_Z\big(1+r^2_q\big)-2r^4_Z\Big]\big|F^u_{qt}\big|^2\,,\\
    \Gamma\big(t\rightarrow hq\big)&=\frac{G_Fm^3_t}{16\pi\sqrt{2}}\lambda\big[r_h,r_q\big] \Big[\big(1+r^2_q\big)^2+4r^2_q-r^2_h\big(1+r^2_q\big)\Big]\big|F^u_{qt}\big|^2\,,
\end{split}\end{equation}
where $r_i=m_i/m_t$ and the function $\lambda$ was defined in eq. \eqref{eq:lambda-function}.
Since $\textrm{BR}(t\rightarrow W^+b)\simeq1$ and $r_q\ll1$, one can employ eq. ~\eqref{eq:rtd-width} to obtain the approximations
\begin{equation}\label{eq:rtd-branchingratio}\begin{split}
    \textrm{BR}\big(t\rightarrow Zq\big)&\simeq\frac{1-3r^4_Z+2r^6_Z}{1-3r^4_W+2r^6_W} \frac{|F^u_{qt}|^2}{2|V_{tb}|^2}\,,\\
    \textrm{BR}\big(t\rightarrow hq\big)&\simeq\frac{(1-r^2_h)^2}{1-3r^4_W+2r^6_W} \frac{|F^u_{qt}|^2}{2|V_{tb}|^2}\,.
\end{split}\end{equation}

In the SM, the values for these branching ratios  are very suppressed
BR$(t \to u Z)_\text{SM} \sim 10^{-16}$ and 
BR$(t \to c Z)_\text{SM} \sim 10^{-14}$~\cite{AguilarSaavedra:2004wm}.
In the small angle approximation, one predicts $\text{Br}(t \to q_i Z) \simeq 0.46\, \theta_{i4}^2 \,\theta_{34}^2$, which for $\mathcal{O}(0.01)$ angles still exceeds the SM contribution by several orders of magnitude.
At present, the strongest bound on these branching ratios is set by the ATLAS collaboration, namely
BR$(t \to u Z)_\text{exp} < 1.7 \times 10^{-4}$ and 
BR$(t \to c Z)_\text{exp} < 2.4 \times 10^{-4}$ ($95\%$ CL)~\cite{Aaboud:2018nyl}.
As for models featuring singlet VLQs, the existing experimental data~\cite{ParticleDataGroup:2020ssz} imposes the following limitation on these expressions:
\begin{equation}\label{eq:rtd-constraint}
    |F^u_{qt}|<(3.337\pm0.099)\times10^{-2}\,.
\end{equation}
It is important to observe that the experimental limit on $|F^u_{qt}|$ will consistently be dominated by one of the decay channels mentioned in eq.~\eqref{eq:rtd-branchingratio}. Currently, the most stringent constraint arises from $t\rightarrow Zq$. For instance, it should be noted that in models incorporating VLQs to address the $W$-mass anomaly, a sizeable branching ratio for $t \rightarrow Zc$ can be produced, as demonstrated in ref. \cite{Crivellin:2022fdf}. \par
When only a single up-type VLQ is introduced, cf. section \ref{sec:quarkup}, this bound becomes
\begin{equation}\label{eq:rtd-constraintU}
    |\Theta_q\Theta^*_t|=\theta_{q4}\theta_{34}<(3.337\pm0.099)\times10^{-2}\,,
\end{equation}
where the notation of eq. \eqref{eq:SVLQ-Tup} was used. Note that these observables provide no constraints to models with only down-type VLQs, cf. section \ref{sec:quarkdown}, as a model with only down-type VLQs does not introduce FCNC in the up sector. \par

When VLQs are present, the flavour-changing radiative decays involving a photon or a gluon in the final state, namely $t \to \gamma q$ and $t \to g q$ ($q=u,c$), can experience significant enhancements, exceeding the predictions of the SM by orders of magnitude \cite{Aguilar-Saavedra:2002lwv, Balaji:2021lpr}. However, it is worth noting that these signals, in general, lie beyond the capabilities of current experiments to detect. \par

\subsubsection{\texorpdfstring{$D^0$-$\overline{D}^0$}{D0-D0bar} Mixing}
\label{sec:D0D0bar}
\begin{figure}[H]
  \centering
        \includegraphics[width=0.45\textwidth]{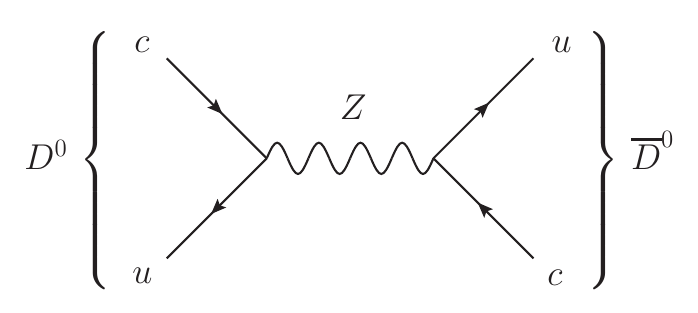}
        \caption{NP contribution to $D^0$-$\overline{D}^0$ mixing via $Z$-mediated FCNC, in the presence of one up-type VLQ.}
        \label{fig:D}
    \end{figure}

Within the $D^0$ system, the SM contribution to neutral meson mixing occurs through box diagrams involving internal down-type quarks and the GIM mechanism effectively renders this contribution negligible, as demonstrated in refs. \cite{Datta:1984jx, Donoghue:1985hh}. In contrast, in models incorporating VLQs, the parameter $M^D_{12}$ is essentially controlled by the New Physics (NP), which in the case $n_d=0$ occurs at tree level and in the case of $n_u=0$ occurs at loop level. Fig. \ref{fig:D} illustrates the leading diagram in the presence of one up-type VLQ.
Neglecting the Higgs-mediated FCNC, the effects of the VLQs are estimated through~\cite{Aguilar-Saavedra:2002phh}
\begin{equation}\label{eq:dm-m12}
    M^D_{12}=B_D\frac{G^2_Ff^2_Dm^2_Wm_D}{12\pi^2}\Delta_{D^0}\,,
\end{equation}
where $f_D=(212.0\pm0.7)~\unit{MeV}$ is the $D^0$ decay constant~\cite{FlavourLatticeAveragingGroupFLAG:2021npn} and its bag factor is estimated to be $B_D=1.0\pm0.3$~\cite{Aguilar-Saavedra:2002phh} and, solely in this subsection, $\lambda^q_{cu}=V_{cq}V^*_{uq}$.
Subsequently, as the quark masses are small, resulting in $S_0(x{d,s,b},x_{D_k})\ll1$, and the CKM matrix ensures that $|\lambda^q_{cu}|\ll1$, the leading-order approximation of the new effects $\Delta_{D^0}$ can be expressed as follows,
\begin{equation}\label{eq:delta-D}\begin{split}
    \Delta_{D^0}&=\sum^{n_d}_{k=1}S_0(x_{D_k})\big(\lambda^{D_k}_{cu}\big)^2 ~,~ n_u =0\,,\\
    \Delta_{D^0}&=\frac{4\pi s^2_W}{\alpha_\text{em}}\eta^D_Z\big(F^u_{cu}\big)^2 ~,~n_d =0\,,
\end{split}\end{equation}
where $\eta^D_Z=0.59$~\cite{Buchalla:1995vs} is the only known QCD correction to this process and the $n_u=0$ expression is for models with only down-type VLQs, and the $n_d=0$ for models with only up-type VLQs, see section \ref{sec:quarkmodels}. $D_k$ in the $n_u=0$ expression refers to all down-type VLQs $D_1,..,D_k$, since the leading diagrams are the ones where these are in the loop. \par
Even with NP, the  $D^0$--$\overline{D^0}$ mixing is, in principle, dominated by non-perturbative long-range effects~\cite{Wolfenstein:1985ft}. In general, models with VLQs must satisfy
\begin{equation}\label{eq:dm-constraintM}
    \big(\Delta m_D\big)_\text{NP}=2\Big|\big(M^D_{12}\big)_\text{NP} \Big|<\big(\Delta m_D\big)_\text{exp}\,.
\end{equation}
By requiring that $\big(\Delta m_D\big)_\text{NP} <\big(\Delta m_D\big)_\text{exp}$ one also controls the NP tree-level contribution to the yet unobserved $\Delta C = 1$ decay $D^0 \to \mu^+\mu^-$ \cite{Golowich:2009ii}.\par
From the current experimental limits~\cite{ParticleDataGroup:2020ssz}, one obtains the most updated bound for $\big|\Delta_{D^0}\big|$,
\begin{equation}\label{eq:dm-constraint}
    \big|\Delta_{D^0}\big|<(5.0\pm2.8)\times10^{-6}\,.
\end{equation}
For models with $n_d=0$, where no down-type VLQs are present, eq. \eqref{eq:dm-constraint} can be converted into a bound on $|F^u_{cu}|$
\begin{equation}\label{eq:dm-constraintU}
    |F^u_{cu}|^2<(2.1\pm1.2)\times10^{-8}\,.
\end{equation}
In the simplest case of
$n_u=1$, as in section \ref{sec:quarkup}, one may also write
\begin{equation}\label{eq:dm-FU}
    F^u_{cu}=\Theta_c\Theta^*_u=\theta_{14}\theta_{24}e^{i(\delta_{14}-\delta_{24})} ~,
\end{equation}
where the notation of eq. \eqref{eq:SVLQ-Tup} was used. In extensions of the SM with a single down-type VLQ $B$, eq. ~\eqref{eq:dm-constraint} looks like
\begin{equation}\label{eq:dm-constraintD}
    S_0(x_B)\big|\lambda^B_{cu}\big|^2<(5.0\pm2.8)\times10^{-6}\,.
\end{equation}
Using the parameterisation introduced in section \ref{sec:quarkdown} yields
\begin{equation}\label{eq:dm-LD}
    \lambda^B_{cu}=\Theta_d\Theta^*_s=\theta_{14}\theta_{24}e^{-i(\delta_{14}-\delta_{24})}\,.
\end{equation}
Hence, eq. \eqref{eq:dm-constraintD} may provide a restriction on the VLQ-SM quark mixing parameter space of models such as the ones in section \ref{sec:quarkdown}. \par
The current experimental bounds on $F_{ij}$, cf. eq. \eqref{eq:dm-FU}, from this process are still orders of magnitude away from reasonable values one obtains in theoretical models such as the ones in section \ref{sec:quarkup} and in ref. \cite{Branco:2021vhs}. Nonetheless, the experimental limits obtained from this process are of paramount importance in constraining the parameter space of models of this type \cite{Branco:2021vhs}.

\subsubsection{\texorpdfstring{$B_{d,s}^0$-$\overline{B}_{d,s}^0$}{B0-B0bar} mixing}
\label{sec:B0q}
In models featuring singlet VLQs, the involvement of Higgs-mediated FCNC in $B^0_q$--$\overline{B^0_q}$ ($q=d,s$) mixing is significantly suppressed by a factor of $m^2_{B_q}/m^2_h\sim10^{-3}$ compared to the contributions arising from $Z$ interactions.
Consequently, at the leading order, the expression is given by~\cite{Aguilar-Saavedra:2002phh}
\begin{equation}\label{eq:bm-m12}
    M^{B_q}_{12}=
    B_{B_q}\frac{G^2_Ff^2_{B_q}m^2_Wm_{B_q}}{12\pi^2}
    \Big[\eta^{B_q}_{tt}S_0(x_t)\big(\lambda^t_{qb}\big)^2+\Delta_{B_q}\Big]\,,
\end{equation}
where the NLO QCD corrections are given by $\eta^{B_d}_{tt}=\eta^{B_s}_{tt}=0.55$~\cite{Buchalla:1993bv}, the bag factors are $B_{B_d}=1.222\pm0.061$ and $B_{B_s}=1.232\pm0.053$, the decay constants are $f_{B_d}=(190.0\pm1.3)~\unit{MeV}$ and $f_{B_s}=(230.3\pm1.3)~\unit{MeV}$, which were determined through lattice computations with $N_f=2+1+1$~\cite{FlavourLatticeAveragingGroupFLAG:2021npn}.\par
\begin{figure}[H]
  \centering
        \includegraphics[width=\textwidth]{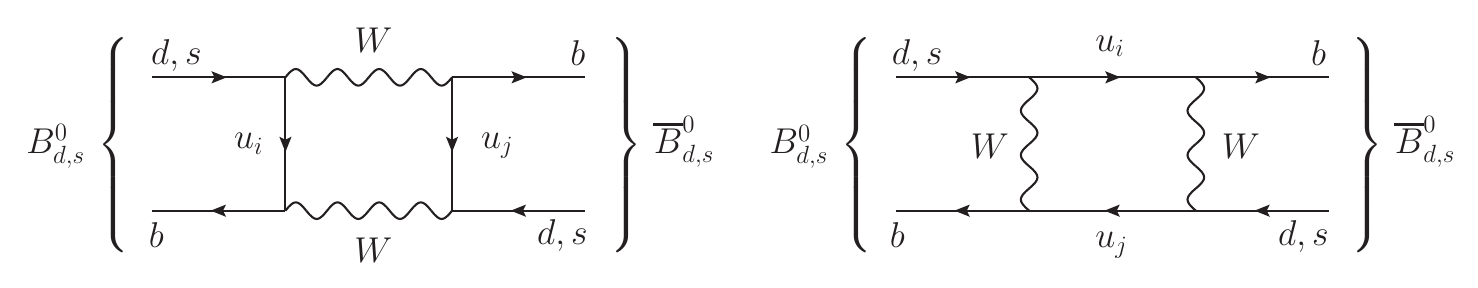}
  \caption{   \label{fig:B}Leading contributions to $B_{d,s}^0$-$\overline{B}_{d,s}^0$ mixing in the presence of one up-type VLQ, including the effect of the new heavy quark, $u_{i,j} = u,c,t,T$.}
\end{figure}

In contrast to section \ref{sec:D0D0bar}, at the leading order, the expression for $\Delta_{B_q}$ when $n_u=0$ contains terms at the tree level, while the expression for $n_d=0$ only contains terms at the loop level, cf. Fig \ref{fig:B}. This distinction arises because the valence quarks of $B^0_{d,s}$ are down-type quarks, unlike the case of the $D^0$ meson, where all valence quarks are up-type quarks.
\begin{equation}\label{eq:delta-Bq}\begin{split}
    \Delta_{B_q}&=                                                                                         \sum^{n_u}_{k=1}\eta^{B_q}_{U_kU_k}S_0(x_{U_k})\big(\lambda^{U_k}_{qb}\big)^2+                        2\sum^{n_u}_{k=1}\eta^{B_q}_{tU_k}S_0(x_t,x_{U_k})\lambda^t_{qb}\lambda^{U_k}_{qb} ~,~ n_d=0\,,\\
    \Delta_{B_q}&=                                                                            -8F^d_{qb}\eta^{B_q}_{tt}Y_0(x_t)\lambda^t_{qb}+\frac{4\pi s^2_W}{\alpha_\text{em}}\eta^{B_q}_Z\big(F^d_{qb}\big)^2  ~,~ n_u=0 ~,
\end{split}\end{equation}
where $\eta^{B_d}_Z=\eta^{B_s}_Z=0.57$~\cite{Buchalla:1995vs}. Note that for VLQs with masses above $m_t$, one can assume $\eta^{B_q}_{U_kU_k}\simeq\eta^{B_q}_{tU_k}\simeq\eta^{B_q}_{tt}$, due to the slow-running nature of QCD. The $U_k$ in the $n_d=0$ expression refers to all up-type VLQs $U_1,..,U_k$, since the leading diagrams are the ones where these are in the loop. \par

The $B^0_q$ systems can be characterized using the perturbative interactions given in equation \eqref{eq:bm-m12}. Consequently, they stand out as the only system where $M_{12}$ can be constrained through data fitting. In ref. \cite{Lenz:2012az}, this fitting process was carried out in a model-independent manner. This was done by introducing the quantity
\begin{equation}\label{eq:bm-constraintD}
    \Delta_q=M^{B_q}_{12}/\big(M^{B_q}_{12}\big)_\text{SM}\equiv
    1+\big(M^{B_q}_{12}\big)_\text{NP}/\big(M^{B_q}_{12}\big)_\text{SM}\,,
\end{equation}
and obtaining
$\text{Re}(\Delta_d)=0.82\pm0.14$, $\text{Im}(\Delta_d)=-0.199\pm0.062$, $\text{Re}(\Delta_s)=0.97\pm0.13$ and $\text{Im}(\Delta_s)=0.00\pm0.10$.
Thus, 
\begin{equation}\label{eq:bm-constraints}\begin{split}
    \text{Re}\Big[\Delta_{B_d}/\eta^{B_d}_{tt}S_0(x_t)\big(\lambda^t_{db}\big)^2\Big]&= -0.18\pm0.14\,,\\
    \text{Im}\Big[\Delta_{B_d}/\eta^{B_d}_{tt}S_0(x_t)\big(\lambda^t_{db}\big)^2\Big]&= -0.199\pm0.062\,,\\
    \text{Re}\Big[\Delta_{B_s}/\eta^{B_s}_{tt}S_0(x_t)\big(\lambda^t_{sb}\big)^2\Big]&= -0.03\pm0.13\,,\\
    \text{Im}\Big[\Delta_{B_s}/\eta^{B_s}_{tt}S_0(x_t)\big(\lambda^t_{sb}\big)^2\Big]&= 0.00\pm0.10 ~,
\end{split}\end{equation}
which must be satisfied by every model with singlet VLQs. \par
In a model with only singlet VLQs of the down-type, $n_u=0$, the constraints in eq. \eqref{eq:bm-constraints} become constraints on $F^d_{ij}$,
\begin{equation}\label{eq:bm-constraintsD}\begin{split}
    \mathrm{Re}\Big\{(0.707-0.707i)F^d_{db}\big[F^d_{db}-(1.58+0.66i)\times10^{-4}\big]\Big\}&=                (-8.1\pm6.3)\times10^{-8}\,,\\
    \mathrm{Im}\Big\{(0.707-0.707i)F^d_{db}\big[F^d_{db}-(1.58+0.66i)\times10^{-4}\big]\Big\}&=                (-8.9\pm2.8)\times10^{-8}\,,\\
    \mathrm{Re}\Big[F^d_{sd}\big(8.37\times10^{-4}+F^d_{sb}\big)\Big]&=(-0.3\pm1.4)\times10^{-6}\,,\\
    \mathrm{Im}\Big[F^d_{sd}\big(8.37\times10^{-4}+F^d_{sb}\big)\Big]&=(0.0\pm1.1)\times10^{-6}\,,
\end{split}\end{equation}
where SM best-fit values for the CKM matrix~\cite{ParticleDataGroup:2020ssz} were used.
When $n_d=1$, cf. section \ref{sec:quarkdown}, one can write
\begin{equation}\label{eq:bm-FD}
    F^d_{qd}=-\Theta_{d_q}\Theta^*_b=-\theta_{q4}\theta_{34}e^{-i\delta_{q4}}
\end{equation}
where the notation of eq. \eqref{eq:SVLQ-Tdown} was used. When one VLQ $T$ of the up-type is introduced into the SM, one should use
\begin{equation}\label{eq:bm-constraintsU}\begin{split}
    \Delta_{B_d}&=\lambda^T_{db}\Big[0.55S_0(x_T)\lambda^T_{db}+(8.70+3.60i)\times10^{-3}S_0(x_t,x_T)\Big]\,,\\
    \Delta_{B_s}&=\lambda^T_{sb}\Big[0.55S_0(x_T)\lambda^T_{sb}-4.60\times10^{-2}S_0(x_t,x_T)\Big]\,,
\end{split}\end{equation}
in eq. \eqref{eq:bm-constraints}, to obtain limits on the mixing between SM quarks and the up VLQ. The SM best-fit values for the CKM matrix~\cite{ParticleDataGroup:2020ssz} were used in eq. \eqref{eq:bm-constraintsU}. Again, using the notation of eq. \eqref{eq:SVLQ-Tup},
\begin{equation}\label{eq:dm-LU}
    \lambda^T_{qb}=\Theta_{u_q}\Theta^*_t=\theta_{q4}\theta_{34}e^{-i\delta_{q4}}\,.
\end{equation}

\subsubsection{\texorpdfstring{$K^0$-$\overline{K}^0$}{K0-K0bar}  mixing}
\label{sec:Kmixing}

As explained in section \ref{sec:quarksgen}, models involving VLQs exhibit tree-level $Z$ and Higgs-mediated FCNC, which contribute to $K^0$--$\overline{K^0}$ mixing. Nevertheless, the Higgs-mediated FCNC contributions are significantly suppressed, approximately by a factor of $m^2_K/m^2_h\sim10^{-5}$.
Therefore, at the leading order, the short-range NP contributions to this specific mixing can be expressed as~\cite{Aguilar-Saavedra:2002phh}
\begin{equation}\label{eq:km-m12}
    M^K_{12}=B_K\frac{G^2_Ff^2_Km^2_Wm_K}{12\pi^2}\Big[ \eta^K_{cc}S_0(x_c)\big(\lambda^c_{ds}\big)^2\!+ \eta^K_{tt}S_0(x_t)\big(\lambda^t_{ds}\big)^2\!+ 2\eta^K_{ct}S_0(x_c,x_t)\lambda^c_{ds}\lambda^t_{ds}+                     \Delta_{K^0}\Big]\,,
\end{equation}
where $\eta^K_{cc}=1.38\pm0.20$, $\eta^K_{ct}=0.47\pm0.04$ and $\eta^K_{tt}=0.57\pm0.01$ are QCD correction factors evaluated at NLO~\cite{Buchalla:1995vs}, while 
$B_K=0.717\pm0.018 \,(\text{stat.})\pm0.016\, (\text{syst.})$ and $f_K=(155.7\pm0.3)~\unit{MeV}$ are, respectively, the kaon bag factor and decay constant determined from lattice computations with $N_f=2+1+1$~\cite{FlavourLatticeAveragingGroupFLAG:2021npn}.
Similarly to section \ref{sec:B0q}, as the valence quarks of $K^0$ are down-type quarks, the main contributions to $\Delta_{K^0}$ in the case $n_u=0$ occur at tree and loop level and in the case of $n_d=0$ occur at loop level
\begin{equation}\label{eq:delta-K}\begin{split}
    \Delta_{K^0}&=                                                                                         \sum^{n_u}_{k=1}\eta^K_{U_kU_k}S_0(x_{U_k})\big(\lambda^{U_k}_{ds}\big)^2+                                2\sum_{q=c,t}\sum^{n_u}_{k=1}\eta^K_{qU_k}S_0(x_q,x_{U_k})\lambda^q_{ds}\lambda^{U_k}_{ds} ~,~n_d=0~,\\
    \Delta_{K^0}&=                                                                          -8F^d_{ds}\Big[\eta^K_ZY_0(x_c)\lambda^c_{ds}+\eta^K_{tt}Y_0(x_t)\lambda^t_{ds}\Big]+                       \frac{4\pi s^2_W}{\alpha_\text{em}}\eta^K_Z\big(F^d_{ds}\big)^2 ~,~ n_u =0,
\end{split}\end{equation}
where $\eta^K_Z=0.60$ is a QCD correction factor estimated at NLO~\cite{Buchalla:1995vs} and the slow-running nature of QCD can be used to assume $\eta^K_{qU_k}\simeq\eta^K_{qt}$ and $\eta^K_{U_kU_k}\simeq\eta^K_{tt}$. The $U_k$ in the $n_d=0$ expression refers to all up-type VLQs $U_1,..,U_k$, since the leading diagrams are the ones where these are in the loop. \par

Due to the presence of unknown long-distance physics, a comprehensive assessment of kaon physics necessitates more information than what is currently available. Consequently, it is not straightforward to directly fit $M^K_{12}$ to the relevant experimental data. Nonetheless, akin to section \ref{sec:D0D0bar}, one can restrict NP scenarios by making the assumption that their short-distance contributions do not reach the experimental bound. In other words, one can impose the condition 
\begin{equation}\label{eq:km-constraintM}
    \big(\Delta m_K\big)_\text{NP}=2\Big|\big(M^K_{12}\big)_\text{NP}\Big|<\big(\Delta m_K\big)_\text{exp}\,.
\end{equation}

Given the current experimental data~\cite{ParticleDataGroup:2020ssz}, eq. \eqref{eq:km-constraintM} translates into the experimental bound
\begin{equation}\label{eq:km-constraint}\begin{split}
    \big|\Delta_{K^0}\big|&<(2.714\pm0.092)\times10^{-5}\,.\\
\end{split}\end{equation}

$\epsilon_K$, an observable connected to CP Violation which is sensitive to the imaginary part of $\big(M^K_{12}\big)_\text{NP}$, discussed in section \ref{sec:epsilon}, provides the most significant constraint in the $K^0$ system.
However, it is important to note that $\Delta m_K$ cannot be disregarded, as $\epsilon_K$ is not sensitive to real (non-imaginary) NP contributions.
To convert eq. \eqref{eq:km-constraint} into bounds for specific models, the SM best-fit values for the CKM matrix were used~\cite{ParticleDataGroup:2020ssz}.\par
 In models with only down-type VLQs, $n_u=0$, the constraint in eq. \eqref{eq:km-constraint} becomes
\begin{equation}\label{eq:km-constraintD}\begin{split}
    \Big|F^d_{ds}\big[F^d_{ds}+(7.07+2.70i)\times10^{-6}\big]\Big|&<(1.136\pm0.039)\times10^{-7}\,.\\
\end{split}\end{equation}
Note that in the case of $n_d=1, n_u=0$, cf. section \ref{sec:quarkdown}, one has
\begin{equation}\label{eq:km-FD}
    F^d_{ds}=-\Theta_d\Theta^*_s=-\theta_{14}\theta_{24}e^{i(\delta_{14}-\delta_{24})} ~,
\end{equation}
written using the parameterisation in section \ref{sec:quarkdown}.

\begin{figure}[H]
    \vskip 2mm
  \centering
        \includegraphics[width=\textwidth]{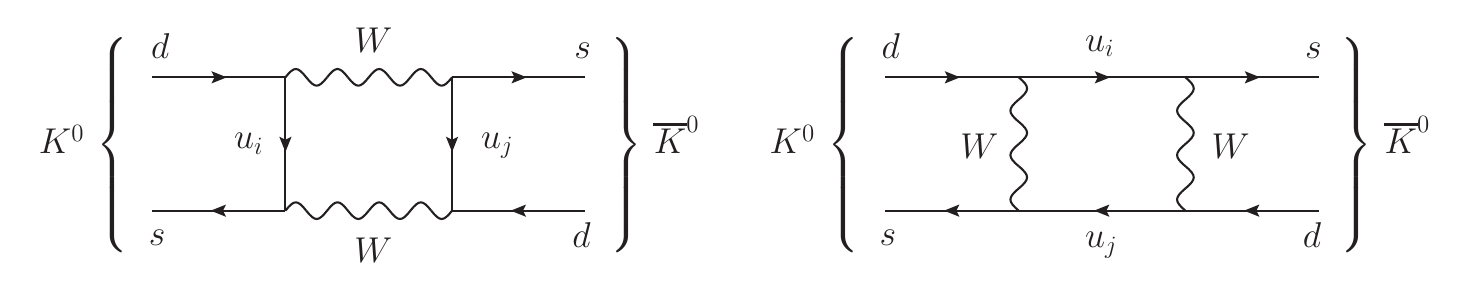}
        \caption{Leading contributions to $K^0$-$\overline{K}^0$ mixing in the presence of one up-type VLQ, including the effect of the new heavy quark, $u_{i,j} = u,c,t,T$.}
        \label{fig:K}
    \end{figure}

In models where only one up-type VLQ singlet $T$ was introduced, $n_u=1 ~,~ n_d=0$, cf. section \ref{sec:quarkup} and ref. \cite{Branco:2021vhs}, the main contribution to eq. \eqref{eq:km-m12} is at loop level as depicted in Fig \ref{fig:K}.

In this case, eq. \eqref{eq:delta-K} becomes
\begin{equation}\label{eq:km-constraintU}
    \Delta_{K^0}= \lambda^T_{ds}\Big[0.57S_0(x_T)\lambda^T_{ds}-(3.77+1.56i)\times10^{-4}S_0(x_t,x_T)-0.206S_0(x_c,x_T)\Big]\,,
\end{equation}
where using the parameterisation introduced in section \ref{sec:quarkup} one has 
\begin{equation}\label{eq:km-constraintU2}
    \lambda^T_{ds}=\Theta_u\Theta^*_c=\theta_{14}\theta_{24}e^{-i(\delta_{14}-\delta_{24})}\,.
\end{equation}
Combining eqs. \eqref{eq:km-constraint} \eqref{eq:km-constraintU} \eqref{eq:km-constraintU2} one obtains an experimental bound on the mixing among SM quarks and VLQs.

The terms in eq. \eqref{eq:km-constraintU} are sensitive to the mass of the new quark $m_T$, via the Inami-Lim functions $S_0(x)$, defined in eq. \eqref{eq:InamiLim}, and to the elements of the fourth row of $V$, via $\lambda^T_{ds}$. The same applies to other mesons with down-type valence quarks, such as $B^0_{d,s}$, cf. section \ref{sec:B0q}.
In the analysis of the model with $n_u=1 ~,~n_d=0$ in section \ref{sec:quarkup} and ref. \cite{Branco:2021vhs}, bounds on the mixing matrix elements were obtained for reference masses of the $T$ VLQ, see Table~\ref{tab:mesonbounds}.
%
\begin{table}
  \centering
  \renewcommand{\arraystretch}{1.5}
  \begin{tabular}{ccc}
    \toprule
    Observable  & $m_T = 1$ TeV & $m_T = 3$ TeV\\
    \midrule
    $\Delta m_K$ & $\big|V_{Td}\big| \big|V_{Ts}\big| < 7.4 \times 10^{-4}$ & $\big|V_{Td}\big| \big|V_{Ts}\big| < 2.7 \times 10^{-4}$\\
    $\Delta m_B$ & $\big|V_{Td}\big| \big|V_{Tb}\big| < 6.7 \times 10^{-4}$ & $\big|V_{Td}\big| \big|V_{Tb}\big| < 3.4 \times 10^{-4}$\\
    $\Delta m_{B_s}$ & $\big|V_{Ts}\big| \big|V_{Tb}\big| < 3.2 \times 10^{-3}$ & $\big|V_{Ts}\big| \big|V_{Tb}\big| < 1.6 \times 10^{-3}$\\
    \bottomrule
  \end{tabular}
  \caption{Constraints from neutral meson observables on products of mixing matrix elements ($\Theta = \arg V_{Ts}^* V_{Td}$) for two benchmark masses of the new heavy top quark.}
  \label{tab:mesonbounds}
\end{table}
%

\subsubsection{Scalar Mediated Box Diagrams: Effects of Very Heavy VLQs}
\begin{figure}[H]
    \vskip 2mm
  \centering
        \includegraphics[width=0.8\textwidth]{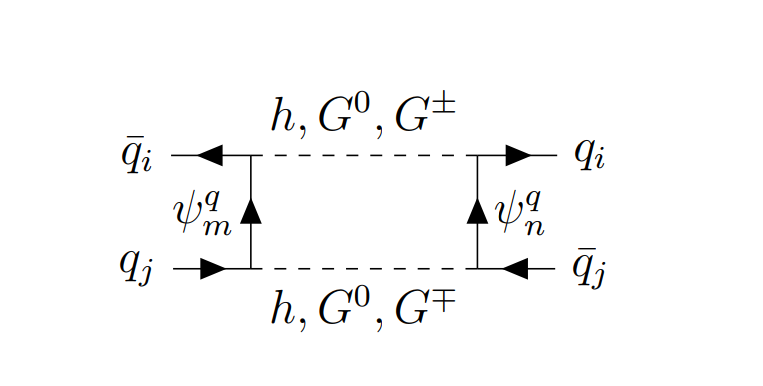}
        \caption{Scalar-box contributions to neutral meson mixing, where $\psi^q$ ($q=u,d$), with $\psi^u = (u,c,t,U_1,\ldots,U_{n_u})^T$ and $\psi^d = (d,s,b,D_1,\ldots,D_{n_d})^T$.}
         \label{fig:hvlq-box}
    \end{figure}

In sections \ref{sec:D0D0bar}, \ref{sec:B0q} and \ref{sec:Kmixing}, the gauge-mediated one-loop contributions to neutral meson mixing alongside the tree-level NP effects were considered. In this section, the emphasis will be on the scalar-mediated box diagrams, illustrated in fig. \ref{fig:hvlq-box}, which become significant for large VLQ masses~\cite{Ishiwata:2015cga,Bobeth:2016llm}.
    
With the notation introduced in section~\ref{sec:quarksgen}, the Lagrangian where these Feynman diagrams are extracted from has the form
\begin{equation}\begin{split}\label{eq:hvlq-lagrangian}
    \mathscr{L}_Y\supset&                                                                                  -\frac{\sqrt{2}}{v}G^+\Big[\bar{\psi}^u\big(VD_d\gamma_R-D_uV\gamma_L\big)\psi^d\Big]                  +\frac{\sqrt{2}}{v}G^-\Big[\bar{\psi}^d\big(V^\dagger D_u\gamma_R-D_dV^\dagger\gamma_L\big)\psi^u\Big]\\
    &-i\epsilon_q\frac{G^0}{v}\Big[\bar{\psi}^q\big(F^qD_q\gamma_R-D_qF^q\gamma_L\big)\psi^q\Big]                 -\frac{h}{v}\Big[\bar{\psi}^q\big(F^qD_q\gamma_R+D_qF^q\gamma_L\big)\psi^q\Big]\,,
\end{split}\end{equation}
where, as in fig. \ref{fig:hvlq-box},  the column vectors $\psi^q$ ($q=u,d$), with $\psi^u = (u,c,t,U_1,\ldots,U_{n_u})^T$ and $\psi^d = (d,s,b,D_1,\ldots,D_{n_d})^T$ and, in the second line of the equation, there are implicit sums over $q=d,u$ with $\epsilon_d=-\epsilon_u=1$. $G^0$ and $G^\pm$ are the unphysical Goldstone Bosons, since the calculation is performed in a general gauge, see the end of section \ref{sec:summary_SM} for a brief discussion on the role of these bosons in the SM.\par
Working in the $R_\xi$ gauge, where $m^2_{G^0}=\xi m^2_Z$ and $m^2_{G^\pm}=\xi m^2_W$, the contributions from the neutral scalars to the effective Hamiltonian can be written in the following manner,
\begin{equation}\label{eq:hvlq-Hneutral}
    \mathcal{H}_{\textrm{eff}}=\frac{m^2_mm^2_n}{64\pi^2v^4}                                                                   \Big[\Upsilon_3\big(m^2_h,m^2_h\big)+\Upsilon_3\big(\xi m^2_Z,\xi m^2_Z\big)+2\Upsilon_3\big(m^2_h,\xi m^2_Z\big)\Big]\Lambda^m_{ij}\Lambda^n_{ij}\big(\bar{q}_i\gamma^\mu\gamma_Lq_j\big)^2\,,
\end{equation}
where there are implicit sums over all fermions $\psi^q_{m,n}$, $\Lambda^m_{ij}=F^{*q}_{mi}F^q_{mj}$ and $\Upsilon_3(m^2_a,m^2_b)\equiv\Upsilon_3[m^2_a,m^2_b,m^2_m,m^2_n]$, where
\begin{equation}\begin{split}\label{eq:hvlq-Ups}
    \Upsilon_3&[m^2_a,m^2_b,m^2_c,m^2_d]\equiv\\
    &\frac{m^4_a\log m^2_a}{2(m^2_a-m^2_b)(m^2_a-m^2_c)(m^2_a-m^2_d)}                                     +\frac{m^4_b\log m^2_b}{2(m^2_b-m^2_a)(m^2_b-m^2_c)(m^2_b-m^2_d)}\\
    +&\frac{m^4_c\log m^2_c}{2(m^2_c-m^2_a)(m^2_c-m^2_b)(m^2_c-m^2_d)}                                    +\frac{m^4_d\log m^2_d}{2(m^2_d-m^2_a)(m^2_d-m^2_b)(m^2_d-m^2_c)}\,.
\end{split}\end{equation}
Also important are the charged scalar contributions to neutral meson mixing, also represented in Fig. \ref{fig:hvlq-box}. These equate to
\begin{equation}\label{eq:hvlq-Hcharged}
      \mathcal{H}_{\textrm{eff}}=\frac{m^2_mm^2_n}{16\pi^2v^4}                                                                        \Upsilon_3\big(\xi m^2_W,\xi m^2_W\big)\lambda^m_{ij}\lambda^n_{ij}\big(\bar{q}_i\gamma^\mu\gamma_Lq_j\big)^2\,,
\end{equation}
with $\lambda^m_{ij}=V^*_{mi}V_{mj}$ ($\lambda^m_{ij}=V_{im}V^*_{jm}$) for neutral mesons with down-type (up-type) valence quarks. 
In models with a heavy VLQ $Q$, where $m_Q\gg m_{W,Z}$, eq. \eqref{eq:hvlq-Ups} becomes
\begin{equation}\label{eq:hvlq-UpsUps}
    \Upsilon_3[m^2_Q,m^2_Q,0,0]=\frac{1}{2m^2_Q}
\end{equation}
at leading order. Thus, the previous Hamiltonians, expressed in eqs. \eqref{eq:hvlq-Hneutral} and \eqref{eq:hvlq-Hcharged}, may be combined into
\begin{equation}\label{eq:hvlq-Heff}
      \mathcal{H}_{\textrm{eff}}=\frac{m^2_Q}{32\pi^2v^4}                            \Big[
        \big(\lambda^Q_{ij}\big)^2
        +\big(\Lambda^Q_{ij}\big)^2        
    \Big]                         \big(\bar{q}_i\gamma^\mu\gamma_Lq_j\big)^2\,,
\end{equation}
which gains importance against other diagrams as $m_Q$ grows and reaches the regime $m_Q \gg v$.
Hence, in theories with a heavy VLQ, eqs. \eqref{eq:dm-m12}, \eqref{eq:bm-m12} and \eqref{eq:km-m12} should be replaced by
\begin{equation}\begin{split}\label{eq:hvlq-m12}
    M^D_{12}&=-\frac{f^2_Dm^2_Qm_D}{96\pi^2v^4}                                    \Big[\big(\Lambda^Q_{cu}\big)^2+\big(\lambda^Q_{cu}\big)^2\Big]\,,\\
    M^{B_q}_{12}&=-\frac{f^2_{B_q}m^2_Qm_{B_q}}{96\pi^2v^4}                        \Big[\big(\Lambda^Q_{qb}\big)^2+\big(\lambda^Q_{qb}\big)^2\Big]\,,\\
     M^K_{12}&=-\frac{f^2_Km^2_Qm_K}{96\pi^2v^4}                                    \Big[\big(\Lambda^Q_{ds}\big)^2+\big(\lambda^Q_{ds}\big)^2\Big]\,.
\end{split}\end{equation}
In ref. \cite{Ishiwata:2015cga}, it was demonstrated that these expressions are favored over the ones presented in sections \ref{sec:D0D0bar}, \ref{sec:B0q}, and \ref{sec:Kmixing} when the VLQ mass is around $m_Q\gtrsim10,\mathrm{TeV}$.
Moreover, it is important to note that the calculations from the previous sections, involving $W$ boson exchange, already encompass the corresponding Goldstone contribution shown in Fig. \ref{fig:hvlq-box}. However, the contributions involving the $Z$ boson and the physical Higgs are missing.

\subsubsection{CP Violation: The $K^0$ system, $\epsilon_K$ and $\epsilon^\prime/\epsilon$}
\label{sec:epsilon}

The parameter $\epsilon_K$ describes the indirect CP violation in the kaon system and may be written in terms of $M_{12}^K$, defined in eq. \eqref{eq:km-m12}, via~\cite{Buras:1997fb}
\begin{equation}
\label{eq:epsilonK}
|\epsilon_K| = 
\frac{\kappa_\epsilon}{\sqrt{2}\,\Delta m_K} \left|\im M_{12}^K \right|\,,
\end{equation}
where $\kappa_\epsilon \simeq 0.92 \pm 0.02$~\cite{Buras:2008nn}. It has been measured to be $|(\epsilon_K)_\text{exp}|=(2.257\pm0.018)\times10^{-3}$ \cite{ParticleDataGroup:2020ssz}, with 
most uncertainties in its determination being cancelled by using the unitarity of the CKM matrix \cite{Brod:2019rzc}. Since the measurement is in good agreement with the SM estimate  $|(\epsilon_K)_\text{SM}|=(2.16\pm0.18)\times10^{-3}$, it is reasonable to demand
\begin{equation}\label{eq:km-constraintE}
    \Big|\big(\epsilon_K\big)_\text{NP}\Big|<              \Big|\big(\epsilon_K\big)_\text{exp}\Big|-\Big|\big(\epsilon_K\big)_\text{SM}\Big|\,.
\end{equation}
Currently, there is a theoretical uncertainty of approximately $10\%$ associated with $(\epsilon_K)_\text{SM}$, which makes it compatible with $(\epsilon_K)_\text{exp}$. Due to this uncertainty, some authors opt to restrict the contributions to $(\epsilon_K)_\text{NP}$ to be within one-tenth of the experimental value $(\epsilon_K)_\text{exp}$~\cite{Brod:2019rzc, Botella:2021uxz}. Given the current experimental data~\cite{ParticleDataGroup:2020ssz}, using eqs. \eqref{eq:km-m12}, \eqref{eq:delta-K}, \eqref{eq:epsilonK} and \eqref{eq:km-constraintE}, one obtains
\begin{equation}\label{eq:km-constraint2}\begin{split}
    \big|\text{Im}\,\Delta_{K^0}\big|&<(0.8\pm1.5)\times10^{-8}\,.
\end{split}\end{equation}
In models with $n_u=0$, the constraints in eq. \eqref{eq:km-constraint2} become
\begin{equation}\label{eq:km-constraintD2}\begin{split}
    \Big|\mathrm{Im}\Big\{F^d_{ds}\big[F^d_{ds}+(7.07+2.70i)\times10^{-6}\big]\Big\}\Big|&<(3.3\pm6.3)\times10^{-11}\,.
\end{split}\end{equation}

In models with $n_d=0$, eqs. \eqref{eq:km-m12}, \eqref{eq:delta-K}, \eqref{eq:epsilonK} \eqref{eq:km-constraintE}, \eqref{eq:km-constraintU} and \eqref{eq:km-constraintU2} may be used to constrain the parameter space. For the case of $n_u=1~,n_d=0$, cf. section \ref{sec:quarkup} and ref. \cite{Branco:2021vhs}, with one $T$ VLQ, one may write the maximum possible value for the NP contribution for $|\epsilon_K|$, see eqs. \eqref{eq:km-m12}, \eqref{eq:delta-K} and \eqref{eq:epsilonK}, as
\begin{equation}
\begin{aligned}
|\epsilon_K|^\text{NP}
    &\,\simeq\, 
\frac{G_F^2 m_W^2\, m_K f^2_K B_K\, \kappa_\epsilon}{12 \sqrt{2}\, \pi^2 \,\Delta m_K}
     \bigg| 
       2\, \eta_{cT}^K  S_0(x_c,x_T) \im\big(\lambda_{ds}^c \lambda^T_{ds}\big) \\
    &\qquad\quad + 2\, \eta^L_{tT} S_0(x_t,x_T) \im\big(\lambda_{ds}^t \lambda^T_{ds}\big)
     +   \eta^K_{TT} S_0(x_T) \im\Big[\big(\lambda_{ds}^{T} \big)^2\Big]  \bigg|\,.
\end{aligned}
\label{eq:epsilonNP}
\end{equation}
This maximum NP contribution depends on $m_T$ and on the angles and phases in $V$. In the analysis in section \ref{sec:quarkup} and ref. \cite{Branco:2021vhs}, it was required that its absolute value does not exceed the measured value, $|\epsilon_K|^\text{NP} < |\epsilon_K|^\text{exp}$ (taking $\eta^K_{iT} = 1$).
For illustrative purposes, a preliminary estimate of the bound on $\lambda^T_{ds}$ can be obtained by assuming that only the last term in equation~\eqref{eq:epsilonNP} contributes significantly, as $S_0(x_c,x_T) \ll S_0(x_t,x_T) < S_0(x_T)$ and $|V_{td}| |V_{ts}| \sim 3\times 10^{-4}$. Denoting by $\Theta$ the phase of $\lambda^T_{ds}$, one has
\begin{equation}
|\epsilon_K|^\text{NP}
    \,\sim\, 0.5 \,
\frac{G_F^2 m_W^2\, m_K f^2_K B_K\, \kappa_\epsilon}{12 \sqrt{2}\, \pi^2 \,\Delta m_K}
       S_0(x_T)  \big|\lambda^T_{ds}\big|^2 |\sin 2 \Theta| < |\epsilon_K|^\text{exp}\,,
\label{eq:boundek}
\end{equation}
where the ad-hoc $1/2$ factor takes into account the fact that the $tT$ term may partly cancel the $TT$ one and leads to a more conservative bound. The consequences of eq.~\eqref{eq:boundek} for the previously considered benchmarks ($m_T = 1,3$ TeV) are shown in Table~\ref{tab:mesonbounds2}.

\begin{table}[H]
  \centering
  \renewcommand{\arraystretch}{1.5}
  \begin{tabular}{ccc}
    \toprule
    Observable  & $m_T = 1$ TeV & $m_T = 3$ TeV\\
    \midrule
      $|\epsilon_K|$ & $\big|V_{Td}\big| \big|V_{Ts}\big| \sqrt{|\sin 2 \Theta|} < 8.8 \times 10^{-5}$ & $\big|V_{Td}\big| \big|V_{Ts}\big| \sqrt{|\sin 2 \Theta|} < 3.1 \times 10^{-5}$ \\
    \bottomrule
  \end{tabular}
  \caption{Constraints from indirect CP Violation in the kaon system on products of mixing matrix elements ($\Theta = \arg \lambda^T_{ds} = \arg( V_{Ts}^* V_{Td})$) for two benchmark masses of the new heavy top quark.}
  \label{tab:mesonbounds2}
\end{table}

\vskip 2mm
Concerning direct CP violation in the $K$ sector, usually it is quantified by the parameter $\epsilon^\prime$. However, its theoretical determination involves several uncertainties, so researchers often prefer to consider the ratio~\cite{Aguilar-Saavedra:2002phh}
\begin{equation}\label{eq:eps-ratio}
    \frac{\epsilon'}{\epsilon}=F_{\epsilon'}(x_t)\mathrm{Im}\,\lambda^t_{sd}+\Delta_{\epsilon'}\,,
\end{equation}
where \cite{Buras:2001pn}
\begin{equation}\label{eq:eps-Fx}
    F_{\epsilon'}(x)=P_0+P_XX_0(x)+P_YY_0(x)+P_ZZ_0(x)+P_EE_0(x)\,.
\end{equation}
In the NDR scheme~\cite{Buras:2000qz}, the coefficients of the Inami-Lim functions above are given by
\begin{equation}\label{eq:eps-FxP}\begin{split}
    P_0&=-3.167+12.409B^{(1/2)}_6+1.262B^{(3/2)}_8\,,\\
    P_X&=0.540+0.023B^{(1/2)}_6\,,\\
    P_Y&=0.387+0.088B^{(1/2)}_6\,,\\
    P_Z&=0.474-0.017B^{(1/2)}_6-10.186B^{(3/2)}_8\,,\\
    P_E&=0.188-1.399B^{(1/2)}_6+0.459B^{(3/2)}_8\,,
\end{split}\end{equation}
with the bag factors $B^{(1/2)}_6=1.36\pm0.23$ and $B^{(3/2)}_8=0.79\pm0.05$, obtained with lattice QCD~\cite{Aebischer:2020jto} methods.
The NP contributions to eq. \eqref{eq:eps-ratio} from VLQ models are
\begin{equation}\label{eq:delta-Eps}\begin{aligned}
    \Delta_{\epsilon'}\! &=   \!                                 \sum^{n_u}_{k=1}\!F_{\epsilon'}(x_{U_k})\mathrm{Im}\,\lambda^{U_k}_{sd}\!+                                     \!(P_X\!+\!P_Y\!+\!P_Z)\!\sum_{i,j}\mathrm{Im}\big[V^*_{is}(F^u\!-\!\id)_{ij}V_{jd}\big]N(x_i,x_j) ~,~ n_d =0 ~,~\\
    \Delta_{\epsilon'}\!&=-\frac{\pi s^2_W}{\alpha_\text{em}}\big(P_X+P_Y+P_Z\big)\mathrm{Im}\,F^d_{sd}\ ~,~ n_u =0 ~.~\\
\end{aligned}\end{equation}
Using the current experimental value  $\epsilon'/\epsilon=(1.66\pm0.23)\times10^{-3}$~\cite{ParticleDataGroup:2020ssz}, one obtains a constraint that all models with VLQ singlets must satisfy
\begin{equation}\label{eq:eps-constraint}
    \Delta_{\epsilon'}=(2.9\pm4.4)\times10^{-4} ~.
\end{equation}
In the case of $n_u=0$, combining eq. \eqref{eq:eps-constraint} with eq. \eqref{eq:delta-Eps} gives
\begin{equation}\label{eq:eps-constraintD}
    \mathrm{Im}\,F^d_{sd}=(4.5\pm6.8)\times10^{-7}\,.
\end{equation}
As done before, when $n_d=1$ this may be written as a constraint on the mixing among SM quarks and the VLQ
\begin{equation}\label{eq:eps-FD}
    F^d_{sd}=-\Theta_s\Theta^*_d=-\theta_{14}\theta_{24}e^{i(\delta_{14}-\delta_{24})}
\end{equation}
by using the parameterisation introduced in section \ref{sec:quarkdown}.

In models with a single up-type VLQ $T$, $n_u=1~,n_d=0$ the constraint in eq. \eqref{eq:eps-constraint} may be approximately, in the limit $F^u\to\diag(1,1,1,0)$,  translated into
\begin{equation}\label{eq:eps-constraintU}
    \mathrm{Im}\,\lambda^T_{sd}\Big[F_{\epsilon'}(x_T)-\big(P_X\!+\!P_Y\!+\!P_Z\big)N(x_T)\Big]=                (2.9\pm4.4)\times10^{-4}\,.
\end{equation}
When generating the entire parameter space and aiming for precise results, one may directly limit the parameter space of $n_d=0$ using eqs. \eqref{eq:eps-constraint} and \eqref{eq:delta-Eps}, since $V$ and $F^u$ are interrelated through equation \eqref{eq:Fud}.

Again, for $n_u=1~,n_d=0$, the parameterisation of section \ref{sec:quarkdown} may be employed
\begin{equation}\label{eq:eps-LU}
    \lambda^T_{sd}=\Theta_c\Theta^*_u=\theta_{14}\theta_{24}e^{i(\delta_{14}-\delta_{24})}\,,
\end{equation}
to write the constraint in terms of the mixing angles among SM quarks and the VLQ.

\subsubsection{\texorpdfstring{$B_q$}{Bq} Decays}

In the SM, the $B_q\rightarrow\mu^+\mu^-$ decays ($q=d,s$) generated at the one-loop level encounter additional suppression due to the effects of the CKM matrix and helicity factors~\cite{Blake:2016olu}, rendering them exceedingly rare. After considering the Next-to-Leading Order (NLO) electroweak corrections~\cite{Bobeth:2013tba} and Next-to-Next-to-Leading Order (NNLO) QCD corrections~\cite{Hermann:2013kca}, the predicted branching ratios are $\textrm{BR}(B_d\rightarrow\mu^+\mu^-)=(1.12\pm0.12)\times10^{-10}$ and $\textrm{BR}(B_s\rightarrow\mu^+\mu^-)=(3.52\pm0.15)\times10^{-9}$~\cite{Blake:2016olu}, providing ample room for NP effects. In extensions of the SM involving singlet VLQs, the contributions to these decays are given by~\cite{Morozumi:2018cnc}
\begin{equation}\label{eq:bd-rBq}
    \textrm{BR}\big(B_q\rightarrow\mu^+\mu^-\big)=                                                                  \tau_{B_q}\frac{G^2_F}{16\pi}\bigg(\frac{\alpha_\text{em}}{\pi s^2_W}\bigg)^2f^2_{B_q}m_{B_q}m^2_\mu                \sqrt{1-\frac{4m^2_\mu}{m^2_{B_q}}}|\eta^2_Y|\big|\lambda^t_{qb}Y_0(x_t)+\Delta^{B_q}_{\mu\mu}\big|^2\,,
\end{equation}
where the decay constants $f_{B_d}=(190.0\pm1.3)~\unit{MeV}$ and $f_{B_s}=(230.3\pm1.3)~\unit{MeV}$ were determined from lattice computations with $N_f=2+1+1$~\cite{FlavourLatticeAveragingGroupFLAG:2021npn}, the factor $\eta_Y=1.0113$ incorporates QCD corrections in a scheme where NLO electroweak corrections can be neglected, as indicated in ref. \cite{Buras:2012ru}, $\tau_{B_q}$ is the meson lifetime and $\Delta^{B_q}_{\mu\mu}$ encodes the NP effects.
Using the current experimental values~\cite{ParticleDataGroup:2020ssz}, eq. \eqref{eq:bd-rBq} leads to
\begin{equation}\label{eq:bd-constraints}\begin{split}
    \big|\lambda^t_{db}Y_0(x_t)+\Delta^{B_d}_{\mu\mu}\big|^2&=(0.8\pm1.0)\times10^{-4}\,,\\
    \big|\lambda^t_{sb}Y_0(x_t)+\Delta^{B_s}_{\mu\mu}\big|^2&=(1.47\pm0.20)\times10^{-3}\,.
\end{split}\end{equation}
The uncertainty in the first constraint arises from the significant uncertainty associated with the experimental measurement of $\textrm{BR}(B_d\rightarrow\mu^+\mu^-)$.
In models with only down-type VLQs, $n_u=0$ the factors $\Delta^{B_q}_{\mu\mu}$ are given by~\cite{Aguilar-Saavedra:2002phh}
\begin{equation}\label{eq:bd-deltaD}
    \Delta^{B_q}_{\mu\mu}=-\frac{\pi s^2_W}{\alpha_\text{em}}F^d_{qb}\,.
\end{equation}
Using this in eq. \eqref{eq:bd-constraints} together with the SM best-fit values for the CKM matrix\cite{ParticleDataGroup:2020ssz} leads to the the constraints
\begin{equation}\label{eq:bd-constraintsD}\begin{split}
    \big|F^d_{db}-(8.19+3.40i)\times10^{-5}\big|^2&=(0.8\pm1.0)\times10^{-8}\,,\\
    \big|F^d_{sb}+4.21\times10^{-4}\big|^2&=(1.48\pm0.20)\times10^{-7}\,.
\end{split}\end{equation}

In models with $n_u=0~,~ n_d=1$, eq. \eqref{eq:bd-constraintsD} can be written in terms of the mixing angles among VLQs and SM quarks with
\begin{equation}\label{eq:bd-FD}
    F^d_{db}=-\Theta_d\Theta^*_b=-\theta_{14}\theta_{34}e^{-i\delta_{14}}\,,\quad                            F^d_{sb}=-\Theta_s\Theta^*_b=-\theta_{24}\theta_{34}e^{-i\delta_{24}}\,,
\end{equation}
where $\Theta_d$ was expanded as $(\Theta_d,\Theta_s,\Theta_b)^T$, as done in section \ref{sec:quarkdown}.

When only up-type VLQs are present, $n_d=0$, the NP contributions are \cite{Nardi:1995fq,Vysotsky:2006fx,Kopnin:2008ca,Picek:2008dd,Botella:2017caf}
\begin{equation}\label{eq:bd-deltaU}
    \Delta^{B_q}_{\mu\mu}=                                                                            \sum^{n_u}_{k=1}\lambda^{U_k}_{qb}Y_0(x_{U_k})+\sum_{i,j}V^*_{iq}\big(F^u -\id\big)_{ij}V_{jb}N(x_i,x_j)\,,
\end{equation}
where the sums over $i,j$ cover every up-type quark in the theory (light-$u$ and heavy-$U$). By considering the limit 
$F^u\to \diag(1,1,1,0)$ of no $3 \times 3$ deviations from unitarity, using the SM best-fit values for the CKM matrix~\cite{ParticleDataGroup:2020ssz}, and using eq. \eqref{eq:bd-deltaU} along with eq. \eqref{eq:bd-constraints}, one can obtain, in good approximation, numerical constraints for a model with $n_u=1$,
\begin{equation}\label{eq:bd-constraintsU}\begin{split}
    \big|\lambda^T_{db}\big[Y_0(x_T)-N(x_T)\big]+(8.15+3.39i)\times10^{-3}\big|^2&=(0.8\pm1.0)\times10^{-4}\,,\\
    \big|\lambda^T_{sb}\big[Y_0(x_T)-N(x_T)\big]-4.19\times10^{-2}\big|^2&=(1.47\pm0.20)\times10^{-3}\,.
\end{split}\end{equation}
When generating the entire parameter space and aiming for precise results, one may directly limit the parameter space of $n_u=1$ using eqs. \eqref{eq:bd-constraints} and \eqref{eq:bd-deltaU}, since $V$ and $F^u$ are interrelated through equation \eqref{eq:Fud}.
As before, eq. \eqref{eq:bd-constraintsU} may be written in terms of the mixing among SM quarks and VLQs, via the parameterisation discussed in section \ref{sec:quarkup}, which establishes
\begin{equation}\label{eq:bd-LU}
    \lambda^T_{db}=\Theta_u\Theta^*_t=\theta_{14}\theta_{34}e^{-i\delta_{14}}\,,\quad                           \lambda^T_{sb}=\Theta_c\Theta^*_t=\theta_{24}\theta_{34}e^{-i\delta_{24}}\,,
\end{equation}
with the vector $\Theta_u$ expanded as $(\Theta_u,\Theta_c,\Theta_t)^T$.\par

\subsubsection{\texorpdfstring{$K$}{K} Decays}
\label{sec:Kdecays}
The rare kaon decays $K^+\rightarrow\pi^+\nu\bar{\nu}$, $K_L\rightarrow\mu^+\mu^-$, and $K_L\rightarrow\pi^0\bar{\nu}\nu$ play a crucial role in imposing significant constraints on the mixing in the $sd$ sector of any model involving singlet VLQs. In this section, these decays and their implications on VLQ models will be discussed by utilizing fractions of their branching ratios, where most theoretical uncertainties are mitigated. To describe the first decay, the isospin symmetry of QCD is typically employed in the following manner~\cite{Aguilar-Saavedra:2002phh},
\begin{equation}\label{eq:kd-rKp}
    \frac{\textrm{BR}(K^+\rightarrow\pi^+\nu\bar{\nu})}{\textrm{BR}(K^+\rightarrow\pi^0e^+\bar{\nu})}=                   \frac{\alpha^2_\text{em}r_{K^+}}{2\pi^2s^4_W|V_{us}|^2}                                                      \sum_{l=e,\mu,\tau}\big|\lambda^c_{sd}X^l_{NL}+\lambda^t_{sd}\eta^X_tX_0(x_t)+\Delta^{K^+}_{\pi\nu\nu}\big|^2\,,
\end{equation}
where $r_{K^+}=0.901$~\cite{Marciano:1996wy} captures all isospin breaking corrections.
At NLO, the charm contributions to this ratio are given by $X^{e,\mu}{NL}=(10.6\pm1.5)\times10^{-4}$ and $X^\tau{NL}=(7.1\pm1.4)\times10^{-4}$\cite{Buchalla:1998ba}. Additionally, the QCD correction to the top contributions is expressed as $\eta^X_t=0.994$\cite{Buchalla:1993bv}.
In models with only up, $n_d=0$, or down, $n_u=0$, singlet VLQs
\cite{Nardi:1995fq,Aguilar-Saavedra:2002phh,Vysotsky:2006fx,Kopnin:2008ca,Picek:2008dd,Botella:2017caf}
\begin{equation}\label{eq:delta-Kp}\begin{split}
    \Delta^{K^+}_{\pi\nu\nu}&=                                                                        \sum^{n_u}_{k=1}\eta^X_{U_k}\lambda^{U_k}_{sd}X_0\big(x_{U_k}\big)+                                             \sum_{i,j}\eta^X_iV^*_{is}(F^u-1)_{ij}V_{jd}N(x_i,x_j)~,~ n_d=0 ~,\\
    \Delta^{K^+}_{\pi\nu\nu}&=-\frac{\pi s^2_W}{\alpha_\text{em}}F^d_{sd}~,~n_u=0 ~,
\end{split}\end{equation}
where the sums over $i,j$ cover every up-type quark (light-$u$ and heavy-$U$) in the theory and the slow-running nature of QCD may be used to motivate the assumption $\eta^X_{U_k}\simeq\eta^X_t$.\par
To analyse the second rare decay, the typical approach involves comparing the short-distance contributions to $K_L\rightarrow\mu^+\mu^-$ with the branching ratio of $K^+\rightarrow\mu^+\nu$~\cite{Aguilar-Saavedra:2002phh},
\begin{equation}\label{eq:kd-rKL}
    \frac{\textrm{BR}(K_L\rightarrow\mu^+\mu^-)_{\rm SD}}{\textrm{BR}(K^+\rightarrow\mu^+\nu)}=                      \frac{\tau_{K_L}}{\tau_{K^+}}\frac{\alpha^2_\text{em}}{\pi^2s^4_W|V_{us}|^2}                                \big[Y_{NL}\,\mathrm{Re}\,\lambda^c_{sd}+\eta^Y_tY_0(x_t)\,\mathrm{Re}\,\lambda^t_{sd}+          \Delta^{K_L}_{\mu\mu}\big]^2\,.
\end{equation}
At NLO, the QCD corrections to the box diagrams, which generate this expression, are given by $Y_{NL}=(2.94\pm0.28)\times10^{-4}$\cite{Buchalla:1998ba}, and the QCD correction to the top contributions is $\eta^Y_t=1.012$\cite{Buchalla:1993bv}.
In this analysis, the conservative bound $\textrm{BR}(K_L\rightarrow \mu^+\mu^-)_{\rm SD}< 2.5\times10^{-9}$~\cite{Isidori:2003ts} will be used. The NP effects provided by models with up, $n_d=0$, or down, $n_u=0$, type VLQs are represented as \cite{Nardi:1995fq, Aguilar-Saavedra:2002phh, Vysotsky:2006fx, Kopnin:2008ca, Picek:2008dd, Botella:2017caf}
\begin{equation}\label{eq:delta-KL}\begin{split}
    \Delta^{K_L}_{\mu\mu}&=                                                                           \sum^{n_u}_{k=1}\eta^Y_{U_k}\mathrm{Re}\big(\lambda^{U_k}_{sd}\big)Y_0\big(x_{U_k}\big)+                \sum_{i,j}\eta^Y_i\mathrm{Re}\Big[V^*_{is}(F^u-\id)_{ij}V_{jd}\Big]N(x_i,x_j)~,~ n_d=0 ~,\\
    \Delta^{K_L}_{\mu\mu}&=-\frac{\pi s^2_W}{\alpha_\text{em}}\mathrm{Re}\,F^d_{sd}~,~ n_u=0 ~,\\
\end{split}\end{equation}
where $\eta^Y_{U_k}\simeq\eta^Y_t$ was again considered for heavy $U_k$.
Finally, the third rare decay, $K_L\rightarrow\pi^0\bar{\nu}\nu$, is frequently compared to the SM predictions~\cite{Botella:2021uxz}
\begin{equation}\label{eq:kd-rKl}
    \frac{\textrm{BR}(K_L\rightarrow\pi^0\bar{\nu}\nu)}{\textrm{BR}(K_L\rightarrow\pi^0\bar{\nu}\nu)_\mathrm{SM}}=              \bigg|\frac{\mathrm{Im}[\lambda^c_{sd}X_0(x_c)+\lambda^t_{sd}X_0(x_t)+\Delta^{K_L}_{\pi\nu\nu}]}          {\mathrm{Im}[\lambda^c_{sd}X_0(x_c)+\lambda^t_{sd}X_0(x_t)]}\bigg|^2\,,
\end{equation}
where $\textrm{BR}(K_L\rightarrow\pi^0\bar{\nu}\nu)_\mathrm{SM}=(3.0\pm0.6)\times10^{-11}$~\cite{Buras:2004uu}.
Due to the significant constraints imposed by the mixing in neutral meson systems on the $Z$-mediated FCNCs in the down sector, this observable is expected to be predominantly influenced by the contributions from the up-type VLQs. Hence, one can consider\cite{Nardi:1995fq,Vysotsky:2006fx,Kopnin:2008ca,Picek:2008dd,Botella:2017caf}
\begin{equation}\label{eq:delta-Kl}
\begin{split}
    &\Delta^{K_L}_{\pi\nu\nu}=                                                         \sum^{n_u}_{k=1}\lambda^{U_k}_{sd}X_0\big(x_{U_k}\big)+\sum_{i,j}V^*_{is}(F^u-\id)_{ij}V_{jd}N(x_i,x_j)~,~n_d=0 ~,\\
    &\Delta^{K_L}_{\pi\nu\nu}=0 ~,~ n_u=0 ~.
\end{split}
\end{equation}
Given current experimental data~\cite{ParticleDataGroup:2020ssz}, eqs. \eqref{eq:kd-rKp}, \eqref{eq:kd-rKL}, \eqref{eq:kd-rKl} constrain models with VLQs in the following manner,
\begin{equation}\label{eq:kd-constraints}\begin{split}
    \sum_{l=e,\mu,\tau}\big|\lambda^c_{sd}X^l_{NL}+\lambda^t_{sd}\eta^X_tX_0(x_t)+\Delta^{K^+}_{\pi\nu\nu}\big|^2&=(3.7\pm2.4)\times10^{-6}\,,\\
    \big[Y_{NL}\,\mathrm{Re}\,\lambda^c_{sd}+\eta^Y_tY_0(x_t)\,\mathrm{Re}\,\lambda^t_{sd}+          \Delta^{K_L}_{\mu\mu}\big]^2&<(4.753\pm0.041)\times10^{-7}\,,\\
    \Big|\mathrm{Im}\big[\lambda^c_{sd}X_0(x_c)+\lambda^t_{sd}X_0(x_t)+\Delta^{K_L}_{\pi\nu\nu}\big]\Big|^2&<  (4.4\pm1.0)\times10^{-6}\,.
\end{split}\end{equation}
It is important to acknowledge that although these observables probe the same couplings, each one plays a crucial role, and none can be disregarded. While the second observable only determines their real part, the others are essential to constrain the imaginary component of NP. In models incorporating additional down (up)-type VLQs, the first (third) observable tends to dominate. However, this intricate scenario may evolve in the future as experimental sensitivity improves for each decay.\par
In models with $n_u=0$, where all VLQs are of the down-type, equation \eqref{eq:kd-constraints} takes the form:
\begin{equation}\label{eq:kd-constraintsD}\begin{split}
    \sum_{l=e,\mu,\tau}\big|F^d_{sd}+(5.06-2.10i)\times10^{-6}+2.22\times10^{-3}X^l_{NL}\big|^2&=               (3.7\pm2.4)\times10^{-10},\\
    \big[\mathrm{Re}\,F^d_{sd}+4.01\times10^{-6}\big]^2&<(4.797\pm0.041)\times10^{-11}\,,
\end{split}\end{equation}
when the best-fit values for the SM CKM matrix are employed\cite{ParticleDataGroup:2020ssz}.
Using the parameterisation introduced in~\ref{sec:quarkdown}, for theories with $n_d=1~,n_u=0$, these bounds become
\begin{equation}\label{eq:kd-FD}
    F^d_{sd}=-\Theta_s\Theta^*_d=-\theta_{14}\theta_{24}e^{i(\delta_{14}-\delta_{24})}\,,
\end{equation}
where $\Theta_s=(\Theta_d)_s$.

In models with $n_u=1~,n_d=0$, one can extract approximate numerical constraints from eq. \eqref{eq:kd-constraints}
\begin{equation}\label{eq:kd-constraintsU}\begin{split}
    \big\{\mathrm{Re}\,\lambda^T_{sd}\big[Y_0(x_T)-N(x_T)\big]-3.94\times10^{-4}\big\}^2&<(4.641\pm0.040)\times10^{-7}\,,\\
    \big|\mathrm{Im}\,\lambda^T_{sd}\big[X_0(x_T)-N(x_T)\big]+2.11\times10^{-4}\big|^2&<(4.4\pm1.0)\times10^{-6}\,,
\end{split}\end{equation}
using the SM best-fit values for the CKM matrix~\cite{ParticleDataGroup:2020ssz} and the limit $F^u\to \diag(1,1,1,0)$.
Eq. \eqref{eq:kd-constraintsU} can be further simplified using the parameterisation in section \ref{sec:quarkup},
Such an expression can be further developed with
\begin{equation}\label{eq:kd-LU}
    \lambda^T_{sd}=\Theta_c\Theta^*_u=\theta_{14}\theta_{24}e^{i(\delta_{14}-\delta_{24})}\,.
\end{equation}

Again, it is important to note that when attempting to generate the complete parameter space and achieve precise results, one can directly constrain the parameter space of $n_u=1$ by employing eq. \eqref{eq:kd-constraints} and eqs. \eqref{eq:delta-Kp}, \eqref{eq:delta-KL} and \eqref{eq:delta-Kl}. This is possible because the matrices $V$ and $F^q$ are related to each other through equation \eqref{eq:Fud}.

\subsubsection{\texorpdfstring{$Z\to bb$: $R_b$}{Z -> bb: Rb}}
\label{sec:Rb}
The inclusion of VLQs into the SM introduces effects beyond  $Z/h$-mediated FCNC. Notably, it alters the diagonal couplings of down-type quarks to the $Z$ boson. These modifications are subject to strong constraints, particularly by the ratio
\begin{equation}\label{eq:Rb-Rdef}
    R_b=\frac{\Gamma(Z\rightarrow b\bar{b})}{\Gamma(Z\rightarrow\text{hadrons})}\,,
\end{equation}
which currently has an experimental value which is in agreement with the SM prediction~\cite{ALEPH:2005ab,Haller:2018nnx}. Hence, one usually employs the approximate formula~\cite{Aguilar-Saavedra:2013qpa}
\begin{equation}\label{eq:Rb-Rconstraint}
    R_b=R^\text{SM}_b(1+0.5118\delta c_{L_d}+0.5118\delta c_{L_s}-1.8178\delta c_{L_b})\,,
\end{equation}
with $R^\text{SM}_b=0.21582\pm0.00011$~\cite{Haller:2018nnx} to constrain these models.
The parameters $\delta c_{L_q}$ are typically defined through the Lagrangian
\begin{equation}\label{eq:Rb-Ldef}
    \mathscr{L}_{Zqq}=-\frac{g}{2c_W}\bar{q}\gamma^\mu(c_{L_q}\gamma_L+c_{R_q}\gamma_R)bZ_\mu\,,
\end{equation}
where $c_{L_q}=c^\text{SM}_{L_q}+\delta c_{L_q}$. Although up-type VLQs offer one-loop contributions to these couplings through top corrections to the effective $Zb_Lb_L$ vertices~\cite{Bernabeu:1987me, Bamert:1996px}, the ensuing constraints do not rival those provided by electroweak observables~\cite{Aguilar-Saavedra:2013qpa}. Thus, only the tree-level result will be considered
\begin{equation}\label{eq:Rb-Cdef}
    \delta c_{L_q}=1-F^d_{qq}\,.
\end{equation}
Using the current experimental value, $R^\text{exp}_b = 0.21629 \pm 0.00066$~\cite{ALEPH:2005ab}, models with down-type VLQs are constrained in the following manner,
\begin{equation}\label{eq:Rb-constraint}
    F^d_{bb}-0.2815(F^d_{dd}+F^d_{ss})=0.4381\pm0.0017\,.
\end{equation}
In models with $n_d=1$, this constraint can be further developed 
\begin{equation}\label{eq:Rb-FD}
    F^d_{qq}=1-|\Theta_q|^2=1-\theta^2_{q4}\,,
\end{equation}
where the parameterisation introduced in section \ref{sec:quarkdown} was used.

\subsubsection{Oblique Parameters}
Any models that introduce modifications to the gauge couplings of quarks must confront numerous constraints imposed by electroweak precision data~\cite{ALEPH:2010aa}. Generally, researchers describe NP in this sector by evaluating the oblique parameters introduced in refs. \cite{Peskin:1990zt, Peskin:1991sw, Maksymyk:1993zm}. Among these parameters, the most significant ones, namely $S$, $T$, and $U$, are expressed in models featuring singlet VLQ as\cite{Lavoura:1992np}
\begin{equation}\label{eq:oblique-parameters}\begin{split}
    S&=\frac{3}{2\pi}\Bigg[                                                       \sum_{\alpha,i}\big|V_{\alpha i}\big|^2\psi(y_\alpha,y_i)-                         \sum_{j<i}\big|F^d_{ij}\big|^2\chi(y_i,y_j)-                              \sum_{\beta<\alpha}\big|F^u_{\alpha\beta}\big|^2\chi(y_\alpha,y_\beta)\Bigg]\,,\\
    T&=\frac{3}{16\pi s^2_Wc^2_W}\Bigg[                                           \sum_{\alpha,i}\big|V_{\alpha i}\big|^2\theta(y_\alpha,y_i)-                       \sum_{j<i}\big|F^d_{ij}\big|^2\theta(y_i,y_j)-                            \sum_{\beta<\alpha}\big|F^u_{\alpha\beta}\big|^2\theta(y_\alpha,y_\beta)\Bigg]\,,\\
    U&=-\frac{3}{2\pi}\Bigg[                                                      \sum_{\alpha,i}\big|V_{\alpha i}\big|^2\chi(y_\alpha,y_i)-                         \sum_{j<i}\big|F^d_{ij}\big|^2\chi(y_i,y_j)-                              \sum_{\beta<\alpha}\big|F^u_{\alpha\beta}\big|^2\chi(y_\alpha,y_\beta)\Bigg]\,,
\end{split}\end{equation}
where sums over Greek (Latin) indices cover every up-type (down-type) quark in the theory and the following functions were introduced
\begin{equation}\label{eq:oblique-functions}\begin{split}
    \psi(y_\alpha,y_i)&\equiv\frac{1}{3}-\frac{1}{9}\log\frac{y_\alpha}{y_i}\,,\\
    \chi(y_1,y_2)&\equiv\frac{5(y^2_1+y^2_2)-22y_1y_2}{9(y_1-y_2)^2}+ \frac{3y_1y_2(y_1+y_2)-(y^3_1+y^3_2)}{3(y_1-y_2)^3}\log\frac{y_1}{y_2}\,,\\
    \theta(y_1,y_2)&\equiv(y_1+y_2)-\frac{2y_1y_2}{y_1-y_2}\log\frac{y_1}{y_2}\,,
\end{split}\end{equation}
with $y_i=m^2_i/m^2_Z$.
In models with $n_d=1$, a single down-type VLQ $B$, which only mixes with the third generation, one typically parameterises the mixing matrix through~\cite{Lavoura:1992qd}
\begin{equation}\label{eq:oblique-parameterisationD}
    V\supset\begin{pmatrix}
        V_{tb}&V_{tB}
    \end{pmatrix}=\begin{pmatrix}
        c&s
    \end{pmatrix}\,,
\end{equation}
where $c=\cos\theta$ and $s=\sin\theta$.
Thus, by using eq. \eqref{eq:oblique-parameters} one can define
\begin{equation}\label{eq:oblique-parametersD}\begin{split}
    \Delta S&=                                                                           \frac{3}{2\pi}s^2\Big[\psi(y_t,y_B)-c^2\chi(y_B,y_b)-\psi(y_t,y_b)\Big]\,,\\
    \Delta T&=                                                                           \frac{3}{16\pi s^2_Wc^2_W}s^2\Big[\theta(y_t,y_B)-c^2\theta(y_B,y_b)-\theta(y_t,y_b)\Big]\,,\\
    \Delta U&=                                                                          -\frac{3}{2\pi}s^2\Big[\chi(y_t,y_B)-c^2\chi(y_B,y_b)-\chi(y_t,y_b)\Big]\,,\\
\end{split}\end{equation}
where $(\Delta S,\Delta T,\Delta U)=(S-S_\text{SM},T-T_\text{SM},U-U_\text{SM})$ with $S_\text{SM}=3\psi(y_t,y_b)/2\pi$, $T_\text{SM}=3\theta(y_t,y_b)/16\pi s^2_Wc^2_W$ and $U_\text{SM}=-3\chi(y_t,y_b)/2\pi$.
In extensions of the SM with $n_u=1$, a single up-type VLQ $T$, which mixes exclusively with the third generation, the situation is similar.
Hence, after parameterising the mixing matrix via
\begin{equation}\label{eq:oblique-parameterisationU}
    V\supset\begin{pmatrix}
        V_{tb}\\V_{Tb}
    \end{pmatrix}=\begin{pmatrix}
        c\\s
    \end{pmatrix}\,,
\end{equation}
and using eq. \eqref{eq:oblique-parameters}, the subsequent constraints follow
\begin{equation}\label{eq:oblique-parametersU}\begin{split}
    \Delta S&=                                                                           \frac{3}{2\pi}s^2\Big[\psi(y_T,y_b)-c^2\chi(y_T,y_t)-\psi(y_t,y_b)\Big]\,,\\
    \Delta T&=                                                                           \frac{3}{16\pi s^2_Wc^2_W}s^2\Big[\theta(y_T,y_b)-c^2\theta(y_T,y_t)-\theta(y_t,y_b)\Big]\,,\\
    \Delta U&=                                                                          -\frac{3}{2\pi}s^2\Big[\chi(y_T,y_b)-c^2\chi(y_T,y_t)-\chi(y_t,y_b)\Big]\,.
\end{split}\end{equation}
Evaluating eq. \eqref{eq:oblique-parametersU} in the heavy VLQ limit, $y_T\gg1$,
\begin{equation}\label{eq:oblique-parametersH}\begin{split}
    \Delta S\simeq-\frac{s^2}{6\pi}\big(1-3c^2\big)\log y_T\,,\quad                      \Delta T\simeq\frac{3y_ts^2c^2}{8\pi s^2_Wc^2_W}\log y_T\,,\quad                     \Delta U\simeq\frac{s^4}{2\pi}\log y_T\,.\\
\end{split}\end{equation}
And using eq. \eqref{eq:perturb-traceU} to establish $s^2\sim y^{-1}_T$, one obtains a very small $\Delta U$,
\begin{equation}\label{eq:oblique-parametersHy}
    \Delta S,\Delta T\sim y^{-1}_T\log y_T\,,\qquad\Delta U\sim y^{-2}_T\log y_T\,.
\end{equation}
Due to this reason, it is prudent to focus solely on $\Delta S$ and $\Delta T$ in the phenomenological analysis of a model featuring singlet VLQs. It is important to highlight that the same rationale is applicable to eq. \eqref{eq:oblique-parametersD}.
The most recent data for  $(\Delta S,\Delta T,\Delta U)=(S-S_\text{SM},T-T_\text{SM},U-U_\text{SM})$ \cite{pdg} gives the constraints
\begin{equation}
\begin{split}
    &\Delta S = -0.02 \pm 0.10 ~,\\
    &\Delta T = 0.03 \pm 0.12 ~,\\
    &\Delta U = 0.01 \pm 0.11 ~.
\end{split}
\end{equation}

\subsection{Future search prospects}
\label{sec:future}
In the majority of recent searches, the reported results are presented for a range of branching ratios (BRs) without imposing the restrictions given by eqs. \eqref{eq:SU2BRs} and \eqref{eq:3rdgen}. In other words, these searches do not assume "$SU(2)$ singlet VLQ BRs" or exclusive decays to third-generation quarks. However, if a VLQ predominantly decays into lighter generations, it might not be detected in a search concentrated on the phenomenology of VLQs that decay into third-generation quarks. This is because the branching ratios of the VLQ in the channels utilized in the third-generation quark search could be very small. Consequently, searches specifically targeting VLQs that primarily decay into lighter generations ~\cite{ATLAS2015lpr, CMS2017asf} offer complementary insights to those where VLQs are assumed to mainly decay into third-generation quarks. \par
According to ref. \cite{Liu:2016jho}, at the High-Luminosity Large Hadron Collider (HL-LHC), a $5\sigma$ discovery reach for the single production of an up-type VLQ with a mass of approximately $1~\unit{TeV}$ can be achieved for couplings to Standard Model (SM) quarks $q$ as low as $|V_{Tq}| \sim 0.2$. In the case of pair-produced down-type VLQs, a $5\sigma$ discovery reach is feasible for masses around $730~\unit{GeV}$\cite{Paul:2020mul}. Additionally, scenarios where an up-type VLQ couples exclusively to first-generation quarks can also be probed\cite{Cui:2022hjg}.\par
As for the Future Circular Collider (FCC), it is argued that a $5\sigma$ discovery reach for a pair-produced down-type VLQ can be extended up to $2980~\unit{GeV}$, considering only the $4l + 2j$ decay channel \cite{Paul:2020mul}. Furthermore, it is demonstrated that the FCC-hh generally requires approximately two orders of magnitude less integrated luminosity than the HL-LHC to discover a down-type VLQ at a given mass. Additionally, with $V_{Tb}\sim 0.3$, up-type VLQs with masses up to $2.1~\unit{TeV}$ can be probed at the $5\sigma$ level at the FCC-hh through single production followed by a decay to $ht$, while 95\% confidence level (C.L.) exclusion limits can extend up to $2.6~\unit{TeV}$~\cite{Tian:2021nmj}.\par
The prospects for VLQ searches at the CLIC are also highly promising, offering an opportunity to investigate these particles through electron-positron interactions. In ref.~\cite{Han:2021lpg}, a study is presented for the $3~\unit{TeV}$ CLIC with an integrated luminosity of $5,\unit{ab}^{-1}$, focusing on a singly-produced down-type VLQ that exclusively decays to third-generation quarks. The VLQ is produced in the process $e e^+ \rightarrow \overline{B}b$ and decays via $B \rightarrow Zb$. For couplings to the third-generation of $|V_{tB}| \sim 0.3$, the discovery region is found to be $m_B \in [1200,1900]~\unit{GeV}$.\par
Finally, in the work presented in ref. \cite{AlAli:2021let}, it is proposed that a future high-energy muon collider offers several advantages compared to other projects currently under consideration. Specifically, the production cross-section of a TeV-mass VLQ can be significantly enhanced when the VLQ is produced through $\mu^-\mu^+$ annihilation, particularly for $\sqrt{s}<100~\unit{TeV}$.\par
\cleardoublepage

\chapter{Conclusions}
\label{chapter:conclusions}

The works that this thesis is based on aimed to explore and classify different regions of the 3$\nu$SM \footnote{SM $+~3$ right-handed neutrinos without imposing Lepton Number conservation.}, and the $n_u$SM$n_d$ \footnote{SM $+~n_d$ down-type VLQs $+~n_u$ up-type VLQs.} parameter space utilizing a newly-developed exact parameterization \cite{Agostinho:2017wfs,Branco:2019avf,Branco:2021vhs,Pereira:2022pqu,Alves:2023ufm}.
This parameterisation is particularly well-suited for models with non-unitary mixing matrices, with the 3$\nu$SM and $n_u$SM$n_d$ serving as straightforward examples.\par
These models are motivated by the sizeable deviations from unitarity one naturally encounters in models with neutrinos with masses in the or below the GeV/TeV scale, which existence may help explain many open problems in neutrino physics \cite{Deppisch:2015qwa, Escrihuela:2015wra, Fernandez-Martinez:2016lgt,Dentler:2018sju}, such as the puzzles in neutrino oscillation experiments \cite{MiniBooNE:2022emn,MINOS:2016viw,Arguelles:2021meu,Hu:2020uvx}, and the CKM Unitarity problem or Cabibbo Angle Anomaly \cite{Czarnecki:2019mwq,Seng:2020wjq,Hayen:2020cxh,Shiells:2020fqp,Belfatto:2019swo,Belfatto:2021jhf}, in quark physics, which may be explained via the introduction of VLQs \cite{Belfatto:2019swo,Belfatto:2021jhf,Branco:2021vhs}.\par
Both heavy neutrinos and vector-like quarks share a crucial characteristic: their mixing with SM leptons/quarks, and consequently the deviations from unitarity in the corresponding 
$3 \times 3$ mixing matrices, is suppressed by their masses. Despite their distinct integration into the SM, they both possess this feature due to an important property they have in common: the existence of two scales that remain invariant under the gauge group. The effective suppression of beyond-the-SM effects, such as Z-mediated and Higgs-mediated FCNC, as required by experimental data, is accomplished due to the presence of the electroweak symmetry breaking scale and the bare mass term scale for vector-like quarks or the Majorana mass term scale for neutrinos. Notably, since these mass terms are not protected by the gauge symmetry, the bare mass term or the Majorana mass term scale, denoted as $M$, can be significantly larger than the electroweak symmetry breaking scale, $v$. As elaborated in chapters \ref{chapter:framework_neutrinos} and \ref{chapter:framework_quarks}, the ratio $v/M$ serves as a reliable measure of the magnitude of $3 \times 3$ deviations from unitarity in each theory, which govern many beyond-the-SM effects, that experiments require to be suppressed. This characteristic renders them straightforward and well-justified additions to the SM. \par
Neutrinos with masses in or below the GeV/TeV scale and Vector-like Quarks hold great promise in explaining various phenomena in Particle Physics. As discussed in the fourth chapter, the results of ongoing experiments are approaching the predictions of many models, and the near future looks very promising in terms of neutrino and meson precision frontier experiments. Moreover, the outlook for the far future is optimistic, with the prospects of new experiments such as the Future Circular Collider (FCC). \par
Nonetheless, this advantageous feature can also become their greatest challenge, as the suppression of their mixing with the SM particles may lead to difficulties in detecting them. Despite this potential difficulty, the search for heavy neutrinos and VLQs is undoubtedly worth the effort, as their discovery could provide crucial insights into the mysteries of the Universe.\par
\vskip 4mm


\cleardoublepage


\phantomsection
\addcontentsline{toc}{chapter}{\bibname}

\bibliographystyle{JHEPwithnote} 

\bibliography{Thesis} 

\cleardoublepage

%





\end{document}